\documentclass[11pt]{article}
\usepackage{amssymb}
\usepackage{amsmath}
\newtheorem{theorem}{Theorem}[section]
\newtheorem{lemma}{Lemma}[section]
\newtheorem{corollary}{Corollary}[section]
\begin{document}

\author{E. M. Cioroianu\thanks{%
e-mail address: manache@central.ucv.ro} , S. C. S\u{a}raru\thanks{%
e-mail address: scsararu@central.ucv.ro} \\
Faculty of Physics, University of Craiova\\
13 A. I. Cuza Str., Craiova, 200585, Romania}
\title{Self-interactions in a topological BF-type model in D=5}
\date{}
\maketitle

\begin{abstract}
All consistent interactions in five spacetime dimensions that can
be added to a free BF-type model involving one scalar field, two
types of one-forms, two sorts of two-forms, and one three-form are
investigated by means of deforming the solution to the master
equation with the help of specific cohomological techniques. The
couplings are obtained on the grounds of smoothness, locality,
(background) Lorentz invariance, Poincar\'{e} invariance, and the
preservation of the number of derivatives on each field.

\end{abstract}

\section{Introduction}

The power of the BRST formalism was strongly increased by its
cohomological development, which allowed, among others, a useful
investigation of many interesting aspects related to the
perturbative renormalization problem~\cite {4a,4b,4c,4d,5},
anomaly-tracking mechanism~\cite{5,6a,6b,6c,6d,6e}, simultaneous
study of local and rigid invariances of a given theory~\cite{7},
as well as to the reformulation of the construction of consistent
interactions in gauge theories~\cite{7a,7b,7c,7d,7e} in terms of
the deformation theory~\cite{8a,17and5,8b,8c,8d}, or, actually, in
terms of the deformation of the solution to the master equation.

The main aim of this paper is to construct the consistent
interactions in five spacetime dimensions that can be added to a
free BF-type model~\cite{12} involving one scalar field, two types
of one-forms, two sorts of two-forms, and one three-form by means
of deforming the solution to the master equation with the help of
specific cohomological techniques. Interacting topological field
theories of BF-type are important in view of their relationship
with Poisson Sigma Models, which are known to explain interesting
aspects of two-dimensional gravity, including the study of
classical
solutions~\cite{psm1,psm2,psm3,psm4,psm5,psm6,psm7,psm8,psm9}.
Various aspects of BF models can be found
in~\cite{bfaspects,bfaspects1,bfaspects2,bfaspects3,bfaspects4}.
The present paper extends our former Hamiltonian
results~\cite{BFhamnoPT,BFhamnoPT1} on BF-type models.

The couplings are obtained on the grounds of smoothness, locality,
(background) Lorentz invariance, Poincar\'{e} invariance, and the
preservation of the number of derivatives on each field. The
starting, free BF model possesses Abelian gauge transformations,
which are off-shell, third-order reducible. The entire Lagrangian
formulation of the interacting theory is obtained from the
computation of the deformed solution to the master equation, order
by order in the coupling constant $g$. Thus, the first-order
deformation of the solution to the master equation is parametrized
by seven arbitrary, smooth functions of the undifferentiated
scalar field. The consistency of the deformation procedure at
order $g^{2}$ imposes some restrictions on the above mentioned
functions, which lead to three kinds of interacting models that
are in a way complementary to each other. In all situations the
fully deformed solution to the master equation that is consistent
to all orders in the coupling constant stops at order one in $g$.
Related to the three types of interacting BF theories, all of them
describe a deformed model with an open gauge algebra, which closes
on-shell (on the stationary surface of deformed field equations).
At the level of reducibility relations, the first coupled model
possesses on-shell first- and second-order reducibility relations,
the second interacting theory exhibits on-shell reducibility
relations to all (the three) orders, while in the last situation
only the first-order reducibility relations close on-shell.

The paper is organized into six sections. Section \ref{free}
introduces the model to be considered and constructs its free
Lagrangian BRST symmetry. Section \ref{defth} briefly reviews the
procedure of adding consistent interactions in gauge theories
based on the deformation of the solution to the master equation.
In Sec. \ref{inter} we construct the Lagrangian interactions for
the starting free system in five dimensions by solving the
deformation equations with the help of standard cohomological
techniques. Section \ref{lagfor} discusses the resulting
interacting models and Sec. \ref{conc} ends the paper with the
main conclusions. The paper also contains seven Appendix sections
including various aspects mentioned in the main text.

\section{Free BRST differential \label{free}}

We start from a free, five-dimensional BF-like theory involving
one scalar field $\varphi $, two types of one-forms $H_{\mu }$ and
$A_{\mu }$, two sorts of two-forms $B_{\mu \nu }$ and $\phi _{\mu
\nu }$, and one three-form $K_{\mu \nu \rho }$
\begin{equation}\label{f1}
S_{0}^{\text{L}}\left[ \varphi ,H,A,B,\phi ,K\right] =\int
d^{5}x\left( H_{\mu }\partial ^{\mu }\varphi +\tfrac{1}{2}B^{\mu
\nu }\partial _{[\mu }A_{\nu ]}+\tfrac{1}{3}K^{\mu \nu \rho
}\partial _{[\mu }\phi _{\nu \rho ]}\right) .
\end{equation}
Action (\ref{f1}) is found invariant under the gauge
transformations
\begin{align}
\delta _{\epsilon ,\xi }A^{\mu }& =\partial ^{\mu }\epsilon ,&
\delta _{\epsilon ,\xi }H^{\mu }& =2\partial _{\nu }\epsilon ^{\mu
\nu },& \delta _{\epsilon ,\xi }\varphi & =0, \label{f2a}
\\
\delta _{\epsilon ,\xi }B^{\mu \nu } & =-3\partial _{\rho
}\epsilon ^{\mu \nu \rho },& \delta _{\epsilon ,\xi }\phi _{\mu
\nu }& =\partial _{[\mu }\xi _{\nu ]},& \delta _{\epsilon ,\xi
}K^{\mu \nu \rho }& =4\partial _{\lambda }\xi ^{\mu \nu \rho
\lambda }, \label{f2b}
\end{align}
where the gauge parameters $\epsilon $, $\epsilon ^{\mu \nu }$,
$\epsilon ^{\mu \nu \rho }$, $\xi _{\mu }$, and $\xi ^{\mu \nu
\rho \lambda }$ are bosonic, with $\epsilon ^{\mu \nu }$,
$\epsilon ^{\mu \nu \rho }$, and $\xi
^{\mu \nu \rho \lambda }$ completely antisymmetric. By means of (\ref{f2a}%
)--(\ref{f2b}) we read the nonvanishing gauge generators (written
in De Witt condensed notations)
\begin{equation}
(Z_{(A)}^{\mu })=\partial ^{\mu },\quad  (Z_{(H)}^{\mu })_{\alpha
\beta }=-\partial _{\left[ \alpha \right. } \delta _{\left. \beta
\right] }^{\mu },\quad (Z_{(B)}^{\mu \nu })_{\alpha \beta \gamma
}=-\tfrac{1}{2}\partial _{\left[ \alpha \right. }\delta _{\beta
}^{\mu }\delta _{\left. \gamma \right] }^{\nu }, \label{f3a}
\end{equation}
\begin{equation}
(Z_{(\phi )}^{\mu \nu })_{\alpha } =\partial ^{[\mu }\delta
_{\alpha }^{\nu ]}, \quad (Z_{(K)}^{\mu \nu \rho })_{\alpha \beta
\gamma \delta }=- \tfrac{1}{6}\partial _{\left[ \alpha \right.
}\delta _{\beta }^{\mu }\delta _{\gamma }^{\nu }\delta _{\left.
\delta \right] }^{\rho },  \label{f3c}
\end{equation}
where we put an extra lower index $(A)$, $(H)$, etc., in order to
indicate with what field is a certain gauge generator associated.
Everywhere in this paper we understand that the notation $[\alpha
\beta \ldots \gamma ]$ signifies complete antisymmetry with
respect to the Lorentz indices between brackets, with the
conventions that the minimum number of terms is always used and
the result is never divided by the number of terms. The above
gauge transformations are Abelian and off-shell, third-order
reducible. More precisely, the gauge generators of the one-form
$H^{\mu }$ are third-order reducible, with the first-, second-,
and respectively third-order reducibility functions
\begin{equation}
(Z_{1}^{\alpha \beta })_{\mu ^{\prime }\nu ^{\prime }\rho ^{\prime
}}=-\tfrac{1}{2}\partial _{\left[ \mu ^{\prime }\right. }\delta
_{\nu ^{\prime }}^{\alpha }\delta _{\left. \rho ^{\prime }\right]
}^{\beta }, \quad (Z_{2}^{\mu ^{\prime }\nu ^{\prime }\rho
^{\prime }})_{\alpha ^{\prime }\beta ^{\prime }\gamma ^{\prime
}\delta ^{\prime }}=- \tfrac{1}{6}\partial _{\left[ \alpha
^{\prime }\right. }\delta _{\beta ^{\prime }}^{\mu ^{\prime
}}\delta _{\gamma ^{\prime }}^{\nu ^{\prime }}\delta _{\left.
\delta ^{\prime }\right] }^{\rho ^{\prime }}, \label{f4a}
\end{equation}
\begin{equation}
(Z_{3}^{\alpha ^{\prime }\beta ^{\prime }\gamma ^{\prime }\delta
^{\prime }})_{\mu ^{\prime \prime }\nu ^{\prime \prime }\rho
^{\prime \prime }\lambda
^{\prime \prime }\sigma ^{\prime \prime }}=-\tfrac{1}{24}%
\partial _{\left[ \mu ^{\prime \prime }\right. }\delta _{\nu ^{\prime
\prime }}^{\alpha ^{\prime }}\delta _{\rho ^{\prime \prime
}}^{\beta ^{\prime }}\delta _{\lambda ^{\prime \prime }}^{\gamma
^{\prime }}\delta _{\left. \sigma ^{\prime \prime }\right]
}^{\delta ^{\prime }},  \label{f4c}
\end{equation}
the gauge generators of the two-form $B^{\mu \nu }$ are
second-order reducible, with the reducibility functions
\begin{equation}
(Z_{1}^{\alpha \beta \gamma })_{\mu ^{\prime }\nu ^{\prime }\rho
^{\prime }\lambda ^{\prime }}=-\tfrac{1}{6}\partial _{\left[ \mu
^{\prime }\right. }\delta _{\nu ^{\prime }}^{\alpha }\delta _{\rho
^{\prime }}^{\beta }\delta _{\left. \lambda ^{\prime }\right]
}^{\gamma }, \quad (Z_{2}^{\mu ^{\prime }\nu ^{\prime }\rho
^{\prime }\lambda ^{\prime }})_{\alpha \beta \gamma \delta
\varepsilon }=-\tfrac{1}{24 }\partial _{\left[ \alpha \right.
}\delta _{\beta }^{\mu ^{\prime }}\delta _{\gamma }^{\nu ^{\prime
}}\delta _{\delta }^{\rho ^{\prime }}\delta _{\left. \varepsilon
\right] }^{\lambda ^{\prime }},  \label{f5a}
\end{equation}
while the gauge generators of the two-form $\phi _{\mu \nu }$ and
of the three-form $K^{\mu \nu \rho }$ are first-order reducible,
with the corresponding reducibility functions respectively of the
form
\begin{equation}
(Z_{1}^{\alpha })=\partial ^{\alpha }, \quad (Z_{1}^{\alpha \beta
\gamma \delta })_{\mu ^{\prime }\nu ^{\prime }\rho ^{\prime
}\lambda ^{\prime }\sigma ^{\prime }}=-\tfrac{1}{24}
\partial _{\left[ \mu ^{\prime }\right. }\delta _{\nu ^{\prime
}}^{\alpha }\delta _{\rho ^{\prime }}^{\beta }\delta _{\lambda
^{\prime }}^{\gamma }\delta _{\left. \sigma ^{\prime }\right]
}^{\delta }.  \label{f6b}
\end{equation}
The concrete form of the reducibility relations written in
condensed De Witt notations are expressed as follows. The
first-order reducibility relations are
\begin{align}
(Z_{(H)}^{\mu })_{\alpha \beta }(Z_{1}^{\alpha \beta })_{\mu
^{\prime }\nu ^{\prime }\rho ^{\prime }}& =0,& (Z_{(B)}^{\mu \nu
})_{\alpha \beta \gamma }(Z_{1}^{\alpha \beta \gamma })_{\mu
^{\prime }\nu ^{\prime }\rho
^{\prime }\lambda ^{\prime }}& =0,  \label{f7a} \\
(Z_{(\phi )}^{\mu \nu })_{\alpha }(Z_{1}^{\alpha })& =0,&
(Z_{(K)}^{\mu \nu \rho })_{\alpha \beta \gamma \delta
}(Z_{1}^{\alpha \beta \gamma \delta })_{\mu ^{\prime }\nu ^{\prime
}\rho ^{\prime }\lambda ^{\prime }\sigma ^{\prime }}& =0,
\label{f7b}
\end{align}
the second-order ones read as
\begin{equation}
(Z_{1}^{\alpha \beta })_{\mu ^{\prime }\nu ^{\prime }\rho ^{\prime
}}(Z_{2}^{\mu ^{\prime }\nu ^{\prime }\rho ^{\prime }})_{\alpha
^{\prime }\beta ^{\prime }\gamma ^{\prime }\delta ^{\prime
}}=0,\quad (Z_{1}^{\alpha \beta \gamma })_{\mu ^{\prime }\nu
^{\prime }\rho ^{\prime }\lambda ^{\prime }}(Z_{2}^{\mu ^{\prime
}\nu ^{\prime }\rho ^{\prime }\lambda ^{\prime }})_{\alpha \beta
\gamma \delta \varepsilon }=0,  \label{f8}
\end{equation}
while the third-order reducibility relations can be written as
\begin{equation}
(Z_{2}^{\mu ^{\prime }\nu ^{\prime }\rho ^{\prime }})_{\alpha
^{\prime }\beta ^{\prime }\gamma ^{\prime }\delta ^{\prime
}}(Z_{3}^{\alpha ^{\prime }\beta ^{\prime }\gamma ^{\prime }\delta
^{\prime }})_{\mu ^{\prime \prime }\nu ^{\prime \prime }\rho
^{\prime \prime }\lambda ^{\prime \prime }\sigma ^{\prime \prime
}}=0.  \label{f9}
\end{equation}
We observe that the BF-like theory under study is a usual linear
gauge theory (its field equations are linear in the fields and
first-order in their spacetime derivatives), whose generating set
of gauge transformations is third-order reducible, such that we
can define in a consistent manner its Cauchy order, which is found
to be equal to five.

In order to construct the BRST symmetry of this free theory, we
introduce the field/ghost and antifield spectra
\begin{align}
\Phi ^{\alpha _{0}}& =\left( A^{\mu },H^{\mu },\varphi ,B^{\mu \nu
},K^{\mu \nu \rho },\phi _{\mu \nu }\right) ,& \Phi _{\alpha
_{0}}^{*}& =\left( A_{\mu }^{*},H_{\mu }^{*},\varphi ^{*},B_{\mu
\nu }^{*},K_{\mu \nu \rho }^{*},\phi ^{*\mu \nu }\right) ,
\label{f10a} \\
\eta ^{\alpha _{1}}& =\left( \eta ,C^{\mu \nu },\eta ^{\mu \nu \rho },%
\mathcal{G}^{\mu \nu \rho \lambda },C_{\mu }\right) ,& \eta
_{\alpha _{1}}^{*}& =\left( \eta ^{*},C_{\mu \nu }^{*},\eta _{\mu
\nu \rho }^{*},\mathcal{G}_{\mu \nu \rho \lambda }^{*},C^{*\mu
}\right) ,
\label{f10d} \\
\eta ^{\alpha _{2}}& =\left( C^{\mu \nu \rho },\eta ^{\mu \nu \rho
\lambda },\mathcal{G}^{\mu \nu \rho \lambda \sigma },\, C\right)
,& \eta _{\alpha _{2}}^{*}& =\left( C_{\mu \nu \rho }^{*},\eta
_{\mu \nu \rho \lambda }^{*},\mathcal{G}_{\mu \nu \rho \lambda
\sigma }^{*},C^{*}\right) ,
\label{f10f} \\
\eta ^{\alpha _{3}}& =\left( C^{\mu \nu \rho \lambda },\eta ^{\mu
\nu \rho \lambda \sigma }\right) ,& \eta _{\alpha _{3}}^{*}&
=\left( C_{\mu \nu \rho \lambda }^{*},\eta _{\mu \nu \rho \lambda
\sigma }^{*}\right) ,  \label{f10g}
\\
\eta ^{\alpha _{4}}& =\left( C^{\mu \nu \rho \lambda \sigma
}\right) ,& \eta _{\alpha _{4}}^{*}& =\left( C_{\mu \nu \rho
\lambda \sigma }^{*}\right) .  \label{f10h}
\end{align}
The fermionic ghosts $\eta ^{\alpha _{1}}$ respectively correspond
to the bosonic gauge parameters $\epsilon ^{\alpha _{1}}$ in
(\ref{f2a})--(\ref{f2b}), the bosonic ghosts for ghosts $\eta
^{\alpha _{2}}$ are due to the first-order reducibility relations
(\ref{f7a})--(\ref{f7b}), the fermionic ghosts for ghosts for
ghosts $\eta ^{\alpha _{3}}$ are required by the second-order
reducibility relations (\ref{f8}), while the bosonic ghosts for
ghosts for ghosts for ghosts $\eta ^{\alpha _{4}}$ are imposed by
the third-order reducibility relations (\ref{f9}). The star
variables represent the antifields of the corresponding
fields/ghosts. Their Grassmann parities are obtained via the usual
rule $\varepsilon \left( \chi _{\Delta }^{*}\right) =\left(
\varepsilon \left( \chi ^{\Delta }\right) +1\right) \mod \;2$,
where we employed the notations
\begin{equation}
\chi ^{\Delta }=\left( \Phi ^{\alpha _{0}},\eta ^{\alpha
_{1}},\eta ^{\alpha _{2}},\eta ^{\alpha _{3}},\eta ^{\alpha
_{4}}\right) ,\quad \chi _{\Delta }^{*}=\left( \Phi _{\alpha
_{0}}^{*},\eta _{\alpha _{1}}^{*},\eta _{\alpha _{2}}^{*},\eta
_{\alpha _{3}}^{*},\eta _{\alpha _{4}}^{*}\right) . \label{notat}
\end{equation}

Since both the gauge generators and the reducibility functions are
field-independent, it follows that the BRST differential reduces to $%
s=\delta +\gamma $, where $\delta $ is the Koszul-Tate differential, and $%
\gamma $ means the exterior longitudinal derivative. The
Koszul-Tate
differential is graded in terms of the antighost number [$\text{agh}$, $%
\text{agh}\left( \delta \right) =-1$, $\text{agh}\left( \gamma
\right) =0$] and enforces a resolution of the algebra of smooth
functions defined on the stationary surface of field equations for
the action (\ref{f1}), $C^{\infty }\left( \Sigma \right) $,
$\Sigma :\delta S_{0}^{\text{L}}/\delta \Phi ^{\alpha _{0}}=0$.
The exterior longitudinal derivative is graded in terms
of the pure ghost number [$\text{pgh}$, $\text{pgh}\left( \gamma \right) =1$%
, $\text{pgh}\left( \delta \right) =0$] and is correlated with the
original
gauge symmetry via its cohomology in pure ghost number zero computed in $%
C^{\infty }\left( \Sigma \right) $, which is isomorphic to the
algebra of physical observables for this free theory. These two
degrees of the generators (\ref{f10a})--(\ref{f10h}) from the BRST
complex are valued like
\begin{align}
\text{pgh}\left( \Phi ^{\alpha _{0}}\right) & =0,&
\text{pgh}\left( \eta ^{\alpha _{k}}\right) & =k,&
\text{pgh}\left( \Phi _{\alpha _{0}}^{*}\right) & =0,&
\text{pgh}\left(
\eta _{\alpha _{k}}^{*}\right) & =0,  \label{f11a} \\
\text{agh}\left( \Phi ^{\alpha _{0}}\right) & =0,&
\text{agh}\left( \eta ^{\alpha _{k}}\right) & =0,&
\text{agh}\left( \Phi _{\alpha _{0}}^{*}\right) & =1,&
\text{agh}\left( \eta _{\alpha _{k}}^{*}\right) & =k+1,
\label{f11d}
\end{align}
for $k=\overline{1,4}$. The actions of the differentials $\delta $ and $%
\gamma $ on the above generators read as
\begin{equation}
\delta \Phi ^{\alpha _{0}}=0,\quad \delta \eta ^{\alpha _{k}}=0\quad (k=%
\overline{1,4}),  \label{f12a}
\end{equation}
\begin{equation}
\delta A_{\mu }^{*}=-\partial ^{\nu }B_{\mu \nu },\quad \delta
H_{\mu }^{*}=-\partial _{\mu }\varphi ,\quad \delta \varphi
^{*}=\partial ^{\mu }H_{\mu },\quad \delta B_{\mu \nu
}^{*}=-\tfrac{1}{2}\partial _{[\mu }A_{\nu ]}, \label{f12b}
\end{equation}
\begin{equation}
\delta \phi _{\mu \nu }^{*}=\partial ^{\rho }K_{\mu \nu \rho
},\quad \delta K_{\mu \nu \rho }^{*}=-\tfrac{1}{3}\partial _{[\mu
}\phi _{\nu \rho ]}, \label{f12c}
\end{equation}
\begin{equation}
\delta \eta ^{*}=-\partial ^{\mu }A_{\mu }^{*},\quad \delta C_{\mu
\nu }^{*}=\partial _{[\mu }H_{\nu ]}^{*},\quad \delta \eta _{\mu
\nu \rho }^{*}=\partial _{[\mu }B_{\nu \rho ]}^{*},  \label{f12d}
\end{equation}
\begin{equation}
\delta C_{\mu }^{*}=2\partial ^{\nu }\phi _{\mu \nu }^{*},\quad
\delta \mathcal{G}_{\mu \nu \rho \lambda }^{*}=\partial _{[\mu
}K_{\nu \rho \lambda ]}^{*},\quad \delta C_{\mu \nu \rho
}^{*}=-\partial _{[\mu }C_{\nu \rho ]}^{*},  \label{f12e}
\end{equation}
\begin{equation}
\delta \eta _{\mu \nu \rho \lambda }^{*}=-\partial _{[\mu }\eta
_{\nu \rho \lambda ]}^{*},\quad \delta C^{*}=\partial ^{\mu
}C_{\mu }^{*},\quad \delta
\mathcal{G}_{\mu \nu \rho \lambda \sigma }^{*}=-\partial _{[\mu }\mathcal{G}%
_{\nu \rho \lambda \sigma ]}^{*},  \label{f12f}
\end{equation}
\begin{equation}
\delta C_{\mu \nu \rho \lambda }^{*}=\partial _{[\mu }C_{\nu \rho
\lambda ]}^{*},\quad \delta \eta _{\mu \nu \rho \lambda \sigma
}^{*}=\partial _{[\mu }\eta _{\nu \rho \lambda \sigma ]}^{*},\quad
\delta C_{\mu \nu \rho \lambda \sigma }^{*}=-\partial _{[\mu
}C_{\nu \rho \lambda \sigma ]}^{*}, \label{f12g}
\end{equation}
\begin{equation}
\gamma \Phi _{\alpha _{0}}^{*}=0,\quad \gamma \eta _{\alpha
_{k}}^{*}=0\quad (k=\overline{1,4}),  \label{f13a}
\end{equation}
\begin{equation}
\gamma A^{\mu }=\partial ^{\mu }\eta ,\quad \gamma H^{\mu
}=2\partial _{\nu }C^{\mu \nu },\quad \gamma B^{\mu \nu
}=-3\partial _{\rho }\eta ^{\mu \nu \rho },\quad \gamma \varphi
=0,  \label{f13b}
\end{equation}
\begin{equation}
\gamma \phi _{\mu \nu }=\partial _{[\mu }C_{\nu ]},\quad \gamma
K^{\mu \nu \rho }=4\partial _{\lambda }\mathcal{G}^{\mu \nu \rho
\lambda },\quad \gamma \eta =0,\quad \gamma C^{\mu \nu
}=-3\partial _{\rho }C^{\mu \nu \rho },  \label{f13c}
\end{equation}
\begin{equation}
\gamma \eta ^{\mu \nu \rho }=4\partial _{\lambda }\eta ^{\mu \nu
\rho
\lambda },\quad \gamma C_{\mu }=\partial _{\mu }C,\quad \gamma \mathcal{G}%
^{\mu \nu \rho \lambda }=-5\partial _{\sigma }\mathcal{G}^{\mu \nu
\rho \lambda \sigma },  \label{f13d}
\end{equation}
\begin{equation}
\gamma C^{\mu \nu \rho }=4\partial _{\lambda }C^{\mu \nu \rho
\lambda },\quad \gamma \eta ^{\mu \nu \rho \lambda }=-5\partial
_{\sigma }\eta ^{\mu \nu \rho \lambda \sigma },\quad \gamma
C=\gamma \mathcal{G}^{\mu \nu \rho \lambda \sigma }=0,
\label{f13e}
\end{equation}
\begin{equation}
\gamma C^{\mu \nu \rho \lambda }=-5\partial _{\sigma }C^{\mu \nu
\rho \lambda \sigma },\quad \gamma \eta ^{\mu \nu \rho \lambda
\sigma }=0,\quad \gamma C^{\mu \nu \rho \lambda \sigma }=0.
\label{f13f}
\end{equation}
The overall degree that grades the BRST complex is named ghost number ($%
\text{gh}$) and is defined like the difference between the pure
ghost number and the antighost number, such that gh$\left( \delta
\right) =$gh$\left( \gamma \right) =\text{gh}\left( s\right) =1$.

The BRST symmetry admits a canonical action $s\cdot =\left( \cdot ,\bar{S}%
\right) $, where its canonical generator [$\text{gh}\left( \bar{S}\right) =0$%
, $\varepsilon \left( \bar{S}\right) =0$] satisfies the classical
master equation $\left( \bar{S},\bar{S}\right) =0$. The symbol
$(,)$ denotes the antibracket, defined by decreeing the
fields/ghosts conjugated with the corresponding antifields. In the
case of the free theory under discussion, the solution to the
master equation takes the form
\begin{equation}\label{f14}
\begin{split}
\bar{S}& =S_{0}^{\text{L}}+\int d^{5}x\left[ A_{\mu }^{*}\partial
^{\mu }\eta +2H_{\mu }^{*}\partial _{\nu }C^{\mu \nu }-3B_{\mu \nu
}^{*}\partial _{\rho }\eta ^{\mu \nu \rho }+\phi ^{*\mu \nu
}\partial _{[\mu }C_{\nu
]}+\right. \\
&\quad +4K_{\mu \nu \rho }^{*}\partial _{\lambda }\mathcal{G}^{\mu
\nu \rho \lambda }-3C_{\mu \nu }^{*}\partial _{\rho }C^{\mu \nu
\rho }+4\eta _{\mu
\nu \rho }^{*}\partial _{\lambda }\eta ^{\mu \nu \rho \lambda }-5\mathcal{G}%
_{\mu \nu \rho \lambda }^{*}\partial _{\sigma }\mathcal{G}^{\mu
\nu \rho
\lambda \sigma } +\\
&\quad \left. +C^{*\mu }\partial _{\mu }C+4C_{\mu \nu \rho
}^{*}\partial _{\lambda }C^{\mu \nu \rho \lambda }-5\left( \eta
_{\mu \nu \rho \lambda }^{*}\partial _{\sigma }\eta ^{\mu \nu \rho
\lambda \sigma }+C_{\mu \nu \rho \lambda }^{*}\partial _{\sigma
}C^{\mu \nu \rho \lambda \sigma }\right) \right] .
\end{split}
\end{equation}
The solution to the master equation encodes all the information on
the gauge structure of a given theory. We remark that in our case
the solution (\ref {f14}) to the master equation breaks into terms
with the antighost number ranging from zero to four. Let us
briefly recall the significance of the various terms present in
the solution to the master equation. Thus, the part with the
antighost number equal to zero is nothing but the Lagrangian
action of the gauge model under study. The components of antighost
number equal to
one are always proportional with the gauge generators [in this situation (%
\ref{f3a})--(\ref{f3c})]. If the gauge algebra were non-Abelian,
then there would appear terms linear in the antighost number two
antifields and quadratic in the pure ghost number one ghosts. The
absence of such terms in
our case shows that the gauge transformations are Abelian. The terms from (%
\ref{f14}) with higher antighost number give us information on the
reducibility functions (\ref{f4a})--(\ref{f6b}). If the
reducibility relations held on-shell, then there would appear
components linear in the ghosts for ghosts (ghosts of pure ghost
number strictly greater than one) and quadratic in the various
antifields. Such pieces are not present in (\ref {f14}), since the
reducibility relations (\ref{f7a})--(\ref{f9}) hold off-shell.
Other possible components in the solution to the master equation
offer information on the higher-order structure functions related
to the
tensor gauge structure of the theory. There are no such terms in (\ref{f14}%
), as a consequence of the fact that all higher-order structure
functions vanish for the theory under study.

\section{Deformation of the master equation: a brief
review\label{defth}}

We begin with a ``free'' gauge theory, described by a Lagrangian action $%
S_{0}\left[ \Phi ^{\alpha _{0}}\right] $, invariant under some
gauge transformations
\begin{equation}
\delta _{\epsilon }\Phi ^{\alpha _{0}}=Z_{\;\;\alpha _{1}}^{\alpha
_{0}}\epsilon ^{\alpha _{1}},\quad \frac{\delta S_{0}}{\delta \Phi
^{\alpha _{0}}}Z_{\;\;\alpha _{1}}^{\alpha _{0}}=0,
\label{bfa2.1}
\end{equation}
and consider the problem of constructing consistent interactions
among the fields $\Phi ^{\alpha _{0}}$ such that the couplings
preserve the field spectrum and the original number of gauge
symmetries. This matter is addressed by means of reformulating the
problem of constructing consistent interactions as a deformation
problem of the solution to the master equation corresponding to
the ``free'' theory~\cite{8a,17and5}. Such a reformulation is
possible due to the fact that the solution to the master equation
contains all the information on the gauge structure of the theory.
If a consistent interacting gauge theory can be constructed, then
the solution $\bar{S}$ to
the master equation associated with the ``free'' theory, $\left( \bar{S},%
\bar{S}\right) =0$, can be deformed into a solution $S$,
\begin{align}
\bar{S}\rightarrow S& =\bar{S}+gS_{1}+g^{2}S_{2}+\cdots =  \nonumber \\
& =\bar{S}+g\int d^{D}x\,a+g^{2}\int d^{D}x\,b+\cdots ,
\label{bfa2.2}
\end{align}
of the master equation for the deformed theory
\begin{equation}
\left( S,S\right) =0,  \label{bfa2.3}
\end{equation}
such that both the ghost and antifield spectra of the initial
theory are preserved. The equation (\ref{bfa2.3}) splits,
according to the various orders in the coupling constant (or
deformation parameter) $g$, into
\begin{align}
\left( \bar{S},\bar{S}\right) & =0,  \label{bfa2.4} \\
2\left( S_{1},\bar{S}\right) & =0,  \label{bfa2.5} \\
2\left( S_{2},\bar{S}\right) +\left( S_{1},S_{1}\right) & =0,
\label{bfa2.6}
\\
\left( S_{3},\bar{S}\right) +\left( S_{1},S_{2}\right) & =0,
\label{bfa2.7}
\\
&\vdots  \nonumber
\end{align}

The equation (\ref{bfa2.4}) is fulfilled by hypothesis. The next
one requires that the first-order deformation of the solution to
the master equation, $S_{1}$, is a cocycle of the ``free'' BRST
differential $s\cdot =\left( \cdot ,\bar{S}\right) $. However,
solely cohomologically nontrivial solutions to (\ref{bfa2.5})
should be taken into account, as the BRST-exact ones can be
eliminated by some (in general nonlinear) field redefinitions.
This means that $S_{1}$ pertains to the ghost number zero
cohomological space of $s$, $H^{0}\left( s\right) $, which is
generically nonempty due to its isomorphism to the space of
physical observables of the ``free'' theory. It has been shown
in~\cite{8a,17and5} (on behalf of the triviality of the
antibracket map in the cohomology of the BRST differential) that
there are no obstructions in finding solutions to the remaining
equations [(\ref {bfa2.6})--(\ref{bfa2.7}), etc.]. However, the
resulting interactions may be nonlocal, and there might even
appear obstructions if one insists on their locality. The analysis
of these obstructions can be done with the help of cohomological
techniques. As it will be seen below, all the interactions in the
case of the model under study turn out to be local.

\section{Determination of consistent interactions\label{inter}}

In this section we determine all consistent interactions that can
be added to the free theory that describes a topological BF-type
model in five spacetime dimensions. This is done by means of
solving the deformation equations (\ref{bfa2.5})--(\ref{bfa2.7}),
etc., by means of specific cohomological techniques in the
presence of certain hypotheses to be discussed below. The
interacting theory and its gauge structure are deduced from the
analysis of the deformed solution to the master equation that is
consistent to all orders in the deformation parameter.

\subsection{Standard material: $H\left( \gamma \right) $ and $H\left( \delta
\vert d\right) $ \label{stand}}

For obvious reasons, we consider only smooth, local, (background)
Lorentz invariant, Poincar\'{e} invariant quantities (i.e., we do
not allow explicit dependence on the spacetime coordinates), and,
moreover, require the preservation of the number of derivatives on
each field with respect to the free theory. The smoothness of the
deformations refers to the fact that the deformed solution to the
master equation (\ref{bfa2.2}) is
smooth in the coupling constant $g$ and reduces to the original solution (%
\ref{f14}) in the free limit ($g=0$). If we make the notation
$S_{1}=\int d^{5}x\,a$, with $a$ a local function, then the
equation (\ref{bfa2.5}), which we have seen that controls the
first-order deformation, takes the local form
\begin{equation}
sa=\partial _{\mu }m^{\mu },\quad \text{gh}\left( a\right)
=0,\quad \varepsilon \left( a\right) =0,  \label{3.1}
\end{equation}
for some local $m^{\mu }$, and it shows that the nonintegrated
density of the first-order deformation pertains to the local
cohomology of $s$ in ghost number zero, $a\in H^{0}\left( s\vert
d\right) $, where $d$ denotes the exterior spacetime differential.
The solution to the equation (\ref{3.1}) is unique up to $s$-exact
pieces plus divergences
\begin{equation}
a\rightarrow a+sb+\partial _{\mu }n^{\mu },\quad \text{gh}\left(
b\right) =-1,\quad \varepsilon \left( b\right) =1,\quad
\text{gh}\left( n^{\mu }\right) =0,\quad \varepsilon \left( n^{\mu
}\right) =0.  \label{3.1a}
\end{equation}
At the same time, if the general solution of (\ref{3.1}) is found
to be completely trivial, $a=sb+\partial _{\mu }n^{\mu }$, then it
can be made to vanish $a=0$.

In order to analyze the equation (\ref{3.1}), we develop $a$
according to the antighost number
\begin{equation}
a=\sum\limits_{i=1}^{I}a_{i},\quad \text{agh}\left( a_{i}\right)
=i,\quad \text{gh}\left( a_{i}\right) =0,\quad \varepsilon \left(
a_{i}\right) =0, \label{3.2}
\end{equation}
and assume, without loss of generality, that the decomposition
(\ref{3.2})
stops at some finite value of $I$. This can be shown, for instance, like in~%
\cite{gen2} (Section 3), under the sole assumption that the
interacting Lagrangian at the first order in the coupling
constant, $a_{0}$, has a finite, but otherwise arbitrary
derivative order. Inserting the decomposition (\ref{3.2}) into the
equation (\ref{3.1}) and projecting it on the various values of
the antighost number, we obtain the tower of equations
\begin{align}
\gamma a_{I}& =\partial _{\mu }\stackrel{\left( I\right) }{m}^{\mu
},
\label{3.3} \\
\delta a_{I}+\gamma a_{I-1}& =\partial _{\mu }\stackrel{\left( I-1\right) }{%
m}^{\mu },  \label{3.4} \\
\delta a_{i}+\gamma a_{i-1}& =\partial _{\mu }\stackrel{\left( i-1\right) }{%
m}^{\mu },\quad 1\leq i\leq I-1,  \label{3.5}
\end{align}
where $\left( \stackrel{\left( i\right) }{m}^{\mu }\right) _{i=\overline{0,I}%
}$ are some local currents with $\text{agh}\left( \stackrel{\left(
i\right) }{m}^{\mu }\right) =i$. The equation (\ref{3.3}) can be
replaced in strictly positive values of the antighost number by
\begin{equation}
\gamma a_{I}=0,\quad I>0.  \label{3.6}
\end{equation}
The proof of this statement is done in Corollary \ref{gaplusdb}
from the
Appendix \ref{hgama}. Due to the second-order nilpotency of $\gamma $ ($%
\gamma ^{2}=0$), the solution to the equation (\ref{3.6}) is
clearly unique up to $\gamma $-exact contributions
\begin{equation}
a_{I}\rightarrow a_{I}+\gamma b_{I},\quad \text{agh}\left(
b_{I}\right) =I,\quad \text{pgh}\left( b_{I}\right) =I-1,\quad
\varepsilon \left( b_{I}\right) =1.  \label{r68}
\end{equation}
Meanwhile, if it turns out that $a_{I}$ exclusively reduces to $\gamma $%
-exact terms, $a_{I}=\gamma b_{I}$, then it can be made to vanish, $a_{I}=0$%
. In other words, the nontriviality of the first-order deformation
$a$ is translated at its highest antighost number component into
the requirement that $a_{I}\in H^{I}\left( \gamma \right) $, where
$H^{I}\left( \gamma
\right) $ denotes the cohomology of the exterior longitudinal derivative $%
\gamma $ in pure ghost number equal to $I$. So, in order to solve
the
equation (\ref{3.1}) [equivalent with (\ref{3.6}) and (\ref{3.4})--(\ref{3.5}%
)], we need to compute the cohomology of $\gamma $, $H\left( \gamma \right) $%
, and, as it will be made clear below, also the local homology of
$\delta $, $H\left( \delta \vert d\right) $.

On behalf of the definitions (\ref{f13a})--(\ref{f13f}) it is
simple to see that $H\left( \gamma \right) $ is spanned by
\begin{equation}
F_{\bar{A}}=\left( \varphi ,\partial ^{\left[ \mu \right.
}A^{\left. \nu \right] },\partial _{\mu }H^{\mu },\partial _{\mu
}B^{\mu \nu },\partial _{[\mu }\phi _{\nu \rho ]},\partial _{\rho
}K^{\mu \nu \rho }\right) , \label{3.7}
\end{equation}
by the antifields $\chi _{\Delta }^{*}$ from (\ref{notat}), by all
of their spacetime derivatives, as well as by the undifferentiated
ghosts
\begin{equation}
\eta ^{\bar{\Upsilon}}=\left( \eta ,C,\mathcal{G}^{\mu \nu \rho
\lambda \sigma },\eta ^{\mu \nu \rho \lambda \sigma },C_{\mu \nu
\rho \lambda \sigma }\right) .  \label{notat1}
\end{equation}
[The derivatives of the ghosts $\eta ^{\bar{\Upsilon}}$ are removed from $%
H\left( \gamma \right) $ since they are $\gamma $-exact, in
agreement with the first relation in (\ref{f13b}), the second
formula in (\ref{f13d}), the last equation in (\ref{f13d}), the
second relation in (\ref{f13e}), and the
first definition from (\ref{f13f}).] If we denote by $e^{M}\left( \eta ^{%
\bar{\Upsilon}}\right) $ the elements with pure ghost number $M$
of a basis in the space of the polynomials in the ghosts
(\ref{notat1}), it follows that the general solution to the
equation (\ref{3.6}) takes the form
\begin{equation}
a_{I}=\alpha _{I}\left( \left[ F_{\bar{A}}\right] ,\left[ \chi
_{\Delta }^{*}\right] \right) e^{I}\left( \eta
^{\bar{\Upsilon}}\right) ,  \label{3.8}
\end{equation}
where $\text{agh}\left( \alpha _{I}\right) =I$ and
$\text{pgh}\left( e^{I}\right) =I$. The notation $f([q])$ means
that $f$ depends on $q$ and its spacetime derivatives up to a
finite order. The objects $\alpha _{I}$ [obviously nontrivial in
$H^{0}\left( \gamma \right) $] will be called ``invariant
polynomials''. The result that we can replace the equation (\ref
{3.3}) with the less obvious one (\ref{3.6}) is a nice consequence
of the fact that the cohomology of the exterior spacetime
differential is trivial in the space of invariant polynomials in
strictly positive antighost numbers. For more details on invariant
polynomials, see the Appendix \ref {hgama}.

Inserting (\ref{3.8}) in (\ref{3.4}) we obtain that a necessary
(but not sufficient) condition for the existence of (nontrivial)
solutions $a_{I-1}$ is that the invariant polynomials $\alpha
_{I}$ are (nontrivial) objects from the local cohomology of
Koszul-Tate differential $H\left( \delta \vert d\right) $ in
antighost number $I>0$ and in pure ghost number zero,
\begin{equation}
\delta \alpha _{I}=\partial _{\mu }\stackrel{\left( I-1\right)
}{j}^{\mu },\quad \text{agh}\left( \stackrel{\left( I-1\right)
}{j}^{\mu }\right) =I-1,\quad \text{pgh}\left( \stackrel{\left(
I-1\right) }{j}^{\mu }\right) =0.  \label{3.10a}
\end{equation}
We recall that the local cohomology $H\left( \delta \vert d\right)
$ is completely trivial in both strictly positive antighost
\textit{and} pure ghost numbers (for instance, see~\cite{gen1},
Theorem 5.4, and~\cite{gen11}). Using the fact that the BF model
under study is a linear gauge theory of Cauchy order equal to five
and the general result from~\cite{gen1,gen11}, according to which
the local cohomology of the Koszul-Tate differential at pure ghost
number zero is trivial in antighost numbers strictly greater than
its Cauchy order, we can state that
\begin{equation}
H_{J}\left( \delta \vert d\right) =0\quad \text{for\ all}\quad
J>5,  \label{3.11}
\end{equation}
where $H_{J}\left( \delta \vert d\right) $ represents the local
cohomology of the Koszul-Tate differential in antighost number $J$
and in zero pure ghost
number. Moreover, if the invariant polynomial $\alpha _{J}$, with $\text{agh}%
\left( \alpha _{J}\right) =J\geq 5$, is trivial in $H_{J}\left(
\delta
\vert d\right) $, then it can be taken to be trivial also in $H_{J}^{\text{inv}%
}\left( \delta \vert d\right) $%
\begin{equation}
\left( \alpha _{J}=\delta b_{J+1}+\partial _{\mu }\stackrel{(J)}{c}^{\mu },%
\text{agh}\left( \alpha _{J}\right) =J\geq 5\right) \Rightarrow
\alpha _{J}=\delta \beta _{J+1}+\partial _{\mu
}\stackrel{(J)}{\gamma }^{\mu }, \label{3.12ax}
\end{equation}
with both $\beta _{J+1}$ and $\stackrel{(J)}{\gamma }^{\mu }$
invariant polynomials. Here, $H_{J}^{\text{inv}}\left( \delta
\vert d\right) $ denotes the invariant characteristic cohomology
in antighost number $J$ (the local cohomology of the Koszul-Tate
differential in the space of invariant polynomials). [An element
of $H_{I}^{\text{inv}}\left( \delta \vert d\right) $ is defined
via an equation of the type (\ref{3.10a}), but with $\alpha _{I}$
and the corresponding current invariant polynomials.] The result
(\ref {3.12ax}) is proved in detail in Theorem \ref{hinvdelta}
from the Appendix \ref{charcoh}. It is important since, together
with (\ref{3.11}), ensures that the entire invariant
characteristic cohomology in antighost numbers strictly greater
than five is trivial
\begin{equation}
H_{J}^{\text{inv}}\left( \delta \vert d\right) =0\quad \text{for
all}\quad J>5. \label{3.12x}
\end{equation}

The nontrivial representatives of $H_{J}\left( \delta \vert
d\right) $ at pure
ghost number zero and of $H_{J}^{\text{inv}}\left( \delta \vert d\right) $ for $%
J\geq 2$ depend neither on $\left( \partial ^{\left[ \mu \right.
}A^{\left. \nu \right] },\partial _{\mu }H^{\mu },\partial _{\mu
}B^{\mu \nu },\partial _{[\mu }\phi _{\nu \rho ]},\partial _{\rho
}K^{\mu \nu \rho }\right) $ nor on the spacetime derivatives of
$F_{\bar{A}}$ defined in (\ref{3.7}), but only on the
undifferentiated scalar field $\varphi $. With the help of the
relations (\ref{f12b})--(\ref{f12g}), it can be shown that both
$H_{5}\left( \delta \vert d\right) $ at pure ghost number zero and
$H_{5}^{\text{inv}}\left( \delta \vert d\right) $ are generated by
the elements
\begin{equation}\label{3.13}
\begin{split}
\left( W\right) _{\mu \nu \rho \lambda \sigma }&
=\frac{dW}{d\varphi }C_{\mu \nu \rho \lambda \sigma
}^{*}+\frac{d^{2}W}{d\varphi ^{2}}\left( H_{[\mu }^{*}C_{\nu \rho
\lambda \sigma ]}^{*}+C_{[\mu \nu }^{*}C_{\rho \lambda
\sigma ]}^{*}\right) +\\
&\quad +\frac{d^{3}W}{d\varphi ^{3}}\left( H_{[\mu }^{*}H_{\left.
\nu \right. }^{*}C_{\rho \lambda \sigma ]}^{*}+H_{[\mu
}^{*}C_{\left. \nu \rho \right. }^{*}C_{\lambda \sigma
]}^{*}\right) +\frac{d^{4}W}{d\varphi ^{4}}H_{[\mu }^{*}H_{\left.
\nu \right. }^{*}H_{\left. \rho \right. }^{*}C_{\lambda
\sigma ]}^{*} +\\
&\quad +\frac{d^{5}W}{d\varphi ^{5}}H_{\mu }^{*}H_{\nu
}^{*}H_{\rho }^{*}H_{\lambda }^{*}H_{\sigma }^{*},
\end{split}
\end{equation}
where $W=W\left( \varphi \right) $ is an arbitrary, smooth
function of the undifferentiated scalar field $\varphi $. Indeed,
direct computation yields
\begin{equation}
\delta \left( W\right) _{\mu \nu \rho \lambda \sigma }=-\partial
_{[\mu }\left( W\right) _{\nu \rho \lambda \sigma ]},\quad
\text{agh}\left( \left( W\right) _{\nu \rho \lambda \sigma
}\right) =4,  \label{3.13a}
\end{equation}
where we made the notation
\begin{equation} \label{3.14}
\begin{split}
\left( W\right) _{\mu \nu \rho \lambda }& =\frac{dW}{d\varphi
}C_{\mu \nu \rho \lambda }^{*}+\frac{d^{2}W}{d\varphi ^{2}}\left(
H_{[\mu }^{*}C_{\nu \rho \lambda ]}^{*}+C_{[\mu \nu }^{*}C_{\rho
\lambda ]}^{*}\right) + \\
&\quad +\frac{d^{3}W}{d\varphi ^{3}}H_{[\mu }^{*}H_{\left. \nu
\right. }^{*}C_{\rho \lambda ]}^{*}+\frac{d^{4}W}{d\varphi
^{4}}H_{\mu }^{*}H_{\nu }^{*}H_{\rho }^{*}H_{\lambda }^{*}.
\end{split}
\end{equation}
Using again the actions of $\delta $ on the BRST generators, it
can be proved that both $H_{4}\left( \delta \vert d\right) $ at
pure ghost number zero and $H_{4}^{\text{inv}}\left( \delta \vert
d\right) $ are spanned by the elements $\left( W\right) _{\mu \nu
\rho \lambda }$ given in (\ref{3.14}) and by the undifferentiated
antifields $\eta _{\mu \nu \rho \lambda \sigma }^{*}$ [the second
definition in (\ref{f12g})]. Related to $\left( W\right) _{\mu \nu
\rho \lambda }$, we have that
\begin{equation}
\delta \left( W\right) _{\mu \nu \rho \lambda }=\partial _{[\mu
}\left( W\right) _{\nu \rho \lambda ]},\quad \text{agh}\left(
\left( W\right) _{\nu \rho \lambda }\right) =3,  \label{3.14a}
\end{equation}
where we employed the convention
\begin{equation}
\left( W\right) _{\mu \nu \rho }=\frac{dW}{d\varphi }C_{\mu \nu \rho }^{*}+%
\frac{d^{2}W}{d\varphi ^{2}}H_{[\mu }^{*}C_{\nu \rho ]}^{*}+\frac{d^{3}W}{%
d\varphi ^{3}}H_{\mu }^{*}H_{\nu }^{*}H_{\rho }^{*}.  \label{3.15}
\end{equation}
On account of the same arguments, it can be shown that the
generators of the
spaces $H_{3}\left( \delta \vert d\right) $ at pure ghost number zero and $H_{3}^{%
\text{inv}}\left( \delta \vert d\right) $ are exactly $\left(
W\right) _{\mu \nu
\rho }$ expressed by (\ref{3.15}), as well the undifferentiated antifields $%
\eta _{\mu \nu \rho \lambda }^{*}$, $\mathcal{G}_{\mu \nu \rho
\lambda \sigma }^{*}$, and $C^{*}$ [see the formula (\ref{f12f})].
For the first element, straightforward calculations produce
\begin{equation}
\delta \left( W\right) _{\mu \nu \rho }=-\partial _{[\mu }\left(
W\right) _{\nu \rho ]},\quad \text{agh}\left( \left( W\right)
_{\nu \rho }\right) =2, \label{3.15a}
\end{equation}
where we used the notation
\begin{equation}
\left( W\right) _{\mu \nu }=\frac{dW}{d\varphi }C_{\mu \nu }^{*}+\frac{d^{2}W%
}{d\varphi ^{2}}H_{\mu }^{*}H_{\nu }^{*}.  \label{3.16}
\end{equation}
Finally, it can be proved that the spaces $H_{2}\left( \delta
\vert d\right) $ at pure ghost number zero and
$H_{2}^{\text{inv}}\left( \delta \vert d\right) $ are spanned by
$\left( W\right) _{\mu \nu }$ defined in (\ref{3.16}) and by the
undifferentiated antifields $\eta _{\mu \nu \rho }^{*}$,
$\mathcal{G}_{\mu \nu \rho \lambda }^{*}$, $C_{\mu }^{*}$, and
$\eta ^{*}$ [see the first and
last relations in (\ref{f12d}), as well as the first two definitions from (%
\ref{f12e})]. Concerning $\left( W\right) _{\mu \nu }$, simple
computation leads to
\begin{equation}
\delta \left( W\right) _{\mu \nu }=\partial _{[\mu }\left(
W\right) _{\nu ]},\quad \text{agh}\left( \left( W\right) _{\nu
}\right) =1,  \label{3.16a}
\end{equation}
with
\begin{equation}
\left( W\right) _{\mu }=\frac{dW}{d\varphi }H_{\mu }^{*}.
\label{3.17}
\end{equation}
In contrast to the spaces $\left( H_{J}\left( \delta \vert
d\right) \right) _{J\geq 2}$ and $\left( H_{J}^{\text{inv}}\left(
\delta \vert d\right) \right) _{J\geq 2}$, which are
finite-dimensional, the cohomology $H_{1}\left( \delta \vert
d\right) $ at pure ghost number zero, that is related to global
symmetries and ordinary conservation laws, is infinite-dimensional
since the theory is free. Fortunately, it will not be needed in
the sequel.

The previous results on $H\left( \delta \vert d\right) $ and $H^{\text{inv}%
}\left( \delta \vert d\right) $ in strictly positive antighost
numbers are important because they control the obstructions to
removing the antifields from the first-order deformation. More
precisely, we can successively eliminate all the pieces of
antighost number strictly greater that five from the nonintegrated
density of the first-order deformation by adding solely trivial
terms, so we can take, without loss of nontrivial objects, the
condition $I\leq 5$ in the decomposition (\ref{3.2}). The proof of
this statement is contained in the Appendix \ref{elim4}. In
addition, the last representative is of the form (\ref{3.8}),
where the invariant polynomial is necessarily a nontrivial object
from $H_{5}^{\text{inv}}\left( \delta \vert d\right) $.

\subsection{First-order deformation}

Using the results stated in the previous subsection, we can assume
that the first-order deformation stops at antighost number five
($I=5$)
\begin{equation}
a=a_{0}+a_{1}+a_{2}+a_{3}+a_{4}+a_{5},  \label{fr1}
\end{equation}
where $a_{5}$ is of the form (\ref{3.8}), with $\alpha _{5}$ from $H_{5}^{%
\text{inv}}\left( \delta \vert d\right) $ [elements of the form
(\ref{3.13}), generated by arbitrarily smooth functions,
exclusively depending on the
undifferentiated scalar field $\varphi $] and $e^{5}\left( \eta ^{\bar{%
\Upsilon}}\right) $ denoting the elements with pure ghost number
five of a
basis in the space of the polynomials in the ghosts $\eta ^{\bar{\Upsilon}}$%
\begin{equation}\label{fr2a}
e^{5}:\left( \eta C^{\mu \nu \rho
\lambda \sigma },C\eta ^{\mu \nu \rho \lambda \sigma },\eta C\mathcal{G}%
^{\mu \nu \rho \lambda \sigma },\eta \mathcal{G}^{\mu \nu \rho
\lambda \sigma }\mathcal{G}^{\mu ^{\prime }\nu ^{\prime }\rho
^{\prime }\lambda ^{\prime }\sigma ^{\prime }},\eta CC,\eta ^{\mu
\nu \rho \lambda \sigma }\mathcal{G}^{\mu ^{\prime }\nu ^{\prime
}\rho ^{\prime }\lambda ^{\prime }\sigma ^{\prime }}\right) .
\end{equation}
In order to couple (\ref{3.13}) to the last three elements from
(\ref{fr2a}) like in (\ref{3.8}) we need some completely
antisymmetric constants, which, by covariance arguments, can only
be proportional with the completely antisymmetric five-dimensional
symbol, $\varepsilon _{\mu \nu \rho \lambda \sigma }$. Thus, the
most general (manifestly covariant) form of the last
representative from the expansion (\ref{fr1}) is given by
\begin{equation}\label{fr3}
\begin{split}
a_{5}& =\left( W_{1}\right) _{\mu \nu \rho \lambda \sigma }\eta
C^{\mu \nu \rho \lambda \sigma }+\left( W_{2}\right) _{\mu \nu
\rho \lambda \sigma }C\eta ^{\mu \nu \rho \lambda \sigma }+\left(
W_{3}\right) _{\mu \nu \rho \lambda \sigma }\eta C\mathcal{G}^{\mu
\nu \rho \lambda \sigma }  -\\
&\quad -\epsilon ^{\alpha \beta \gamma \delta \varepsilon }\left(
\left( W_{4}\right) _{\mu \nu \rho \lambda \sigma }\eta
\mathcal{G}^{\mu \nu \rho
\lambda \sigma }\mathcal{G}_{\alpha \beta \gamma \delta \varepsilon }-\tfrac{%
1}{5!}\left( W_{5}\right) _{\alpha \beta \gamma \delta \varepsilon
}\eta
CC\right.  +\\
&\quad \quad \quad \quad \quad \left. +\left( W_{6}\right) _{\mu
\nu \rho \lambda \sigma }\eta ^{\mu \nu \rho \lambda \sigma
}\mathcal{G}_{\alpha \beta \gamma \delta \varepsilon }\right) ,
\end{split}
\end{equation}
where each of the elements $\left( \left( W_{k}\right) _{\mu \nu
\rho
\lambda \sigma }\right) _{k=\overline{1,6}}$ is expressed like in (\ref{3.13}%
), being generated by an arbitrary smooth function of the
undifferentiated scalar field, $W_{k}\left( \varphi \right) $.

Inserting (\ref{fr3}) into the equation (\ref{3.4}) for $I=5$ and
using the definitions (\ref{f12a})--(\ref{f13f}), after some
computation we obtain the piece with antighost number equal to
four from the first-order deformation like
\begin{equation}\label{fr4c}
\begin{split}
a_{4}& =\left( W_{1}\right) _{\mu \nu \rho \lambda }\left(
5A_{\sigma }C^{\mu \nu \rho \lambda \sigma }-\eta C^{\mu \nu \rho
\lambda }\right) -\left( W_{2}\right) _{\mu \nu \rho \lambda
}\left( 5C_{\sigma }\eta ^{\mu
\nu \rho \lambda \sigma }+C\eta ^{\mu \nu \rho \lambda }\right) + \\
&\quad +\left( W_{3}\right) _{\mu \nu \rho \lambda }\left( 4A_{\sigma }C\mathcal{G%
}^{\mu \nu \rho \lambda \sigma }+4\eta C_{\sigma }\mathcal{G}^{\mu
\nu \rho \lambda \sigma }-\eta C\mathcal{G}^{\mu \nu \rho \lambda
}\right) +\\
&\quad +\epsilon ^{\alpha \beta \gamma \delta \varepsilon }\left[
-\left( \left( W_{4}\right) _{[\mu \nu \rho \lambda }A_{\sigma
]}\mathcal{G}^{\mu \nu \rho
\lambda \sigma }-2\left( W_{4}\right) _{\mu \nu \rho \lambda }\eta \mathcal{G%
}^{\mu \nu \rho \lambda }\right) \mathcal{G}_{\alpha \beta \gamma
\delta
\varepsilon }+\right.  \\
&\quad \quad \quad \quad \quad +\tfrac{1}{5!}\left( \left(
W_{5}\right) _{[\alpha \beta \gamma \delta }A_{\varepsilon
]}C-2\left( W_{5}\right) _{[\alpha \beta \gamma \delta
}C_{\varepsilon ]}\eta \right) C+ \\
&\quad \quad \quad \quad \quad \left. +\left( W_{6}\right) _{\mu
\nu \rho \lambda }\left( \eta ^{\mu \nu
\rho \lambda }\mathcal{G}_{\alpha \beta \gamma \delta \varepsilon }-\mathcal{%
G}^{\mu \nu \rho \lambda }\eta _{\alpha \beta \gamma \delta
\varepsilon
}\right) \right] + \\
&\quad +2\left( \left( W_{1}\right) _{[\mu \nu \rho }B_{\lambda
\sigma
]}^{*}+\left( W_{1}\right) _{[\mu \nu }\eta _{\rho \lambda \sigma ]}^{*}+%
\frac{dW_{1}}{d\varphi }H_{[\mu }^{*}\eta _{\nu \rho \lambda
\sigma ]}^{*}+W_{1}\eta _{\mu \nu \rho \lambda
\sigma }^{*}\right) C^{\mu \nu \rho \lambda \sigma } +\\
&\quad +2\left( \left( W_{3}\right) _{[\mu \nu \rho }B_{\lambda
\sigma ]}^{*}+\left( W_{3}\right) _{[\mu \nu }\eta _{\rho \lambda
\sigma ]}^{*}+\frac{dW_{3}}{d\varphi }H_{[\mu }^{*}\eta _{\nu \rho
\lambda \sigma ]}^{*}+W_{3}\eta _{\mu \nu \rho \lambda \sigma
}^{*}\right) C\mathcal{G}^{\mu \nu \rho \lambda \sigma } -\\
&\quad -2\epsilon ^{\alpha \beta \gamma \delta \varepsilon }\left[
\left( \left( W_{4}\right) _{[\mu \nu \rho }B_{\lambda
\sigma ]}^{*}+\left( W_{4}\right) _{[\mu \nu }\eta _{\rho \lambda \sigma ]}^{*}+%
\frac{dW_{4}}{d\varphi }H_{[\mu }^{*}\eta _{\nu \rho \lambda
\sigma
]}^{*}+W_{4}\eta _{\mu \nu \rho \lambda \sigma }^{*}\right) \times \right. \\
&\quad \quad \quad \quad \quad \quad \times \mathcal{G}^{\mu \nu
\rho \lambda \sigma }\mathcal{G}_{\alpha \beta \gamma \delta
\varepsilon }-\tfrac{1}{5!}\left( \left( W_{5}\right) _{[\alpha
\beta \gamma }B_{\delta \varepsilon ]}^{*}+\left( W_{5}\right)
_{[\alpha \beta }\eta _{\gamma \delta \varepsilon ]}^{*}+\right. \\
&\quad \quad \quad \quad \quad \quad \quad \quad \quad \quad \quad
\quad \quad \quad \quad \left. \left. +\frac{dW_{5}}{d\varphi
}H_{[\alpha }^{*}\eta _{\beta \gamma \delta \varepsilon
]}^{*}+W_{5}\eta _{\alpha \beta \gamma \delta \varepsilon
}^{*}\right) CC\right] .
\end{split}
\end{equation}
Here and in the sequel $\left( W_{k}\right) _{\mu \nu \rho \lambda }$, $%
\left( W_{k}\right) _{\mu \nu \rho }$, and $\left( W_{k}\right) _{\mu \nu }$%
are written like in (\ref{3.14}), (\ref{3.15}), and respectively (\ref{3.16}%
), with $W\left( \varphi \right) $ replaced by the corresponding $%
W_{k}\left( \varphi \right) $.

From now on we will need the relations (\ref{3.14a}),
(\ref{3.15a}), and (\ref{3.16a}).
Substituting the solution (\ref{fr4c}) into the equation (\ref{3.5}) for $%
i=4 $ and employing the same definitions like before, we derive
the terms of antighost number three from the first-order
deformation as
\begin{equation}\label{fr5c}
\begin{split}
a_{3}& =-\left( W_{1}\right) _{\mu \nu \rho }\left( 4A_{\lambda
}C^{\mu \nu \rho \lambda }-\eta C^{\mu \nu \rho }\right) +\left(
W_{2}\right) _{\mu \nu \rho }\left( 4C_{\lambda }\eta ^{\mu \nu
\rho \lambda }+C\eta ^{\mu \nu \rho
}\right) - \\
&\quad -2\left( \left( W_{1}\right) _{[\mu \nu }B_{\rho \lambda
]}^{*}+\frac{dW_{1}}{d\varphi }H_{[\mu }^{*}\eta _{\nu \rho
\lambda ]}^{*}+W_{1}\eta _{\mu \nu \rho \lambda }^{*}\right)
C^{\mu \nu \rho \lambda } -\left( W_{2}\right) _{[\mu \nu \rho
}\phi
_{\lambda \sigma ]}\eta ^{\mu \nu \rho \lambda \sigma } +\\
&\quad +\left( \left( W_{3}\right) _{[\mu \nu \rho }A_{\lambda
}C_{\sigma ]}+2\left( W_{3}\right) _{[\mu \nu }B_{\rho \lambda
}^{*}C_{\sigma ]}+\right.  \\
&\quad \quad \quad \left. +2\frac{dW_{3}}{d\varphi }H_{[\mu
}^{*}\eta _{\nu \rho \lambda }^{*}C_{\sigma ]}+2W_{3}\eta _{[\mu
\nu \rho \lambda }^{*}C_{\sigma ]}-\left( W_{3}\right) _{[\mu \nu
\rho }\phi _{\lambda \sigma ]}\eta \right)
\mathcal{G}^{\mu \nu \rho \lambda \sigma } - \\
&\quad -\left[ \left( \left( W_{3}\right) _{[\mu \nu \rho
}A_{\lambda ]}+2\left(
W_{3}\right) _{[\mu \nu }B_{\rho \lambda ]}^{*}+2\frac{dW_{3}}{d\varphi }%
H_{[\mu }^{*}\eta _{\nu \rho \lambda ]}^{*}+2W_{3}\eta _{\mu \nu
\rho \lambda }^{*}\right) C-\right. \\
&\quad \quad \quad \left. -\left( W_{3}\right) _{[\mu \nu \rho
}C_{\lambda ]}\eta \right] \mathcal{G}^{\mu \nu \rho \lambda
}+\left( W_{3}\right) _{\mu \nu \rho }\eta CK^{\mu \nu \rho }+
\\
&\quad +\epsilon ^{\alpha \beta \gamma \delta \varepsilon }\left[
4\left( \tfrac{1}{2}\left( W_{4}\right) _{[\mu \nu \rho
}A_{\lambda ]}+\left( W_{4}\right) _{[\mu \nu }B_{\rho \lambda
]}^{*}+\frac{dW_{4}}{d\varphi }H_{[\mu }^{*}\eta _{\nu \rho
\lambda ]}^{*}+W_{4}\eta _{\mu \nu \rho \lambda }^{*}\right)
\times \right.  \\
&\quad \quad \quad \quad \quad \times \mathcal{G}^{\mu \nu \rho
\lambda }\mathcal{G}_{\alpha \beta \gamma \delta \varepsilon
}+\left( \tfrac{1}{3!}\left( W_{4}\right) _{\alpha \beta \gamma
}\epsilon _{\delta \mu \nu \rho \lambda }\mathcal{G}^{\mu \nu \rho
\lambda }\epsilon
_{\varepsilon \mu ^{\prime }\nu ^{\prime }\rho ^{\prime }\lambda ^{\prime }}%
\mathcal{G}^{\mu ^{\prime }\nu ^{\prime }\rho ^{\prime }\lambda
^{\prime
}}-\right.  \\
&\quad \quad \quad \quad \quad \quad \quad \quad \quad \quad \quad
\quad \quad \left. -2\left( W_{4}\right) _{\mu \nu \rho }K^{\mu
\nu \rho }\mathcal{G}_{\alpha \beta \gamma \delta
\varepsilon }\right) \eta +\\
&\quad \quad \quad \quad \quad +\tfrac{4}{5!}\left(
\tfrac{1}{2}\left( W_{5}\right) _{[\alpha \beta \gamma }A_{\delta
}C_{\varepsilon ]}+\left( W_{5}\right) _{[\alpha \beta }B_{\gamma
\delta
}^{*}C_{\varepsilon ]}+\right.  \\
&\quad \quad \quad \quad \quad \quad \quad \quad \left.
+\frac{dW_{5}}{d\varphi }H_{[\alpha }^{*}\eta _{\beta \gamma
\delta }^{*}C_{\varepsilon ]}+W_{5}\eta _{[\alpha \beta \gamma
\delta }^{*}C_{\varepsilon ]}\right)
C+\\
&\quad \quad \quad \quad \quad +\tfrac{2}{5!}\left( \left(
W_{5}\right) _{[\alpha \beta \gamma }C_{\delta }C_{\varepsilon
]}-\left( W_{5}\right) _{[\alpha \beta \gamma
}\phi _{\delta \varepsilon ]}C\right) \eta -\\
&\quad \quad \quad \quad \quad \left. -\left( W_{6}\right) _{\mu
\nu \rho }\left( \eta ^{\mu \nu \rho }\mathcal{G}_{\alpha \beta
\gamma \delta \varepsilon }+K^{\mu \nu \rho }\eta _{\alpha \beta
\gamma \delta \varepsilon }-\eta ^{\mu \nu \rho \lambda }\sigma
_{\lambda \alpha }\mathcal{G}_{\beta \gamma \delta \varepsilon
}\right) \right] -
\\
&\quad -3\left[ \left( \left( W_{2}\right) _{[\mu \nu }K_{\rho
\lambda \sigma ]}^{*}+\frac{dW_{2}}{d\varphi }H_{[\mu
}^{*}\mathcal{G}_{\nu \rho \lambda \sigma ]}^{*}
+W_{2}\mathcal{G}_{\mu \nu \rho \lambda \sigma }^{*}\right) \eta
^{\mu \nu \rho \lambda \sigma }+\right.
\\
&\quad \quad \quad +\left( \left( W_{3}\right) _{[\mu \nu }K_{\rho
\lambda \sigma ]}^{*}+\frac{dW_{3}}{d\varphi }H_{[\mu
}^{*}\mathcal{G}_{\nu \rho \lambda \sigma
]}^{*}+W_{3}\mathcal{G}_{\mu \nu \rho \lambda \sigma }^{*}\right)
\eta \mathcal{G}^{\mu \nu \rho \lambda
\sigma } +\\
&\quad \quad \quad \left. +\left( \left( W_{3}\right) _{\mu \nu
}\phi ^{*\mu \nu }+\frac{dW_{3}}{d\varphi }H_{\mu }^{*}C^{*\mu
}-W_{3}C^{*}\right) \eta C  \right] +\\
&\quad +\epsilon ^{\alpha \beta \gamma \delta \varepsilon }\left[
6\left( \left( W_{4}\right) _{\mu \nu }\phi ^{*\mu \nu
}+\frac{dW_{4}}{d\varphi }H_{\mu }^{*}C^{*\mu }-W_{4}C^{*}\right)
\eta \mathcal{G}_{\alpha \beta \gamma
\delta \varepsilon }-\right.  \\
&\quad \quad \quad \quad \quad -\tfrac{1}{20}\left( \left(
W_{5}\right) _{[\alpha \beta }K_{\gamma \delta \varepsilon
]}^{*}+\frac{dW_{5}}{d\varphi }H_{[\alpha }^{*}\mathcal{G}_{\beta
\gamma \delta \varepsilon ]}^{*}+W_{5}\mathcal{G}_{\alpha \beta
\gamma
\delta \varepsilon }^{*}\right) \eta C  +\\
&\quad \quad \quad \quad \quad \left. +3\left( \left( W_{6}\right)
_{\mu \nu }\phi ^{*\mu \nu }+\frac{dW_{6}}{d\varphi }H_{\mu
}^{*}C^{*\mu }-W_{6}C^{*}\right) \eta _{\alpha \beta \gamma \delta
\varepsilon }\right] .
\end{split}
\end{equation}

The component with the antighost number equal to two results as
solution to the equation (\ref{3.5}) for $i=3$, by relying on the
formula (\ref{fr5c}) and the definitions
(\ref{f12a})--(\ref{f13f}), and takes the form
\begin{equation}\label{fr6c}
\begin{split}
a_{2}& =\left( W_{1}\right) _{\mu \nu }\left( 3A_{\rho }C^{\mu \nu
\rho }-\eta C^{\mu \nu }\right) -\left( W_{2}\right) _{\mu \nu
}\left( 3C_{\rho
}\eta ^{\mu \nu \rho }+B^{\mu \nu }C\right) + \\
&\quad +2\left( \frac{dW_{1}}{d\varphi }H_{[\mu }^{*}B_{\nu \rho
]}^{*}+W_{1}\eta _{\mu \nu \rho }^{*}\right) C^{\mu \nu \rho
} +\\
&\quad +\left( \left( W_{2}\right) _{[\mu \nu }\phi _{\rho \lambda
]}+3\frac{dW_{2}}{d\varphi }H_{[\mu }^{*}K_{\nu \rho \lambda
]}^{*}+3W_{2}\mathcal{G}_{\mu \nu \rho \lambda }^{*}\right) \eta
^{\mu \nu \rho \lambda }-\\
&\quad -\left[ \left( W_{3}\right) _{[\mu \nu }A_{\rho }\phi
_{\lambda \sigma ]}+2\frac{dW_{3}}{d\varphi }H_{[\mu }^{*}B_{\nu
\rho
}^{*}\phi _{\lambda \sigma ]}+\right.  \\
&\quad \quad \quad \left. +3\frac{dW_{3}}{d\varphi }H_{[\mu
}^{*}K_{\nu \rho \lambda }^{*}A_{\sigma ]}+W_{3}\left( 2\eta
_{[\mu \nu \rho }^{*}\phi _{\lambda \sigma ]}+3\mathcal{G}_{[\mu
\nu \rho \lambda }^{*}A_{\sigma ]}\right)
\right] \mathcal{G}^{\mu \nu \rho \lambda \sigma } -\\
&\quad -\left[ \left( W_{3}\right) _{[\mu \nu }A_{\rho }C_{\lambda
]}-\left( W_{3}\right) _{[\mu \nu }\phi _{\rho \lambda ]}\eta
+2\frac{dW_{3}}{d\varphi
}H_{[\mu }^{*}B_{\nu \rho }^{*}C_{\lambda ]}-\right.  \\
&\quad \quad \quad \left. -3\frac{dW_{3}}{d\varphi }H_{[\mu
}^{*}K_{\nu \rho \lambda
]}^{*}\eta +W_{3}\left( 2\eta _{[\mu \nu \rho }^{*}C_{\lambda ]}-3\mathcal{G}%
_{\mu \nu \rho \lambda }^{*}\eta \right) \right] \mathcal{G}^{\mu
\nu \rho
\lambda } + \\
&\quad +\left( \left( W_{3}\right) _{[\mu \nu }A_{\rho
]}+2\frac{dW_{3}}{d\varphi }H_{[\mu }^{*}B_{\nu \rho
]}^{*}+2W_{3}\eta _{\mu \nu \rho }^{*}\right)
K^{\mu \nu \rho }C + \\
&\quad +3\left( 2\frac{dW_{3}}{d\varphi }H_{\mu }^{*}\phi ^{*\mu
\nu
}-W_{3}C^{*\nu }\right) A_{\nu }C + \\
&\quad +3\left( \left( W_{3}\right) _{\mu \nu }K^{\mu \nu \rho }+2\frac{dW_{3}}{%
d\varphi }H_{\mu }^{*}\phi ^{*\mu \rho }-W_{3}C^{*\rho }\right)
\eta C_{\rho
} + \\
&\quad +\epsilon ^{\alpha \beta \gamma \delta \varepsilon }\left[
-2\left( \left( W_{4}\right) _{[\mu \nu }A_{\rho
]}+2\frac{dW_{4}}{d\varphi }H_{[\mu }^{*}B_{\nu \rho
]}^{*}+2W_{4}\eta _{\mu \nu \rho }^{*}\right) K^{\mu \nu \rho
}\mathcal{G}_{\alpha \beta \gamma \delta \varepsilon }+\right.
\\
&\quad \quad \quad \quad \quad +\tfrac{1}{3}\left(
\tfrac{1}{2}\left( W_{4}\right) _{[\alpha \beta }A_{\gamma ]}+
\frac{dW_{4}}{d\varphi }H_{[\alpha }^{*}B_{\beta \gamma
]}^{*}\right. \left. +W_{4}\eta _{\alpha \beta \gamma }^{*}\right) \times\\
&\quad \quad \quad \quad \quad \quad \quad \quad \times \epsilon
_{\delta \mu \nu \rho \lambda }\mathcal{G}^{\mu \nu \rho \lambda
}\epsilon
_{\varepsilon \mu ^{\prime }\nu ^{\prime }\rho ^{\prime }\lambda ^{\prime }}%
\mathcal{G}^{\mu ^{\prime }\nu ^{\prime }\rho ^{\prime }\lambda
^{\prime }}+
\\
&\quad \quad \quad \quad \quad +6\left( \left( W_{4}\right) _{\mu
\nu }K^{\mu \nu \rho }+2\frac{dW_{4}}{d\varphi }H_{\mu }^{*}\phi
^{*\mu \rho }-W_{4}C^{*\rho }\right) \sigma _{\rho \alpha }\eta
\mathcal{G}_{\beta \gamma \delta
\varepsilon } - \\
&\quad \quad \quad \quad \quad -6\left( \frac{dW_{4}}{d\varphi
}H_{[\mu }^{*}A_{\nu ]}\phi ^{*\mu \nu }-W_{4}C^{*\mu }A_{\mu
}\right) \mathcal{G}_{\alpha \beta \gamma \delta
\varepsilon } - \\
&\quad \quad \quad \quad \quad -\tfrac{2}{5!}\left( \left(
W_{5}\right) _{[\alpha \beta }A_{\gamma }\phi _{\delta \varepsilon
]}+2\frac{dW_{5}}{d\varphi }H_{[\alpha }^{*}B_{\beta \gamma
}^{*}\phi _{\delta \varepsilon ]}+2W_{5}\eta _{[\alpha \beta
\gamma
}^{*}\phi _{\delta \varepsilon ]}\right) C - \\
&\quad \quad \quad \quad \quad -\tfrac{1}{20}\left(
\frac{dW_{5}}{d\varphi }H_{[\alpha }^{*}K_{\beta \gamma \delta
}^{*}A_{\varepsilon ]}+W_{5}\mathcal{G}_{[\alpha \beta \gamma
\delta }^{*}A_{\varepsilon ]}\right) C + \\
&\quad \quad \quad \quad \quad +\tfrac{1}{20}\left(
\tfrac{1}{3}\left( W_{5}\right) _{[\alpha \beta }\phi
_{\gamma \delta }C_{\varepsilon ]}+\tfrac{1}{2}\frac{dW_{5}}{d\varphi }%
H_{[\alpha }^{*}K_{\beta \gamma \delta }^{*}C_{\varepsilon ]}+W_{5}\mathcal{G%
}_{[\alpha \beta \gamma \delta }^{*}C_{\varepsilon ]}\right) \eta + \\
&\quad \quad \quad \quad \quad +\tfrac{2}{5!}\left( \left(
W_{5}\right) _{[\alpha \beta }A_{\gamma }C_{\delta }C_{\varepsilon
]}+2\frac{dW_{5}}{d\varphi }H_{[\alpha }^{*}B_{\beta \gamma
}^{*}C_{\delta }C_{\varepsilon ]}+2W_{5}\eta _{[\alpha
\beta \gamma }^{*}C_{\delta }C_{\varepsilon ]}\right) + \\
&\quad \quad \quad \quad \quad +\left( W_{6}\right) _{\mu \nu
}\left( B^{\mu \nu }\mathcal{G}_{\alpha \beta \gamma \delta
\varepsilon }+3\eta ^{\mu \nu \rho }\sigma _{\rho \alpha
}\mathcal{G}_{\beta \gamma \delta \varepsilon }\right) + \\
&\quad \quad \quad \quad \quad \left. +3\left( \left( W_{6}\right)
_{\mu \nu }K^{\mu \nu \rho }+2\frac{dW_{6}}{d\varphi }H_{\mu
}^{*}\phi ^{*\mu \rho }-W_{6}C^{*\rho }\right) \sigma _{\rho
\alpha }\eta _{\beta \gamma \delta \varepsilon }\right] -
\\
&\quad -2\left( \frac{dW_{2}}{d\varphi }H_{\mu }^{*}A^{*\mu
}-W_{2}\eta ^{*}\right) C-6W_{3}\left( \phi ^{*\mu \nu }B_{\mu \nu
}^{*}C+K_{[\mu \nu \rho }^{*}B_{\lambda \sigma
]}^{*}\mathcal{G}^{\mu \nu \rho \lambda \sigma
}\right) + \\
&\quad +\epsilon ^{\alpha \beta \gamma \delta \varepsilon }\left[
12W_{4}\phi ^{*\mu \nu }B_{\mu \nu }^{*}\mathcal{G}_{\alpha \beta
\gamma \delta \varepsilon }+\tfrac{1}{10}W_{5}B_{[\alpha \beta
}^{*}K_{\gamma \delta \varepsilon ]}^{*}C+\right. \\
&\quad \quad \quad \quad \quad \left. +2\left(
\frac{dW_{6}}{d\varphi }H_{\mu }^{*}A^{*\mu }-W_{6}\eta
^{*}\right) \mathcal{G}_{\alpha \beta \gamma \delta \varepsilon
}\right] .
\end{split}
\end{equation}

Replacing now the expression (\ref{fr6c}) into the equation (\ref{3.5}) for $%
i=2$ and recalling the same definitions like in the above, we
obtain that the piece of antighost number one in the first-order
deformation is written like
\begin{equation}\label{fr7c}
\begin{split}
a_{1}& =-\frac{dW_{1}}{d\varphi }H_{\mu }^{*}\left( 2A_{\nu
}C^{\mu \nu }-H^{\mu }\eta \right) +\frac{dW_{2}}{d\varphi }H_{\mu
}^{*}\left( 2B^{\mu
\nu }C_{\nu }-3\phi _{\nu \rho }\eta ^{\mu \nu \rho }\right) - \\
&\quad -2W_{1}B_{\mu \nu }^{*}C^{\mu \nu }-W_{2}\left( 3K_{\mu \nu
\rho }^{*}\eta
^{\mu \nu \rho }+2A^{*\mu }C_{\mu }\right) + \\
&\quad +\left[ \frac{dW_{3}}{d\varphi }H_{[\mu }^{*}A_{\nu }\phi
_{\rho \lambda ]}+W_{3}\left( 2B_{[\mu \nu }^{*}\phi _{\rho
\lambda ]}+3K_{[\mu \nu \rho }^{*}A_{\lambda ]}\right) \right]
\mathcal{G}^{\mu \nu \rho \lambda } +
\\
&\quad +3\left[ \frac{dW_{3}}{d\varphi }H_{\mu }^{*}\left( 2A_{\nu
}C_{\rho }-\phi _{\nu \rho }\eta \right) +W_{3}\left( 2B_{\mu \nu
}^{*}C_{\rho
}-K_{\mu \nu \rho }^{*}\eta \right) \right] K^{\mu \nu \rho } - \\
&\quad -3W_{3}\phi _{\mu \nu }^{*}\left( A^{[\mu }C^{\nu ]}-\phi
^{\mu \nu }\eta \right) +\\
&\quad +\epsilon ^{\alpha \beta \gamma \delta \varepsilon }\left[
6\left( \frac{dW_{4}}{d\varphi }H_{[\mu }^{*}A_{\nu
]}+2W_{4}B_{\mu \nu }^{*}\right) K^{\mu \nu \rho }\sigma _{\rho
\alpha }\mathcal{G}_{\beta \gamma \delta \varepsilon }\right. + \\
&\quad \quad \quad \quad \quad +\tfrac{1}{4}\frac{%
dW_{4}}{d\varphi }H_{\alpha }^{*}\epsilon _{\beta \gamma \mu \nu
\rho }K^{\mu \nu \rho }\epsilon _{\delta \varepsilon \mu ^{\prime
}\nu ^{\prime }\rho ^{\prime }}K^{\mu ^{\prime }\nu ^{\prime }\rho
^{\prime }}+
\\
&\quad \quad \quad \quad \quad +6W_{4}\left( \phi _{\alpha \beta
}^{*}K_{\gamma \delta \varepsilon }\eta -2\phi ^{*\mu \nu }A_{\mu
}\sigma _{\nu \alpha }\mathcal{G}_{\beta \gamma
\delta \varepsilon }\right) -\\
&\quad \quad \quad \quad \quad -\tfrac{1}{5!}\left( 2\frac{dW_{5}}{d\varphi }%
H_{[\alpha }^{*}A_{\beta }\phi _{\gamma \delta }C_{\varepsilon ]}
+4W_{5}B_{[\alpha \beta }^{*}\phi _{\gamma \delta }C_{\varepsilon
]}+6W_{5}K_{[\alpha \beta \gamma }^{*}A_{\delta }C_{\varepsilon ]}\right) +\\
&\quad \quad \quad \quad \quad +\tfrac{1}{5!}\left(
2\frac{dW_{5}}{d\varphi }H_{[\alpha }^{*}\phi _{\beta \gamma }\phi
_{\delta \varepsilon ]}+6W_{5}K_{[\alpha \beta \gamma }^{*}\phi
_{\delta \varepsilon ]}\right) \eta + \\
&\quad \quad \quad \quad \quad +2\left( \frac{dW_{6}}{d\varphi
}H_{\mu }^{*}B^{\mu \nu }-W_{6}A^{*\nu }\right) \sigma _{\nu
\alpha }\mathcal{G}_{\beta \gamma
\delta \varepsilon } - \\
&\quad \quad \quad \quad \quad \left. -3\left(
\frac{dW_{6}}{d\varphi }H^{*\mu }K_{\mu \alpha \beta }-W_{6}\phi
_{\alpha \beta }^{*}\right) \eta _{\gamma \delta \varepsilon
}\right] -W_{1}\varphi ^{*}\eta .
\end{split}
\end{equation}

In the last step we solve the equation (\ref{3.5}) for $i=1$ with
the help of the relation (\ref{fr7c}) and the definitions of
$\gamma $ acting on the BRST generators, whose solution reads as
\begin{equation}\label{fr8}
\begin{split}
a_{0}& =W_{1}A_{\mu }H^{\mu }+W_{2}B_{\mu \nu }\phi ^{\mu \nu
}-W_{3}\phi
_{[\mu \nu }A_{\rho ]}K^{\mu \nu \rho } + \\
&\quad +\epsilon ^{\alpha \beta \gamma \delta \varepsilon }\left( \tfrac{1}{4}%
W_{4}A_{\alpha }\epsilon _{\beta \gamma \mu \nu \rho }K^{\mu \nu
\rho }\epsilon _{\delta \varepsilon \mu ^{\prime }\nu ^{\prime
}\rho ^{\prime
}}K^{\mu ^{\prime }\nu ^{\prime }\rho ^{\prime }}+\right.  \\
&\quad \quad \quad \quad \quad \left. +\tfrac{1}{4}W_{5}A_{\alpha
}\phi _{\beta \gamma }\phi _{\delta \varepsilon }+W_{6}B_{\alpha
\beta }K_{\gamma \delta \varepsilon }\right) +\bar{M}(\varphi ),
\end{split}
\end{equation}
and represents nothing but the interacting Lagrangian at order one
in the coupling constant. The solution $\bar{M}(\varphi )$
represents the general solution to the `homogeneous' equation
\begin{equation}
\gamma \bar{a}_{0}=\partial _{\mu }j_{0}^{\mu },  \label{aze1in}
\end{equation}
which is determined in Appendix \ref{homog}. [The solutions to
this `homogeneous' equation come from $\bar{a}_{1}=0$, and hence
they bring contributions only to the deformed lagrangian density
at order one in the coupling constant. ]

We emphasize that the solutions $\left( a_{m}\right)
_{m=\overline{0,4}}$ obtained in the above also include the
solutions corresponding to the associated `homogeneous' equations
$\gamma \bar{a}_{m}=0$. In order to simplify the exposition we
avoided the discussion regarding the selection procedure of these
solutions such as to comply with obtaining some consistent
components of the first-order deformation at each value of the
antighost number. It is however interesting to note that this
procedure
allows no new functions of the scalar fields beside $\left( W_{k}\right) _{k=%
\overline{1,6}}$ and $\bar{M}\left( \varphi \right) $ to enter
$\left( a_{m}\right) _{m=\overline{0,4}}$. In consequence, we
succeeded in finding the complete form of the nonintegrated
density of the first-order deformation of the solution to the
master equation for the model under study (\ref{fr1}), which
reduces to a sum of terms with antighost numbers ranging from zero
to five, namely, the right-hand sides of the formulas
(\ref{fr3})--(\ref{fr8}), and is parametrized in terms of seven
arbitrary, smooth functions of the undifferentiated scalar field
$\varphi $.

\subsection{Higher-order deformations\label{highord}}

Next, we investigate the equations responsible for higher-order
deformations. The second-order deformation is governed by the
equation (\ref {bfa2.6}). Making use of the first-order
deformation derived in the previous subsection, after some
computation we organize the second term in the left-hand side of
(\ref{bfa2.6}) like
\begin{equation}
\left( S_{1},S_{1}\right) =\int d^{5}x\left( Y^{\left( 0 \right)
}X^{\left( 0 \right) }+
\sum\limits_{a=0}^{5}\sum\limits_{i=1}^{8}\frac{d^{a}Y^{\left( i\right) }}{%
d\varphi ^{a}}X_{a}^{\left( i\right) }\right) ,  \label{pt13}
\end{equation}
where
\begin{gather}
Y^{\left( 0 \right) }\left( \varphi \right) =\frac{d\bar{M}\left(
\varphi \right) }{d\varphi }W_{1}\left( \varphi \right) ,
\label{pt14a} \\
Y^{\left( 1\right) }\left( \varphi \right) =W_{1}\left( \varphi
\right)
W_{2}\left( \varphi \right) ,  \label{pt14} \\
Y^{\left( 2\right) }\left( \varphi \right) =W_{1}\left( \varphi
\right) \frac{dW_{2}\left( \varphi \right) }{d\varphi
}-3W_{2}\left( \varphi \right) W_{3}\left( \varphi \right)
+6W_{5}\left( \varphi \right) W_{6}\left(
\varphi \right) ,  \label{pt15} \\
Y^{\left( 3\right) }\left( \varphi \right) =W_{2}\left( \varphi
\right) W_{3}\left( \varphi \right) +W_{5}\left( \varphi \right)
W_{6}\left( \varphi
\right) ,  \label{pt16} \\
Y^{\left( 4\right) }\left( \varphi \right) =W_{1}\left( \varphi
\right) \frac{dW_{6}\left( \varphi \right) }{d\varphi
}+3W_{3}\left( \varphi \right) W_{6}\left( \varphi \right)
-6W_{2}\left( \varphi \right) W_{4}\left(
\varphi \right) ,  \label{pt17} \\
Y^{\left( 5\right) }\left( \varphi \right) =W_{1}\left( \varphi
\right)
W_{6}\left( \varphi \right) ,  \label{pt18} \\
Y^{\left( 6\right) }\left( \varphi \right) =W_{2}\left( \varphi
\right) W_{4}\left( \varphi \right) +W_{3}\left( \varphi \right)
W_{6}\left( \varphi
\right) ,  \label{pt19} \\
Y^{\left( 7\right) }\left( \varphi \right) =W_{2}\left( \varphi
\right)
W_{5}\left( \varphi \right) ,  \label{pt19'} \\
Y^{\left( 8\right) }\left( \varphi \right) =W_{4}\left( \varphi
\right) W_{6}\left( \varphi \right) ,  \label{pt20}
\end{gather}
while the remaining objects, namely, $X^{\left( 0 \right) } $ and
$\left( X_{a}^{\left( i\right) }\right)
_{a=\overline{1,5},\;i=\overline{1,8}}$ can be found in the
Appendix \ref{A}. On the one hand, $X^{\left( 0 \right) } $ and
all $\left( X_{a}^{\left( i\right) }\right)
_{a=\overline{1,5},\;i=\overline{1,8}}$ are polynomials of ghost
number one involving only the \emph{undifferentiated}
fields/ghosts and antifields. On the other hand, the
equation (\ref{bfa2.6}) requires that $\left( S_{1},S_{1}\right) $ is $s$%
-exact. However, since none of the terms present in the right-hand side of (%
\ref{pt13}) can be brought to such a form, the nonintegrated density of $%
\left( S_{1},S_{1}\right) $ must vanish. This takes place if and
only if the following equations are simultaneously obeyed
\begin{equation}
Y^{\left( k\right) }\left( \varphi \right) =0,\;k=\overline{0,8}.
\label{eqn}
\end{equation}
Using the above results, we can further take $S_{2}=0$, the
remaining higher-order deformation equations being satisfied with
the choice
\begin{equation}
S_{k}=0,\;k>2.  \label{sup}
\end{equation}

In this way the complete deformation of the solution to the master
equation, consistent to all orders in the coupling constant,
simply reduces to the sum between the `free' solution (\ref{f14})
and the first-order deformation
\begin{equation}
S=\bar{S}+gS_{1}=\bar{S}+g\int d^{5}x\left(
\sum\limits_{k=0}^{5}a_{k}\right) ,  \label{soldef}
\end{equation}
where the components $\left( a_{k}\right) _{k=\overline{0,5}}$ are given in (%
\ref{fr3})--(\ref{fr8}) and the functions $\bar{M}\left( \varphi
\right) $ and $\left( W_{k}\left( \varphi \right) \right)
_{k=\overline{1,6}}$ are no longer arbitrary; they must satisfy
the equations (\ref{eqn}).

\section{Lagrangian formulation of the deformed gauge theory\label{lagfor}}

By virtue of the discussion from the end of Sec. \ref{free} on the
significance of terms with various antighost numbers in the
solution to the master equation, at this stage we can extract all
the information on the
gauge structure of the coupled model. The antifield-independent piece in (%
\ref{soldef}) provides the expression of the overall Lagrangian
action of the interacting gauge theory
\begin{equation}\label{actlgrdef}
\begin{split}
\tilde{S}\left[ \varphi ,H,A,B,\phi ,K\right] & =\int d^{5}x\left[
H_{\mu }\left(
\partial ^{\mu
}\varphi +gW_{1}A^{\mu }\right) +g\bar{M}\left( \varphi
\right) +\right. \\
&\quad \quad \quad \quad +\tfrac{1}{2}B^{\mu \nu }\left(
\partial _{[\mu }A_{\nu
]}+2gW_{2}\phi _{\mu \nu }\right) +\\
&\quad \quad \quad \quad
+\tfrac{1}{3}K^{\mu \nu \rho }\left(
\partial _{[\mu
}\phi _{\nu \rho ]}-3gW_{3}\phi _{[\mu \nu }A_{\rho ]}\right) + \\
&\quad \quad \quad \quad +g\varepsilon ^{\mu \nu \rho \lambda
\sigma }\left( \tfrac{1}{4} W_{4}A_{\mu }\varepsilon _{\nu \rho
\alpha \beta \gamma }K^{\alpha \beta \gamma }\varepsilon _{\lambda
\sigma \alpha ^{\prime }\beta ^{\prime }\gamma ^{\prime
}}K^{\alpha ^{\prime }\beta ^{\prime }\gamma ^{\prime }}+\right.
\\
&\quad \quad \quad \quad \quad \quad \quad \quad \quad \left.
\left. +\tfrac{1}{4}W_{5}A_{\mu }\phi _{\nu \rho }\phi _{\lambda
\sigma }+W_{6}B_{\mu \nu }K_{\rho \lambda \sigma }\right) \right]
,
\end{split}
\end{equation}
while from the components with antighost number one we conclude
that it is invariant under the gauge transformations
\begin{equation}
\bar{\delta}_{\epsilon ,\xi }A^{\mu }=\partial ^{\mu }\epsilon
-2gW_{2}\xi ^{\mu }-2gW_{6}\varepsilon ^{\mu \nu \rho \lambda
\sigma }\xi _{\nu \rho \lambda \sigma },  \label{i2}
\end{equation}
\begin{equation}\label{i3}
\begin{split}
\bar{\delta}_{\epsilon ,\xi }H^{\mu }& =2D_{\nu }\epsilon ^{\mu
\nu }+g\left( \frac{dW_{1}}{d\varphi }H^{\mu
}-3\frac{dW_{3}}{d\varphi }K^{\mu
\nu \rho }\phi _{\nu \rho }\right) \epsilon - \\
&\quad -3g\frac{dW_{2}}{d\varphi }\phi _{\nu \rho }\epsilon ^{\mu
\nu \rho
}+2g\left( \frac{dW_{2}}{d\varphi }B^{\mu \nu }-3\frac{dW_{3}}{d\varphi }%
K^{\mu \nu \rho }A_{\rho }\right) \xi _{\nu } + \\
&\quad +12g\frac{dW_{3}}{d\varphi }A_{\nu }\phi _{\rho \lambda
}\xi ^{\mu \nu \rho \lambda }+2g\frac{dW_{6}}{d\varphi }B^{\mu \nu
}\varepsilon _{\nu \alpha \beta \gamma \delta }\xi ^{\alpha \beta
\gamma \delta } +\\
&\quad +3gK^{\mu \nu \rho }\left( 4\frac{dW_{4}}{d\varphi }A_{\nu
}\varepsilon _{\rho \alpha \beta \gamma \delta }\xi ^{\alpha \beta
\gamma \delta }-\frac{dW_{6}}{d\varphi }\varepsilon _{\nu \rho
\alpha \beta \gamma }\epsilon ^{\alpha \beta \gamma }\right) +
 \\
&\quad +g\varepsilon ^{\mu \nu \rho \lambda \sigma }\left[
\tfrac{1}{4}\frac{dW_{4}}{d\varphi }\varepsilon _{\nu \rho \alpha
\beta \gamma }K^{\alpha \beta \gamma }\varepsilon _{\lambda \sigma
\alpha ^{\prime }\beta ^{\prime }\gamma ^{\prime }}K^{\alpha
^{\prime }\beta ^{\prime }\gamma ^{\prime
}}\epsilon - \right.  \\
&\quad \quad \quad \quad \quad \quad \left.
-\frac{dW_{5}}{d\varphi }\phi _{\nu \rho }\left( A_{\lambda }\xi
_{\sigma }-\tfrac{1}{4}\phi _{\lambda \sigma }\epsilon \right)
\right] ,
\end{split}
\end{equation}
\begin{equation}
\bar{\delta}_{\epsilon ,\xi }\varphi =-gW_{1}\epsilon ,
\label{i4}
\end{equation}
\begin{equation}\label{i5}
\begin{split}
\bar{\delta}_{\epsilon ,\xi }B^{\mu \nu }& =-3\partial _{\rho
}\epsilon ^{\mu \nu \rho }-2gW_{1}\epsilon ^{\mu \nu
}+6gW_{3}\left( 2\phi _{\rho \lambda }\xi ^{\mu \nu \rho \lambda
}+K^{\mu \nu \rho }\xi _{\rho }\right) +
\\
&\quad +g\left( 12W_{4}K^{\mu \nu \rho }\varepsilon _{\rho \alpha
\beta \gamma \delta }\xi ^{\alpha \beta \gamma \delta
}-W_{5}\varepsilon ^{\mu \nu \rho \lambda \sigma }\phi _{\rho
\lambda }\xi _{\sigma }\right) ,
\end{split}
\end{equation}
\begin{equation}\label{i6}
\begin{split}
\bar{\delta}_{\epsilon ,\xi }\phi _{\mu \nu }& =D_{[\mu }^{\left(
-\right) }\xi _{\nu ]}+3g\left( W_{3}\phi _{\mu \nu }\epsilon
-2W_{4}A_{[\mu }\varepsilon _{\nu ]\alpha \beta \gamma \delta
}\xi ^{\alpha \beta \gamma \delta }\right) +\\
&\quad +3g\varepsilon _{\mu \nu \rho \lambda \sigma }\left(
2W_{4}K^{\rho \lambda \sigma }\epsilon +W_{6}\epsilon ^{\rho
\lambda \sigma }\right) ,
\end{split}
\end{equation}
\begin{equation}\label{i7}
\begin{split}
\bar{\delta}_{\epsilon ,\xi }K^{\mu \nu \rho }& =4D_{\lambda
}^{\left( +\right) }\xi ^{\mu \nu \rho \lambda }-3g\left(
W_{2}\epsilon ^{\mu \nu \rho
}+W_{3}K^{\mu \nu \rho }\epsilon \right)  -\\
&\quad -g\varepsilon ^{\mu \nu \rho \lambda \sigma }W_{5}\left(
A_{\lambda }\xi _{\sigma }-\tfrac{1}{2}\phi _{\lambda \sigma
}\epsilon \right) ,
\end{split}
\end{equation}
where we employed the notations
\begin{equation}
D_{\nu }=\partial _{\nu }-g\frac{dW_{1}}{d\varphi }A_{\nu },\quad
D_{\nu }^{\left( \pm \right) }=\partial _{\nu }\pm 3gW_{3}A_{\nu
}.  \label{i8}
\end{equation}
The commutators among the deformed gauge transformations, as well
as the accompanying reducibility relations, result from the
analysis of the structure of terms with antighost numbers greater
than one in (\ref{soldef})
and are listed in the Appendices \ref{defgauge} and respectively \ref{defred}%
.

However, the functions $\left( W_{k}\right) _{k=\overline{1,6}}$
and $\bar{M}\left( \varphi \right) $ are no longer arbitrary
smooth functions of the undifferentiated scalar field. They are
required to fulfill the equations (\ref{eqn}) in order to ensure
the consistency of the deformed solution to the master equation to
all orders in the coupling constant. Let us analyze now the
solutions to the system (\ref {eqn}). It it easy to see that
(\ref{eqn}) is equivalent with the equations
\begin{align}
\frac{d\bar{M}\left( \varphi \right) }{d\varphi }W_{1}\left(
\varphi \right) & =0, \label{s1a} \\
W_{1}\left( \varphi \right) W_{2}\left( \varphi \right) & =0,  \label{s1} \\
W_{1}\left( \varphi \right) \frac{dW_{2}\left( \varphi \right) }{d\varphi }%
-9W_{2}\left( \varphi \right) W_{3}\left( \varphi \right) & =0,
\label{s2}
\\
W_{2}\left( \varphi \right) W_{3}\left( \varphi \right)
+W_{5}\left( \varphi
\right) W_{6}\left( \varphi \right) & =0,  \label{s3} \\
W_{1}\left( \varphi \right) \frac{dW_{6}\left( \varphi \right) }{d\varphi }%
+9W_{3}\left( \varphi \right) W_{6}\left( \varphi \right) & =0,
\label{s4}
\\
W_{1}\left( \varphi \right) W_{6}\left( \varphi \right) & =0,  \label{s5} \\
W_{2}\left( \varphi \right) W_{4}\left( \varphi \right)
+W_{3}\left( \varphi
\right) W_{6}\left( \varphi \right) & =0,  \label{s6} \\
W_{2}\left( \varphi \right) W_{5}\left( \varphi \right) & =0,  \label{s7} \\
W_{4}\left( \varphi \right) W_{6}\left( \varphi \right) & =0.
\label{s8}
\end{align}
There are three different types of solutions to
(\ref{s1a})--(\ref{s8}). The first type is described by the choice
\begin{equation}
W_{1}\left( \varphi \right) =W_{3}\left( \varphi \right)
=W_{4}\left( \varphi \right) =W_{5}\left( \varphi \right) =0,
\label{sol1}
\end{equation}
with $\bar{M}\left( \varphi \right) $, $W_{2}\left( \varphi
\right) $, and $W_{6}\left( \varphi \right) $ arbitrary smooth
functions of the undifferentiated scalar field. The second kind of
solutions is pictured by the pick
\begin{equation}
\bar{M}\left( \varphi \right) =W_{2}\left( \varphi \right)
=W_{6}\left( \varphi \right) =0, \label{sol2}
\end{equation}
with $W_{1}\left( \varphi \right) $, $W_{3}\left( \varphi \right) $, $%
W_{4}\left( \varphi \right) $, and $W_{5}\left( \varphi \right) $
arbitrary smooth functions of $\varphi $. Finally, the third sort
of solutions is parametrized by
\begin{equation}
W_{1}\left( \varphi \right) =W_{2}\left( \varphi \right)
=W_{6}\left( \varphi \right) =0, \label{sol3}
\end{equation}
while $\bar{M}\left( \varphi \right) $, $W_{3}\left( \varphi
\right) $, $W_{4}\left( \varphi \right) $, and $W_{5}\left(
\varphi \right) $ remain arbitrary smooth functions of the
undifferentiated scalar field. If we particularize the general
results on the Lagrangian formulation of the interacting BF model
contained in this section, and also the formulas related to the
commutators among the deformed gauge transformations and to the
accompanying reducibility relations contained in the Appendices
\ref{defgauge} and \ref{defred} to the above solutions, we obtain
three interacting theories that are in a way complementary to each
other. More precisely, the first solution produces a deformed
interacting BF theory with an open gauge algebra that closes
on-shell in an Abelian way, but on-shell first- and second-order
reducibility relations. The second one leads to a coupled
topological BF model displaying an open gauge algebra, which
closes on-shell in a non-Abelian manner, and on-shell reducibility
relations for all the three levels. The last case yields an
interacting BF model with an open gauge algebra, on-shell
first-order reducibility relations, but off-shell second- and
third-order reducibility relations (the second- and third-order
reducibility functions are not modified by the deformation
procedure).

\section{Conclusion\label{conc}}

In conclusion, in this paper we have generated the consistent
Lagrangian interactions in five spacetime dimensions that can be
introduced among one scalar field, two types of one-forms, two
sorts of two-forms, and one three-form, pictured in the free limit
by an Abelian topological field theory of BF-type, with Abelian
gauge transformations, which are off-shell, third-order reducible.
Our treatment is mainly based on the Lagrangian BRST deformation
procedure, that relies on the construction of the consistent
deformations of the solution to the master equation with the help
of some cohomological techniques. The couplings are obtained under
the hypotheses of smoothness, locality, (background) Lorentz
invariance, Poincar\'{e} invariance, and the preservation of the
number of derivatives on each field. As a result, we obtain three
sorts of coupled models that are in a way complementary to each
other. All of them underlies a deformed interacting BF theory with
an open gauge algebra, which only closes on-shell, where on-shell
means on the stationary surface of field equations for the coupled
model. However, for the first situation it closes according to an
Abelian algebra, while for the second and third model it produces
on-shell a non-Abelian algebra. Related to the reducibility
relations, we remark that the first model outputs on-shell first-
and second-order reducibility relations, but off-shell third-order
reducibility, the second describes a coupled topological BF model
displaying on-shell reducibility relations to all the three
levels, while the third coupled theory exhibits on-shell
first-order reducibility relations, but off-shell second- and
third-order redundancy relations.

\section*{Acknowledgment}

The authors are partially supported by the European Commission FP6
program MRTN-CT-2004-005104. E.M.C. also acknowledges partial
support from the grant AT35/2004 with the Romanian National
Council for Academic Scientific Research (C.N.C.S.I.S.) and the
Romanian Ministry of Education and Research (M.E.C.).

\appendix

\section{Cohomology of $\gamma $ and related matters\label{hgama}}

In this section we study the first ingredient implied in the local
BRST cohomology $H\left( s\vert d\right) $, namely, the cohomology
algebra of the exterior longitudinal derivative $H\left( \gamma
\right) $. Let $a$ be an element of $H\left( \gamma \right) $ with
definite pure ghost number, antighost number and form degree
($\deg $)
\begin{equation}
\gamma a=0,\quad \text{pgh}\left( a\right) =l\geq 0,\quad
\text{agh}\left( a\right) =k\geq 0,\quad \deg \left( a\right)
=p\leq 5.  \label{e1}
\end{equation}
Extending the analysis realized in Sec. \ref{stand} for
nonintegrated densities (0-forms) to objects that may have
nonvanishing form degrees, we can state that the general, local
solution to the equation (\ref{e1}) (up to trivial, $\gamma
$-exact contributions) is of the type
\begin{equation}
a=\sum_{J}\alpha _{J}\left( \left[ \chi _{\Delta }^{*}\right]
,\left[ F_{A}\right] \right) e^{J}\left( \eta ^{\Upsilon }\right)
,  \label{e2}
\end{equation}
where $F_{A}$ denotes the set of $\gamma $-closed
(gauge-invariant) quantities that can be constructed out of the
original fields
\begin{equation}
F_{A}=\left\{ \varphi ,\partial ^{\left[ \mu \right. }A^{\left.
\nu \right] },\partial _{\mu }H^{\mu },\partial ^{[\mu
}\tilde{B}^{\nu \rho \lambda ]},\partial _{[\mu }\phi _{\nu \rho
]},\partial ^{[\mu }\tilde{K}^{\nu \rho ]}\right\} ,  \label{e3}
\end{equation}
and $\chi _{\Delta }^{*}$ is explained in (\ref{notat}). The notation $%
e^{J}\left( \eta ^{\Upsilon }\right) $ in (\ref{e2}) signifies
here the elements of pure ghost number equal to $l$ of a basis in
the space of polynomials in the undifferentiated ghosts
\begin{equation}
\eta ^{\Upsilon }=\left( \eta ,C,\stackrel{\sim }{\mathcal{G}},\tilde{\eta},%
\tilde{C}\right) ,  \label{notat2}
\end{equation}
so they have the properties
\begin{equation}
\text{pgh}\left( e^{J}\right) =l>0,\quad \text{agh}\left(
e^{J}\right) =0,\quad \deg \left( e^{J}\right) =0.  \label{e5}
\end{equation}
By contrast to Sec. \ref{stand}, here we work with slightly
modified quantities $F_{A}$ and $\eta ^{\Upsilon }$ instead of
(\ref{3.7}) and (\ref {notat1}), such as to include tilde
quantities, defined like the Hodge duals of the untilded ones
\begin{equation}
\stackrel{\sim }{\Psi }^{\mu _{1}\mu _{2}\ldots \mu _{k}}\mathcal{=}\tfrac{1%
}{\left( 5-k\right) !}\varepsilon ^{\mu _{1}\ldots \mu _{k}\nu
_{1}\ldots \nu _{5-k}}\Psi _{\nu _{1}\ldots \nu _{5-k}}.
\label{e4}
\end{equation}
In fact, (\ref{e2}) is nothing but the analogue of (\ref{3.8}) in
form degree $p$ and written in terms of the newly defined ghosts
and gauge-invariant quantities. Here, the objects $\alpha _{J}$
[obviously nontrivial in $H^{0}\left( \gamma \right) $] are
$p$-forms and were taken to have a finite antighost number and a
bounded number of derivatives, so they are local $p$-forms with
coefficients that are polynomials in the antifields $\chi _{\Delta
}^{*}$, in the quantities $F_{A}$ (excluding the undifferentiated
scalar field $\varphi $), and also in their spacetime derivatives
(including the derivatives of $\varphi $). However, $\alpha _{J}$
may contain infinite, formal series in the undifferentiated scalar field $%
\varphi $. Due to their $\gamma $-closeness, $\gamma \alpha
_{I}=0$, and to their (partial) polynomial character, $\alpha
_{J}$ will be called `invariant polynomials'. In agreement with
(\ref{e1}), they display the properties
\begin{equation}
\text{pgh}\left( \alpha _{J}\right) =0,\quad \text{agh}\left(
\alpha _{J}\right) =k\geq 0,\quad \deg \left( \alpha _{J}\right)
=p\leq 5. \label{e6}
\end{equation}
In antighost number zero the invariant polynomials are local
$p$-forms with coefficients that are polynomials in $F_{A}$
(excluding the undifferentiated scalar field $\varphi $) and also
in their spacetime derivatives (including the derivatives of
$\varphi $), with coefficients that may be infinite, formal series
in the undifferentiated scalar field $\varphi $.

In order to establish that just $H\left( \gamma \right) $ is
required at the
computation of $H\left( s\vert d\right) $, and not the local cohomology of $%
\gamma $, we need the cohomology of the exterior spacetime
differential $d$ in the space of invariant polynomials, as well as
other interesting properties, which are addressed below.

\begin{theorem}
\label{hginv}The cohomology of $d$ in form degree strictly less
than $5$ is trivial in the space of invariant polynomials with
strictly positive antighost number. This means that the conditions
\begin{equation}
\gamma \alpha =0,\quad d\alpha =0,\quad \text{agh}\left( \alpha
\right)
>0,\quad \deg (\alpha )<5,\quad \alpha =\alpha \left( \left[ \chi _{\Delta
}^{*}\right] ,\left[ F_{A}\right] \right) ,  \label{e7}
\end{equation}
imply
\begin{equation}
\alpha =d\beta ,  \label{e8}
\end{equation}
for some invariant polynomial $\beta \left( \left[ \chi _{\Delta
}^{*}\right] ,\left[ F_{A}\right] \right) $.
\end{theorem}

\emph{Proof} In order to prove the theorem, we decompose $d$ like
\begin{equation}
d=d_{0}+d_{1},  \label{e9}
\end{equation}
where $d_{1}$ acts solely on the antifields $\chi _{\Delta }^{*}$
and their derivatives, while $d_{0}$ acts exclusively on the
$\gamma $-invariant objects $F_{A}$ and on their derivatives
\begin{equation}
d_{0}=\partial _{\mu _{1}}^{0}dx^{\mu _{1}},\quad d_{1}=\partial
_{\mu _{1}}^{1}dx^{\mu _{1}},  \label{e10}
\end{equation}
with
\begin{align}
\partial _{\mu _{1}}^{0} & =F_{A,\mu _{1}}\frac{\partial }{\partial F_{A}}%
+F_{A,\mu _{1}\mu _{2}}\frac{\partial }{\partial F_{A,\mu
_{2}}}+\cdots ,
\label{e11} \\
\partial _{\mu _{1}}^{1} & =\chi _{\Delta ,\mu _{1}}^{*}\frac{\partial ^{L}}{%
\partial \chi _{\Delta }^{*}}+\chi _{\Delta ,\mu _{1}\mu _{2}}^{*}\frac{%
\partial ^{L}}{\partial \chi _{\Delta ,\mu _{2}}^{*}}+\cdots .  \label{e12}
\end{align}
We used the common convention $f_{,\mu _{1}}\equiv \partial
f/\partial x^{\mu _{1}}$. Obviously, $d^{2}=0$ on invariant
polynomials is equivalent with the nilpotency and anticommutation
of its components acting on invariant polynomials
\begin{equation}
d_{0}^{2}=0=d_{1}^{2},\quad d_{0}d_{1}+d_{1}d_{0}=0.  \label{e13}
\end{equation}
The action of $d_{0}$ on a given invariant polynomial with say $l$
derivatives of $F_{A}$ and $j$ derivatives of $\chi _{\Delta
}^{*}$ results in an invariant polynomial with $\left( l+1\right)
$ derivatives of $F_{A}$ and $j$ derivatives of $\chi _{\Delta
}^{*}$, while the action of $d_{1}$ on
the same object leads to an invariant polynomial with $l$ derivatives of $%
F_{A}$ and $\left( j+1\right) $ derivatives of $\chi _{\Delta
}^{*}$. In particular, $d_{0}$ gives zero when acting on an
invariant polynomial that does not involve any of the objects
$F_{A}$ or of their derivatives, and the same is valid with
respect to $d_{1}$ acting on an invariant polynomial that does not
depend on any of the antifields $\chi _{\Delta }^{*}$ or on their
derivatives. With the help of the relations
(\ref{e11})--(\ref{e12}) we observe that
\begin{equation}
\text{agh}\left( d_{0}\right) =\text{agh}\left( d_{1}\right) =\text{agh}%
\left( d\right) =0,  \label{e14}
\end{equation}
such that neither of them changes the antighost number of the
objects on which any of them acts.

For convenience, the antifields $\chi _{\Delta }^{*}$ will be
called ``foreground'' fields, and the $\gamma $-invariant objects
$F_{A}$ will be named ``background'' fields. So, $d_{0}$ acts just
on the background fields and their derivatives, while $d_{1}$ acts
solely on the foreground fields and their derivatives. According
to the proposition from page 363 in~\cite {dubplb}, we have that
the entire cohomology of $d_{1}$ in form degree strictly less than
$5$ is trivial in the space of invariant polynomials with strictly
positive antighost number. This means that
\begin{equation}
\alpha =\alpha \left( \left[ \chi _{\Delta }^{*}\right] ,\left[
F_{A}\right] \right) ,\quad \text{agh}\left( \alpha \right)
=k>0,\quad \deg \left( \alpha \right) =p<5,\quad d_{1}\alpha =0,
\label{e15}
\end{equation}
implies that
\begin{equation}
\alpha =d_{1}\beta ,  \label{e16}
\end{equation}
with
\begin{equation}
\beta =\beta \left( \left[ \chi _{\Delta }^{*}\right] ,\left[
F_{A}\right] \right) ,\quad \text{agh}\left( \beta \right)
=k>0,\quad \deg \left( \beta \right) =p-1.  \label{e17}
\end{equation}
In particular, we have that if an invariant polynomial (of form
degree $p<5$ and with strictly positive antighost number)
depending only on the undifferentiated antifields is
$d_{1}$-closed, then it vanishes
\begin{equation}
\left( \bar{\alpha}=\bar{\alpha}\left( \chi _{\Delta }^{*},\left[
F_{A}\right] \right) ,\quad \text{agh}\left( \bar{\alpha}\right)
>0,\quad \deg \left( \bar{\alpha}\right) =p<5,\quad d_{1}\bar{\alpha}%
=0\right)\quad \Rightarrow \quad \bar{\alpha}=0.  \label{e18}
\end{equation}
Just $d_{0}$ has nontrivial cohomology. For instance, any form
exclusively depending on the antifields and their derivatives is
$d_{0}$-closed, but it is clearly not $d_{0}$-exact.

Next, assume that $\alpha $ is a homogeneous form of degree $p<5$
and antighost number $k>0$ that satisfies the conditions
(\ref{e7}). We decompose $\alpha $ according to the number of
derivatives of the antifields
\begin{equation}
\alpha =\stackrel{(0)}{\alpha }+\stackrel{(1)}{\alpha }+\cdots +\stackrel{(s)%
}{\alpha },\quad \text{agh}\left( \stackrel{(i)}{\alpha }\right)
=k>0,\quad \deg \left( \stackrel{(i)}{\alpha }\right) =p<5,
\label{e19}
\end{equation}
where $\stackrel{(i)}{\alpha }$ signifies the component from $\alpha $ with $%
i$ derivatives of the antifields. (The decomposition contains a
finite number of terms since $\alpha $ is by assumption local.) As
$\alpha $ is an invariant polynomial of form degree $p<5$ and
strictly positive antighost number, each component $\left(
\stackrel{(i)}{\alpha }\right) _{0\leq i\leq s}$ is an invariant
polynomial with the same form degree and strictly positive
antighost number. The proof of the theorem is realized in $\left(
s+1\right) $ steps.

Step 1. Taking into account the splitting (\ref{e9}), the
projection of the equation
\begin{equation}
d\alpha =0  \label{e20}
\end{equation}
on the maximum number of derivatives of the antifields ($s+1$)
produces
\begin{equation}
d_{1}\stackrel{(s)}{\alpha }=0,  \label{e21}
\end{equation}
and hence the triviality of the cohomology of $d_{1}$ ensures that
\begin{equation}
\stackrel{(s)}{\alpha }=d_{1}\stackrel{(s-1)}{\beta },\quad
\text{agh}\left(
\stackrel{(s-1)}{\beta }\right) =k>0,\quad \deg \left( \stackrel{(s-1)}{%
\beta }\right) =p-1,  \label{e22}
\end{equation}
where $\stackrel{(s-1)}{\beta }$ is an invariant polynomial of form degree ($%
p-1$), with strictly positive antighost number and containing only
($s-1$) derivatives of the antifields. If we introduce the
$p$-form
\begin{equation}
\alpha _{1}=\alpha -d\stackrel{(s-1)}{\beta },  \label{e23}
\end{equation}
then the equation (\ref{e20}) together with the nilpotency of $d$
further yield
\begin{equation}
d\alpha _{1}=0.  \label{e24}
\end{equation}
It is by construction an invariant polynomial of form degree $p$
and of strictly positive antighost number and, most important, the
maximum number of derivatives of the antifields from $\alpha _{1}$
is equal to ($s-1$). Indeed, if we replace (\ref{e22}) in
(\ref{e19}) and then in (\ref{e23}), we get that
\begin{equation}
\alpha _{1}=\stackrel{(0)}{\alpha }+\stackrel{(1)}{\alpha }+\cdots +%
\stackrel{(s-2)}{\alpha }+\stackrel{(s-1)}{\alpha }-d_{0}\stackrel{(s-1)}{%
\beta }.  \label{e25}
\end{equation}
Then, the maximum number of derivatives of the antifields from the
first $s$
terms in the right-hand side of (\ref{e25}) is contained in $\stackrel{(s-1)%
}{\alpha }$, being equal to ($s-1$), while
$d_{0}\stackrel{(s-1)}{\beta }$
has the same number of derivatives of the antifields like $\stackrel{(s-1)}{%
\beta }$, which is again ($s-1$).

Step 2. If we project now the equation (\ref{e24}) on the maximum
number of derivatives of the antifields ($s$), we infer that
\begin{equation}
d_{1}\left( \stackrel{(s-1)}{\alpha }-d_{0}\stackrel{(s-1)}{\beta
}\right) =0,  \label{e26}
\end{equation}
with $\stackrel{(s-1)}{\alpha }-d_{0}\stackrel{(s-1)}{\beta }$ an
invariant polynomial of form degree $p$ and of strictly positive
antighost number. Using again the triviality of the cohomology of
$d_{1}$, we deduce that
\begin{equation}
\stackrel{(s-1)}{\alpha }-d_{0}\stackrel{(s-1)}{\beta }=d_{1}\stackrel{(s-2)%
}{\beta },\quad \text{agh}\left( \stackrel{(s-2)}{\beta }\right)
=k>0,\quad \deg \left( \stackrel{(s-2)}{\beta }\right) =p-1,
\label{e27}
\end{equation}
where $\stackrel{(s-2)}{\beta }$ is an invariant polynomial of form degree ($%
p-1$), with strictly positive antighost number and containing only
($s-2$) derivatives of the antifields. At this stage, we define
the $p$-form
\begin{equation}
\alpha _{2}=\alpha -d\left( \stackrel{(s-1)}{\beta }+\stackrel{(s-2)}{\beta }%
\right) .  \label{e28}
\end{equation}
The equation (\ref{e20}) together with the nilpotency of $d$
further yield
\begin{equation}
d\alpha _{2}=0.  \label{e29}
\end{equation}
Clearly, $\alpha _{2}$ is an invariant polynomial of form degree
$p$ and of strictly positive antighost number. It is essential to
remark that the maximum number of derivatives of the antifields
from $\alpha _{2}$ is equal to ($s-2$). This results by inserting
(\ref{e22}) and (\ref{e27}) in (\ref {e19}) and consequently in
(\ref{e28}), which then gives
\begin{equation}
\alpha _{2}=\stackrel{(0)}{\alpha }+\stackrel{(1)}{\alpha }+\cdots +%
\stackrel{(s-3)}{\alpha }+\stackrel{(s-2)}{\alpha }-d_{0}\stackrel{(s-2)}{%
\beta }.  \label{e30}
\end{equation}
etc.

Step $s$. Proceeding in the same manner, at the $s$-th step we
obtain an invariant polynomial of form degree $p$ and with
strictly positive antighost number, which contains only the
undifferentiated antifields
\begin{gather}
\alpha _{s}=\alpha -d\left( \stackrel{(s-1)}{\beta }+\cdots +\stackrel{(0)}{%
\beta }\right) =\stackrel{(0)}{\alpha }-d_{0}\stackrel{(0)}{\beta
},
\label{e31} \\
\text{agh}\left( \stackrel{(j)}{\beta }\right) =k>0,\quad \deg
\left( \stackrel{(j)}{\beta }\right) =p-1,\quad 0\leq j\leq s-1.
\label{e32}
\end{gather}
[All $\left( \stackrel{(j)}{\beta }\right) 0\leq j\leq s-1$ are
invariant polynomials.] The equation (\ref{e20}) and the
nilpotency of $d$ lead to the equation
\begin{equation}
d\alpha _{s}=0.  \label{e33}
\end{equation}

Step ($s+1$). The projection of (\ref{e33}) on the maximum number
of derivatives of the antifields (one) is
\begin{equation}
d_{1}\left( \stackrel{(0)}{\alpha }-d_{0}\stackrel{(0)}{\beta
}\right)
=0,\quad \text{agh}\left( \stackrel{(0)}{\alpha }-d_{0}\stackrel{(0)}{\beta }%
\right) =k>0.  \label{e34}
\end{equation}
Taking into account the relations (\ref{e34}) and (\ref{e18}) (with $\bar{%
\alpha}$ replaced by $\stackrel{(0)}{\alpha
}-d_{0}\stackrel{(0)}{\beta }$) we get that
\begin{equation}
\stackrel{(0)}{\alpha }-d_{0}\stackrel{(0)}{\beta }=0,
\label{e35}
\end{equation}
which substituted in (\ref{e31}) finally allows us to write that
\begin{equation}
\alpha =d\beta ,  \label{e36}
\end{equation}
with
\begin{equation}
\beta =\left( \stackrel{(s-1)}{\beta }+\cdots
+\stackrel{(0)}{\beta }\right) ,\quad \text{agh}\left( \beta
\right) =k>0,\quad \deg \left( \beta \right) =p-1,  \label{e37}
\end{equation}
and this proves the theorem since $\beta $ is an invariant
polynomial of
form degree ($p-1$) and with strictly positive antighost number. $%
\blacksquare $

In form degree $5$ the Theorem \ref{hginv} is replaced with: let
$\alpha =\rho dx^{0}\wedge \cdots \wedge dx^{4}$ be a $d$-exact
invariant polynomial
of form degree $5$ and of strictly positive antighost number, $\text{agh}%
\left( \alpha \right) =k>0$, $\deg \left( \alpha \right) =5$,
$\alpha =d\beta $. Then, one can take the $4$-form $\beta $ to be
an invariant
polynomial (of antighost number $k$). In dual notations, this means that if $%
\rho $ with $\text{agh}\left( \rho \right) =k>0$ is an invariant
polynomial whose Euler-Lagrange derivatives are all vanishing,
$\rho =\partial _{\mu }j^{\mu }$, then $j^{\mu }$ can be taken to
be also invariant. Theorem \ref {hginv} can be generalized as
follows.

\begin{theorem}
\label{hdinhg}The cohomology of $d$ computed in $H\left( \gamma
\right) $ is trivial in form degree strictly less than $5$ and in
strictly positive antighost number
\begin{equation}
H_{p}^{g,k}\left( d,H\left( \gamma \right) \right) =0,\quad
k>0,\quad p<5, \label{e38}
\end{equation}
where $p$ is the form degree, $k$ is the antighost number and $g$
is the ghost number.
\end{theorem}

\emph{Proof} An element $a$ from $H_{p}^{g,k}\left( d,H\left(
\gamma \right)
\right) $ is a $p$-form of definite ghost number $g$ and antighost number $k$%
, pertaining to the cohomology of $\gamma $, which is $d$-closed modulo $%
\gamma $%
\begin{equation}
\gamma a=0,\;da=\gamma \mu ,\quad \text{agh}\left( a\right) =k,\quad \text{gh%
}\left( a\right) =g,\quad \deg \left( a\right) =p.  \label{e39}
\end{equation}
The theorem states that if $a$ satisfies the conditions (\ref{e39}) with $%
p<5 $ and $k>0$, then $a$ is trivial in $H_{p}^{g,k}\left(
d,H\left( \gamma
\right) \right) $%
\begin{equation}
a=d\nu +\gamma \rho ,\quad \gamma \nu =0,  \label{e40}
\end{equation}
where
\begin{gather}
\text{agh}\left( \nu \right) =\text{agh}\left( \rho \right) =k>0,\quad \text{%
gh}\left( \nu \right) =g,\quad \text{gh}\left( \rho \right) =g-1,
\label{e41} \\
\deg \left( \nu \right) =p-1,\quad \deg \left( \rho \right) =p<5.
\label{e42}
\end{gather}
Since $g=l^{\prime }-k$, with $l^{\prime }$ the pure ghost number
of $a$, and $l^{\prime }$ takes positive values $l^{\prime }\geq
0$, it follows that $g$ is restricted to fulfill the condition
$g\geq -k$. Thus, if $g<-k$, then $a=0$. The theorem is thus
trivially obeyed for $g<-k$.

We consider a nontrivial element $a$ from $H\left( \gamma \right)
$ of form degree $p<5$, of antighost number $k>0$
\begin{equation}
\gamma a=0,\;\text{agh}\left( a\right) =k>0,\quad \deg \left(
a\right) =p<5. \label{e43}
\end{equation}
In agreement with the previous results, $a$ can be expressed, up to $\gamma $%
-exact contributions, like
\begin{equation}
a=\sum_{J}\alpha _{J}e^{J}.  \label{e44}
\end{equation}
We will use in extenso the following obvious properties
\begin{gather}
\gamma ^{2}=0,\quad d^{2}=0,\quad \gamma d+d\gamma =0,\quad
\text{pgh}\left(
d\right) =0,\quad \deg \left( \gamma \right) =0,  \label{e45} \\
\sum \alpha _{J}e^{J}=\gamma \left( \text{something}\right)
\Leftrightarrow
\left( \alpha _{J}=0\quad \text{for all}\quad J\right),  \label{e46} \\
d\alpha _{J}=\alpha _{J}^{\prime },  \label{e47}
\end{gather}
where
\begin{equation}
\text{agh}\left( \alpha _{J}^{\prime }\right) =\text{agh}\left(
\alpha _{J}\right) ,\quad \deg \left( \alpha _{J}^{\prime }\right)
=\deg \left( \alpha _{J}\right) +1.  \label{e48}
\end{equation}
By applying the exterior spacetime differential on $a$ of the form (\ref{e44}%
), we infer that
\begin{equation}
da=\pm \sum_{J}\left( d\alpha _{J}\right) e^{J}+\sum_{J}\alpha
_{J}\left( de^{J}\right) .  \label{e49}
\end{equation}
By means of the relations (\ref{f13b})--(\ref{f13f}), we get
\begin{gather}
d\eta =\partial _{\mu }\eta dx^{\mu }=\gamma \left( -A_{\mu
}dx^{\mu }\right) ,\quad dC=\partial _{\mu }Cdx^{\mu }=\gamma
\left( -C_{\mu }dx^{\mu
}\right) ,  \label{e50a} \\
d\stackrel{\sim }{\mathcal{G}}=\partial _{\mu }\stackrel{\sim }{\mathcal{G}}%
dx^{\mu }=\gamma \left( \tfrac{1}{5}\stackrel{\sim
}{\mathcal{G}}_{\mu
}dx^{\mu }\right) ,  \label{e50b} \\
d\tilde{\eta}=\partial _{\mu }\tilde{\eta}dx^{\mu }=\gamma \left( \tfrac{1}{5%
}\tilde{\eta}_{\mu }dx^{\mu }\right) ,\quad d\tilde{C}=\partial _{\mu }%
\tilde{C}dx^{\mu }=\gamma \left( \tfrac{1}{5}\tilde{C}_{\mu
}dx^{\mu }\right) ,  \label{e50c}
\end{gather}
which allow us to write
\begin{equation}
de^{J}=\gamma \hat{e}^{J},  \label{e52}
\end{equation}
where $\hat{e}^{J}$ depend in general on $A_{\mu }$, $C_{\mu }$, $\stackrel{%
\sim }{\mathcal{G}}_{\mu }$, $\tilde{\eta}_{\mu }$, and
$\tilde{C}_{\mu }$. Substituting (\ref{e52}) in (\ref{e49}), it
follows that
\begin{equation}
da=\pm \sum_{J}\left( d\alpha _{J}\right) e^{J}+\gamma \left(
\sum_{J}\alpha _{J}\hat{e}^{J}\right) .  \label{e53}
\end{equation}
Since $a$ is a $d$-closed modulo $\gamma $ $p$-form, the equations (\ref{e39}%
) and (\ref{e53}) yield
\begin{equation}
\pm \sum_{J}\left( d\alpha _{J}\right) e^{J}=\gamma \mu ^{\prime
}. \label{e54}
\end{equation}
With the help of the property (\ref{e46}), from (\ref{e54}) we
arrive to
\begin{equation}
d\alpha _{J}=0.  \label{e55}
\end{equation}
Theorem \ref{hginv} then implies that
\begin{equation}
\alpha _{J}=d\beta _{J},  \label{e56}
\end{equation}
with $\beta _{J}$ an invariant polynomial. Inserting $\alpha _{J}$
of the form (\ref{e56}) in (\ref{e44}), we obtain that
\begin{equation}
a=\sum_{J}d\beta _{J}e^{J}=\pm d\left( \sum_{J}\beta
_{J}e^{J}\right) \mp \gamma \left( \sum_{J}\beta
_{J}\hat{e}^{J}\right) ,  \label{e57}
\end{equation}
which proves the theorem. $\blacksquare $

Theorem \ref{hdinhg} is one of the main tools needed for the computation of $%
H\left( s\vert d\right) $. In particular, it implies that there is
no nontrivial descent for $H\left( \gamma \vert d\right) $ in
strictly positive antighost number.

\begin{corollary}
\label{gaplusdb}If $a$ with
\begin{equation}
\text{agh}\left( a\right) =k>0,\quad \text{gh}\left( a\right)
=g\geq -k,\quad \deg \left( a\right) =p\leq 5,  \label{e58}
\end{equation}
satisfies the equation
\begin{equation}
\gamma a+db=0,  \label{e59}
\end{equation}
where
\begin{equation}
\text{agh}\left( b\right) =k>0,\quad \text{gh}\left( b\right)
=g+1>-k,\quad \deg \left( b\right) =p-1<5,  \label{e60}
\end{equation}
then one can always redefine $a$%
\begin{equation}
a\rightarrow a^{\prime }=a+d\nu ,  \label{e61}
\end{equation}
so that
\begin{equation}
\gamma a^{\prime }=0.  \label{e62}
\end{equation}
\end{corollary}

\emph{Proof} We construct the descent associated with the equation (\ref{e59}%
). Acting with $\gamma $ on (\ref{e59}) and using the first and
the third relations in (\ref{e45}), we find that
\begin{equation}
d\left( -\gamma b\right) =0,  \label{e63}
\end{equation}
such that the triviality of the cohomology of $d$ implies that
\begin{equation}
\gamma b+dc=0,  \label{e64}
\end{equation}
where
\begin{equation}
\text{agh}\left( c\right) =k>0,\quad \text{gh}\left( c\right)
=g+2,\quad \deg \left( c\right) =p-2.  \label{e65}
\end{equation}
Going on in the same way, we get the next equation from the
descent
\begin{equation}
\gamma c+de=0,  \label{e66}
\end{equation}
with
\begin{equation}
\text{agh}\left( e\right) =k>0,\quad \text{gh}\left( e\right)
=g+3,\quad \deg \left( e\right) =p-3,  \label{e67}
\end{equation}
and so on. The descent stops after a finite number of steps with
the last equations
\begin{align}
\gamma t+du & =0,  \label{e68} \\
\gamma u+dv & =0,  \label{e69} \\
\gamma v & =0,  \label{e70}
\end{align}
either because $v$ is a zero-form or because we stopped at a
higher form-degree with a $\gamma $-closed term. It is essential
to remark that irrespective of the step at which the descent is
cut, we have that
\begin{equation}
\text{agh}\left( v\right) =k>0,\quad \text{gh}\left( v\right)
=g^{\prime }>-k,\quad \deg \left( v\right) =p^{\prime }<5.
\label{e71}
\end{equation}
[The earliest step where the descent may terminate is $v=b$ and,
according
to (\ref{e60}), we have that $\deg \left( b\right) =p-1<5$ and $\text{gh}%
\left( b\right) =g+1>-k$.]

The equations (\ref{e69})--(\ref{e70}) together with the
conditions (\ref {e71}) tell us that $v$ belongs to $H_{p^{\prime
}}^{g^{\prime },k}\left( d,H\left( \gamma \right) \right) $ for
$k>0$, $p^{\prime }<5$ and $g^{\prime
}>-k$, so Theorem \ref{hdinhg} guarantees that $v$ is trivial in $%
H_{p^{\prime }}^{g^{\prime },k}\left( d,H\left( \gamma \right) \right) $%
\begin{equation}
v=d\nu ^{\prime }+\gamma \rho ^{\prime },\quad \gamma \nu ^{\prime
}=0, \label{e72}
\end{equation}
which substituted in (\ref{e69}) allows us, due to the
anticommutation between $d$ and $\gamma $, to replace it with the
equivalent equation
\begin{equation}
\gamma u^{\prime }=0,  \label{e73}
\end{equation}
where
\begin{equation}
u^{\prime }=u-d\rho ^{\prime }.  \label{e74}
\end{equation}
In the meantime, (\ref{e74}) and the nilpotency of $d$ induces that $%
du^{\prime }=du$, such that the equation (\ref{e68}) becomes
\begin{equation}
\gamma t+du^{\prime }=0.  \label{e75}
\end{equation}
[Note that if the descent stops in form degree zero, $\deg \left(
v\right) =0 $, then the proof remains valid with the sole
modification $\nu ^{\prime
}=0$ in (\ref{e72}).] Reprising the same argument in relation with (\ref{e73}%
) and the last equation, we find that (\ref{e75}) can be replaced
with
\begin{equation}
\gamma t^{\prime }=0,  \label{e76}
\end{equation}
where
\begin{equation}
t^{\prime }=t-d\rho ^{\prime \prime },  \label{e77}
\end{equation}
and $\rho ^{\prime \prime }$ comes from
\begin{equation}
u^{\prime }=d\nu ^{\prime \prime }+\gamma \rho ^{\prime \prime
},\quad \gamma \nu ^{\prime \prime }=0.  \label{e78}
\end{equation}
Performing exactly the same operations for the remaining equations
from the descent, we finally infer that (\ref{e59}) is equivalent
with
\begin{equation}
\gamma a^{\prime }=0,  \label{e79}
\end{equation}
where
\begin{equation}
a^{\prime }=a-d\rho ^{\prime \prime \prime },  \label{e80}
\end{equation}
and $\rho ^{\prime \prime \prime }$ appears in
\begin{equation}
b^{\prime }=d\nu ^{\prime \prime \prime }+\gamma \rho ^{\prime
\prime \prime },\quad \gamma \nu ^{\prime \prime \prime }=0.
\label{e81}
\end{equation}
The corollary is now demonstrated once we perform the
identification
\begin{equation}
\nu =-\rho ^{\prime \prime \prime },  \label{e82}
\end{equation}
between (\ref{e80}) and (\ref{e61}). Meanwhile, it is worth noticing that $%
b^{\prime }=b-dg$, with $\gamma g$ nonvanishing in general, so
from (\ref {e81}) we can also state that
\begin{equation}
b=\gamma \rho ^{\prime \prime \prime }+df,\quad f=\nu ^{\prime
\prime \prime }+g,  \label{e83}
\end{equation}
with $\gamma f\neq 0$ in general. $\blacksquare $

\section{Some results on the (invariant) characteristic cohomology \label%
{charcoh}}

We have argued in Sec. \ref{stand} that the characteristic
cohomology for the model under study is trivial in antighost
numbers strictly greater than five, $H_{k}\left( \delta \vert
d\right) =0$ for all $k>5$. It appears the natural question if
this result is still valid in the space of invariant polynomials,
or, in other words, at the level of the invariant characteristic
cohomology $H^{\text{inv}}\left( \delta \vert d\right) $. The
answer is affirmative and is proved below, in Theorem
\ref{hinvdelta}. Actually, we prove that if $\alpha _{k}$ is
trivial in $H_{k}\left( \delta
\vert d\right) $, then it can be taken to be trivial also in $H_{k}^{\text{inv}%
}\left( \delta \vert d\right) $. We consider only the case $k\geq
5$ since our main scope is to argue the triviality of
$H^{\text{inv}}\left( \delta \vert d\right) $ in antighost number
strictly greater than five. To this end, we firstly need the
following lemma.

\begin{lemma}
\label{trivinv}Let $\alpha $ be a $\delta $-exact invariant
polynomial
\begin{equation}
\alpha =\delta \beta .  \label{A72}
\end{equation}
Then, $\beta $ can also be taken to be an invariant polynomial.
\end{lemma}

\emph{Proof} Let $v$ be a function of $\left[ \chi _{\Delta }^{*}\right] $, $%
\left[ \varphi \right] $, $\left[ A_{\mu }\right] $, $\left[
H_{\mu }\right]
$, $\left[ B_{\mu \nu }\right] $, $\left[ \phi _{\mu \nu }\right] $, and $%
\left[ K_{\mu \nu \rho }\right] $. The dependence of $v$ on
$\left[ \varphi \right] $, $\left[ A_{\mu }\right] $, $\left[
H_{\mu }\right] $, $\left[ B_{\mu \nu }\right] $, $\left[ \phi
_{\mu \nu }\right] $, and $\left[ K_{\mu \nu \rho }\right] $ can
be reorganized as a dependence on $\left[ F_{A}\right] $ and on
$\bar{A}_{\mu }=\left\{ A_{\mu },\partial A_{\mu },\ldots \right\}
$, $\bar{H}_{\mu }=\left\{ H_{\mu },\partial H_{\mu },\ldots
\right\} $, $\bar{B}_{\mu \nu }=\left\{ B_{\mu \nu },\partial
B_{\mu \nu },\ldots \right\} $,\ $\bar{\phi}_{\mu \nu }=\left\{
\phi _{\mu \nu },\partial \phi _{\mu \nu },\ldots \right\} $,
$\bar{K}_{\mu \nu \rho }=\left\{ K_{\mu \nu \rho },\partial K_{\mu
\nu \rho },\ldots \right\} $,
where, $\bar{A}_{\mu }$, $\bar{H}_{\mu }$, $\bar{B}_{\mu \nu }$,\ $\bar{\phi}%
_{\mu \nu }$, $\bar{K}_{\mu \nu \rho }$ are not $\gamma
$-invariant. If $v$
is $\gamma $-invariant, then it does not involve, $\bar{A}_{\mu }$, $\bar{H}%
_{\mu }$, $\bar{B}_{\mu \nu }$,\ $\bar{\phi}_{\mu \nu }$,
$\bar{K}_{\mu \nu
\rho }$, i.e., $v=\left. v\right\vert  _{\bar{A}_{\mu }=0,\bar{H}_{\mu }=0,\bar{B}%
_{\mu \nu }=0,\bar{\phi}_{\mu \nu }=0,\bar{K}_{\mu \nu \rho }=0}$,
so we have by hypothesis that
\begin{equation}
\alpha =\left. \alpha \right\vert  _{\bar{A}_{\mu }=0,\bar{H}_{\mu }=0,\bar{B}%
_{\mu \nu }=0,\bar{\phi}_{\mu \nu }=0,\bar{K}_{\mu \nu \rho }=0}.
\label{A73}
\end{equation}
On the other hand, $\beta $ depends in general on $\left[ \chi
_{\Delta }^{*}\right] $, $\left[ F_{A}\right] $ and, $\bar{A}_{\mu
}$, $\bar{H}_{\mu } $, $\bar{B}_{\mu \nu }$,\ $\bar{\phi}_{\mu \nu
}$, $\bar{K}_{\mu \nu \rho
} $. Making $\bar{A}_{\mu }=0$, $\bar{H}_{\mu }=0$, $\bar{B}_{\mu \nu }=0$, $%
\bar{\phi}_{\mu \nu }=0$, $\bar{K}_{\mu \nu \rho }=0$ in (\ref{A72}), using (%
\ref{A73}) and taking into account the fact that $\delta $
commutes with the
operation of setting the `fields' $\bar{A}_{\mu }$, $\bar{H}_{\mu }$, $\bar{B%
}_{\mu \nu }$,\ $\bar{\phi}_{\mu \nu }$, and $\bar{K}_{\mu \nu
\rho }$ equal to zero, we find that
\begin{equation}
\alpha =\delta \left( \left. \beta \right\vert  _{\bar{A}_{\mu
}=0,\bar{H}_{\mu }=0,\bar{B}_{\mu \nu }=0,\bar{\phi}_{\mu \nu
}=0,\bar{K}_{\mu \nu \rho }=0}\right) ,  \label{A74}
\end{equation}
with $\left. \beta \right\vert  _{\bar{A}_{\mu }=0,\bar{H}_{\mu
}=0,\bar{B}_{\mu \nu }=0,\bar{\phi}_{\mu \nu }=0,\bar{K}_{\mu \nu
\rho }=0}$ invariant. This proves the lemma. $\blacksquare $

Now, we have the necessary tools for proving the next theorem.

\begin{theorem}
\label{hinvdelta}Let $\alpha _{k}^{p}$ be an invariant polynomial
with $\deg \left( \alpha _{k}^{p}\right) =p$ and $\text{agh}\left(
\alpha
_{k}^{p}\right) =k\geq 5$, which is $\delta $-exact modulo $d$%
\begin{equation}
\alpha _{k}^{p}=\delta \lambda _{k+1}^{p}+d\lambda
_{k}^{p-1},\quad k\geq 5. \label{g1}
\end{equation}
Then, we can choose $\lambda _{k+1}^{p}$ and $\lambda _{k}^{p-1}$
to be invariant polynomials.
\end{theorem}

\emph{Proof} Initially, by successively acting with $d$ and $\delta $ on (%
\ref{g1}) we obtain a tower of equations of the same type. Indeed,
acting with $d$ on (\ref{g1}) we find that $d\alpha
_{k}^{p}=-\delta \left( d\lambda _{k+1}^{p}\right) $. On the other
hand, as $d\alpha _{k}^{p}$ is invariant, by means of Lemma
\ref{trivinv} we obtain that $d\alpha _{k}^{p}=-\delta \alpha
_{k+1}^{p+1}$, with $\alpha _{k+1}^{p+1}$ invariant. The last two
relations imply that $\delta \left( \alpha _{k+1}^{p+1}-d\lambda
_{k+1}^{p}\right) =0$. As $\delta $ is acyclic at strictly
positive antighost numbers, the last relation implies that
\begin{equation}
\alpha _{k+1}^{p+1}=\delta \lambda _{k+2}^{p+1}+d\lambda
_{k+1}^{p}. \label{g2}
\end{equation}
Starting now with (\ref{g2}) and reprising the same operations
like those performed between the formulas (\ref{g1}) and
(\ref{g2}), we obtain a descent that stops in form degree $5$ with
the equation $\alpha _{k+5-p}^{5}=\delta \lambda
_{k+5-p+1}^{5}+d\lambda _{k+5-p}^{4}$. Now, we act with $\delta $
on (\ref{g1}) and deduce that $\delta \alpha _{k}^{p}=-d\delta
\lambda _{k}^{p-1}$. As $\delta \alpha _{k}^{p}$ is invariant, in
the case $k>1$, due to the Theorem \ref{hginv}, we obtain that
$\delta \alpha _{k}^{p}=-d\alpha _{k-1}^{p-1}$, where $\alpha
_{k-1}^{p-1}$ is invariant. Using the last two relations we get
that $d\left( \alpha _{k-1}^{p-1}-\delta \lambda _{k}^{p-1}\right)
=0$, such that it follows that
\begin{equation}
\alpha _{k-1}^{p-1}=\delta \lambda _{k}^{p-1}+d\lambda
_{k-1}^{p-2}. \label{g3}
\end{equation}
If $k=5$ in (\ref{g1}), we cannot go down since by assumption
$k\geq 5$, and so the bottom of the tower is (\ref{g1}) for $k=5$.
Starting from (\ref{g3}) and reprising the same procedure we reach
a descent that ends at either form degree zero or antighost number
five, hence the last equation respectively takes the form
\begin{equation}
\alpha _{k-p}^{0}=\delta \lambda _{k-p+1}^{0},  \label{g4}
\end{equation}
for $k-p\geq 5$ or
\begin{equation}
\alpha _{5}^{p-k+5}=\delta \lambda _{6}^{p-k+5}+d\lambda
_{5}^{p-k+4}, \label{g5}
\end{equation}
for $k-p<5$. In consequence, the procedure described in the above
leads to the chain
\begin{align}
\alpha _{k+5-p}^{5}& =\delta \lambda _{k+6-p}^{5}+d\lambda
_{k+5-p}^{4},
\nonumber \\
& \vdots  \nonumber \\
\alpha _{k+1}^{p+1}& =\delta \lambda _{k+2}^{p+1}+d\lambda
_{k+1}^{p},
\nonumber \\
\alpha _{k}^{p}& =\delta \lambda _{k+1}^{p}+d\lambda _{k}^{p-1},
\label{g6}
\\
\alpha _{k-1}^{p-1}& =\delta \lambda _{k}^{p-1}+d\lambda
_{k-1}^{p-2},
\nonumber \\
& \vdots  \nonumber \\
\alpha _{k-p}^{0}& =\delta \lambda _{k-p+1}^{0}\quad
\text{or}\quad \alpha _{5}^{p-k+5}=\delta \lambda
_{6}^{p-k+5}+d\lambda _{5}^{p-k+4}.  \nonumber
\end{align}
All the $\alpha $'s in the descent (\ref{g6}) are invariant.

Now, we show that if one of the $\lambda $'s in (\ref{g6}) is
invariant, then all the other $\lambda $'s can be taken to be also
invariant. Indeed, let $\lambda _{B}^{A-1}$ be invariant. It is
involved in two of the equations from (\ref{g6}), namely
\begin{align}
\alpha _{B}^{A}& =\delta \lambda _{B+1}^{A}+d\lambda _{B}^{A-1},
\label{g7}
\\
\alpha _{B-1}^{A-1}& =\delta \lambda _{B}^{A-1}+d\lambda
_{B-1}^{A-2}. \label{g8}
\end{align}
The relation (\ref{g7}) yields that $\alpha _{B}^{A}-d\lambda
_{B}^{A-1}$ is invariant. Then, in agreement with Lemma
\ref{trivinv} the object $\lambda
_{B+1}^{A}$ can be chosen to be invariant. Using (\ref{g8}), we have that $%
\alpha _{B-1}^{A-1}-\delta \lambda _{B}^{A-1}$ is invariant, such
that Theorem \ref{hginv} ensures that $\lambda _{B-1}^{A-2}$ is
also invariant. On the other hand, $\lambda _{B+1}^{A}$ and
$\lambda _{B-1}^{A-2}$ are involved in other two sets of equations
from the descent. (For instance, the former element appears in the
equations $\alpha _{B+1}^{A+1}=\delta \lambda
_{B+2}^{A+1}+d\lambda _{B+1}^{A}$ and $\alpha _{B+2}^{A+2}=\delta
\lambda _{B+3}^{A+2}+d\lambda _{B+2}^{A+1}$.) Going on in the same
fashion, we find that all the $\lambda $'s are invariant. In the
case where $\lambda _{B}^{A-1}$ appears at the top or at the
bottom of the descent, we act in a similar way, but only with
respect to a single equation. The above considerations emphasize
that it is enough to verify the theorem in form degree $5$ and for
all the values $k\geq 5$ of the antighost number.

If $k\geq 10$ (and hence $k-p\geq 5$), the last equation from the descent (%
\ref{g6}) for $p=5$ reads as
\begin{equation}
\alpha _{k-5}^{0}=\delta \lambda _{k-4}^{0}.  \label{g9}
\end{equation}
Using Lemma \ref{trivinv}, it results that $\lambda _{k-4}^{0}$
can be taken to be invariant, such that the above arguments lead
to the conclusion that all the $\lambda $'s from the descent can
be chosen invariant. As a consequence, in the first equation from
the descent in this situation, namely, $\alpha _{k}^{5}=\delta
\lambda _{k+1}^{5}+d\lambda _{k}^{4}$, we have that both $\lambda
_{k+1}^{5}$ and $\lambda _{k}^{4}$ are invariant. Therefore, the
theorem is true in form degree $5$ and in all antighost
numbers $k\geq 10$, so it remains to be proved that it holds in form degree $%
5$ and in all antighost numbers $5\leq k<10$. This is done below.

In the sequel we consider the case $p=5$ and $5\leq k<10$. The top
equation from (\ref{g6}), written in dual notations, takes the
form
\begin{equation}
\alpha _{k}=\delta \lambda _{k+1}+\partial _{\mu }\lambda
_{k}^{\mu },\quad 5\leq k<10.  \label{g10}
\end{equation}
On the other hand, we can express $\alpha _{k}$ in terms of its
E.L. derivatives by means of the homotopy formula
\begin{equation}
\alpha _{k}=\int\nolimits_{0}^{1}d\tau \left( \frac{\delta ^{R}\alpha _{k}}{%
\delta \chi _{\Delta }^{*}}\left( \tau \right) \chi _{\Delta }^{*}+\frac{%
\delta \alpha _{k}}{\delta \Phi ^{\alpha _{0}}}\left( \tau \right)
\Phi ^{\alpha _{0}}\right) +\partial _{\mu }j_{k}^{\mu },
\label{g10a}
\end{equation}
where $\frac{\delta ^{R}\alpha _{k}}{\delta \chi _{\Delta
}^{*}}\left( \tau \right) =\frac{\delta ^{R}\alpha _{k}}{\delta
\chi _{\Delta }^{*}}\left( \tau \left[ F_{A}\right] ,\tau \left[
\chi _{\Delta }^{*}\right] \right) $ and similarly for the other
terms. For further convenience, we denote the E.L. derivatives of
$\lambda _{k+1}$ by
\begin{gather}
\frac{\delta ^{R}\lambda _{k+1}}{\delta C_{\mu \nu \rho \lambda \sigma }^{*}}%
=G_{k-4}^{\mu \nu \rho \lambda \sigma },\quad \frac{\delta ^{R}\lambda _{k+1}%
}{\delta C_{\mu \nu \rho \lambda }^{*}}=G_{k-3}^{\mu \nu \rho
\lambda
},\quad \frac{\delta ^{R}\lambda _{k+1}}{\delta C_{\mu \nu \rho }^{*}}%
=G_{k-2}^{\mu \nu \rho },  \label{g11} \\
\frac{\delta ^{R}\lambda _{k+1}}{\delta C_{\mu \nu
}^{*}}=G_{k-1}^{\mu \nu },\quad \frac{\delta ^{R}\lambda
_{k+1}}{\delta H_{\mu }^{*}}=G_{k}^{\mu },\quad \frac{\delta
\lambda _{k+1}}{\delta H_{\mu }}=\bar{G}_{k+1}^{\mu },
\label{g11a} \\
\frac{\delta \lambda _{k+1}}{\delta \varphi ^{*}}=M_{k},\quad
\frac{\delta
\lambda _{k+1}}{\delta \varphi }=\bar{M}_{k+1},  \label{g11b} \\
\frac{\delta ^{R}\lambda _{k+1}}{\delta \eta ^{*}}=N_{k-1},\quad \frac{%
\delta ^{R}\lambda _{k+1}}{\delta A_{\mu }^{*}}=N_{k}^{\mu },\quad \frac{%
\delta ^{R}\lambda _{k+1}}{\delta A_{\mu }}=\bar{N}_{k+1}^{\mu },
\label{g12} \\
\frac{\delta ^{R}\lambda _{k+1}}{\delta
\tilde{\eta}^{*}}=Q_{k-3},\quad
\frac{\delta ^{R}\lambda _{k+1}}{\delta \tilde{\eta}_{\mu }^{*}}%
=Q_{k-2}^{\mu },\quad \frac{\delta ^{R}\lambda _{k+1}}{\delta \tilde{\eta}%
_{\mu \nu }^{*}}=Q_{k-1}^{\mu \nu },  \label{g13a} \\
\frac{\delta ^{R}\lambda _{k+1}}{\delta \tilde{B}_{\mu \nu \rho }^{*}}%
=Q_{k}^{\mu \nu \rho },\quad \frac{\delta ^{R}\lambda _{k+1}}{\delta \tilde{B%
}_{\mu \nu \rho }}=\bar{Q}_{k+1}^{\mu \nu \rho },  \label{g13b} \\
\frac{\delta ^{R}\lambda _{k+1}}{\delta C^{*}}=L_{k-2},\quad
\frac{\delta ^{R}\lambda _{k+1}}{\delta C_{\mu }^{*}}=L_{k-1}^{\mu
},\quad \frac{\delta ^{R}\lambda _{k+1}}{\delta \phi _{\mu \nu
}^{*}}=L_{k}^{\mu \nu },\quad \frac{\delta ^{R}\lambda
_{k+1}}{\delta \phi _{\mu \nu }}=\bar{L}_{k+1}^{\mu \nu },
\label{g14} \\
\frac{\delta ^{R}\lambda _{k+1}}{\delta \stackrel{\sim
}{\mathcal{G}}^{*}} =P_{k-2},\quad \frac{\delta ^{R}\lambda
_{k+1}}{\delta \stackrel{\sim }{\mathcal{G}}_{\mu
}^{*}}=P_{k-1}^{\mu },\quad \frac{\delta ^{R}\lambda
_{k+1}}{\delta \tilde{K}_{\mu \nu }^{*}}=P_{k}^{\mu \nu },\quad
\frac{\delta ^{R}\lambda _{k+1}}{\delta \tilde{K}_{\mu \nu
}}=\bar{P}_{k+1}^{\mu \nu }.  \label{g15}
\end{gather}
Using (\ref{g10}), as well as the homotopy formula for $\lambda
_{k+1}$, together with the notations (\ref{g11})--(\ref{g11b}), we
determine the
relationship between the E.L. derivatives of $\alpha _{k}$ and those of $%
\lambda _{k+1}$ in the $(H^{\mu },\varphi )$-field/antifield
sector like
\begin{align}
\frac{\delta ^{R}\alpha _{k}}{\delta C_{\mu \nu \rho \lambda \sigma }^{*}}%
& =-\delta G_{k-4}^{\mu \nu \rho \lambda \sigma },& \frac{\delta
^{R}\alpha _{k}}{\delta C_{\mu \nu \rho \lambda }^{*}}& =\delta
G_{k-3}^{\mu \nu \rho \lambda }+5\partial _{\sigma }G_{k-4}^{\mu
\nu \rho \lambda \sigma
},  \label{g16a} \\
\frac{\delta ^{R}\alpha _{k}}{\delta C_{\mu \nu \rho }^{*}}&
=-\delta G_{k-2}^{\mu \nu \rho }+4\partial _{\lambda }G_{k-3}^{\mu
\nu \rho \lambda },& \frac{\delta ^{R}\alpha _{k}}{\delta C_{\mu
\nu }^{*}}& =\delta G_{k-1}^{\mu \nu }+3\partial _{\rho
}G_{k-2}^{\mu \nu \rho },  \label{g16b}
\\
\frac{\delta ^{R}\alpha _{k}}{\delta H_{\mu }^{*}}& =-\delta
G_{k}^{\mu }+2\partial _{\nu }G_{k-1}^{\mu \nu },& \frac{\delta
\alpha _{k}}{\delta
H_{\mu }}& =\delta \bar{G}_{k+1}^{\mu }-\partial ^{\mu }M_{k},  \label{g16c} \\
\frac{\delta \alpha _{k}}{\delta \varphi ^{*}}& =-\delta M_{k},&
\frac{\delta \alpha _{k}}{\delta \varphi }& =\delta \bar{M}_{k+1}.
\label{g16d}
\end{align}
Due to Lemma \ref{trivinv} ($k>0$), the equations (\ref{g16d})
allow us to state that
\begin{equation}
\frac{\delta \alpha _{k}}{\delta \varphi ^{*}}=-\delta
M_{k}^{\prime },\quad \frac{\delta \alpha _{k}}{\delta \varphi
}=\delta \bar{M}_{k+1}^{\prime }, \label{g17a}
\end{equation}
with both $M_{k}^{\prime }$ and $\bar{M}_{k+1}^{\prime }$
invariant
polynomials. Applying a similar reasoning in connection with the descent (%
\ref{g16a})--(\ref{g16d}) from bottom to top, we obtain that
\begin{align}
\frac{\delta ^{R}\alpha _{k}}{\delta C_{\mu \nu \rho \lambda \sigma }^{*}}%
& =-\delta G_{k-4}^{\prime \mu \nu \rho \lambda \sigma },&
\frac{\delta ^{R}\alpha _{k}}{\delta C_{\mu \nu \rho \lambda
}^{*}}& =\delta G_{k-3}^{\prime \mu \nu \rho \lambda }+5\partial
_{\sigma }G_{k-4}^{\prime
\mu \nu \rho \lambda \sigma },  \label{g17b} \\
\frac{\delta ^{R}\alpha _{k}}{\delta C_{\mu \nu \rho }^{*}}&
=-\delta G_{k-2}^{\prime \mu \nu \rho }+4\partial _{\lambda
}G_{k-3}^{\prime \mu \nu
\rho \lambda },& \frac{\delta ^{R}\alpha _{k}}{\delta C_{\mu \nu }^{*}}%
& =\delta G_{k-1}^{\prime \mu \nu }+3\partial _{\rho
}G_{k-2}^{\prime \mu \nu
\rho },  \label{g17c} \\
\frac{\delta ^{R}\alpha _{k}}{\delta H_{\mu }^{*}}& =-\delta
G_{k}^{\prime \mu
}+2\partial _{\nu }G_{k-1}^{\prime \mu \nu },& \frac{\delta \alpha _{k}}{%
\delta H_{\mu }}& =\delta \bar{G}_{k+1}^{\prime \mu }-\partial
^{\mu }M_{k}^{\prime },  \label{g17d}
\end{align}
where all the `prime' quantities are invariant polynomials. On the
other hand, since $\alpha _{k}$ is an invariant polynomial that
depends on $H_{\mu }$ just through $\partial _{\mu }H^{\mu }$ and
its spacetime derivatives, we get that
\begin{equation}
\frac{\delta \alpha _{k}}{\delta H_{\mu }}=\partial ^{\mu }\Delta
_{k}. \label{g19}
\end{equation}
Using now the last equation from (\ref{g17d}) together with
(\ref{g19}), we arrive at
\begin{equation}
\delta \bar{G}_{k+1}^{\prime \mu }=\partial ^{\mu }\left(
M_{k}^{\prime }+\Delta _{k}\right) ,  \label{g20}
\end{equation}
which indicates that $\bar{G}_{k+1}^{\prime \mu }$ belongs to $%
H_{k+1}^{1}\left( \delta \vert d\right) $. As $H_{k+1}^{1}\left(
\delta \vert d\right) \simeq H_{k+2}^{2}\left( \delta \vert
d\right) \simeq H_{k+3}^{3}\left( \delta \vert d\right) \simeq
H_{k+4}^{4}\left( \delta \vert d\right) \simeq H_{k+5}^{5}\left(
\delta \vert d\right) $ and $H_{k+5}^{5}\left( \delta \vert
d\right) \simeq 0$, the equation (\ref{g20}) further implies
\begin{equation}
\bar{G}_{k+1}^{\prime \mu }=\delta \bar{G}_{k+2}^{\prime \prime
\mu }+\partial ^{\mu }M_{k+1}^{\prime \prime }.  \label{g21}
\end{equation}

We will prove the theorem for $5\leq k<10$ by induction. More
precisely, we
will assume that the theorem holds in antighost numbers $\left( k+2\right) $%
, $\left( k+3\right) $, $\left( k+4\right) $, $\left( k+5\right)
$, and in form degree $5$, and will prove that it is also valid in
antighost number $k$ and in form degree $5$. In agreement with
this inductive hypothesis [more precisely, that the theorem is
satisfied in antighost number $\left(
k+5\right) $ and in form degree $5$], we can take both $\bar{G}%
_{k+2}^{\prime \prime \mu }$ and $M_{k+1}^{\prime \prime }$ in
(\ref{g21}) to be invariant polynomials.

From (\ref{g10}), the homotopy formula for $\lambda _{k+1}$ and
the definitions (\ref{g12})--(\ref{g13b}) we deduce the
relationship between the
E.L. derivatives of $\alpha _{k}$ and those of $\lambda _{k+1}$ in the ($%
B^{\mu \nu },A^{\mu }$)-field/antifield sector. It is more
convenient to work, instead of the field $B^{\mu \nu }$ and of the
antifields of the ghosts associated with its gauge invariance and
the accompanying reducibility relations, with their Hodge duals,
denoted by tilde variables. In terms of these new fields we have
that
\begin{gather}
\frac{\delta ^{R}\alpha _{k}}{\delta \eta ^{*}}=\delta N_{k-1},\quad \frac{%
\delta ^{R}\alpha _{k}}{\delta A_{\mu }^{*}}=-\delta N_{k}^{\mu
}+\partial
^{\mu }N_{k-1},  \label{g22a} \\
\frac{\delta ^{R}\alpha _{k}}{\delta A_{\mu }}=\delta \bar{N}_{k+1}^{\mu }-%
\tfrac{1}{8}\sigma ^{\mu \alpha }\varepsilon _{\alpha \beta \gamma
\delta \varepsilon }\partial ^{[\beta }Q_{k}^{\gamma \delta
\varepsilon ]},
\label{g22b} \\
\frac{\delta ^{R}\alpha _{k}}{\delta \tilde{\eta}^{*}}=\delta
Q_{k-3},\quad \frac{\delta ^{R}\alpha _{k}}{\delta
\tilde{\eta}_{\mu }^{*}}=-\delta
Q_{k-2}^{\mu }-\partial ^{\mu }Q_{k-3},  \label{g22c} \\
\frac{\delta ^{R}\alpha _{k}}{\delta \tilde{\eta}_{\mu \nu
}^{*}}=\delta
Q_{k-1}^{\mu \nu }-\tfrac{1}{2}\partial ^{[\mu }Q_{k-2}^{\nu ]},\quad \frac{%
\delta ^{R}\alpha _{k}}{\delta \tilde{B}_{\mu \nu \rho
}^{*}}=-\delta Q_{k}^{\mu \nu \rho }-\tfrac{1}{3}\partial ^{[\mu
}Q_{k-1}^{\nu \rho ]},
\label{g22d} \\
\frac{\delta ^{R}\alpha _{k}}{\delta \tilde{B}_{\mu \nu \rho }}=\delta \bar{Q%
}_{k+1}^{\mu \nu \rho }+\tfrac{1}{12}\varepsilon ^{\mu \nu \rho
\lambda \sigma }\sigma _{\lambda \alpha }\sigma _{\sigma \beta
}\partial ^{[\alpha }N_{k}^{\beta ]}.  \label{g22e}
\end{gather}
Due to the Lemma \ref{trivinv} and on account of the fact that the
E.L. derivatives of invariant polynomials are also invariant, the
first equation in (\ref{g22a}) and the former relation from
(\ref{g22c}) can be written like
\begin{equation}
\frac{\delta ^{R}\alpha _{k}}{\delta \eta ^{*}}=\delta
N_{k-1}^{\prime },\quad \frac{\delta ^{R}\alpha _{k}}{\delta
\tilde{\eta}^{*}}=\delta Q_{k-3}^{\prime },  \label{g23a}
\end{equation}
where both $N_{k-1}^{\prime }$ and $Q_{k-3}^{\prime }$ are
invariant.
Reprising the same arguments for the remaining equations in (\ref{g22a})--(%
\ref{g22e}), we infer that
\begin{gather}
\frac{\delta ^{R}\alpha _{k}}{\delta A_{\mu }^{*}}=-\delta
N_{k}^{\prime \mu
}+\partial ^{\mu }N_{k-1}^{\prime },  \label{g23b} \\
\frac{\delta ^{R}\alpha _{k}}{\delta A_{\mu }}=\delta
\bar{N}_{k+1}^{\prime \mu }-\tfrac{1}{8}\sigma ^{\mu \alpha
}\varepsilon _{\alpha \beta \gamma \delta \varepsilon }\partial
^{[\beta }Q_{k}^{\prime \gamma \delta
\varepsilon ]},  \label{g23c} \\
\frac{\delta ^{R}\alpha _{k}}{\delta \tilde{\eta}_{\mu
}^{*}}=-\delta
Q_{k-2}^{\prime \mu }-\partial ^{\mu }Q_{k-3}^{\prime },  \label{g23d} \\
\frac{\delta ^{R}\alpha _{k}}{\delta \tilde{\eta}_{\mu \nu
}^{*}}=\delta Q_{k-1}^{\prime \mu \nu }-\tfrac{1}{2}\partial
^{[\mu }Q_{k-2}^{\prime \nu
]},\quad \frac{\delta ^{R}\alpha _{k}}{\delta \tilde{B}_{\mu \nu \rho }^{*}}%
=-\delta Q_{k}^{\prime \mu \nu \rho }-\tfrac{1}{3}\partial ^{[\mu
}Q_{k-1}^{\prime \nu \rho ]},  \label{g23e} \\
\frac{\delta ^{R}\alpha _{k}}{\delta \tilde{B}_{\mu \nu \rho }}=\delta \bar{Q%
}_{k+1}^{\prime \mu \nu \rho }+\tfrac{1}{12}\varepsilon ^{\mu \nu
\rho \lambda \sigma }\sigma _{\lambda \alpha }\sigma _{\sigma
\beta }\partial ^{[\alpha }N_{k}^{\prime \beta ]},  \label{g23f}
\end{gather}
where all the prime quantities are invariant. Let us analyze the relations (%
\ref{g23c}) and (\ref{g23f}). Since the invariant quantity $\alpha
_{k}$ depends on $A_{\mu }$ through the combination $\partial
_{[\alpha }A_{\beta ]}$ and on $\tilde{B}_{\mu \nu \rho }$ via the
expression $\partial _{[\alpha }\tilde{B}_{\beta \gamma \delta
]}$, it follows that there exist some elements $\Delta _{k}^{\mu
\nu }$ and $\Delta _{k}^{\mu \nu \rho \lambda }$, completely
antisymmetric in their Lorentz indices, such that
\begin{align}
\frac{\delta ^{R}\alpha _{k}}{\delta A_{\mu }}& =\partial _{\nu
}\Delta
_{k}^{\mu \nu },  \label{g24a} \\
\frac{\delta ^{R}\alpha _{k}}{\delta \tilde{B}_{\mu \nu \rho }}&
=\partial _{\lambda }\Delta _{k}^{\mu \nu \rho \lambda }.
\label{g24b}
\end{align}
Inserting (\ref{g24a}) in (\ref{g23c}) and (\ref{g24b})
respectively in (\ref {g23f}), we obtain the relations
\begin{align}
\delta \bar{N}_{k+1}^{\prime \mu }& =\partial _{\nu }\left( \Delta
_{k}^{\mu \nu }+3\tilde{Q}_{k}^{\prime \mu \nu }\right) ,  \label{g25a} \\
\delta \bar{Q}_{k+1}^{\prime \mu \nu \rho }& =\partial _{\lambda
}\left( \Delta _{k}^{\mu \nu \rho \lambda
}-\tfrac{1}{6}\tilde{N}_{k}^{\prime \mu \nu \rho \lambda }\right)
,  \label{g25b}
\end{align}
where $\tilde{Q}_{k}^{\prime \mu \nu }$ is the Hodge dual of
$Q_{k}^{\prime \mu \nu \rho }$, while $\tilde{N}_{k}^{\prime \mu
\nu \rho \lambda }$ represents the Hodge dual of $N_{k}^{\prime
\mu }$. In dual language, the
equation (\ref{g25a}) shows that $\bar{N}_{k+1}^{\prime \mu }$ belongs to $%
H_{k+1}^{4}\left( \delta \vert d\right) $. But $H_{k+1}^{4}\left(
\delta \vert d\right) \simeq H_{k+2}^{5}\left( \delta \vert
d\right) $ and $H_{k+2}^{5}\left( \delta \vert d\right) \simeq 0$,
so the inductive hypothesis [in this case the validity of the
theorem in antighost number $\left( k+2\right) $ and in form
degree $5$] allows us to conclude that there exist some invariant
polynomials $\bar{N}_{k+2}^{\prime \prime \mu }$ and $\tilde{Q}%
_{k+1}^{\prime \prime \mu \nu }$, in terms of which we have that
\begin{equation}
\bar{N}_{k+1}^{\prime \mu }=\delta \bar{N}_{k+2}^{\prime \prime
\mu }+\partial _{\nu }\tilde{Q}_{k+1}^{\prime \prime \mu \nu }.
\label{g26a}
\end{equation}
In the same dual language, from (\ref{g25b}) we read that $\bar{Q}%
_{k+1}^{\prime \mu \nu \rho }\in H_{k+1}^{2}\left( \delta \vert
d\right) \simeq H_{k+2}^{3}\left( \delta \vert d\right) \simeq
H_{k+3}^{4}\left( \delta \vert d\right) \simeq H_{k+4}^{5}\left(
\delta \vert d\right) \simeq 0$. Using again the
inductive hypothesis [here, that the theorem holds in antighost number $%
\left( k+4\right) $ and in form degree $5$], we then get the
existence of some invariant polynomials $\bar{Q}_{k+2}^{\prime
\prime \mu \nu \rho }$ and $\tilde{N}_{k+1}^{\prime \prime \mu \nu
\rho \lambda }$ with the property
\begin{equation}
\bar{Q}_{k+1}^{\prime \mu \nu \rho }=\delta \bar{Q}_{k+2}^{\prime
\prime \mu \nu \rho }+\partial _{\lambda }\tilde{N}_{k+1}^{\prime
\prime \mu \nu \rho \lambda }.  \label{g26b}
\end{equation}

Now, we invoke once more the relation (\ref{g10}) and the homotopy
formula for $\lambda _{k+1}$, which, combined with the definitions
(\ref{g14})--(\ref {g15}), provides the relationship between the
E.L. derivatives of $\alpha
_{k}$ and of $\lambda _{k+1}$ in the ($K^{\mu \nu \rho },\phi ^{\mu \nu }$%
)-field/antifield sector. Instead of $K^{\mu \nu \rho }$ and of
the antifields corresponding to the ghosts associated with the
gauge invariance of this field, we work with their Hodge duals,
and deduce
\begin{align}
\frac{\delta ^{R}\alpha _{k}}{\delta C^{*}}& =-\delta L_{k-2},& \frac{%
\delta ^{R}\alpha _{k}}{\delta C_{\mu }^{*}}& =\delta L_{k-1}^{\mu
}-\partial
^{\mu }L_{k-2},  \label{g27a} \\
\frac{\delta ^{R}\alpha _{k}}{\delta \phi _{\mu \nu }^{*}}&
=-\delta L_{k}^{\mu \nu }+\partial ^{[\mu }L_{k-1}^{\nu ]},&
\frac{\delta
^{R}\alpha _{k}}{\delta \phi _{\mu \nu }}& =\delta \bar{L}_{k+1}^{\mu \nu }+%
\tfrac{1}{3}\partial _{\rho }\tilde{P}_{k}^{\mu \nu \rho },  \label{g27b} \\
\frac{\delta ^{R}\alpha _{k}}{\delta \stackrel{\sim }{\mathcal{G}}^{*}}%
& =-\delta P_{k-2},& \frac{\delta ^{R}\alpha _{k}}{\delta \stackrel{\sim }{%
\mathcal{G}}_{\mu }^{*}}& =\delta P_{k-1}^{\mu }+\partial ^{\mu
}P_{k-2},
\label{g27c} \\
\frac{\delta ^{R}\alpha _{k}}{\delta \tilde{K}_{\mu \nu }^{*}}&
=-\delta
P_{k}^{\mu \nu }+\tfrac{1}{2}\partial ^{[\mu }P_{k-1}^{\nu ]},& \frac{%
\delta ^{R}\alpha _{k}}{\delta \tilde{K}_{\mu \nu }}& =\delta \bar{P}%
_{k+1}^{\mu \nu }+\partial _{\rho }\tilde{L}_{k}^{\mu \nu \rho },
\label{g27d}
\end{align}
where $\tilde{P}_{k}^{\mu \nu \rho }$ and $\tilde{L}_{k}^{\mu \nu
\rho }$ are dual to $P_{k}^{\mu \nu }$ and respectively to
$L_{k}^{\mu \nu }$. The Lemma \ref{trivinv} and the fact that the
E.L. derivatives of any invariant polynomial is also invariant
allow us to write the former equations in (\ref {g27a}) and
respectively in (\ref{g27c}) like
\begin{equation}
\frac{\delta ^{R}\alpha _{k}}{\delta C^{*}}=-\delta
L_{k-2}^{\prime },\quad
\frac{\delta ^{R}\alpha _{k}}{\delta \stackrel{\sim }{\mathcal{G}}^{*}}%
=-\delta P_{k-2}^{\prime },  \label{g28a}
\end{equation}
with both $L_{k-2}^{\prime }$ and $P_{k-2}^{\prime }$ invariant.
The same
reasoning, extended to the remaining equations in (\ref{g27a})--(\ref{g27d}%
), produces
\begin{gather}
\frac{\delta ^{R}\alpha _{k}}{\delta C_{\mu }^{*}}=\delta
L_{k-1}^{\prime
\mu }-\partial ^{\mu }L_{k-2}^{\prime },\quad \frac{\delta ^{R}\alpha _{k}}{%
\delta \phi _{\mu \nu }^{*}}=-\delta L_{k}^{\prime \mu \nu
}+\partial ^{[\mu
}L_{k-1}^{\prime \nu ]},  \label{g28b} \\
\frac{\delta ^{R}\alpha _{k}}{\delta \phi _{\mu \nu }}=\delta \bar{L}%
_{k+1}^{\prime \mu \nu }+\tfrac{1}{3}\partial _{\rho
}\tilde{P}_{k}^{\prime
\mu \nu \rho },  \label{g28c} \\
\frac{\delta ^{R}\alpha _{k}}{\delta \stackrel{\sim }{\mathcal{G}}_{\mu }^{*}%
}=\delta P_{k-1}^{\prime \mu }+\partial ^{\mu }P_{k-2}^{\prime },\quad \frac{%
\delta ^{R}\alpha _{k}}{\delta \tilde{K}_{\mu \nu }^{*}}=-\delta
P_{k}^{\prime \mu \nu }+\tfrac{1}{2}\partial ^{[\mu
}P_{k-1}^{\prime \nu ]},
\label{g28d} \\
\frac{\delta ^{R}\alpha _{k}}{\delta \tilde{K}_{\mu \nu }}=\delta \bar{P}%
_{k+1}^{\prime \mu \nu }+\partial _{\rho }\tilde{L}_{k}^{\prime
\mu \nu \rho },  \label{g28e}
\end{gather}
where all the prime objects from (\ref{g28b})--(\ref{g28e}) are
invariant polynomials. Let us focus on the relations (\ref{g28c})
and (\ref{g28e}). Due to the fact that the invariant element
$\alpha _{k}$ depends on $\phi _{\mu \nu }$ through the
combination $\partial _{[\alpha }\phi _{\beta \gamma ]}$ and on
$\tilde{K}_{\mu \nu }$ via the expression $\partial _{[\alpha
}\tilde{K}_{\beta \gamma ]}$, it results that there exist some
objects $\Omega _{k}^{\mu \nu \rho }$ and $\Gamma _{k}^{\mu \nu
\rho }$, completely antisymmetric in their Lorentz indices, such
that
\begin{align}
\frac{\delta ^{R}\alpha _{k}}{\delta \phi _{\mu \nu }}& =\partial
_{\rho
}\Omega _{k}^{\mu \nu \rho },  \label{g29a} \\
\frac{\delta ^{R}\alpha _{k}}{\delta \tilde{K}_{\mu \nu }}&
=\partial _{\rho }\Gamma _{k}^{\mu \nu \rho }.  \label{g29b}
\end{align}
Putting together the equation (\ref{g29a}) with (\ref{g28c}) and
respectively the relation (\ref{g29b}) with (\ref{g28e}), we find
that
\begin{align}
\delta \bar{L}_{k+1}^{\prime \mu \nu }& =\partial _{\rho }\left(
\Omega _{k}^{\mu \nu \rho }-\tfrac{1}{3}\tilde{P}_{k}^{\prime \mu
\nu \rho }\right)
,  \label{g30a} \\
\delta \bar{P}_{k+1}^{\prime \mu \nu }& =\partial _{\rho }\left(
\Gamma _{k}^{\mu \nu \rho }-\tilde{L}_{k}^{\prime \mu \nu \rho
}\right) , \label{g30b}
\end{align}
which indicates that both the invariant polynomials
$\bar{L}_{k+1}^{\prime
\mu \nu }$ and $\bar{P}_{k+1}^{\prime \mu \nu }$ pertain to $%
H_{k+1}^{3}\left( \delta \vert d\right) \simeq H_{k+2}^{4}\left(
\delta \vert d\right) \simeq H_{k+3}^{5}\left( \delta \vert
d\right) \simeq 0$. The inductive
hypothesis [more exactly, that the theorem is fulfilled in antighost number $%
\left( k+3\right) $ and in form degree $5$] ensures now the
existence of two
sets of invariant polynomials, $\bar{L}_{k+2}^{\prime \prime \mu \nu \rho }$%
, $\tilde{P}_{k+1}^{\prime \prime \mu \nu \rho }$ and
$\bar{P}_{k+2}^{\prime \prime \mu \nu }$, $\tilde{L}_{k}^{\prime
\prime \mu \nu \rho }$, with the help of which we can write
\begin{align}
\bar{L}_{k+1}^{\prime \mu \nu }& =\delta \bar{L}_{k+2}^{\prime
\prime \mu \nu \rho }+\partial _{\rho }\tilde{P}_{k+1}^{\prime
\prime \mu \nu \rho },
\label{g31a} \\
\bar{P}_{k+1}^{\prime \mu \nu }& =\delta \bar{P}_{k+2}^{\prime
\prime \mu \nu }+\partial _{\rho }\tilde{L}_{k}^{\prime \prime \mu
\nu \rho }. \label{g31b}
\end{align}

Introducing in (\ref{g10a}) the E.L. derivatives of the invariant
polynomial $\alpha _{k}$, expressed via the relations
(\ref{g17a})--(\ref{g17d}), (\ref {g23a})--(\ref{g23f}), and
(\ref{g28a})--(\ref{g28e}), we finally determine that
\begin{equation}\label{g32}
\begin{split}
\alpha _{k}& =\delta \left[ \int_{0}^{1}d\tau \left(
G_{k-4}^{\prime \mu \nu \rho \lambda \sigma }\left( \tau \right)
C_{\mu \nu \rho \lambda \sigma }^{*}+G_{k-3}^{\prime \mu \nu \rho
\lambda }\left( \tau \right) C_{\mu \nu \rho \lambda
}^{*}+G_{k-2}^{\prime \mu \nu \rho }\left( \tau \right) C_{\mu
\nu \rho }^{*}+\right. \right.  \\
&\quad \quad \quad \quad \quad \quad +G_{k-1}^{\prime \mu \nu
}\left( \tau \right) C_{\mu \nu }^{*}+G_{k}^{\prime \mu }\left(
\tau \right) H_{\mu }^{*}+M_{k}^{\prime }\left( \tau \right)
\varphi ^{*}+\bar{M}_{k+1}^{\prime }\left( \tau \right) \varphi -\\
&\quad \quad \quad \quad \quad \quad -M_{k+1}^{\prime \prime
}\left( \tau \right) \partial ^{\mu }H_{\mu } +N_{k-1}^{\prime
}\left( \tau \right) \eta ^{*}+Q_{k-3}^{\prime }\left( \tau
\right) \tilde{\eta}^{*}+N_{k}^{\prime \mu }\left( \tau \right)
A_{\mu }^{*}+\\
&\quad \quad \quad \quad \quad \quad +Q_{k-2}^{\prime \mu }\left(
\tau \right) \tilde{\eta}_{\mu }^{*}+Q_{k-1}^{\prime \mu \nu
}\left( \tau \right) \tilde{\eta}_{\mu \nu }^{*}
+\tfrac{1}{4}\tilde{N}_{k+1}^{\prime \prime \mu \nu \rho \lambda
}\left( \tau \right) \partial _{[\mu }\tilde{B}_{\nu \rho \lambda ]}+\\
&\quad \quad \quad \quad \quad \quad +Q_{k}^{\prime \mu \nu \rho
}\left( \tau \right) \tilde{B}_{\mu \nu \rho }^{*}
+\tfrac{1}{2}\tilde{Q}_{k+1}^{\prime \prime \mu \nu }\left( \tau
\right) \partial _{[\mu }A_{\nu ]}+L_{k-1}^{\prime \mu }\left(
\tau \right) C_{\mu }^{*}+\\
&\quad \quad \quad \quad \quad \quad +P_{k-2}^{\prime }\left( \tau
\right) \stackrel{\sim }{\mathcal{G}}^{*}+L_{k}^{\prime \mu \nu
}\left( \tau \right) \phi _{\mu \nu }^{*}+P_{k-1}^{\prime \mu
}\left( \tau \right) \stackrel{\sim }{\mathcal{G}}_{\mu
}^{*}+P_{k}^{\prime \mu
\nu }\left( \tau \right) \tilde{K}_{\mu \nu }^{*}+\\
&\quad \quad \quad \quad \quad \quad \left. \left.
+L_{k-2}^{\prime }\left( \tau \right)
C^{*}-\tfrac{1}{3}\tilde{P}_{k+1}^{\prime \prime \mu \nu \rho
}\left( \tau \right) \partial _{[\mu }\phi _{\nu \rho
]}-\tfrac{1}{3}\tilde{L}_{k}^{\prime \prime \mu \nu \rho }\left(
\tau \right) \partial _{[\mu }\tilde{K}_{\nu \rho ]}\right)
\right] +\\
&\quad +\partial _{\mu }\psi _{k}^{\mu }.
\end{split}
\end{equation}
We observe that all the terms from the integrand are invariant. In
order to prove that the current $\psi _{k}^{\mu }$ can also be
taken invariant, we switch (\ref{g32}) to the original form
notation
\begin{equation}
\alpha _{k}^{5}=\delta \lambda _{k+1}^{5}+d\lambda _{k}^{4},
\label{g33}
\end{equation}
(where $\lambda _{k}^{4}$ is dual to $\psi _{k}^{\mu }$). As
$\alpha _{k}^{5} $ is by assumption invariant and we have shown
that $\lambda _{k+1}^{5}$ can be taken invariant, (\ref{g33})
becomes
\begin{equation}
\beta _{k}^{5}=d\lambda _{k}^{4}.  \label{A107a}
\end{equation}
It states that the invariant polynomial $\beta _{k}^{5}=\alpha
_{k}^{5}-\delta \lambda _{k+1}^{5}$, of form degree $5$ and of
strictly positive antighost number, is $d$-exact. Then, in
agreement with the Theorem \ref{hginv} in form degree $5$ (see the
paragraph following this theorem), we can take $\lambda _{k}^{4}$
(or, which is the same, $\psi _{k}^{\mu }$) to be invariant. In
conclusion, the induction hypothesis in antighost
numbers $\left( k+2\right) $, $\left( k+3\right) $, $\left( k+4\right) $, $%
\left( k+5\right) $, and form degree $5$ leads to the same
property for
antighost number $k$ and form degree $5$, which proves the theorem for all $%
k\geq 5$ since we have shown that it holds for $k\geq 10$.
$\blacksquare $

The most important consequence of the last theorem is the validity
of the result (\ref{3.12x}) on the triviality of
$H^{\text{inv}}\left( \delta \vert d\right) $ in antighost number
strictly greater than five.

\section{Local cohomology of $s$, $H\left( s\vert d\right) $\label{elim4}}

Now, we have all the necessary tools for the study of the local cohomology $%
H\left( s\vert d\right) $ in form degree $5$. We will show that it
is always possible to remove the components of antighost number
strictly greater than five from any co-cycle of $H_{5}^{g}\left(
s\vert d\right) $ in form degree five only by trivial
redefinitions.

We consider a co-cycle from $H_{5}^{g}\left( s\vert d\right) $, $sa+db=0$, with $%
\deg \left( a\right) =5$, $\text{gh}\left( a\right) =g$, $\deg
\left( b\right) =4$, $\text{gh}\left( b\right) =g+1$. Trivial
redefinitions of $a$
and $b$ mean the simultaneous transformations $a\rightarrow a+sc+de$ and $%
b\rightarrow b+df+se$. We expand $a$ and $b$ according to the
antighost number and ask that $a_{0}$ is local, such that each
expansion stops at some
finite antighost number~\cite{gen2}, $a=\sum\nolimits_{k=0}^{I}a_{k}$, $%
b=\sum\nolimits_{k=0}^{M}b_{k}$, $\text{agh}\left( a_{k}\right) =k=\text{agh}%
\left( b_{k}\right) $. Due to the splitting $s=\delta +\gamma $,
the equation $sa+db=0$ is equivalent to the tower of equations
\begin{align*}
\delta a_{1}+\gamma a_{0}+db_{0}& =0, \\
& \vdots \\
\delta a_{I}+\gamma a_{I-1}+db_{I-1}& =0, \\
& \vdots
\end{align*}
The form of the last equation depends on the values of $I$ and
$M$, but we can assume, without loss of generality, that $M=I-1$.
Indeed, if $M>I-1$, the last $\left( M-I\right) $ equations read
as $db_{k}=0$, $I<k\leq M$, which imply that $b_{k}=df_{k}$, $\deg
\left( f_{k}\right) =3$. We can thus absorb all the pieces $\left(
df_{k}\right) _{I<k\leq M}$ in a trivial redefinition of $b$, such
that the new ``current'' stops at antighost number $I$.
Accordingly, the bottom equation becomes $\gamma a_{I}+db_{I}=0$,
so
the Corollary \ref{gaplusdb} ensures that we can make a redefinition $%
a_{I}\rightarrow a_{I}-d\rho _{I}$ such that $\gamma \left(
a_{I}-d\rho _{I}\right) =0$. Meanwhile, the same corollary [see
the formula (\ref{e83})] leads to $b_{I}=dg_{I}+\gamma \rho _{I}$,
where $\deg \left( \rho _{I}\right) =4$, $\deg \left( g_{I}\right)
=3$, $\text{agh}\left( \rho _{I}\right) =\text{agh}\left(
g_{I}\right) =I$, $\text{gh}\left( \rho _{I}\right) =g$,
$\text{gh}\left( g_{I}\right) =g+1$. Then, it follows that
we can make the trivial redefinitions $a\rightarrow a-d\rho _{I}$ and $%
b\rightarrow b-dg_{I}-s\rho _{I}$, such that the new ``current''
stops at antighost number $\left( I-1\right) $, while the last
component of the co-cycle from $H_{5}^{g}\left( s\vert d\right) $
is $\gamma $-closed.

In consequence, we obtained the equation $sa+db=0$, with
\begin{equation}
a=\sum\limits_{k=0}^{I}a_{k},\quad b=\sum\limits_{k=0}^{I-1}b_{k},
\label{exp}
\end{equation}
where $\text{agh}\left( a_{k}\right) =k$ for $0<k<I$ and
$\text{agh}\left( b_{k}\right) =k$ for $0<k<I-1$. All $a_{k}$ are
$5$-forms of ghost number $g$
and all $b_{k}$ are $4$-forms of ghost number $\left( g+1\right) $, with $%
\text{pgh}\left( a_{k}\right) =g+k$ for $0<k<I$ and
$\text{pgh}\left( b_{k}\right) =g+k+1$ for $0<k<I-1$. The equation
$sa+db=0$ is now equivalent with the tower of equations [where
some $\left( b_{k}\right) _{0\leq k\leq I-1}$ could vanish]
\begin{align}
\delta a_{1}+\gamma a_{0}+db_{0}& =0,  \label{A110} \\
&\vdots  \nonumber \\
\delta a_{k+1}+\gamma a_{k}+db_{k}& =0,  \label{A110ab} \\
&\vdots  \nonumber \\
\delta a_{I}+\gamma a_{I-1}+db_{I-1}& =0,  \label{A109} \\
\gamma a_{I}& =0.  \label{A108}
\end{align}
\emph{Next, we show that we can eliminate all the terms }$\left(
a_{k}\right) _{k>5}$\emph{\ and }$\left( b_{k}\right)
_{k>4}$\emph{\ from the expansions (\ref{exp}) by trivial
redefinitions only.}

We can thus assume, without loss of generality, that any co-cycle $a$ from $%
H_{5}^{g}\left( s\vert d\right) $ can be taken to stop at a value
$I>5$ of the antighost number. The last equation from the system
equivalent with $sa+db=0$ takes the form (\ref{A108}), with
$\text{pgh}\left( a_{I}\right) =g+I=L$, so $a_{I}\in H^{L}\left(
\gamma \right) $. In agreement with the general results on
$H\left( \gamma \right) $ (see Sec. \ref{hgama}) it follows that
\begin{equation}
a_{I}=\sum_{J}\alpha _{J}e^{J}+\gamma \bar{a}_{I},  \label{A110a}
\end{equation}
where $\alpha _{J}$ are invariant polynomials satisfying the
properties
\begin{equation}
\text{agh}\left( \alpha _{J}\right) =I,\quad \deg \left( \alpha
_{J}\right) =5,  \label{A110c}
\end{equation}
and $e^{J}$ are the elements of pure ghost number equal to $L$ of
a basis in
the space of polynomials in the ghosts $\eta $, $C$, $\stackrel{\sim }{%
\mathcal{G}}$, $\tilde{\eta}$, $\tilde{C}$. By acting with the operator $%
\gamma $ from the left on (\ref{A109}) and taking into account its
second-order nilpotency, as well as its anticommutation with the
exterior spacetime differential $\gamma d+d\gamma =0$, one obtains
that $-d\left( \gamma b_{I-1}\right) =0$. The triviality of the
cohomology of the differential $d$ in the space of local forms in
form degree equal to $4$ leads to
\begin{equation}
\gamma b_{I-1}+dc_{I-1}=0.  \label{A111}
\end{equation}
By means of the Corollary \ref{gaplusdb}, it follows (as $I>5$ by
assumption, so $I-1>0$) that we can make a trivial redefinition such that (%
\ref{A111}) is replaced with the equation
\begin{equation}
\gamma b_{I-1}=0.  \label{A112}
\end{equation}
In agreement with (\ref{A112}), $b_{I-1}$ belongs to $H^{L}\left(
\gamma \right) $, so we can take
\begin{equation}
b_{I-1}=\sum_{J}\beta _{J}e^{J}+\gamma \bar{b}_{I-1},
\label{A113}
\end{equation}
where $\beta _{J}$ are invariant polynomials with
\begin{equation}
\text{agh}\left( \beta _{J}\right) =I-1,\quad \deg \left( \beta
_{J}\right) =4,  \label{A115}
\end{equation}
and $e^{J}$ is the same notation like in (\ref{A110a}). Inserting
(\ref {A110a}) and (\ref{A113}) in (\ref{A109}) one infers
\begin{equation}
\pm \sum\limits_{J}\left( \delta \alpha _{J}+d\beta _{J}\right)
e^{J}=-\gamma \left( a_{I-1}+\sum\limits_{J}\beta
_{J}\hat{e}^{J}-\delta \bar{a}_{I}-d\bar{b}_{I-1}\right) ,
\label{A116}
\end{equation}
where $\hat{e}^{J}$ has been previously defined via the relation (\ref{e52}%
). Since the left-hand side of (\ref{A116}) is a nontrivial object from $%
H^{L}\left( \gamma \right) $, the equation (\ref{A116}) implies
\begin{equation}\label{A117a}
\delta \alpha _{J}=-d\beta _{J}\quad \text{for all}\quad J.
\end{equation}
The relation (\ref{A117a}) shows that the invariant polynomials
$\alpha _{J}$ belong to the space $H_{I}^{5}\left( \delta \vert
d\right) $. As $I>5$ by assumption and $H_{I}^{5}\left( \delta
\vert d\right) =0$ for $I>5$, it follows that all the invariant
polynomials $\alpha _{J}$ are trivial
\begin{equation}
\alpha _{J}=\delta \lambda _{I+1,J}^{5}+d\lambda _{I,J}^{4},
\label{A119}
\end{equation}
where $\lambda _{I+1,J}^{5}$ are $5$-forms of antighost number
$\left( I+1\right) $, while $\lambda _{I,J}^{4}$ are $4$-forms
with the antighost number equal to $I$. Theorem \ref{hinvdelta}
then ensures that we can also take $\lambda _{I+1,J}^{5}$ and
$\lambda _{I,J}^{4}$ to be invariant
polynomials, and thus $\alpha _{J}$ are in fact trivial elements of $H_{I}^{%
\text{inv}5}\left( \delta \vert d\right) $. Replacing (\ref{A119}) in (\ref{A117a}%
) and using the relations $\delta ^{2}=0$ and $\delta d+d\delta
=0$, we deduce that $d\left( -\delta \lambda _{I,J}^{4}+\beta
_{J}\right) =0$. Because both $\lambda _{I,J}^{4}$ and $\beta
_{J}$ are invariant polynomials with strictly positive values of
the antighost number and with the form degree equal to $4$,
Theorem \ref{hginv} yields $-\delta \lambda _{I,J}^{4}+\beta
_{J}=d\lambda _{I-1,J}^{3}$, where $\lambda _{I-1,J}^{3}$ are also
invariant polynomials with $\text{agh}\left( \lambda
_{I-1,J}^{3}\right) =I-1$ and $\deg \left( \lambda
_{I-1,J}^{3}\right) =3$, such that we can write
\begin{equation}
\beta _{J}=\delta \lambda _{I,J}^{4}+d\lambda _{I-1,J}^{3}.
\label{A120}
\end{equation}
Substituting (\ref{A119}) in (\ref{A110a}), after some computation
we get that $a_{I}$ is expressed in the form
\begin{align}
a_{I}& =\sum_{J}\left( \delta \lambda _{I+1,J}^{5}+d\lambda
_{I,J}^{4}\right) e^{J} = \nonumber \\
& =s\left( \pm \sum_{J}\lambda _{I+1,J}^{5}e^{J}\right) +d\left(
\pm \sum_{J}\lambda _{I,J}^{4}e^{J}\right) \mp \sum_{J}\left(
\lambda _{I,J}^{4}de^{J}\right) .  \label{A121}
\end{align}
Due to the fact that $de^{J}=\gamma \hat{e}^{J}$ and $\gamma
\lambda _{I,J}^{4}=0$, we consequently have that
\begin{equation}
a_{I}=s\left( \pm \sum_{J}\lambda _{I+1,J}^{5}e^{J}\right)
+d\left( \pm \sum_{J}\lambda _{I,J}^{4}e^{J}\right) \mp \gamma
\left( \sum_{J}\lambda _{I,J}^{4}\hat{e}^{J}\right) . \label{A122}
\end{equation}
In a similar manner, with the help of the relation (\ref{A120}) inserted in (%
\ref{A113}), we arrive to
\begin{equation}
b_{I-1}=s\left( \pm \sum_{J}\lambda _{I,J}^{4}e^{J}\right)
+d\left( \pm \sum_{J}\lambda _{I-1,J}^{3}e^{J}\right) \mp \gamma
\left( \sum_{J}\lambda _{I-1,J}^{3}\hat{e}^{J}\right) .
\label{A123}
\end{equation}
Now, if we simultaneously perform some trivial redefinitions of
$a_{I}$ and of the `current' $b_{I-1}$ like
\begin{eqnarray}
a_{I}^{\prime } &=&a_{I}-s\left( \pm \sum_{J}\lambda
_{I+1,J}^{5}e^{J}\right) -d\left( \pm \sum_{J}\lambda
_{I,J}^{4}e^{J}\right)
,  \label{A124a} \\
b_{I-1}^{\prime } &=&b_{I-1}-s\left( \pm \sum_{J}\lambda
_{I,J}^{4}e^{J}\right) -d\left( \pm \sum_{J}\lambda
_{I-1,J}^{3}e^{J}\right) ,  \label{A124b}
\end{eqnarray}
and, meanwhile, fix $\bar{a}_{I}$ and $\bar{b}_{I-1}$ from
(\ref{A110a}) and respectively from (\ref{A113}) as
\begin{eqnarray}
\bar{a}_{I} &=&\pm \sum_{J}\lambda _{I,J}^{4}\hat{e}^{J},  \label{A124c} \\
\bar{b}_{I-1} &=&\pm \sum_{J}\lambda _{I-1,J}^{3}\hat{e}^{J},
\label{A124d}
\end{eqnarray}
then both $a_{I}^{\prime }$ and $b_{I-1}^{\prime }$ become equal
to zero. Reprising now exactly the same procedure like before, but
for the antighost number equal to $\left( I-1\right) $, we find
that we can take $a_{I-1}=0=b_{I-2}$ by trivial redefinitions
only. The elimination procedure stops at $k=5$ in the tower
(\ref{A110})--(\ref{A108}) since for $k=5$ we cannot pass from
(\ref{A117a}) to (\ref{A119}) as $H_{5}\left( \delta \vert
d\right) \neq 0$. In conclusion, we can replace $a_{k}$ with
$a_{k}=0$ by trivial redefinitions for all $k>5$ in the tower
(\ref{A110})--(\ref{A108}), such that the first-order deformation
can always be taken to end at $I=5$ (formula (\ref{fr1})), with
$a_{5}$ from $H\left( \gamma \right) $, $\gamma a_{5}=0$.
Furthermore, the above arguments show that $a_{5}$ can be assumed
to involve only non-trivial elements from $H^{\mathrm{inv}}\left(
\delta \vert d\right) $.

\section{Solution to the `homogeneous' equation (\ref{aze1in})\label{homog}}

In the sequel we consider the consistent interactions that do not
modify the original gauge transformations, which are solutions to
the equation
\begin{equation}
\gamma \bar{a}_{0}=\partial _{\mu }j_{0}^{\mu },  \label{aze1}
\end{equation}
where $\bar{a}_{0}$ has the ghost number and the pure ghost number
equal to
zero. The solutions to this `homogeneous' equation come from $\bar{a}_{1}=0$%
, and hence they bring contributions only to the deformed
lagrangian density at order one in the coupling constant. We
maintain all the hypotheses introduced in the beginning of Section
\ref{inter} (smoothness, locality,
etc.), including the condition on the maximum derivative order of $\bar{a}%
_{0}$ being equal to one. There are two main types of solutions to
this equation.

The first type, to be denoted by $\bar{a}_{0}^{\prime }$, corresponds to $%
j_{0}^{\mu }=0$, and is given by gauge-invariant, nonintegrated
densities constructed out of the original fields and their
spacetime derivatives, which, according to (\ref{3.7}), are of the
form
\begin{equation}
\bar{a}_{0}^{\prime }=\bar{a}_{0}^{\prime }\left( \left[
F_{\bar{A}}\right] \right) ,  \label{3.41}
\end{equation}
where $\bar{a}_{0}^{\prime }$ may contain at most one derivative
of the fields. The sole possibility that complies with all the
hypotheses on the interactions mentioned before is
\begin{equation}
\bar{a}_{0}^{\prime }=\bar{M}\left( \varphi \right) +N\left(
\varphi \right)
\partial _{\mu }H^{\mu },  \label{3.41a}
\end{equation}
where $\bar{M}\left( \varphi \right) $ and $N\left( \varphi
\right) $\ are smooth, arbitrary real functions depending only on
the undifferentiated
scalar field. The second term in the right-hand side of (\ref{3.41a}) is $%
\delta $-exact
\[
N\left( \varphi \right) \partial _{\mu }H^{\mu }=\delta \left(
N\left( \varphi \right) \varphi ^{*}\right) ,
\]
and hence produces trivial interactions, that can be eliminated
via field redefinitions. This is due to the isomorphism
$H^{k}\left( s\vert d\right) \simeq H^{k}\left( \gamma \vert
d,H_{0}\left( \delta \right) \right) $ in all positive values of
the ghost number and respectively of the pure ghost number
\cite{gen1}, which at $k=0$ allows one to state that any solution
of the homogeneous equation (\ref{aze1}) that is $\delta $-exact
modulo $d$ is in fact a trivial co-cycle from $H^{0}\left( s\vert
d\right) $. In conclusion, the only nontrivial solution to
(\ref{aze1}) for $j_{0}^{\mu }=0$ is represented by
\begin{equation}
\bar{a}_{0}^{\prime }=\bar{M}\left( \varphi \right) .
\label{aze7}
\end{equation}
Although this solution does not contribute to the deformed gauge
transformations of the interacting model, however it is important
since it is involved in the consistency of the first-order
deformation, as we have seen in the subsection \ref{highord} [see
formula (\ref{fr8})].

The second kind of solutions, to be denoted by $\bar{a}_{0}^{\prime \prime }$%
, is associated with $j_{0}^{\mu }\neq 0$ in (\ref{aze1}), being
understood that we discard the divergence-like quantities and work
under the hypothesis on the maximum derivative order of the
interacting lagrangian being equal to one. Then,
$\bar{a}_{0}^{\prime \prime }$ can be decomposed like
\begin{equation}
\bar{a}_{0}^{\prime \prime \left( \text{int}\right) }=\omega
_{0}+\omega _{1}  \label{ww60}
\end{equation}
where $\left( \omega _{i}\right) _{i=0,1}$ contains $i$
derivatives of the fields. Due to the different number of
derivatives in the components $\omega _{0}$ and $\omega _{1}$, the
equation (\ref{aze1}) leads to two independent equations
\begin{align}
\gamma \omega _{0} & =\partial _{\mu }m_{0}^{\mu },  \label{ww62} \\
\gamma \omega _{1} & =\partial _{\mu }m_{1}^{\mu }.  \label{wwxy}
\end{align}
Since $\omega _{0}$ is derivative-free, we find that
\begin{equation}\label{aze9}
\begin{split}
\gamma \omega _{0} & =\frac{\partial \omega _{0}}{\partial H^{\mu
}}\left( 2\partial _{\nu }C^{\mu \nu }\right) +\frac{\partial
\omega _{0}}{\partial
A_{\mu }}\left( \partial _{\mu }\eta \right) +\frac{\partial \omega _{0}}{%
\partial B^{\mu \nu }}\left( -3\partial _{\rho }\eta ^{\mu \nu \rho
}\right) +\\
& \quad +\frac{\partial \omega _{0}}{\partial \phi _{\mu \nu
}}\left(
\partial _{[\mu }C_{\nu ]}\right) +\frac{\partial \omega
_{0}}{\partial K^{\mu \nu \rho }}\left( 4\partial _{\lambda
}\mathcal{G}^{\mu \nu \rho \lambda }\right) .
\end{split}
\end{equation}
The right-hand side of the last equation reduces to a full
derivative if the following conditions are simultaneously
satisfied
\begin{align}
\partial _{[\nu }\left( \frac{\partial \omega _{0}}{%
\partial H^{\mu ]}}\right) & =0,&
\partial _{\mu }\left( \frac{\partial
\omega _{0}}{\partial A_{\mu }}\right) &=0,& \partial _{[\mu
}\left( \frac{\partial \omega _{0}}{\partial B^{\nu \rho ]}}
\right) & =0,  \label{aze10a} \\
\partial _{\mu }\left( \frac{\partial \omega _{0}}{\partial
\phi _{\mu \nu }}\right) & =0,& \partial _{[\mu }\left(
\frac{\partial \omega _{0}}{\partial K^{\nu \rho \lambda
]}}\right) & =0.  \label{aze10c}
\end{align}
Since neither of the functions present in (\ref{aze10a}) and
(\ref{aze10c}) on which the derivatives act contain spacetime
derivatives, these equations possess the solutions
\begin{align}
\frac{\partial \omega _{0}}{\partial H^{\mu }} & =c_{\mu },&\frac{%
\partial \omega _{0}}{\partial A_{\mu }}& =k^{\mu },&
\frac{\partial \omega _{0}}{\partial B^{\mu \nu }} & =c_{\mu \nu
}, \label{we1} \\
\frac{\partial \omega _{0}}{\partial \phi _{\mu \nu }}& =k^{\mu
\nu },& \frac{\partial \omega _{0}}{\partial K^{\mu \nu \rho }} &
=c_{\mu \nu \rho }, \label{we3}
\end{align}
where $c_{\mu }$, $k^{\mu }$, $c_{\mu \nu }$, $k^{\mu \nu }$, and
$c_{\mu \nu \rho }$ are real, non-derivative constants, the last
three sets being completely antisymmetric. Since there are no such
constants, they must vanish, and therefore we have that $\omega
_{0}=\omega _{0}\left( \varphi \right) $, which is nothing but the
most general solution to (\ref{aze1}) for $j_{0}^{\mu }=0$, given
in (\ref{aze7}). In conclusion, there is no consistent solution to
the equation (\ref{aze1}) for $j_{0}^{\mu }\neq 0$ that contains
no derivatives of the fields, so we can take, without loss of
generality, $\omega _{0}=0$ in the expansion (\ref{ww60}).

The last step is to determine the consistent solutions $\omega
_{1}$ to (\ref{wwxy}) for $m_{1}^{\mu }\neq 0$ with just one
spacetime derivative. In view of this, the general representation
of $\omega _{1}$ is $\omega _{1}([\varphi ],[H^{\mu }],[B^{\mu \nu
}],[A_{\mu }],[\phi _{\mu \nu }],[K^{\mu \nu \rho }])$. However,
one can always move the single derivative present in $\omega _{1}$
to act on all the fields but $\varphi $, so we will work with
\begin{equation}
\omega _{1}\left( \varphi ,\left[ H^{\mu }\right] ,\left[ B^{\mu
\nu }\right] ,\left[ A_{\mu }\right] ,\left[ \phi _{\mu \nu
}\right] ,\left[ K^{\mu \nu \rho }\right] \right) .
\label{omega1}
\end{equation}
Taking into account the definitions of $\gamma $, we obtain that
\begin{equation}\label{aze12}
\begin{split}
\gamma \omega _{1}& =\frac{\partial \omega _{1}}{\partial H^{\mu
}}\left( 2\partial _{\nu }C^{\mu \nu }\right) +\frac{\partial
\omega _{1}}{\partial \left( \partial ^{\alpha }H^{\mu }\right)
}\left( 2\partial _{\nu }\partial
^{\alpha }C^{\mu \nu }\right) + \\
&\quad +\frac{\partial \omega _{1}}{\partial B^{\mu \nu }}\left(
-3\partial _{\rho }\eta ^{\mu \nu \rho }\right) +\frac{\partial
\omega _{1}}{\partial \left( \partial ^{\alpha }B^{\mu \nu
}\right) }\left( -3\partial _{\rho
}\partial ^{\alpha }\eta ^{\mu \nu \rho }\right) + \\
&\quad +\frac{\partial \omega _{1}}{\partial A_{\mu }}\left(
\partial _{\mu }\eta \right) +\frac{\partial \omega _{1}}{\partial
\left(
\partial ^{\alpha }A_{\mu }\right) }\left( \partial _{\mu
}\partial ^{\alpha }\eta \right) + \\
&\quad +\frac{\partial \omega _{1}}{\partial \phi _{\mu \nu
}}\left(
\partial _{[\mu }C_{\nu ]}\right) +\frac{\partial \omega
_{1}}{\partial \left(
\partial ^{\alpha }\phi _{\mu \nu }\right) }\left( \partial ^{\alpha
}\partial _{[\mu }C_{\nu ]}\right) + \\
&\quad +\frac{\partial \omega _{1}}{\partial K^{\mu \nu \rho
}}\left( 4\partial _{\lambda }\mathcal{G}^{\mu \nu \rho \lambda
}\right) +\frac{\partial \omega _{1}}{\partial \left( \partial
^{\alpha }K^{\mu \nu \rho }\right) }\left( 4\partial _{\lambda
}\partial ^{\alpha }\mathcal{G}^{\mu \nu \rho \lambda }\right) .
\end{split}
\end{equation}
By successively moving all the derivatives from the ghosts, we
observe that the right-hand side of (\ref{aze12}) reduces to a
total divergence if
\begin{align}
\partial _{[\mu }\left( \frac{\delta \omega _{1}}{\delta
H^{\nu ]}}\right)  & =0,&
\partial _{\mu }\left( \frac{\delta \omega _{1}}{\delta A_{\mu }}\right)
& =0,& \partial _{[\mu }\left( \frac{\delta
\omega _{1}}{\delta B^{\nu \rho ]}}\right)  & =0, \label{aze13a} \\
\partial _{\mu }\left( \frac{\delta \omega _{1}}{\delta \phi _{\mu \nu }}%
\right)  & =0,& \partial _{[\mu }\left( \frac{\delta \omega
_{1}}{\delta K^{\nu \rho \lambda ]}}\right)  & =0.  \label{aze13e}
\end{align}
The general solutions to the equations (\ref{aze13a}) and
(\ref{aze13e}) are
\begin{align}
\frac{\delta \omega _{1}}{\delta H^{\mu }} & =\partial _{\mu }D,&
\frac{\delta \omega _{1}}{\delta A_{\mu }} & =\partial _{\nu
}D^{\mu \nu },& \frac{\delta \omega _{1}}{\delta B^{\mu \nu }} &
=\partial _{[\mu }D_{\nu ]},
\label{aze21a} \\
\frac{\delta \omega _{1}}{\delta \phi _{\mu \nu }} & =\partial
_{\rho }E^{\mu \nu \rho },& \frac{\delta \omega _{1}}{\delta
K^{\mu \nu \rho }} & =\partial _{[\mu }E_{\nu \rho ]},
\label{aze21e}
\end{align}
where in our case the functions $D$, $D_{\mu }$, $D^{\mu \nu }$,
$E^{\mu \nu
\rho }$, and $E_{\mu \nu }$ \emph{depend only on the undifferentiated fields}%
, with $D^{\mu \nu }$, $E^{\mu \nu \rho }$, and $E_{\mu \nu }$
completely antisymmetric. In order to analyze the structure of
these functions, it is convenient to introduce an operator $N$
that counts all the fields excepting
$\varphi $%
\begin{equation}
\bar{\Phi}^{\alpha _{0}}=\left( H^{\mu },A_{\mu },B^{\mu \nu
},\phi _{\mu \nu },K^{\mu \nu \rho }\right) ,  \label{xw1}
\end{equation}
and their derivatives, defined through
\begin{equation}
N=\sum\limits_{k\geq 0}\left( \partial _{\mu _{1}\cdots \mu _{k}}\bar{\Phi}%
^{\alpha _{0}}\right) \frac{\partial }{\partial \left( \partial
_{\mu _{1}\cdots \mu _{k}}\bar{\Phi}^{\alpha _{0}}\right) }.
\label{num}
\end{equation}
Then, it is simple to see that, for every nonintegrated density
$\chi $ depending on $\bar{\Phi}^{\alpha _{0}}$, their
derivatives, and \emph{the undifferentiated scalar field} $\varphi
$, we have that
\begin{equation}
N\chi =\bar{\Phi}^{\alpha _{0}}\frac{\delta \chi }{\delta
\bar{\Phi}^{\alpha _{0}}}+\partial ^{\mu }j_{\mu },  \label{rel}
\end{equation}
where $j_{\mu }$ are some local currents. Let $\chi ^{\left(
k\right) }$ be a homogeneous polynomial of order $k>0$ in the
fields $\bar{\Phi}^{\alpha _{0}}$ and their derivatives. Then, it
follows that
\begin{equation}
N\chi ^{\left( k\right) }=k\chi ^{\left( k\right) }.  \label{zw1}
\end{equation}
Using the solutions (\ref{aze21a}) and (\ref{aze21e}) in
(\ref{rel}), we get that
\begin{equation}\label{yw1}
\begin{split}
N\omega _{1} & =-D\partial _{\mu }H^{\mu }+\frac{1}{2}D^{\mu \nu
}\partial
_{[\mu }A_{\nu ]}-2D_{\mu }\partial _{\nu }B^{\nu \mu }- \\
&\quad -\frac{1}{3}E^{\mu \nu \rho }\partial _{[\mu }\phi _{\nu
\rho ]}-3E_{\mu \nu }\partial _{\rho }K^{\mu \nu \rho }+\partial
^{\mu }v_{\mu }.
\end{split}
\end{equation}
At this stage, we decompose $\omega _{1}$ under the form
\begin{equation}
\omega _{1}=\sum\limits_{k>0}\omega _{1}^{(k)},  \label{desc}
\end{equation}
[the value $k=0$ is excluded due to the fact that we work with
$\omega _{1}$ like in (\ref{omega1})], where $N$ acts on the
component $\omega _{1}^{(k)}$ via
\begin{equation}
N\omega _{1}^{(k)}=k\omega _{1}^{(k)}.  \label{valp}
\end{equation}
Substituting (\ref{valp}) in the expansion (\ref{desc}), we infer
that
\begin{equation}
N\omega _{1}=\sum\limits_{k>0}k\omega _{1}^{(k)}.  \label{rez}
\end{equation}
Comparing (\ref{yw1}) with (\ref{rez}), we conclude that the decomposition (%
\ref{desc}) induces a similar decomposition with respect to the functions $D$%
, $D_{\mu }$, $D^{\mu \nu }$, $E_{\mu \nu }$, and $E^{\mu \nu \rho
}$, i.e.
\begin{align}
D & =\sum\limits_{k>0}D^{(k-1)},& D_{\mu } &
=\sum\limits_{k>0}D_{\mu }^{(k-1)},& D^{\mu \nu } & =
\sum\limits_{k>0}D^{(k-1)\mu \nu },\label{wwx1} \\
E_{\mu \nu } & =\sum\limits_{k>0}E_{\mu \nu }^{(k-1)},& E^{\mu \nu
\rho } & =\sum\limits_{k>0}E^{(k-1)\mu \nu \rho }. \label{wwx5}
\end{align}
Inserting now the outcomes (\ref{wwx1}) and (\ref{wwx5}) in
(\ref{yw1}) and then comparing the corresponding result with
(\ref{rez}), we deduce that
\begin{equation}\label{aze24}
\begin{split}
\omega _{1}^{(k)} & =-\frac{1}{k}\left( D^{(k-1)}\partial _{\mu }H^{\mu }-%
\frac{1}{2}D^{(k-1)\mu \nu }\partial _{[\mu }A_{\nu ]}+2D_{\mu
}^{(k-1)}\partial _{\nu }B^{\nu \mu }\right. +\\
&\quad \left. +\frac{1}{3}E^{(k-1)\mu \nu \rho }\partial _{[\mu
}\phi _{\nu \rho ]}+3E_{\mu \nu }^{(k-1)}\partial _{\rho }K^{\mu
\nu \rho }\right) +\partial ^{\mu }v_{\mu }^{\left( k\right) }.
\end{split}
\end{equation}
Putting together the relations (\ref{aze24}) for the various
values of $k$ like in (\ref{desc}), we are able to reconstruct
$\omega _{1}$ like
\begin{equation}\label{aze25}
\begin{split}
\omega _{1} & =-\bar{D}\partial _{\mu }H^{\mu
}+\frac{1}{2}\bar{D}^{\mu \nu }\partial _{[\mu }A_{\nu
]}-2\bar{D}_{\mu }\partial _{\nu }B^{\nu \mu }-\\
&\quad -\frac{1}{3}\bar{E}^{\mu \nu \rho }\partial _{[\mu }\phi _{\nu \rho ]}-3%
\bar{E}_{\mu \nu }\partial _{\rho }K^{\mu \nu \rho }+\partial ^{\mu }\bar{v}%
_{\mu },
\end{split}
\end{equation}
where
\begin{align}
\bar{D} & =\sum\limits_{k>0}\frac{1}{k}D^{(k-1)},& \bar{D}_{\mu }
& =\sum\limits_{k>0}\frac{1}{k}D_{\mu }^{(k-1)},& \bar{D}^{\mu \nu
} & =\sum\limits_{k>0}\frac{1}{k}D^{(k-1)\mu \nu
},\label{aw1} \\
\bar{E}_{\mu \nu } & =\sum\limits_{k>0}\frac{1}{k}E_{\mu \nu
}^{(k-1)},& \bar{E}^{\mu \nu \rho } &
=\sum\limits_{k>0}\frac{1}{k}E^{(k-1)\mu \nu \rho }.  \label{aw5}
\end{align}
It is clear from the definitions of $\delta $ that (\ref{aze25}) is in fact $%
\delta $-exact modulo $d$%
\begin{equation}
\omega _{1}=\delta \left[ -\left( \bar{D}\varphi ^{*}+\bar{D}^{\mu
\nu }B_{\mu \nu }^{*}+2\bar{D}_{\mu }A^{*\mu }-\bar{E}^{\mu \nu
\rho }K_{\mu \nu \rho }^{*}+3\bar{E}_{\mu \nu }\phi ^{*\mu \nu
}\right) \right] +\partial ^{\mu }\bar{v}_{\mu }.  \label{aze27}
\end{equation}
By virtue of the above discussion on trivial interactions, we can
state that $\omega _{1}$ is trivial, and therefore we can take
$\omega _{1}=0$.

In conclusion, the general, nontrivial, consistent solution to the equation (%
\ref{aze1}) takes the simple form
\begin{equation}
\bar{a}_{0}=\bar{M}\left( \varphi \right) .  \label{aze26}
\end{equation}

\section{Some notations made in the body of the paper\label{A}}

The various notations used in (\ref{pt13}) are listed below. Thus,
the objects denoted by $X^{\left( 0\right) }$ and $\left(
X_{a}^{\left( 1\right) }\right) _{a=\overline{0,5}}$ are expressed
by
\begin{equation}
X^{\left( 0\right) }=-2\eta , \label{hnou}
\end{equation}
\begin{equation}
X_{0}^{\left( 1\right) }=-4\left[ 3\left( \mathcal{G}_{\mu \nu
\rho \lambda \sigma }^{*}C^{\mu \nu \rho \lambda \sigma
}+\mathcal{G}_{\mu \nu \rho \lambda }^{*}C^{\mu \nu \rho \lambda
}+K_{\mu \nu \rho }^{*}C^{\mu \nu \rho }\right) +\phi _{\mu \nu
}C^{\mu \nu }\right] ,  \label{h5}
\end{equation}
\begin{equation}\label{h6}
\begin{split}
X_{1}^{\left( 1\right) }& =4\left[ \left( C_{\mu \nu \rho \lambda
\sigma }^{*}C-C_{[\mu \nu \rho \lambda }^{*}C_{\sigma ]}-C_{[\mu
\nu \rho }^{*}\phi _{\lambda \sigma ]}-3C_{[\mu \nu }^{*}K_{\rho
\lambda \sigma ]}^{*}-3H_{[\mu }^{*}\mathcal{G}_{\nu \rho \lambda
\sigma
]}^{*}\right) C^{\mu \nu \rho \lambda \sigma } \right. +\\
&\quad \quad +\left( C_{\mu \nu \rho \lambda }^{*}C-C_{[\mu \nu
\rho }^{*}C_{\lambda ]}-C_{[\mu \nu }^{*}\phi _{\rho \lambda
]}-3H_{[\mu }^{*}K_{\nu \rho \lambda ]}^{*}\right)
C^{\mu \nu \rho \lambda }+ \\
&\quad \quad \left. +\left( C_{\mu \nu \rho }^{*}C-C_{[\mu \nu
}^{*}C_{\rho ]}-H_{[\mu }^{*}\phi _{\nu \rho ]}\right) C^{\mu \nu
\rho }+\left( C_{\mu \nu }^{*}C-H_{[\mu }^{*}C_{\nu ]}\right)
C^{\mu \nu }\right] ,
\end{split}
\end{equation}
\begin{equation}\label{h7}
\begin{split}
X_{2}^{\left( 1\right) }& =4\left\{ \left[ \left( H_{[\mu
}^{*}C_{\nu \rho \lambda \sigma ]}^{*}+C_{[\mu \nu }^{*}C_{\rho
\lambda \sigma ]}^{*}\right) C-5\left( H_{[\mu }^{*}C_{\nu \rho
\lambda ]}^{*}+C_{[\mu \nu }^{*}C_{\rho
\lambda ]}^{*}\right) C_{\sigma }\right. \right. - \\
&\quad \quad \quad \left. -H_{[\mu }^{*}C_{\nu \rho }^{*}\phi
_{\lambda \sigma ]}-3H_{[\mu }^{*}H_{\nu }^{*}K_{\rho \lambda
\sigma ]}^{*}\right] C^{\mu \nu \rho \lambda \sigma }+\\
& \quad \quad +\left[ \left( H_{[\mu }^{*}C_{\nu \rho \lambda
]}^{*}+C_{[\mu \nu }^{*}C_{\rho \lambda ]}^{*}\right) C-H_{[\mu
}^{*}C_{\nu \rho }^{*}C_{\lambda ]}-H_{[\mu
}^{*}H_{\nu }^{*}\phi _{\rho \lambda ]}\right] C^{\mu \nu \rho \lambda }+\\
&\quad \quad \left. +\left( H_{[\mu }^{*}C_{\nu \rho
]}^{*}C-H_{[\mu }^{*}H_{\nu }^{*}C_{\rho ]}\right) C^{\mu \nu \rho
}+H_{\mu }^{*}H_{\nu }^{*}CC^{\mu \nu }\right\} ,
\end{split}
\end{equation}
\begin{equation}\label{h8}
\begin{split}
X_{3}^{\left( 1\right) }& =4\left\{ \left[ \left( H_{[\mu
}^{*}H_{\nu }^{*}C_{\rho \lambda \sigma ]}^{*}+H_{[\mu }^{*}C_{\nu
\rho }^{*}C_{\lambda \sigma ]}^{*}\right) C-H_{[\mu }^{*}H_{\nu
}^{*}C_{\rho \lambda
}^{*}C_{\sigma ]}-\right. \right. \\
&\quad \quad \quad \left. -H_{[\mu }^{*}H_{\nu }^{*}H_{\rho
}^{*}\phi _{\lambda \sigma ]}\right] C^{\mu \nu \rho \lambda
\sigma }+\left( H_{[\mu }^{*}H_{\nu }^{*}C_{\rho \lambda
]}^{*}C-H_{[\mu }^{*}H_{\nu }^{*}H_{\rho
}^{*}C_{\lambda ]}\right) C^{\mu \nu \rho \lambda } +\\
&\quad \quad \left. +H_{[\mu }^{*}H_{\nu }^{*}H_{\rho
]}^{*}CC^{\mu \nu \rho }\right\} ,
\end{split}
\end{equation}
\begin{equation}
X_{4}^{\left( 1\right) }=4H_{\mu }^{*}H_{\nu }^{*}H_{\rho
}^{*}\left[ 5\left( 2C_{\lambda \sigma }^{*}C-H_{\lambda
}^{*}C_{\sigma }\right) C^{\mu \nu \rho \lambda \sigma
}+H_{\lambda }^{*}CC^{\mu \nu \rho \lambda }\right] , \label{h9}
\end{equation}
\begin{equation}
X_{5}^{\left( 1\right) }=4H_{\mu }^{*}H_{\nu }^{*}H_{\rho
}^{*}H_{\lambda }^{*}H_{\sigma }^{*}CC^{\mu \nu \rho \lambda
\sigma }.  \label{h10}
\end{equation}

The notations $\left( X_{a}^{\left( 2\right) }\right)
_{a=\overline{0,5}}$ signify the functions
\begin{equation}\label{h11}
\begin{split}
X_{0}^{\left( 2\right) }& =4\left\{ \left( \eta _{\mu \nu \rho
\lambda \sigma }^{*}C+\eta _{[\mu \nu \rho \lambda }^{*}C_{\sigma
]}-\eta _{[\mu \nu \rho }^{*}\phi _{\lambda \sigma ]}+3B_{[\mu \nu
}^{*}K_{\rho \lambda \sigma
]}^{*}-\right. \right.  \\
&\quad \quad \quad \left. -\tfrac{3}{2}A_{[\mu }\mathcal{G}_{\nu
\rho \lambda \sigma ]}^{*}+\tfrac{1}{2}\mathcal{G}_{\mu \nu \rho
\lambda \sigma }^{*}\eta \right) \eta ^{\mu \nu \rho \lambda
\sigma }+\\
&\quad \quad +\left[ \eta _{\mu \nu \rho \lambda }^{*}C+\eta
_{[\mu \nu \rho }^{*}C_{\lambda ]}-B_{[\mu \nu }^{*}\phi _{\rho
\lambda ]}+\tfrac{3}{2}\left( A_{[\mu }K_{\nu \rho \lambda
]}^{*}-\mathcal{G}_{\mu \nu \rho \lambda }^{*}\eta \right)
\right] \eta ^{\mu \nu \rho \lambda } +\\
&\quad \quad +\left( \eta _{\mu \nu \rho }^{*}C+B_{[\mu \nu
}^{*}C_{\rho ]}-\tfrac{1}{2}A_{[\mu }\phi _{\nu \rho
]}-\tfrac{3}{2}K_{\mu \nu \rho }^{*}\eta \right) \eta ^{\mu \nu
\rho }-\eta ^{*}\eta C-\\
&\quad \quad \left. -A_{\mu }^{*}\left( A^{\mu }C+\eta C^{\mu
}\right) +\left( B_{\mu \nu }^{*}C+\tfrac{1}{2}A_{[\mu }C_{\nu
]}-\tfrac{1}{2}\phi _{\mu \nu }\eta \right) B^{\mu \nu }\right\} ,
\end{split}
\end{equation}
\begin{equation}\label{h12}
\begin{split}
X_{1}^{\left( 2\right) }& =2\left\{ \left[ C_{\mu \nu \rho \lambda
\sigma }^{*}\eta C+5C_{\mu \nu \rho \lambda }^{*}\left( A_{\sigma
}C+\eta C_{\sigma }\right) +20C_{\mu \nu \rho }^{*}\left(
B_{\lambda \sigma }^{*}C+A_{\lambda }C_{\sigma }-\tfrac{1}{2}\phi
_{\lambda \sigma }\eta \right)+
\right. \right.  \\
&\quad \quad \quad +20C_{\mu \nu }^{*}\left( \eta _{\rho \lambda
\sigma }^{*}C+B_{[\rho \lambda }^{*}C_{\sigma
]}-\tfrac{1}{2}A_{[\rho }\phi _{\lambda \sigma ]}
-\tfrac{3}{2}K_{\rho \lambda \sigma }^{*}\eta
\right) + \\
&\quad \quad \quad \left. +10H_{\mu }^{*}\left( \eta _{\nu \rho
\lambda \sigma }^{*}C+\eta _{[\nu \rho \lambda }^{*}C_{\sigma
]}-B_{[\nu \rho }^{*}\phi _{\lambda \sigma ]}+3A_{[\nu }K_{\rho
\lambda \sigma ]}^{*}-\tfrac{3}{2}\mathcal{G}_{\nu \rho \lambda
\sigma }^{*}\eta \right) \right] \eta ^{\mu \nu \rho \lambda
\sigma }+
\\
&\quad \quad +\left[ C_{\mu \nu \rho \lambda }^{*}\eta C+4C_{\mu
\nu \rho }^{*}\left( A_{\lambda }C+\eta C_{\lambda }\right)
+12C_{\mu \nu }^{*}\left( B_{\rho \lambda }^{*}C+A_{[\rho
}C_{\lambda ]}-\tfrac{1}{2}\eta \phi _{\rho \lambda
}\right)+\right.
\\
&\quad \quad \quad \left. +8H_{\mu }^{*}\left( \eta _{\nu \rho
\lambda }^{*}C+B_{[\nu \rho }^{*}C_{\lambda ]}-\tfrac{1}{2}A_{[\nu
}\phi _{\rho \lambda ]}-\tfrac{3}{2}K_{\nu \rho \lambda }^{*}\eta
\right) \right] \eta ^{\mu \nu \rho \lambda }+\\
&\quad \quad +\left[ C_{\mu \nu \rho }^{*}\eta C+3C_{\mu \nu
}^{*}\left( A_{\rho }C+\eta C_{\rho }\right) +6H_{\mu }^{*}\left(
B_{\nu \rho }^{*}C+\tfrac{1}{2}\left( A_{[\nu }C_{\rho ]}-\phi
_{\nu \rho }\eta \right) \right) \right] \eta ^{\mu \nu \rho } + \\
&\quad \quad \left. +\left[ C_{\mu \nu }^{*}\eta C+2H_{\mu
}^{*}\left( A_{\nu }C+\eta C_{\nu }\right) \right] B^{\mu \nu
}+2H_{\mu }^{*}A^{*\mu }\eta C\right\} ,
\end{split}
\end{equation}
\begin{equation}\label{h13}
\begin{split}
X_{2}^{\left( 2\right) }& =2\left\{ \left[ \left( H_{[\mu
}^{*}C_{\nu \rho \lambda \sigma ]}^{*}+C_{[\mu \nu }^{*}C_{\rho
\lambda \sigma ]}^{*}\right) \eta C+5\left( H_{[\mu }^{*}C_{\nu
\rho \lambda ]}^{*}+C_{[\mu \nu }^{*}C_{\rho \lambda ]}^{*}\right)
\left( A_{\sigma }C+\eta C_{\sigma }\right) +\right. \right. \\
&\quad \quad \quad +20H_{[\mu }^{*}C_{\nu \rho ]}^{*}\left(
B_{\lambda \sigma }^{*}C+\tfrac{1}{2}\left( A_{[\lambda }C_{\sigma
]}-\phi _{\lambda \sigma }\eta \right) \right) + \\
&\quad \quad \quad \left.+20H_{\mu }^{*}H_{\nu }^{*}\left( \eta
_{\rho \lambda \sigma }^{*}C+B_{[\rho \lambda }^{*}C_{\sigma
]}-\tfrac{1}{2}\left( A_{[\rho }\phi _{\lambda \sigma ]}+3K_{\rho
\lambda \sigma }^{*}\eta \right) \right) \right] \eta ^{\mu \nu
\rho \lambda \sigma }+\\
&\quad \quad +\left[ \left( H_{[\mu }^{*}C_{\nu \rho \lambda
]}^{*}+C_{[\mu \nu }^{*}C_{\rho \lambda ]}^{*}\right) \eta
C+4H_{[\mu }^{*}C_{\nu \rho ]}^{*}\left( A_{\lambda }C+\eta
C_{\lambda }\right) \right. +
\\
&\quad \quad \quad \left. +12H_{\mu }^{*}H_{\nu }^{*}\left(
\tfrac{1}{2}\left( A_{[\rho }C_{\lambda ]}-\phi _{\rho \lambda
}\eta \right) +B_{\rho \lambda }^{*}C\right) \right] \eta ^{\mu
\nu \rho \lambda }+\\
&\quad \quad \left. +\left[ H_{[\mu }^{*}C_{\nu \rho ]}^{*}\eta
C+3H_{\mu }^{*}H_{\nu }^{*}\left( A_{\rho }C+\eta C_{\rho }\right)
\right] \eta ^{\mu \nu \rho }+H_{\mu }^{*}H_{\nu }^{*}B^{\mu \nu
}\eta C\right\} ,
\end{split}
\end{equation}
\begin{equation}\label{h14}
\begin{split}
X_{3}^{\left( 2\right) }& =2\left\{ \left[ \left( H_{[\mu
}^{*}H_{\nu }^{*}C_{\rho \lambda \sigma ]}^{*}+H_{[\mu }^{*}C_{\nu
\rho }^{*}C_{\lambda \sigma ]}^{*}\right) \eta C+5H_{[\mu
}^{*}H_{\nu }^{*}C_{\rho \lambda ]}^{*}\left( A_{\sigma }C+\eta
C_{\sigma }\right) +\right. \right. \\
&\quad \quad \quad \left. +20H_{\mu }^{*}H_{\nu }^{*}H_{\rho
}^{*}\left( \tfrac{1}{2}\left( A_{[\lambda }C_{\sigma ]}-\phi
_{\lambda \sigma }\eta \right) +B_{\lambda \sigma }^{*}C\right)
\right] \eta ^{\mu \nu \rho \lambda \sigma }+\\
&\quad \quad \left. +\left[ H_{[\mu }^{*}H_{\nu }^{*}C_{\rho
\lambda ]}^{*}\eta C+4H_{\mu }^{*}H_{\nu }^{*}H_{\rho }^{*}\left(
A_{\lambda }C+\eta C_{\lambda }\right) \right] \eta ^{\mu \nu \rho
\lambda }+H_{\mu }^{*}H_{\nu }^{*}H_{\rho }^{*}\eta C\eta ^{\mu
\nu \rho }\right\} ,
\end{split}
\end{equation}
\begin{equation}
X_{4}^{\left( 2\right) }=20H_{\mu }^{*}H_{\nu }^{*}H_{\rho
}^{*}\left[ \left( C_{\lambda \sigma }^{*}\eta
C+\tfrac{1}{2}H_{\lambda }^{*}\left( A_{\sigma }C+\eta C_{\sigma
}\right) \right) \eta ^{\mu \nu \rho \lambda \sigma
}+\tfrac{1}{10}H_{\lambda }^{*}\eta C\eta ^{\mu \nu \rho \lambda
}\right] ,  \label{h15}
\end{equation}
\begin{equation}
X_{5}^{\left( 2\right) }=2H_{\mu }^{*}H_{\nu }^{*}H_{\rho
}^{*}H_{\lambda }^{*}H_{\sigma }^{*}\eta C\eta ^{\mu \nu \rho
\lambda \sigma }.  \label{h16}
\end{equation}

The elements $\left( X_{a}^{\left( 3\right) }\right)
_{a=\overline{0,5}}$ read as
\begin{equation}\label{h17}
\begin{split}
X_{0}^{\left( 3\right) }& =12\left[ \left( C^{*}C+2C_{\mu
}^{*}C^{\mu }+2\phi _{\mu \nu }^{*}\phi ^{\mu \nu }-2K_{\mu \nu
\rho }^{*}K^{\mu \nu \rho }-2\mathcal{G}_{\mu \nu \rho \lambda
}^{*}\mathcal{G}^{\mu \nu \rho \lambda }\right) C+\right.  \\
&\quad \quad +2\left( \mathcal{G}_{[\mu \nu \rho \lambda
}^{*}C_{\sigma ]}+K_{[\mu \nu \rho }^{*}\phi _{\lambda \sigma
]}-\mathcal{G}_{\mu \nu \rho \lambda \sigma }^{*}C\right)
\mathcal{G}^{\mu \nu \rho \lambda \sigma }+
\\
&\quad \quad \left. +2\left( K_{[\mu \nu \rho }^{*}C_{\lambda
]}+\tfrac{1}{3}\phi _{[\mu \nu }\phi _{\lambda \rho ]}\right)
\mathcal{G}^{\mu \nu \rho \lambda }+2\left( -\phi ^{*\mu \nu
}C_{\mu }C_{\nu }+K^{\mu \nu \rho }\phi _{\mu \nu }C_{\rho
}\right) \right] ,
\end{split}
\end{equation}
\begin{equation}\label{h18}
\begin{split}
X_{1}^{\left( 3\right) }& =4\left\{ \left[ \left( C_{\mu \nu \rho
\lambda \sigma }^{*}C-2C_{[\mu \nu \rho \lambda }^{*}C_{\sigma
]}\right) C-20C_{\mu \nu \rho }^{*}\left( \phi _{\lambda \sigma
}C-C_{\lambda }C_{\sigma }\right) -\right. \right.
\\
&\quad \quad \quad -60C_{\mu \nu }^{*}\left( K_{\rho \lambda
\sigma }^{*}C-\tfrac{1}{3}\phi _{[\rho \lambda }C_{\sigma
]}\right) +
\\
&\quad \quad \quad \left. +30H_{\mu }^{*}\left( K_{[\nu \rho
\lambda }^{*}C_{\sigma ]}-\mathcal{G}_{\nu \rho \lambda \sigma
}^{*}C+\tfrac{1}{3}\phi _{[\nu \rho }\phi _{\lambda \sigma
]}\right) \right] \mathcal{G}^{\mu \nu \rho \lambda \sigma
}+\\
&\quad \quad +\left[ \left( C_{\mu \nu \rho \lambda
}^{*}C-2C_{[\mu \nu \rho }^{*}C_{\lambda ]}\right) C+12C_{\mu \nu
}^{*}\left( C_{\rho }C_{\lambda
}-\phi _{\rho \lambda }C\right) -\right.  \\
&\quad \quad \quad \left. -12H_{\mu }^{*}\left( K_{\nu \rho
\lambda }^{*}C-\tfrac{2}{3}\phi _{[\nu \rho }C_{\lambda ]}\right)
\right] \mathcal{G}^{\mu \nu \rho \lambda }+
\\
&\quad \quad +\left[ \left( C_{\mu \nu \rho }^{*}C-2C_{[\mu \nu
}^{*}C_{\rho ]}\right) C-3H_{\mu }^{*}\left( \phi _{\nu \rho
}C-4C_{\nu }C_{\rho }\right) \right]
K^{\mu \nu \rho }  -\\
&\quad \quad \left. -3\left[ C_{\mu \nu }^{*}\phi ^{*\mu \nu
}C+H_{\mu }^{*}\left( C^{*\mu }C+2\phi ^{*\mu \nu }C_{\nu }\right)
\right] C\right\} ,
\end{split}
\end{equation}
\begin{equation}\label{h19}
\begin{split}
X_{2}^{\left( 3\right) }& =4\left\{ \left[ \left( H_{[\mu
}^{*}C_{\nu \rho \lambda \sigma ]}^{*}+C_{[\mu \nu }^{*}C_{\rho
\lambda \sigma ]}^{*}\right) CC-10\left( H_{[\mu }^{*}C_{\nu \rho
\lambda ]}^{*}+C_{[\mu \nu }^{*}C_{\rho \lambda ]}^{*}\right)
C_{\sigma }C-\right. \right.  \\
&\quad \quad \quad \left. -20H_{[\mu }^{*}C_{\nu \rho ]}^{*}\left(
\phi _{\lambda \sigma }C-C_{\lambda }C_{\sigma }\right) -60H_{\mu
}^{*}H_{\nu }^{*}\left( K_{\rho \lambda \sigma }^{*}C-
\tfrac{1}{3}C_{[\rho }\phi _{\lambda \sigma ]}\right) \right]
\mathcal{G}^{\mu \nu \rho \lambda \sigma }+\\
&\quad \quad +\left[ \left( H_{[\mu }^{*}C_{\nu \rho \lambda
]}^{*}+C_{[\mu \nu }^{*}C_{\rho \lambda ]}^{*}\right) CC-12H_{[\mu
}^{*}C_{\nu \rho ]}^{*}C_{\lambda }C\right. +
\\
&\quad \quad \quad \left. +12H_{\mu }^{*}H_{\nu }^{*}\left(
C_{\rho }C_{\lambda }-\phi _{\rho \lambda }C\right)
\right] \mathcal{G}^{\mu \nu \rho \lambda } + \\
&\quad \quad \left. +\left( H_{[\mu }^{*}C_{\nu \rho
]}^{*}C-6H_{\mu }^{*}H_{\nu }^{*}C_{\rho }\right) K^{\mu \nu \rho
}C-3H_{\mu }^{*}H_{\nu }^{*}\phi ^{*\mu \nu }CC\right\} ,
\end{split}
\end{equation}
\begin{equation}\label{h20}
\begin{split}
X_{3}^{\left( 3\right) }=& 4\left\{ \left[ \left( H_{[\mu
}^{*}H_{\nu }^{*}C_{\rho \lambda \sigma ]}^{*}+H_{[\mu }^{*}C_{\nu
\rho }^{*}C_{\lambda \sigma ]}^{*}\right) C-2H_{[\mu }^{*}H_{\nu
}^{*}C_{\rho \lambda
}^{*}C_{\sigma ]}-\right. \right.  \\
&\quad \quad \quad \left. -20H_{\mu }^{*}H_{\nu }^{*}H_{\rho
}^{*}\left( \phi _{\lambda \sigma }C-C_{\lambda }C_{\sigma
}\right) \right] \mathcal{G}^{\mu \nu \rho \lambda \sigma
}+ \\
&\quad \quad \left. +H_{[\mu }^{*}H_{\nu }^{*}C_{\rho \lambda
]}^{*}\mathcal{G} ^{\mu \nu \rho \lambda }CC+H_{\mu }^{*}H_{\nu
}^{*}H_{\rho }^{*}\left( K^{\mu \nu \rho }C-8C_{\lambda
}\mathcal{G}^{\mu \nu \rho \lambda }\right) C\right\} ,
\end{split}
\end{equation}
\begin{equation}
X_{4}^{\left( 3\right) }=4\left[ H_{[\mu }^{*}H_{\nu }^{*}H_{\rho
}^{*}C_{\lambda \sigma ]}^{*}CC\mathcal{G}^{\mu \nu \rho \lambda
\sigma
}+H_{\mu }^{*}H_{\nu }^{*}H_{\rho }^{*}H_{\lambda }^{*}C\left( C%
\mathcal{G}^{\mu \nu \rho \lambda }-10C_{\sigma }\mathcal{G}^{\mu
\nu \rho \lambda \sigma }\right) \right] ,  \label{h21}
\end{equation}
\begin{equation}
X_{5}^{\left( 3\right) }=4H_{\mu }^{*}H_{\nu }^{*}H_{\rho
}^{*}H_{\lambda }^{*}H_{\sigma }^{*}CC\mathcal{G}^{\mu \nu \rho
\lambda \sigma }. \label{h22}
\end{equation}

The quantities $\left( X_{a}^{\left( 4\right) }\right)
_{a=\overline{0,5}}$ mean
\begin{equation}\label{pt21}
\begin{split}
X_{0}^{\left( 4\right) }& =12\epsilon ^{\alpha \beta \gamma \delta
\varepsilon }\left[ -\tfrac{1}{3}\left( \eta _{\mu \nu \rho
\lambda \sigma }^{*}\eta ^{\mu \nu \rho \lambda \sigma }+\eta
_{\mu \nu \rho \lambda }^{*}\eta ^{\mu \nu \rho \lambda }+\eta
_{\mu \nu \rho }^{*}\eta ^{\mu \nu
\rho }-\right. \right.  \\
&\quad \quad \quad \quad \quad \quad \quad \quad \left. -\eta
^{*}\eta +B_{\mu \nu }^{*}B^{\mu \nu }-A_{\mu }^{*}A^{\mu
}\right) \mathcal{G}_{\alpha \beta \gamma \delta \varepsilon }+\\
&\quad \quad \quad \quad \quad \quad +\left( \tfrac{1}{3}\eta
_{\mu \nu \rho \lambda }^{*}\mathcal{G}^{\mu \nu \rho \lambda }-
\tfrac{1}{3}\eta _{\mu \nu \rho }^{*}K^{\mu \nu \rho }\right. - \\
&\quad \quad \quad \quad \quad \quad \quad \quad \left.
-\tfrac{1}{2}C^{*}\eta +\tfrac{1}{2}C^{*\mu }A_{\mu }-B_{\mu \nu
}^{*}\phi ^{*\mu \nu }\right) \eta _{\alpha \beta \gamma \delta
\varepsilon } +\\
&\quad \quad \quad \quad \quad \quad +\left( \tfrac{4}{3}\eta
_{\mu \nu \rho }^{*}\eta ^{\mu \nu \rho \lambda }-B_{\mu \nu
}^{*}\eta ^{\mu \nu \lambda }+\tfrac{1}{3}A_{\mu }B^{\mu \lambda
}-\tfrac{1}{3}A^{*\lambda }\eta \right) \sigma _{\lambda
\alpha }\mathcal{G}_{\beta \gamma \delta \varepsilon } + \\
&\quad \quad \quad \quad \quad \quad +\left( \tfrac{1}{2}C^{*\rho
}\eta -B_{\mu \nu }^{*}K^{\mu \nu \rho }+A_{\mu }\phi ^{*}{}^{\mu
\rho }\right) \sigma _{\rho \alpha }\eta _{\beta
\gamma \delta \varepsilon } + \\
&\quad \quad \quad \quad \quad \quad \left. +\tfrac{1}{2}\left(
\phi ^{*\nu \rho }\eta -A_{\mu }K^{\mu \nu \rho }\right) \sigma
_{\nu \alpha }\sigma _{\rho \beta }\eta _{\gamma \delta
\varepsilon }-\tfrac{5}{3}B_{[\alpha \beta }K_{\gamma \delta
\varepsilon ]}\eta \right] ,
\end{split}
\end{equation}
\begin{equation}\label{pt22}
\begin{split}
X_{1}^{\left( 4\right) }& =12\epsilon ^{\alpha \beta \gamma \delta
\varepsilon }\left[ \tfrac{1}{3}\left( -H_{[\mu }^{*}\eta _{\nu
\rho \lambda \sigma ]}^{*}\eta ^{\mu \nu \rho \lambda \sigma
}-H_{[\mu }^{*}\eta _{\nu
\rho \lambda ]}^{*}\eta ^{\mu \nu \rho \lambda }\right. \right. - \\
&\quad \quad \quad \quad \quad \quad \quad \left. -H_{[\mu
}^{*}B_{\nu \rho ]}^{*}\eta ^{\mu \nu \rho }-H_{\mu }^{*}A^{*\mu
}\eta -\tfrac{1}{2}H_{[\mu }^{*}A_{\nu ]}B^{\mu \nu }\right)
\mathcal{G}_{\alpha \beta \gamma \delta \varepsilon } + \\
&\quad \quad \quad \quad \quad +\tfrac{1}{3}\left(
-\tfrac{3}{2}H_{[\mu }^{*}A_{\nu ]}\eta ^{\mu \nu \rho
}+H_{\mu }^{*}B^{\mu \rho }\eta \right) \sigma _{\rho \alpha }\mathcal{G}%
_{\beta \gamma \delta \varepsilon } + \\
&\quad \quad \quad \quad \quad +\tfrac{1}{3}\left( H_{[\mu
}^{*}\eta _{\nu \rho \lambda ]}^{*}\eta ^{\mu
\nu \rho \lambda }-H_{[\mu }^{*}B_{\nu \rho ]}^{*}K^{\mu \nu \rho }\right. -\\
&\quad \quad \quad \quad \quad \quad \quad \left. -\tfrac{3%
}{2}H_{[\mu }^{*}A_{\nu ]}\phi ^{*\mu \nu }+\tfrac{3}{2}H_{\mu
}^{*}C^{*\mu }\eta \right) \mathcal{\eta }_{\alpha \beta \gamma
\delta \varepsilon }+
\\
&\quad \quad \quad \quad \quad +2\left( -\tfrac{2}{3}H_{[\mu
}^{*}B_{\nu \rho ]}^{*}\mathcal{G}^{\mu \nu \rho \lambda
}-\tfrac{1}{4}H_{[\mu }^{*}A_{\nu ]}K^{\mu \nu \lambda }-H_{\mu
}^{*}\phi ^{*\mu \lambda }\eta \right) \sigma _{\lambda \alpha
}\eta _{\beta \gamma \delta \varepsilon }+
\\
&\quad \quad \quad \quad \quad \left. +\tfrac{1}{2}%
H_{\mu }^{*}K^{\mu \nu \rho }\sigma _{\nu \alpha }\sigma _{\rho
\beta }\eta \eta _{\gamma \delta \varepsilon }\right] ,
\end{split}
\end{equation}
\begin{equation}\label{pt23}
\begin{split}
X_{2}^{\left( 4\right) }& =12\epsilon ^{\alpha \beta \gamma \delta
\varepsilon }\left\{ -\tfrac{1}{3}\left[ \left(C_{[\mu \nu
}^{*}\eta _{\rho \lambda \sigma ]}^{*}+H_{[\mu }^{*}H_{\nu
}^{*}\eta _{\rho \lambda \sigma ]}^{*}\right) \eta ^{\mu \nu \rho
\lambda \sigma }+\right. \right. \\
&\quad \quad \quad \quad \quad \quad \quad \quad +\left( C_{[\mu
\nu }^{*}B_{\rho \lambda ]}^{*}+H_{[\mu }^{*}H_{\nu }^{*}B_{\rho
\lambda ]}^{*}\right) \eta ^{\mu \nu \rho \lambda
}+\\
&\quad \quad \quad \quad \quad \quad \quad \quad +
\tfrac{1}{2}\left( C_{[\mu \nu }^{*}A_{\rho ]}
+H_{[\mu }^{*}H_{\nu }^{*}A_{\rho ]}\right) \eta ^{\mu \nu \rho } +\\
&\quad \quad \quad \quad \quad \quad \quad \quad \left.
+\tfrac{1}{2}\left( C_{\mu \nu }^{*}+H_{\mu }^{*}H_{\nu
}^{*}\right) B^{\mu \nu }\eta \right] \mathcal{G}_{\alpha \beta
\gamma \delta \varepsilon }+ \\
&\quad \quad \quad \quad \quad \quad +\tfrac{1}{2}\left[
\tfrac{4}{3}\left( C_{[\mu \nu }^{*}A_{\rho ]}+H_{[\mu }^{*}H_{\nu
}^{*}A_{\rho ]}\right) \eta ^{\mu \nu \rho \lambda }+\right.
\\
&\quad \quad \quad \quad \quad \quad \quad \quad \left. +\left(
C_{\mu \nu }^{*}+H_{\mu }^{*}H_{\nu }^{*}\right) \eta ^{\mu \nu
\lambda }\eta \right] \sigma _{\lambda \alpha }\mathcal{G}_{\beta
\gamma \delta \varepsilon
}+\\
&\quad \quad \quad \quad \quad \quad +\tfrac{1}{6}\left[
\tfrac{1}{2}\left( C_{[\mu \nu }^{*}B_{\rho \lambda ]}^{*}
+H_{[\mu }^{*}H_{\nu }^{*}B_{\rho \lambda ]}^{*}\right)
\mathcal{G}^{\mu \nu \rho \lambda }-\right.\\
&\quad \quad \quad \quad \quad \quad \quad \quad -\left( C_{[\mu
\nu }^{*}A_{\rho ]}+H_{[\mu }^{*}H_{\nu }^{*}A_{\rho ]}\right)
K^{\mu \nu \rho }+
\\
&\quad \quad \quad \quad \quad \quad \quad \quad \left.
+\tfrac{1}{3}\left( C_{\mu \nu }^{*}+H_{\mu }^{*}H_{\nu
}^{*}\right) \phi ^{*\mu \nu }\eta \right] \eta _{\alpha \beta
\gamma \delta \varepsilon } -\\
&\quad \quad \quad \quad \quad \quad \left. -\tfrac{1}{2}\left(
C_{\mu \nu }^{*}+H_{\mu }^{*}H_{\nu }^{*}\right) K^{\mu \nu \rho
}\sigma _{\rho \alpha }\eta \eta _{\beta \gamma \delta \varepsilon
}\right\} ,
\end{split}
\end{equation}
\begin{equation}\label{pt24}
\begin{split}
X_{3}^{\left( 4\right) }& =4\epsilon ^{\alpha \beta \gamma \delta
\varepsilon }\left\{ -\left[ \left( C_{[\mu \nu \rho
}^{*}B_{\lambda \sigma ]}^{*}+C_{[\mu \nu }^{*}H_{\rho
}^{*}B_{\lambda \sigma ]}^{*}+H_{[\mu }^{*}H_{\nu }^{*}H_{\rho
}^{*}B_{\lambda \sigma ]}^{*}\right) \eta ^{\mu \nu \rho \lambda
\sigma }+\right.
\right. \\
&\quad \quad \quad \quad \quad \quad \quad +\tfrac{1}{2}\left(
C_{[\mu \nu }^{*}H_{\rho }^{*}A_{\lambda ]}+H_{[\mu }^{*}H_{\nu
}^{*}H_{\rho }^{*}A_{\lambda ]}+C_{[\mu \nu \rho }^{*}A_{\lambda
]}\right)\eta
^{\mu \nu \rho \lambda } -\\
&\quad \quad \quad \quad \quad \quad \quad \left.
-\tfrac{1}{2}\left( C_{\mu \nu \rho }^{*}+C_{[\mu \nu }^{*}H_{\rho
]}^{*}+H_{\mu }^{*}H_{\nu }^{*}H_{\rho }^{*}\right) \eta ^{\mu \nu
\rho }\eta \right] \mathcal{G}_{\alpha \beta \gamma \delta
\varepsilon }+
\\
&\quad \quad \quad \quad \quad +\tfrac{1}{2}\left[ \left( C_{[\mu
\nu \rho }^{*}A_{\lambda ]}+C_{[\mu \nu }^{*}H_{\rho
}^{*}A_{\lambda ]}+H_{[\mu }^{*}H_{\nu }^{*}H_{\rho
}^{*}A_{\lambda ]}\right) \mathcal{G}
^{\mu \nu \rho \lambda }\right.  -\\
&\quad \quad \quad \quad \quad \quad \quad \left. -\left( C_{\mu
\nu \rho }^{*}+C_{[\mu \nu }^{*}H_{\rho ]}^{*}+H_{\mu }^{*}H_{\nu
}^{*}H_{\rho }^{*}\right) K^{\mu \nu \rho }\eta
\right] \eta _{\alpha \beta \gamma \delta \varepsilon } +\\
&\quad \quad \quad \quad \quad \left. +2\left( C_{\mu \nu \rho
}^{*}+C_{[\mu \nu }^{*}H_{\rho ]}^{*}+H_{\mu }^{*}H_{\nu
}^{*}H_{\rho }^{*}\right) \eta ^{\mu \nu \rho \lambda }\sigma
_{\lambda \alpha }\eta \mathcal{G}_{\beta \gamma \delta
\varepsilon }\right\} ,
\end{split}
\end{equation}
\begin{equation}\label{pt25}
\begin{split}
X_{4}^{\left( 4\right) }& =-2\epsilon ^{\alpha \beta \gamma \delta
\varepsilon }\left\{ \left[ \left( C_{[\mu \nu \rho \lambda
}^{*}A_{\sigma ]}+C_{[\mu \nu \rho }^{*}H_{\lambda }^{*}A_{\sigma
]}+C_{[\mu \nu
}^{*}C_{\rho \lambda }^{*}A_{\sigma ]}+\right. \right. \right. \\
&\quad \quad \quad \quad \quad \quad \quad \left. +C_{[\mu \nu
}^{*}H_{\rho }^{*}H_{\lambda }^{*}A_{\sigma ]}+H_{[\mu }^{*}H_{\nu
}^{*}H_{\rho }^{*}H_{\lambda }^{*}A_{\sigma
]}\right) \eta ^{\mu \nu \rho \lambda \sigma }+\\
&\quad \quad \quad \quad \quad \quad +\left( C_{\mu \nu \rho
\lambda }^{*}+C_{[\mu \nu
\rho }^{*}H_{\lambda ]}^{*}+C_{[\mu \nu }^{*}C_{\rho \lambda ]}^{*}+\right.  \\
&\quad \quad \quad \quad \quad \quad \quad \left. \left. +C_{[\mu
\nu }^{*}H_{\rho }^{*}H_{\lambda ]}^{*}+H_{\mu }^{*}H_{\nu
}^{*}H_{\rho
}^{*}H_{\lambda }^{*}\right) \eta ^{\mu \nu \rho \lambda }\right] \mathcal{G}%
_{\alpha \beta \gamma \delta \varepsilon } +\\
&\quad \quad \quad \quad \quad +\left( C_{\mu \nu \rho \lambda
}^{*}+C_{[\mu \nu \rho }^{*}H_{\lambda
]}^{*}+C_{[\mu \nu }^{*}C_{\rho \lambda ]}^{*}+\right. \\
&\quad \quad \quad \quad \quad \quad \left. \left. +C_{[\mu \nu
}^{*}H_{\rho }^{*}H_{\lambda ]}^{*}+H_{\mu }^{*}H_{\nu
}^{*}H_{\rho }^{*}H_{\lambda }^{*}\right) \mathcal{G}^{\mu \nu
\rho \lambda }\eta \eta _{\alpha \beta \gamma \delta \varepsilon
}\right\} ,
\end{split}
\end{equation}
\begin{equation}\label{pt26}
\begin{split}
X_{5}^{\left( 4\right) }& =2\epsilon ^{\alpha \beta \gamma \delta
\varepsilon }\left( C_{\mu \nu \rho \lambda \sigma }^{*}+C_{[\mu
\nu \rho \lambda }^{*}H_{\sigma ]}^{*}+C_{[\mu \nu \rho
}^{*}C_{\lambda \sigma ]}^{*}+C_{[\mu \nu \rho }^{*}H_{\lambda
}^{*}H_{\sigma ]}^{*}+\right.
\\
&\quad \quad \quad \quad \quad +C_{[\mu \nu }^{*}C_{\rho \lambda
}^{*}H_{\sigma ]}^{*}+C_{[\mu \nu
}^{*}H_{\rho }^{*}H_{\lambda }^{*}H_{\sigma ]}^{*}+\\
&\quad \quad \quad \quad \quad \left. +H_{\mu }^{*}H_{\nu
}^{*}H_{\rho }^{*}H_{\lambda }^{*}H_{\sigma }^{*}\right) \eta
^{\mu \nu \rho \lambda \sigma }\eta \mathcal{G}_{\alpha \beta
\gamma \delta \varepsilon }.
\end{split}
\end{equation}

The components $\left( X_{a}^{\left( 5\right) }\right)
_{a=\overline{0,5}}$ are given by
\begin{equation}\label{pt27}
\begin{split}
X_{0}^{\left( 5\right) }& =24\epsilon ^{\alpha \beta \gamma \delta
\varepsilon }\left( -\tfrac{1}{2}C^{*}C_{\alpha \beta \gamma
\delta \varepsilon }+\tfrac{1}{10}C_{[\alpha }^{*}C_{\beta \gamma
\delta \varepsilon ]}+\tfrac{1}{20}\phi _{[\alpha \beta
}^{*}C_{\gamma \delta
\varepsilon ]}-\right. \\
&\quad \quad \quad \quad \quad \left. -\tfrac{1}{60}K_{[\alpha
\beta \gamma }C_{\delta \varepsilon ]}-
\tfrac{1}{30}H_{[\alpha }\mathcal{G}_{\beta \gamma \delta \varepsilon ]}+%
\tfrac{1}{6}\varphi ^{*}\mathcal{G}_{\alpha \beta \gamma \delta
\varepsilon }\right) ,
\end{split}
\end{equation}
\begin{equation}\label{pt28}
\begin{split}
X_{1}^{\left( 5\right) }& =12\epsilon ^{\alpha \beta \gamma \delta
\varepsilon }\left[ \left( H_{\mu }^{*}C^{*\mu }C_{\alpha \beta
\gamma \delta \varepsilon }-\tfrac{1}{3}H_{\mu }^{*}H^{\mu
}\mathcal{G}_{\alpha
\beta \gamma \delta \varepsilon }\right) +\right.  \\
&\quad \quad \quad \quad \quad +2\sigma _{\rho \alpha }\left(
-H_{\mu }^{*}\phi ^{*\mu \rho
}C_{\beta \gamma \delta \varepsilon }+\tfrac{1}{3}H_{\mu }^{*}C^{\mu \rho }%
\mathcal{G}_{\beta \gamma \delta \varepsilon }\right) -\\
&\quad \quad \quad \quad \quad \left. -H^{*\mu }C_{\mu \alpha
\beta }K_{\gamma \delta \varepsilon }\right] ,
\end{split}
\end{equation}
\begin{equation}\label{pt29}
\begin{split}
X_{2}^{\left( 5\right) }& =12\epsilon ^{\alpha \beta \gamma \delta
\varepsilon }\left[ \left( C_{\mu \nu }^{*}+H_{\mu }^{*}H_{\nu
}^{*}\right)
\left( \phi ^{*\mu \nu }C_{\alpha \beta \gamma \delta \varepsilon }-\tfrac{1%
}{3}C^{\mu \nu }\mathcal{G}_{\alpha \beta \gamma \delta
\varepsilon }\right)
-\right. \\
&\quad \quad \quad \quad \quad \left. -\left( C_{\mu \nu
}^{*}+H_{\mu }^{*}H_{\nu }^{*}\right) \sigma _{\rho \alpha }\left(
K^{\mu \nu \rho }C_{\beta \gamma \delta \varepsilon }+C^{\mu \nu
\rho }\mathcal{G}_{\beta \gamma \delta \varepsilon }\right)
\right] ,
\end{split}
\end{equation}
\begin{equation}\label{pt30}
\begin{split}
X_{3}^{\left( 5\right) }& =-4\epsilon ^{\alpha \beta \gamma \delta
\varepsilon }\left[ \left( C_{\mu \nu \rho }^{*}+C_{[\mu \nu
}^{*}H_{\rho ]}^{*}+H_{\mu }^{*}H_{\nu }^{*}H_{\rho }^{*}\right)
\times \right.
\\
&\quad \quad \quad \quad \quad \quad \times \left( K^{\mu \nu \rho
}C_{\alpha \beta \gamma \delta \varepsilon }+C^{\mu \nu \rho
}\mathcal{G}_{\alpha \beta \gamma \delta \varepsilon
}\right) -\\
&\quad \quad \quad \quad \quad \quad \left. -4\left( C_{\mu \nu
\rho }^{*}+C_{[\mu \nu }^{*}H_{\rho ]}^{*}+H_{\mu }^{*}H_{\nu
}^{*}H_{\rho }^{*}\right) C^{\mu \nu \rho \lambda }\sigma
_{\lambda \alpha }\mathcal{G}_{\beta \gamma \delta \varepsilon
}\right] ,
\end{split}
\end{equation}
\begin{equation}\label{pt31}
\begin{split}
X_{4}^{\left( 5\right) }& =-4\epsilon ^{\alpha \beta \gamma \delta
\varepsilon }\left( C_{\mu \nu \rho \lambda }^{*}+C_{[\mu \nu \rho
}^{*}H_{\lambda ]}^{*}+C_{[\mu \nu }^{*}C_{\rho \lambda
]}^{*}+C_{[\mu \nu
}^{*}H_{\rho }^{*}H_{\lambda ]}^{*}+\right. \\
&\quad \quad \quad \quad \quad \quad \left. +H_{\mu }^{*}H_{\nu
}^{*}H_{\rho }^{*}H_{\lambda }^{*}\right) \left( \mathcal{G}^{\mu
\nu \rho \lambda }C_{\alpha \beta \gamma \delta \varepsilon
}+C^{\mu \nu \rho \lambda }\mathcal{G}_{\alpha \beta \gamma \delta
\varepsilon }\right) ,
\end{split}
\end{equation}
\begin{equation}\label{pt32}
\begin{split}
X_{5}^{\left( 5\right) }& =-4\epsilon ^{\alpha \beta \gamma \delta
\varepsilon }\left( C_{\mu \nu \rho \lambda \sigma }^{*}+C_{[\mu
\nu \rho \lambda }^{*}H_{\sigma ]}^{*}+C_{[\mu \nu \rho
}^{*}C_{\lambda \sigma ]}^{*}+C_{[\mu \nu \rho }^{*}H_{\lambda
}^{*}H_{\sigma ]}^{*}+\right.
\\
&\quad \quad \quad \quad \quad \quad +C_{[\mu \nu }^{*}C_{\rho
\lambda }^{*}H_{\sigma ]}^{*}+C_{[\mu
\nu }^{*}H_{\rho }^{*}H_{\lambda }^{*}H_{\sigma ]}^{*}+\\
&\quad \quad \quad \quad \quad \quad \left. +H_{\mu }^{*}H_{\nu
}^{*}H_{\rho }^{*}H_{\lambda }^{*}H_{\sigma }^{*}\right) C^{\mu
\nu \rho \lambda \sigma }\mathcal{G}_{\alpha \beta \gamma \delta
\varepsilon }.
\end{split}
\end{equation}

The terms denoted by $\left( X_{a}^{\left( 6\right) }\right) _{a=\overline{%
0,5}}$ are of the form
\begin{equation}\label{pt33}
\begin{split}
X_{0}^{\left( 6\right) }& =24\epsilon ^{\alpha \beta \gamma \delta
\varepsilon }\left[ \left( 2\mathcal{G}_{\mu \nu \rho \lambda \sigma }^{*}%
\mathcal{G}^{\mu \nu \rho \lambda \sigma }+\mathcal{G}_{\mu \nu
\rho \lambda
}^{*}\mathcal{G}^{\mu \nu \rho \lambda }-C^{*}C-\right. \right.  \\
&\quad \quad \quad \quad \quad \quad \left. -C_{\mu }^{*}C^{\mu
}+K_{\mu \nu \rho }^{*}K^{\mu \nu \rho }-\phi _{\mu \nu }^{*}\phi
^{\mu \nu }\right) \mathcal{G}_{\alpha \beta \gamma
\delta \varepsilon } + \\
&\quad \quad \quad \quad \quad +\left( C^{*\lambda }C+\phi _{\mu
\nu }K^{\mu \nu \lambda }-2\phi ^{*\mu \lambda }C_{\mu }\right)
\sigma _{\lambda \alpha }\mathcal{G}_{\beta \gamma
\delta \varepsilon } -\\
&\quad \quad \quad \quad \quad -\tfrac{1}{12}K_{\alpha \beta
\gamma }^{*}\epsilon _{\delta \mu \nu \rho \lambda
}\mathcal{G}^{\mu \nu \rho \lambda }\epsilon _{\varepsilon \mu
^{\prime }\nu ^{\prime }\rho ^{\prime }\lambda ^{\prime
}}\mathcal{G}^{\mu
^{\prime }\nu ^{\prime }\rho ^{\prime }\lambda ^{\prime }} +\\
&\quad \quad \quad \quad \quad \left. +\phi _{\alpha \beta
}^{*}K_{\gamma \delta \varepsilon }C-\tfrac{1}{24}C_{\alpha
}\epsilon _{\beta \gamma \mu \nu \rho }K^{\mu \nu \rho }\epsilon
_{\delta \varepsilon \mu ^{\prime }\nu ^{\prime }\rho ^{\prime
}}K^{\mu ^{\prime }\nu ^{\prime }\rho ^{\prime }}\right] ,
\end{split}
\end{equation}
\begin{equation}\label{pt34}
\begin{split}
X_{1}^{\left( 6\right) }& =24\epsilon ^{\alpha \beta \gamma \delta
\varepsilon }\left[ \left( H_{\mu }^{*}C^{*\mu }C+2H_{\mu
}^{*}\phi ^{*\mu \nu }C_{\nu }+\tfrac{1}{2}H_{[\mu
}^{*}\mathcal{G}_{\nu \rho \lambda \sigma
]}^{*}\mathcal{G}^{\mu \nu \rho \lambda \sigma }+\right. \right. \\
&\quad \quad \quad \quad \quad \quad \left. +\tfrac{1}{3}H_{[\mu
}^{*}\phi _{\nu \rho ]}K^{\mu \nu \rho }+H_{[\mu }^{*}K_{\nu \rho
\lambda ]}^{*}\mathcal{G}^{\mu \nu \rho \lambda
}\right) \mathcal{G}_{\alpha \beta \gamma \delta \varepsilon } +\\
&\quad \quad \quad \quad \quad +\sigma _{\rho \alpha }\left(
2H_{\mu }^{*}\phi ^{*\mu \rho }C-H_{[\mu }^{*}C_{\nu ]}K^{\mu \nu
\rho }\right) \mathcal{G}_{\beta \gamma \delta
\varepsilon }  +\\
&\quad \quad \quad \quad \quad +\tfrac{1}{24}H_{\alpha
}^{*}C\epsilon _{\beta \gamma \mu \nu \rho }K^{\mu \nu \rho
}\epsilon _{\delta \varepsilon \mu ^{\prime }\nu ^{\prime }\rho
^{\prime }}K^{\mu ^{\prime }\nu ^{\prime }\rho ^{\prime }} -\\
&\quad \quad \quad \quad \quad \left. -\tfrac{1}{36}H_{[\alpha
}^{*}\phi _{\beta \gamma ]}\epsilon _{\delta \mu \nu \rho \lambda
}\mathcal{G}^{\mu \nu \rho \lambda }\epsilon
_{\varepsilon \mu ^{\prime }\nu ^{\prime }\rho ^{\prime }\lambda ^{\prime }}%
\mathcal{G}^{\mu ^{\prime }\nu ^{\prime }\rho ^{\prime }\lambda
^{\prime }}\right] ,
\end{split}
\end{equation}
\begin{equation}\label{pt35}
\begin{split}
X_{2}^{\left( 6\right) }& =24\epsilon ^{\alpha \beta \gamma \delta
\varepsilon }\left\{ \left[ \tfrac{2}{3}\left( C_{[\mu \nu
}^{*}C_{\rho ]}+H_{[\mu }^{*}H_{\nu }^{*}C_{\rho ]}\right) K^{\mu
\nu \rho }+\left( C_{\mu \nu }^{*}+H_{\mu }^{*}H_{\nu }^{*}\right)
\phi ^{*\mu \nu }C+\right. \right. \\
&\quad \quad \quad \quad \quad \quad \quad +\tfrac{1}{2}\left(
C_{[\mu \nu }^{*}K_{\rho \lambda \sigma ]}^{*}+H_{[\mu }^{*}H_{\nu
}^{*}K_{\rho \lambda \sigma ]}^{*}\right) \mathcal{G}^{\mu \nu
\rho \lambda \sigma }  +\\
&\quad \quad \quad \quad \quad \quad \quad \left.
+\tfrac{1}{3}\left( C_{[\mu \nu }^{*}\phi _{\rho \lambda
]}+H_{[\mu }^{*}H_{\nu }^{*}\phi _{\rho \lambda ]}\right)
\mathcal{G}^{\mu \nu \rho \lambda }\right] \mathcal{G}_{\alpha
\beta \gamma \delta \varepsilon }
-\\
&\quad \quad \quad \quad \quad \quad -\sigma _{\rho \alpha }\left(
C_{\mu \nu }^{*}+H_{\mu }^{*}H_{\nu }^{*}\right) K^{\mu \nu \rho
}\mathcal{G}_{\beta \gamma \delta \varepsilon }
-\\
&\quad \quad \quad \quad \quad \quad \left. -\tfrac{1}{6\cdot
4!}\left( C_{[\alpha \beta }^{*}C_{\gamma ]}+H_{[\alpha
}^{*}H_{\beta }^{*}C_{\gamma ]}\right) \epsilon _{\delta \mu \nu
\rho \lambda }\mathcal{G}^{\mu \nu \rho \lambda }\epsilon
_{\varepsilon
\mu ^{\prime }\nu ^{\prime }\rho ^{\prime }\lambda ^{\prime }}\mathcal{G}%
^{\mu ^{\prime }\nu ^{\prime }\rho ^{\prime }\lambda ^{\prime
}}\right\} ,
\end{split}
\end{equation}
\begin{equation}\label{pt36}
\begin{split}
X_{3}^{\left( 6\right) }& =-8\epsilon ^{\alpha \beta \gamma \delta
\varepsilon }\left[ \left( C_{\mu \nu \rho }^{*}+C_{[\mu \nu
}^{*}H_{\rho
]}^{*}+H_{\mu }^{*}H_{\nu }^{*}H_{\rho }^{*}\right) K^{\mu \nu \rho }C%
\mathcal{G}_{\alpha \beta \gamma \delta \varepsilon }+\right. \\
&\quad \quad \quad \quad \quad \quad +\left( C_{[\mu \nu \rho
}^{*}A_{\lambda ]}+C_{[\mu \nu }^{*}H_{\rho
}^{*}A_{\lambda ]}+\right. \\
&\quad \quad \quad \quad \quad \quad \quad \quad \left. +C_{[\mu
\nu }^{*}H_{\rho }^{*}A_{\lambda ]}+H_{[\mu }^{*}H_{\nu
}^{*}H_{\rho }^{*}A_{\lambda ]}\right) \mathcal{G}^{\mu \nu \rho \lambda }%
\mathcal{G}_{\alpha \beta \gamma \delta \varepsilon } -\\
&\quad \quad \quad \quad \quad \quad \left. -\tfrac{1}{12}\left(
C_{[\alpha \beta }^{*}C_{\gamma ]}+H_{[\alpha }^{*}H_{\beta
}^{*}C_{\gamma ]}\right) \epsilon _{\delta \mu \nu \rho \lambda
}\mathcal{G}^{\mu \nu \rho \lambda }\epsilon _{\varepsilon \mu
^{\prime }\nu ^{\prime }\rho ^{\prime }\lambda ^{\prime
}}\mathcal{G}^{\mu ^{\prime }\nu ^{\prime }\rho ^{\prime }\lambda
^{\prime }}\right] ,
\end{split}
\end{equation}
\begin{equation}\label{pt37}
\begin{split}
X_{4}^{\left( 6\right) }& =4\epsilon ^{\alpha \beta \gamma \delta
\varepsilon }\left[ \left( C_{[\mu \nu \rho \lambda }^{*}C_{\sigma
]}+C_{[\mu \nu \rho }^{*}H_{\lambda }^{*}C_{\sigma ]}+C_{[\mu \nu
}^{*}C_{\rho \lambda }^{*}C_{\sigma ]}+\right. \right. \\
&\quad \quad \quad \quad \quad \quad \left. +C_{[\mu \nu
}^{*}H_{\rho }^{*}H_{\lambda }^{*}C_{\sigma ]}+H_{[\mu
}^{*}H_{\nu }^{*}H_{\rho }^{*}H_{\lambda }^{*}C_{\sigma ]}\right) \mathcal{G}%
^{\mu \nu \rho \lambda \sigma }  -\\
&\quad \quad \quad \quad \quad -\left( C_{\mu \nu \rho \lambda
}^{*}+C_{[\mu \nu \rho }^{*}H_{\lambda
]}^{*}+C_{[\mu \nu }^{*}C_{\rho \lambda ]}^{*}\right. +\\
&\quad \quad \quad \quad \quad \quad \left. \left. +C_{[\mu \nu
}^{*}H_{\rho }^{*}H_{\lambda ]}^{*}+H_{\mu }^{*}H_{\nu
}^{*}H_{\rho }^{*}H_{\lambda }^{*}\right) \mathcal{G}^{\mu \nu
\rho \lambda }C\right] \mathcal{G}_{\alpha \beta \gamma \delta
\varepsilon },
\end{split}
\end{equation}
\begin{equation}\label{pt38}
\begin{split}
X_{5}^{\left( 6\right) }& =-4\epsilon ^{\alpha \beta \gamma \delta
\varepsilon }\left( C_{\mu \nu \rho \lambda \sigma }^{*}+C_{[\mu
\nu \rho \lambda }^{*}H_{\sigma ]}^{*}+C_{[\mu \nu \rho
}^{*}C_{\lambda \sigma ]}^{*}+C_{[\mu \nu \rho }^{*}H_{\lambda
}^{*}H_{\sigma ]}^{*}+\right.
\\
&\quad \quad \quad \quad \quad \quad +C_{[\mu \nu }^{*}C_{\rho
\lambda }^{*}H_{\sigma ]}^{*}+C_{[\mu \nu }^{*}H_{\rho
}^{*}H_{\lambda
}^{*}H_{\sigma ]}^{*} +\\
&\quad \quad \quad \quad \quad \quad \left. +H_{\mu }^{*}H_{\nu
}^{*}H_{\rho }^{*}H_{\lambda }^{*}H_{\sigma }^{*}\right)
\mathcal{G}^{\mu \nu \rho \lambda \sigma }\mathcal{G}_{\alpha
\beta \gamma \delta \varepsilon }.
\end{split}
\end{equation}

The pieces $\left( X_{a}^{\left( 7\right) }\right)
_{a=\overline{0,5}}$ are
\begin{equation}\label{pt39}
\begin{split}
X_{0}^{\left( 7\right) }& =\tfrac{3}{5}\epsilon ^{\alpha \beta
\gamma \delta \varepsilon }\left( -\tfrac{1}{2}\mathcal{G}_{\alpha
\beta \gamma \delta \varepsilon }^{*}CC+\mathcal{G}_{[\alpha \beta
\gamma \delta }^{*}C_{\varepsilon ]}C+K_{[\alpha \beta \gamma
}^{*}\phi _{\delta
\varepsilon ]}C-\right. \\
&\quad \quad \quad \quad \quad \left. -K_{[\alpha \beta \gamma
}^{*}C_{\delta }C_{\varepsilon ]}-\tfrac{1}{3}\phi _{[\alpha \beta
}\phi _{\gamma \delta }C_{\varepsilon ]}\right) ,
\end{split}
\end{equation}
\begin{equation}\label{pt40}
\begin{split}
X_{1}^{\left( 7\right) }& =\tfrac{3}{5}\epsilon ^{\alpha \beta
\gamma \delta \varepsilon }\left( -\tfrac{1}{2}H_{[\alpha
}^{*}\mathcal{G}_{\beta \gamma \delta \varepsilon
]}^{*}CC+H_{[\alpha }^{*}K_{\beta \gamma \delta
}^{*}C_{\varepsilon ]}C-\right. \\
&\quad \quad \quad \quad \quad \left. -H_{[\alpha }^{*}\phi
_{\beta \gamma }C_{\delta }C_{\varepsilon ]}+
\tfrac{1}{3}H_{[\alpha }^{*}\phi _{\beta \gamma }\phi _{\delta
\varepsilon ]}\right) ,
\end{split}
\end{equation}
\begin{equation}\label{pt41}
\begin{split}
X_{2}^{\left( 7\right) }& =\tfrac{1}{5}\epsilon ^{\alpha \beta
\gamma \delta \varepsilon }\left[ -\tfrac{3}{2}\left( C_{[\alpha
\beta }^{*}K_{\gamma \delta \varepsilon ]}^{*}+H_{[\alpha
}^{*}H_{\beta }^{*}K_{\gamma \delta \varepsilon ]}^{*}\right)
CC+\right. \\
&\quad \quad \quad \quad \quad +\left( C_{[\alpha \beta }^{*}\phi
_{\gamma \delta }C_{\varepsilon ]}+H_{[\alpha }^{*}H_{\beta
}^{*}\phi _{\gamma \delta }C_{\varepsilon ]}\right)
C-\\
&\quad \quad \quad \quad \quad \left. -\left( C_{[\alpha \beta
}^{*}C_{\gamma }C_{\delta }C_{\varepsilon ]}+H_{[\alpha
}^{*}H_{\beta }^{*}C_{\gamma }C_{\delta }C_{\varepsilon ]}\right)
\right] ,
\end{split}
\end{equation}
\begin{equation}\label{pt42}
\begin{split}
X_{3}^{\left( 7\right) }& =\tfrac{1}{5}\epsilon ^{\alpha \beta
\gamma \delta \varepsilon }\left[ \left( C_{[\alpha \beta \gamma
}^{*}C_{\delta }C_{\varepsilon ]}+C_{[\alpha \beta }^{*}H_{\gamma
}^{*}C_{\delta }C_{\varepsilon ]}+H_{[\alpha }^{*}H_{\beta
}^{*}H_{\gamma }^{*}C_{\delta }C_{\varepsilon
]}\right) -\right. \\
&\quad \quad \quad \quad \quad \left. -\tfrac{1}{2}\left(
C_{[\alpha \beta \gamma }^{*}\phi _{\delta \varepsilon
]}+C_{[\alpha \beta }^{*}H_{\gamma }^{*}\phi _{\delta \varepsilon
]}+H_{[\alpha }^{*}H_{\beta }^{*}H_{\gamma }^{*}\phi _{\delta
\varepsilon ]}\right) C\right] C,
\end{split}
\end{equation}
\begin{equation}\label{pt43}
\begin{split}
X_{4}^{\left( 7\right) }& =-\tfrac{1}{10}\epsilon ^{\alpha \beta
\gamma \delta \varepsilon }\left( C_{[\alpha \beta \gamma \delta
}^{*}C_{\varepsilon ]}+C_{[\alpha \beta \gamma }^{*}H_{\delta
}^{*}C_{\varepsilon ]}+C_{[\alpha \beta
}^{*}C_{\gamma \delta }^{*}C_{\varepsilon ]}+\right.  \\
&\quad \quad \quad \quad \quad \quad \left. +C_{[\alpha \beta
}^{*}H_{\gamma }^{*}H_{\delta }^{*}C_{\varepsilon ]}+H_{[\alpha
}^{*}H_{\beta }^{*}H_{\gamma }^{*}H_{\delta }^{*}C_{\varepsilon
]}\right) CC,
\end{split}
\end{equation}
\begin{equation}\label{pt44}
\begin{split}
X_{5}^{\left( 7\right) }& =\tfrac{1}{30}\epsilon ^{\alpha \beta
\gamma \delta \varepsilon }\left( C_{\alpha \beta \gamma \delta
\varepsilon }^{*}+C_{[\alpha \beta \gamma \delta
}^{*}H_{\varepsilon ]}^{*}+C_{[\alpha \beta \gamma }^{*}C_{\delta
\varepsilon ]}^{*}+C_{[\alpha \beta \gamma
}^{*}H_{\delta }^{*}H_{\varepsilon ]}^{*}+\right.  \\
&\quad \quad \quad \quad \quad \left. +C_{[\alpha \beta
}^{*}C_{\gamma \delta }^{*}H_{\varepsilon ]}^{*}+C_{[\alpha \beta
}^{*}H_{\gamma }^{*}H_{\delta }^{*}H_{\varepsilon ]}^{*}+H_{\alpha
}^{*}H_{\beta }^{*}H_{\gamma }^{*}H_{\delta }^{*}H_{\varepsilon
}^{*}\right) CCC.
\end{split}
\end{equation}

Finally, the coefficients $\left( X_{a}^{\left( 8\right) }\right) _{a=%
\overline{0,5}}$ are given by
\begin{equation}\label{pt45}
\begin{split}
X_{0}^{\left( 8\right) }& =-\tfrac{1}{200}\left[ 60\left(
C^{*}\mathcal{G} ^{\alpha \beta \gamma \delta \varepsilon
}-\tfrac{2}{25}C^{*[\alpha }\mathcal{G}^{\beta \gamma \delta
\varepsilon ]}+\tfrac{1}{5}\phi ^{*[\alpha \beta }K^{\gamma \delta
\varepsilon ]}\right) \mathcal{G}_{\alpha \beta
\gamma \delta \varepsilon }+\right. \\
&\quad \quad \quad \quad +\phi ^{*\alpha \beta }\epsilon _{\alpha
\mu \nu \rho \lambda }\mathcal{G}^{\mu \nu \rho \lambda }\epsilon
_{\beta \mu ^{\prime }\nu ^{\prime }\rho ^{\prime }\lambda
^{\prime }}\mathcal{G}^{\mu
^{\prime }\nu ^{\prime }\rho ^{\prime }\lambda ^{\prime }}-\\
&\quad \quad \quad \quad \left. -\mathcal{G}^{\alpha \beta \gamma
\delta }\epsilon _{\alpha \beta \mu \nu \rho }K^{\mu \nu \rho
}\epsilon _{\gamma \delta \mu ^{\prime }\nu ^{\prime }\rho
^{\prime }}K^{\mu ^{\prime }\nu ^{\prime }\rho ^{\prime }}\right]
,
\end{split}
\end{equation}
\begin{equation}\label{pt46}
\begin{split}
X_{1}^{\left( 8\right) }& =\tfrac{1}{20}\epsilon _{\beta \mu \nu
\rho \lambda }H_{\alpha }^{*}\left( \tfrac{1}{10}K^{\alpha \beta
\gamma }\mathcal{G}^{\mu \nu \rho \lambda }\epsilon _{\gamma \mu
^{\prime }\nu ^{\prime }\rho ^{\prime }\lambda ^{\prime
}}\mathcal{G}^{\mu ^{\prime }\nu ^{\prime }\rho ^{\prime }\lambda
^{\prime }} -\tfrac{1}{3}\mathcal{G}^{\alpha \beta \lambda \delta
\varepsilon }K^{\mu \nu \rho }\epsilon _{\delta \varepsilon \mu
^{\prime }\nu ^{\prime }\rho ^{\prime }}K^{\mu
^{\prime }\nu ^{\prime }\rho ^{\prime }}\right) +\\
&\quad +\tfrac{3}{5}H_{\mu }^{*}\left( -2\phi ^{*\mu \alpha
}\mathcal{G}^{\beta \gamma \delta \varepsilon
}+\tfrac{1}{2}C^{*\mu }\mathcal{G}^{\alpha \beta \gamma \delta
\varepsilon }\right) \mathcal{G}_{\alpha \beta \gamma \delta
\varepsilon },
\end{split}
\end{equation}
\begin{equation}\label{pt47}
\begin{split}
X_{2}^{\left( 8\right) }& =-\tfrac{2}{\left( 5!\right)
^{2}}\epsilon ^{\alpha \beta \gamma \delta \varepsilon }\left(
C_{\alpha \beta }^{*}+H_{\alpha }^{*}H_{\beta }^{*}\right)
\epsilon _{\gamma \mu \nu \rho \lambda }\mathcal{G}^{\mu \nu \rho
\lambda }\epsilon _{\delta \mu ^{\prime }\nu ^{\prime }\rho
^{\prime }\lambda ^{\prime }}\mathcal{G}^{\mu ^{\prime }\nu
^{\prime }\rho ^{\prime }\lambda ^{\prime }}\epsilon _{\varepsilon
\mu ^{\prime \prime }\nu ^{\prime \prime }\rho ^{\prime \prime
}\lambda ^{\prime \prime }}\mathcal{G}^{\mu ^{\prime \prime }\nu
^{\prime \prime }\rho ^{\prime \prime }\lambda ^{\prime
\prime }} +\\
&\quad +\tfrac{3}{10}\left( C_{\mu \nu }^{*}+H_{\mu }^{*}H_{\nu
}^{*}\right) \left( -2K^{\mu \nu \alpha }\mathcal{G}^{\beta \gamma
\delta \varepsilon }+\phi ^{*\mu \nu }\mathcal{G}^{\alpha \beta
\gamma \delta \varepsilon }\right) \mathcal{G}_{\alpha \beta
\gamma \delta \varepsilon },
\end{split}
\end{equation}
\begin{equation}\label{pt48}
\begin{split}
X_{3}^{\left( 8\right) }& =\tfrac{1}{60}\left( C_{\alpha \beta
\gamma }^{*}+C_{[\alpha \beta }^{*}H_{\gamma ]}^{*}+H_{\alpha
}^{*}H_{\beta }^{*}H_{\gamma }^{*}\right) \mathcal{G}^{\alpha
\beta \gamma \delta \varepsilon }\epsilon _{\delta \mu \nu \rho
\lambda }\mathcal{G}^{\mu \nu \rho \lambda }\epsilon _{\varepsilon
\mu ^{\prime }\nu ^{\prime }\rho ^{\prime }\lambda ^{\prime
}}\mathcal{G}^{\mu ^{\prime }\nu ^{\prime }\rho ^{\prime
}\lambda ^{\prime }} -\\
&\quad -\tfrac{1}{10}\left( C_{\mu \nu \rho }^{*}+C_{[\mu \nu
}^{*}H_{\rho
]}^{*}+H_{\mu }^{*}H_{\nu }^{*}H_{\rho }^{*}\right) K^{\mu \nu \rho }%
\mathcal{G}^{\alpha \beta \gamma \delta \varepsilon
}\mathcal{G}_{\alpha \beta \gamma \delta \varepsilon },
\end{split}
\end{equation}
\begin{equation}\label{pt49}
\begin{split}
X_{4}^{\left( 8\right) }& =-\tfrac{1}{20}\left( C_{\mu \nu \rho
\lambda }^{*}+C_{[\mu \nu \rho }^{*}H_{\lambda ]}^{*}+C_{[\mu \nu
}^{*}C_{\rho \lambda ]}^{*}+C_{[\mu \nu }^{*}H_{\rho
}^{*}H_{\lambda
]}^{*}+\right. \\
&\quad \quad \quad \quad \left. +H_{\mu }^{*}H_{\nu
}^{*}H_{\rho }^{*}H_{\lambda }^{*}\right) \mathcal{G}^{\mu \nu \rho \lambda }%
\mathcal{G}^{\alpha \beta \gamma \delta \varepsilon
}\mathcal{G}_{\alpha \beta \gamma \delta \varepsilon },
\end{split}
\end{equation}
\begin{equation}\label{pt50}
\begin{split}
X_{5}^{\left( 8\right) }& =-\tfrac{1}{60}\left( C_{\mu \nu \rho
\lambda \sigma }^{*}+C_{[\mu \nu \rho \lambda }^{*}H_{\sigma
]}^{*}+C_{[\mu \nu \rho }^{*}C_{\lambda \sigma ]}^{*}+C_{[\mu \nu
\rho }^{*}H_{\lambda }^{*}H_{\sigma ]}^{*}+C_{[\mu \nu
}^{*}C_{\rho \lambda
}^{*}H_{\sigma ]}^{*}+\right. \\
&\quad \quad \quad \quad \left. +C_{[\mu \nu }^{*}H_{\rho
}^{*}H_{\lambda }^{*}H_{\sigma ]}^{*}+H_{\mu }^{*}H_{\nu
}^{*}H_{\rho }^{*}H_{\lambda }^{*}H_{\sigma }^{*}\right)
\mathcal{G}^{\mu \nu \rho \lambda \sigma }\mathcal{G}^{\alpha
\beta \gamma \delta \varepsilon }\mathcal{G}_{\alpha \beta \gamma
\delta \varepsilon }.
\end{split}
\end{equation}

\section{Commutators among the deformed gauge transformations \label%
{defgauge}}

We have seen in Sec. \ref{lagfor} that the terms of antighost
number one in the deformed solution to the master equation
(\ref{soldef}) provide the gauge transformations
(\ref{i2})--(\ref{i7}) of the interacting theory. On behalf of
these gauge transformations we are able to identify the nontrivial
gauge generators of all fields. In terms of the notations
(\ref{f10a})--(\ref {f10h}) and taking into account the values
(\ref{f11a})--(\ref{f11d}) of the BRST generators, we observe that
the terms of antighost number one appearing
in the deformed solution (\ref{soldef}) may be generically written like $%
\Phi _{\alpha _{0}}^{*}Z_{\;\;\alpha _{1}}^{\alpha _{0}}\eta
^{\alpha _{1}}$ in De Witt condensed notations, where the
functions $Z_{\;\;\alpha _{1}}^{\alpha _{0}}$ are the precisely
the gauge generators of the deformed gauge transformations.
Identifying the functions $Z_{\;\;\alpha _{1}}^{\alpha _{0}}$ for
each BF field, we initially display the concrete form of the
nonvanishing, deformed gauge
generators. In this way we determine the gauge generators of the one-form $%
A^{\mu }$ (written for convenience in De Witt condensed notations)
like
\begin{equation}
(\tilde{Z}_{(A)}^{\mu })=(Z_{(A)}^{\mu })=\partial ^{\mu }, \;
(\tilde{Z}_{(A)}^{\mu })_{\alpha }=-2gW_{2}\left( \varphi \right)
\delta _{\alpha }^{\mu }, \; (\tilde{Z}_{(A)}^{\mu })_{\alpha
\beta \gamma \delta }=-2gW_{6}\left( \varphi \right) \sigma ^{\mu
\nu }\varepsilon _{\nu \alpha \beta \gamma \delta }, \label{i9c}
\end{equation}
while for the other one-form, $H^{\mu }$, we can write
\begin{equation}
(\tilde{Z}_{(H)}^{\mu })_{\alpha \beta }=-D_{[\alpha }\delta
_{\beta ]}^{\mu }, \label{i10a}
\end{equation}
\begin{eqnarray}
&&\left( \tilde{Z}_{\left( H\right) }^{\mu }\right) =g\left[
\frac{dW_{1}}{d\varphi }\left( \varphi \right) H^{\mu }
-3\frac{dW_{1}}{d\varphi }\left( \varphi \right) K^{\mu
\nu \rho }\phi _{\nu \rho }\right. +\nonumber \\
&&\left. +\tfrac{1}{4}\varepsilon ^{\mu \nu \rho \lambda \sigma
}\left( \frac{dW_{4}}{d\varphi }\left( \varphi \right) \varepsilon
_{\nu \rho \alpha \beta \gamma }K^{\alpha \beta \gamma
}\varepsilon _{\lambda \sigma \alpha ^{\prime }\beta ^{\prime
}\gamma ^{\prime }}K^{\alpha ^{\prime }\beta ^{\prime }\gamma
^{\prime }} +\frac{dW_{5}}{d\varphi }\left( \varphi \right) \phi
_{\nu \rho }\phi _{\lambda \sigma }\right) \right] , \label{i10b}
\end{eqnarray}
\begin{equation}\label{i10c}
\left( \tilde{Z}_{\left( H\right) }^{\mu }\right) _{\alpha \beta
\gamma }=-g\left( \frac{dW_{2}}{d\varphi }\left( \varphi  \right)
\delta _{[\alpha }^{\mu }\phi _{\beta \gamma
]}+3\frac{dW_{6}}{d\varphi }\left( \varphi \right) K^{\mu \nu \rho
}\varepsilon _{\nu \rho \alpha \beta \gamma }\right) ,
\end{equation}
\begin{equation}
\left( \tilde{Z}_{\left( H\right) }^{\mu }\right) ^{\alpha
}=g\left[ 2\left( \frac{dW_{2}}{d\varphi }\left( \varphi \right)
B^{\mu \alpha }+3\frac{dW_{3}}{d\varphi }\left( \varphi \right)
K^{\mu \nu \alpha }A_{\nu }\right) -\frac{dW_{5}}{d\varphi }\left(
\varphi \right) \varepsilon ^{\mu \nu \rho \lambda \alpha }\phi
_{\nu \rho }A_{\lambda }\right] , \label{i10d}
\end{equation}
\begin{equation}
\left( \tilde{Z}_{\left( H\right) }^{\mu }\right) _{\alpha \beta
\gamma \delta }=g\left( \frac{dW_{3}}{d\varphi }\left( \varphi
\right) \delta _{[\alpha }^{\mu }A_{\beta }\phi _{\gamma \delta
]}+12\frac{dW_{4}}{d\varphi }\left( \varphi \right) K^{\mu \nu
\rho }A_{\nu }\varepsilon _{\rho \alpha \beta \gamma \delta }
+2\frac{dW_{6}}{d\varphi }\left( \varphi \right) B^{\mu \nu
}\varepsilon _{\nu \alpha \beta \gamma \delta }\right) .
\label{i10e}
\end{equation}
There is a single nontrivial, deformed gauge generator associated
with the scalar field $\varphi $, which reads as
\begin{equation}
(\tilde{Z}_{(\varphi )})=-gW_{1}\left( \varphi \right) .
\label{i11a}
\end{equation}
Along the same line, we obtain the nonvanishing gauge generators
of the two-forms $B^{\mu \nu }$ and $\phi _{\mu \nu }$ of the type
\begin{gather}
(\tilde{Z}_{(B)}^{\mu \nu })_{\alpha \beta \gamma }=(Z_{(B)}^{\mu
\nu })_{\alpha \beta \gamma }=-\tfrac{1}{2}\partial _{[\alpha
}\delta _{\beta }^{\mu }\delta _{\gamma ]}^{\nu },  \quad
(\tilde{Z}_{(B)}^{\mu \nu })_{\alpha \beta }=-gW_{1}\left( \varphi
\right) \delta _{[\alpha }^{\mu }\delta _{\beta ]}^{\nu },
\label{i12b} \\
(\tilde{Z}_{(B)}^{\mu \nu })_{\alpha \beta \gamma \delta }=g\left(
W_{3}\left( \varphi \right) \delta _{[\alpha }^{\mu }\delta
_{\beta }^{\nu }\phi _{\gamma \delta ]} +12W_{4}\left( \varphi
\right) K^{\mu \nu \rho }\varepsilon _{\rho \alpha \beta \gamma
\delta }\right) , \label{i12b1} \\
(\tilde{Z}_{(B)}^{\mu \nu })^{\alpha }=g\left( 6W_{3}\left(
\varphi \right) K^{\mu \nu \alpha }-W_{5}\left( \varphi \right)
\varepsilon ^{\mu \nu \rho \lambda \alpha }\phi _{\rho \lambda
}\right) ,  \label{i12d} \\
(\tilde{Z}_{(\phi )}^{\mu \nu })_{\alpha }=D^{\left( -\right) [\mu
}\delta _{\alpha }^{\nu ]}, \quad (\tilde{Z}_{(\phi )}^{\mu \nu
})=3g\left( W_{3}\left( \varphi \right) \phi ^{\mu \nu
}+2W_{4}\left( \varphi \right) \varepsilon ^{\mu \nu \rho \lambda
\sigma }K_{\rho \lambda \sigma }\right) ,  \label{i13b} \\
(\tilde{Z}_{(\phi )}^{\mu \nu })^{\alpha \beta \gamma
}=3gW_{6}\left( \varphi \right) \varepsilon ^{\mu \nu \alpha \beta
\gamma },  \quad (\tilde{Z}_{(\phi )}^{\mu \nu })^{\alpha \beta
\gamma \delta }=-6gW_{4}\left( \varphi \right) A^{[\mu }
\varepsilon ^{\nu ]\alpha \beta \gamma \delta }. \label{i13d}
\end{gather}
Finally, the three-form $K^{\mu \nu \rho }$ displays the following
nontrivial, deformed gauge generators:
\begin{gather}
(\tilde{Z}_{(K)}^{\mu \nu \rho })_{\alpha \beta \gamma \delta
}=-\tfrac{1}{6}D_{[\alpha }^{\left( +\right) }\delta _{\beta
}^{\mu }\delta _{\gamma }^{\nu }\delta _{\delta ]}^{\rho }, \quad
(\tilde{Z}_{(K)}^{\mu \nu \rho })^{\alpha }=-gW_{5}\left( \varphi
\right) \varepsilon ^{\mu \nu \rho \lambda \alpha }A_{\lambda
},\label{i14a}\\
(\tilde{Z}_{(K)}^{\mu \nu \rho })=g\left( -3W_{3}\left( \varphi
\right) K^{\mu \nu \rho }+\tfrac{1}{2}\varepsilon ^{\mu \nu \rho
\lambda \sigma }W_{5}\left( \varphi \right) \phi _{\lambda \sigma
}\right) , \label{i14b} \\
(\tilde{Z}_{(K)}^{\mu \nu \rho })_{\alpha \beta \gamma }=-
\tfrac{1}{2}gW_{2}\left( \varphi \right) \delta _{[\alpha }^{\mu
}\delta _{\beta }^{\nu }\delta _{\gamma ]}^{\rho }.  \label{i14c}
\end{gather}

Maintaining the condensed notations introduced in the beginning of
this section, we observe that the deformed solution (\ref{soldef})
contains
pieces of antighost number two generically written like $\left( \frac{1}{2}%
\eta _{\alpha _{1}}^{*}C_{\;\;\;\;\beta _{1}\gamma _{1}}^{\alpha _{1}}-\frac{%
1}{4}M_{\beta _{1}\gamma _{1}}^{\alpha _{0}\beta _{0}}\Phi
_{\alpha
_{0}}^{*}\Phi _{\beta _{0}}^{*}\right) \eta ^{\beta _{1}}\eta ^{\gamma _{1}}$%
. The coefficients $C_{\;\;\;\;\beta _{1}\gamma _{1}}^{\alpha _{1}}$ and $%
M_{\beta _{1}\gamma _{1}}^{\alpha _{0}\beta _{0}}$ represent the
deformed structure functions of order one corresponding to the
gauge algebra of the interacting theory. These structure functions
determine the type of gauge
algebra via the commutators among the new gauge transformations: $%
Z_{\;\;\alpha _{1}}^{\beta _{0}}\frac{\delta Z_{\;\;\beta _{1}}^{\alpha _{0}}%
}{\delta \Phi ^{\beta _{0}}}-Z_{\;\;\beta _{1}}^{\beta
_{0}}\frac{\delta Z_{\;\;\alpha _{1}}^{\alpha _{0}}}{\delta \Phi
^{\beta _{0}}} =C_{\;\;\;\;\alpha _{1}\beta _{1}}^{\gamma
_{1}}Z_{\;\;\gamma _{1}}^{\alpha
_{0}}+M_{\alpha _{1}\beta _{1}}^{\alpha _{0}\beta _{0}}\frac{\delta \tilde{S}%
}{\delta \Phi ^{\beta _{0}}}$. Thus, if at least one coefficient
$M_{\alpha _{1}\beta _{1}}^{\alpha _{0}\beta _{0}}$ is
nonvanishing, then the gauge algebra is open, or, in other words,
only closes on-shell. In the opposite situation the gauge algebra
is closed, but it may be Abelian (all the functions
$C_{\;\;\;\;\alpha _{1}\beta _{1}}^{\gamma _{1}}$ are vanishing)
or non-Abelian. After analyzing all the terms from the deformed
solution to the master equation that contribute to the gauge
algebra relations, we are able to write the expressions of all
commutators corresponding to the interacting BF model. In order to
keep at minimum the number of these relations we omit the Abelian
commutators from the list. In this manner we obtain some
nonvanishing commutators involving the gauge generators of the one
form $A^{\mu }$ like
\begin{equation}
(\tilde{Z}_{(\varphi )})\frac{\delta (\tilde{Z}_{\left( A\right)
}^{\mu })_{\alpha }}{\delta \varphi
}=-3gW_{3}(\tilde{Z}_{(A)}^{\mu })_{\alpha }
-\tfrac{1}{4}gW_{5}\varepsilon ^{\omega \theta \varphi \pi \tau
}\sigma _{\tau \alpha }(\tilde{Z}_{(A)}^{\mu })_{\omega \theta
\varphi \pi }, \label{i15a}
\end{equation}
\begin{equation}
(\tilde{Z}_{(\varphi )})\frac{\delta (\tilde{Z}_{(A)}^{\mu
})_{\alpha \beta \gamma \delta }}{\delta \varphi
}=-6gW_{4}\varepsilon _{\alpha \beta \gamma \delta \eta }\sigma
^{\eta \omega }(\tilde{Z}_{(A)}^{\mu })_{\omega }
+3gW_{3}(\tilde{Z}_{(A)}^{\mu })_{\alpha \beta \gamma \delta },
\label{i15b}
\end{equation}
\begin{equation}
(\tilde{Z}_{(A)}^{\omega })_{\alpha \beta \gamma \delta }\frac{\delta (%
\tilde{Z}_{\left( K\right) }^{\mu \nu \rho })^{\varepsilon
}}{\delta
A^{\omega }}-(\tilde{Z}_{\left( A\right) }^{\omega })^{\varepsilon }\frac{%
\delta (\tilde{Z}_{\left( K\right) }^{\mu \nu \rho })_{\alpha
\beta \gamma \delta }}{\delta A^{\omega }}=
\tfrac{1}{3}gW_{3}\delta _{[\alpha }^{\alpha ^{\prime }}\delta
_{\beta }^{\beta ^{\prime }}\delta _{\gamma }^{\gamma ^{\prime
}}\delta _{\delta ]}^{\varepsilon }(\tilde{Z}_{\left( K\right)
}^{\mu \nu \rho })_{\alpha ^{\prime }\beta ^{\prime }\gamma
^{\prime }},  \label{i16a}
\end{equation}
\begin{equation}
(\tilde{Z}_{\left( A\right) }^{\omega })_{\alpha \beta \gamma \delta }%
\frac{\delta (\tilde{Z}_{\left( K\right) }^{\mu \nu \rho
})_{\theta \varphi \pi \tau }}{\delta A^{\omega
}}-(\tilde{Z}_{\left( A\right) }^{\omega })_{\theta \varphi \pi
\tau }\frac{\delta (\tilde{Z}_{\left( K\right) }^{\mu \nu \rho
})_{\alpha \beta \gamma \delta }}{\delta A^{\omega }}=
\tfrac{1}{3}gW_{4}\delta _{[\alpha }^{\alpha ^{\prime }}\delta
_{\beta }^{\beta ^{\prime }}\delta _{\gamma }^{\gamma ^{\prime
}}\varepsilon _{\delta ]\theta \varphi \pi \tau
}(\tilde{Z}_{\left( K\right) }^{\mu \nu \rho })_{\alpha ^{\prime
}\beta ^{\prime }\gamma ^{\prime }},  \label{i16b}
\end{equation}
\begin{equation}
(\tilde{Z}_{\left( A\right) }^{\omega })^{\alpha }\frac{\delta (\tilde{Z}%
_{\left( K\right) }^{\mu \nu \rho })^{\beta }}{\delta A^{\omega }}-(\tilde{Z}%
_{\left( A\right) }^{\omega })^{\beta }\frac{\delta
(\tilde{Z}_{\left( K\right) }^{\mu \nu \rho })^{\alpha }}{\delta
A^{\omega }}= \tfrac{1}{3}gW_{5}\varepsilon ^{\alpha \beta \alpha
^{\prime }\beta ^{\prime }\gamma ^{\prime }}(\tilde{Z}_{\left(
K\right) }^{\mu \nu \rho })_{\alpha ^{\prime }\beta ^{\prime
}\gamma ^{\prime }},  \label{i16c}
\end{equation}

\begin{eqnarray}
&&(\tilde{Z}_{\left( A\right) }^{\omega })\frac{\delta
(\tilde{Z}_{\left(
K\right) }^{\mu \nu \rho })_{\alpha \beta \gamma \delta }}{\delta A^{\omega }%
}+(\tilde{Z}_{\left( \varphi \right) })\frac{\delta
(\tilde{Z}_{\left( K\right) }^{\mu \nu \rho })_{\alpha \beta
\gamma \delta }}{\delta \varphi }-(\tilde{Z}_{\left( K\right)
}^{\omega \theta \pi })_{\alpha \beta \gamma \delta }\frac{\delta
(\tilde{Z}_{\left( K\right) }^{\mu \nu \rho })}{\delta K^{\omega
\theta \pi }}-(\tilde{Z}_{\left( \phi \right) }^{\omega \theta
})_{\alpha \beta \gamma \delta }\frac{\delta (\tilde{Z}_{\left(
K\right)
}^{\mu \nu \rho })}{\delta \phi ^{\omega \theta }}=  \nonumber \\
&&=\tfrac{1}{8}gW_{3}\delta _{[\alpha }^{\alpha ^{\prime }}\delta
_{\beta }^{\beta ^{\prime }}\delta _{\gamma }^{\gamma ^{\prime
}}\delta _{\delta ]}^{\delta ^{\prime }}(\tilde{Z}_{\left(
K\right) }^{\mu \nu \rho })_{\alpha ^{\prime }\beta ^{\prime
}\gamma ^{\prime }\delta ^{\prime }}-6gW_{4}\varepsilon _{\alpha
\beta \gamma \delta \alpha ^{\prime }}(\tilde{Z}_{\left( K\right)
}^{\mu \nu \rho })^{\alpha ^{\prime
}}-\nonumber\\
&&\quad -\tfrac{1}{2}g\frac{dW_{3}}{d\varphi }\frac{\delta
\tilde{S}}{\delta H^{\alpha ^{\prime }}}\delta _{[\alpha }^{\mu
}\delta _{\beta }^{\nu }\delta _{\gamma }^{\rho }\delta _{\delta
]}^{\alpha ^{\prime }},  \label{i16d}
\end{eqnarray}
\begin{eqnarray}
&&(\tilde{Z}_{\left( A\right) }^{\omega })\frac{\delta
(\tilde{Z}_{\left(
K\right) }^{\mu \nu \rho })^{\alpha }}{\delta A^{\omega }}+(\tilde{Z}%
_{\left( \varphi \right) })\frac{\delta (\tilde{Z}_{\left(
K\right) }^{\mu \nu \rho })^{\alpha }}{\delta \varphi }
-(\tilde{Z}_{\left( K\right) }^{\omega \theta \pi })^{\alpha
}\frac{\delta (\tilde{Z}_{\left( K\right) }^{\mu \nu \rho
})}{\delta K^{\omega \theta \pi } }-(\tilde{Z}_{\left( \phi
\right) }^{\omega \theta })^{\alpha }\frac{\delta (
\tilde{Z}_{\left( K\right) }^{\mu \nu \rho })}{\delta \phi
^{\omega \theta }}
=  \nonumber \\
&&=-\tfrac{1}{4}gW_{5}\varepsilon ^{\alpha \alpha ^{\prime }\beta
^{\prime }\gamma ^{\prime }\delta ^{\prime }}(\tilde{Z}_{\left(
K\right) }^{\mu \nu \rho })_{\alpha ^{\prime }\beta ^{\prime
}\gamma ^{\prime }\delta ^{\prime }} -3gW_{3}\delta _{\alpha
^{\prime }}^{\alpha }(\tilde{Z}_{\left( K\right) }^{\mu \nu \rho
})^{\alpha ^{\prime }}-g\frac{dW_{3}}{d\varphi }\frac{\delta
\tilde{S}}{\delta H^{\alpha ^{\prime }}}\varepsilon ^{\mu \nu \rho
\alpha \alpha ^{\prime }},  \label{i16e}
\end{eqnarray}
\begin{equation}
(\tilde{Z}_{\left( A\right) }^{\omega })_{\alpha }\frac{\delta (\tilde{Z}%
_{\left( \phi \right) }^{\mu \nu })_{\beta }}{\delta A^{\omega }}-(\tilde{Z}%
_{\left( A\right) }^{\omega })_{\beta }\frac{\delta
(\tilde{Z}_{\left( \phi \right) }^{\mu \nu })_{\alpha }}{\delta
A^{\omega }}=\tfrac{1}{3}gW_{5}\varepsilon _{\alpha \beta \alpha
^{\prime }\beta ^{\prime }\gamma ^{\prime }}(\tilde{Z}_{\left(
\phi \right) }^{\mu \nu })^{\alpha ^{\prime }\beta ^{\prime
}\gamma ^{\prime }},  \label{i17a}
\end{equation}
\begin{equation}
(\tilde{Z}_{\left( A\right) }^{\omega })_{\alpha \beta \gamma \delta }%
\frac{\delta (\tilde{Z}_{\left( \phi \right) }^{\mu \nu })_{\theta
\varphi \pi \tau }}{\delta A^{\omega }}-(\tilde{Z}_{\left(
A\right) }^{\omega })_{\theta \varphi \pi \tau }\frac{\delta
(\tilde{Z}_{\left( \phi \right) }^{\mu \nu })_{\alpha \beta \gamma
\delta }}{\delta A^{\omega }}= \tfrac{1}{3}gW_{4}\delta _{[\alpha
}^{\alpha ^{\prime }}\delta _{\beta }^{\beta ^{\prime }}\delta
_{\gamma }^{\gamma ^{\prime }}\varepsilon _{\delta ]\theta \varphi
\pi \tau }(\tilde{Z}_{\left( \phi \right) }^{\mu \nu })_{\alpha
^{\prime }\beta ^{\prime }\gamma ^{\prime }},  \label{i17b}
\end{equation}
\begin{equation}
(\tilde{Z}_{\left( A\right) }^{\omega })^{\alpha \beta \gamma \delta }%
\frac{\delta (\tilde{Z}_{\left( \phi \right) }^{\mu \nu })_{\varepsilon }}{%
\delta A^{\omega }}-(\tilde{Z}_{\left( A\right) }^{\omega })_{\varepsilon }%
\frac{\delta (\tilde{Z}_{\left( \phi \right) }^{\mu \nu })^{\alpha
\beta \gamma \delta }}{\delta A^{\omega }}=
\tfrac{1}{3}gW_{3}\delta _{\alpha ^{\prime }}^{[\alpha }\delta
_{\beta ^{\prime }}^{\beta }\delta _{\gamma ^{\prime }}^{\gamma
}\delta _{\varepsilon }^{\delta ]}(\tilde{Z}_{\left( \phi \right)
}^{\mu \nu })^{\alpha ^{\prime }\beta ^{\prime }\gamma ^{\prime
}},  \label{i17c}
\end{equation}
\begin{eqnarray}
&&(\tilde{Z}_{\left( A\right) }^{\omega })\frac{\delta
(\tilde{Z}_{\left( \phi \right) }^{\mu \nu })_{\alpha }}{\delta
A^{\omega }}+(\tilde{Z}_{\left( \varphi \right) })\frac{\delta
(\tilde{Z}_{\left( \phi \right) }^{\mu \nu })_{\alpha }}{\delta
\varphi }-(\tilde{Z}_{\left( K\right) }^{\omega \theta \pi
})_{\alpha }\frac{\delta (\tilde{Z}_{\left( \phi \right) }^{\mu
\nu \rho })}{\delta K^{\omega \theta
\pi }}-(\tilde{Z}_{\left( \phi \right) }^{\omega \theta })_{\alpha }\frac{%
\delta (\tilde{Z}_{\left( \phi \right) }^{\mu \nu \rho })}{\delta
\phi
^{\omega \theta }}=  \nonumber \\
&&=-3gW_{3}\delta _{\alpha }^{\alpha ^{\prime }}(\tilde{Z}_{\left(
\phi \right) }^{\mu \nu })_{\alpha ^{\prime }}
-\tfrac{1}{4}gW_{5}\varepsilon _{\alpha \alpha ^{\prime }\beta
^{\prime }\gamma ^{\prime }\delta ^{\prime }}(\tilde{Z}_{\left(
\phi \right) }^{\mu \nu })^{\alpha ^{\prime }\beta ^{\prime
}\gamma ^{\prime }\delta ^{\prime }}+
\nonumber \\
&&\quad +3g\frac{\delta W_{3}}{\delta \varphi }\left( \frac{\delta \tilde{S}}{%
\delta H_{\mu }}\delta _{\alpha }^{\nu }-\frac{\delta
\tilde{S}}{\delta H_{\nu }}\delta _{\alpha }^{\mu }\right) ,
\label{i17d}
\end{eqnarray}
\begin{eqnarray}
&&(\tilde{Z}_{\left( A\right) }^{\omega })\frac{\delta
(\tilde{Z}_{\left(
\phi \right) }^{\mu \nu })^{\alpha \beta \gamma \delta }}{\delta A^{\omega }}%
+(\tilde{Z}_{\left( \varphi \right) })\frac{\delta
(\tilde{Z}_{\left( \phi \right) }^{\mu \nu })^{\alpha \beta \gamma
\delta }}{\delta \varphi }-(\tilde{Z}_{\left( K\right) }^{\omega
\theta \pi })^{\alpha \beta \gamma \delta }\frac{\delta
(\tilde{Z}_{\left( \phi \right) }^{\mu \nu })}{\delta K^{\omega
\theta \pi }}-(\tilde{Z}_{\left( \phi \right) }^{\omega \theta
})^{\alpha \beta \gamma \delta }\frac{\delta (\tilde{Z}_{\left(
\phi \right)
}^{\mu \nu })}{\delta \phi ^{\omega \theta }}=  \nonumber \\
&&=-6gW_{4}\varepsilon ^{\alpha \beta \gamma \delta \alpha ^{\prime }}(\tilde{%
Z}_{\left( \phi \right) }^{\mu \nu })_{\alpha ^{\prime }}
+\tfrac{1}{8}gW_{3}\delta _{\alpha ^{\prime }}^{[\alpha }\delta
_{\beta ^{\prime }}^{\beta }\delta _{\gamma ^{\prime }}^{\gamma
}\delta _{\delta ^{\prime }}^{\delta ]}(\tilde{Z}_{\left( \phi
\right) }^{\mu \nu })^{\alpha
^{\prime }\beta ^{\prime }\gamma ^{\prime }\delta ^{\prime }} + \nonumber \\
&&\quad +6g\frac{\delta W_{4}}{\delta \varphi }\left( \frac{\delta \tilde{S}}{%
\delta H_{\mu }}\varepsilon ^{\nu \alpha \beta \gamma \delta
}-\frac{\delta \tilde{S}}{\delta H_{\nu }}\varepsilon ^{\mu \alpha
\beta \gamma \delta }\right) .  \label{i17e}
\end{eqnarray}

Other nonvanishing commutators implying the gauge generators of the fields $%
\varphi $ and $K_{\mu \nu \rho }$ are given by
\begin{eqnarray}
&&(\tilde{Z}_{\left( \varphi \right) })\frac{\delta
(\tilde{Z}_{\left( B\right) }^{\mu \nu })^{\alpha }}{\delta
\varphi }+(\tilde{Z}_{\left( K\right) }^{\omega \theta \pi
})\frac{\delta (\tilde{Z}_{\left( B\right) }^{\mu \nu })^{\alpha
}}{\delta K^{\omega \theta \pi }}+(\tilde{Z}_{\left( \phi \right)
}^{\omega \theta })\frac{\delta (\tilde{Z}_{\left( B\right)
}^{\mu \nu })^{\alpha }}{\delta \phi ^{\omega \theta }}=  \nonumber \\
&&=-3gW_{3}\delta _{\alpha ^{\prime }}^{\alpha }(\tilde{Z}_{\left(
B\right) }^{\mu \nu })^{\alpha ^{\prime
}}+3g\frac{dW_{3}}{d\varphi }K^{\alpha \alpha ^{\prime }\beta
^{\prime }}(\tilde{Z}_{\left( B\right) }^{\mu \nu })_{\alpha
^{\prime }\beta ^{\prime }} - \nonumber \\
&&\quad -\tfrac{1}{4}gW_{5}\varepsilon ^{\alpha \alpha ^{\prime
}\beta ^{\prime }\gamma ^{\prime }\delta ^{\prime
}}(\tilde{Z}_{\left( B\right) }^{\mu \nu })_{\alpha ^{\prime
}\beta ^{\prime }\gamma ^{\prime }\delta ^{\prime }}
-\tfrac{1}{2}g\frac{dW_{5}}{d\varphi }\varepsilon ^{\alpha \alpha
^{\prime }\beta ^{\prime }\gamma ^{\prime }\delta ^{\prime }}\phi
_{\alpha ^{\prime }\beta ^{\prime }}(\tilde{Z}_{\left( B\right)
}^{\mu \nu })_{\gamma ^{\prime }\delta ^{\prime }},  \label{i18a}
\end{eqnarray}
\begin{equation}
(\tilde{Z}_{\left( \varphi \right) })\frac{\delta
(\tilde{Z}_{\left(
B\right) }^{\mu \nu })_{\alpha \beta }}{\delta \varphi }=-\tfrac{1}{2}%
gW_{1}\delta _{[\alpha }^{\alpha ^{\prime }}\delta _{\beta
]}^{\beta ^{\prime }}(\tilde{Z}_{\left( B\right) }^{\mu \nu
})_{\alpha ^{\prime }\beta ^{\prime }},  \label{i18b}
\end{equation}
\begin{eqnarray}
&&(\tilde{Z}_{\left( K\right) }^{\omega \theta \pi })_{\alpha
\beta \gamma \delta }\frac{\delta (\tilde{Z}_{\left( B\right)
}^{\mu \nu })_{\varphi \eta \varsigma \tau }}{\delta K^{\omega
\theta \pi }}+(\tilde{Z}_{\left( \phi \right) }^{\omega \theta
})_{\alpha \beta \gamma \delta }\frac{\delta ( \tilde{Z}_{\left(
B\right) }^{\mu \nu })_{\varphi \eta \varsigma \tau }}{
\delta \phi ^{\omega \theta }} - \nonumber \\
&&-(\tilde{Z}_{\left( K\right) }^{\omega \theta \pi })_{\varphi
\eta \varsigma \tau }\frac{\delta (\tilde{Z}_{\left( B\right)
}^{\mu \nu })_{\alpha \beta \gamma \delta }}{\delta K^{\omega
\theta \pi }}-(\tilde{Z} _{\left( \phi \right) }^{\omega \theta
})_{\varphi \eta \varsigma \tau } \frac{\delta (\tilde{Z}_{\left(
B\right) }^{\mu \nu })_{\alpha \beta \gamma
\delta }}{\delta \phi ^{\omega \theta }}=  \nonumber \\
&&=\tfrac{1}{3}gW_{4}\delta _{[\alpha }^{\alpha ^{\prime }}\delta
_{\beta }^{\beta ^{\prime }}\delta _{\gamma }^{\gamma ^{\prime
}}\varepsilon _{\delta ]\varphi \eta \varsigma \tau
}(\tilde{Z}_{\left( B\right) }^{\mu
\nu })_{\alpha ^{\prime }\beta ^{\prime }\gamma ^{\prime }}  
+\tfrac{1}{2}g\frac{dW_{4}}{d\varphi }\delta _{[\alpha }^{\alpha
^{\prime }}\delta _{\beta }^{\beta ^{\prime }}A_{\gamma
}\varepsilon _{\delta ]\varphi \eta \varsigma \tau
}(\tilde{Z}_{\left( B\right) }^{\mu \nu
})_{\alpha ^{\prime }\beta ^{\prime }}  +\nonumber \\
&&\quad +g\frac{dW_{4}}{d\varphi }\frac{\delta \tilde{S}}{\delta
H^{\alpha ^{\prime }}}\delta _{[\alpha }^{\mu }\delta _{\beta
}^{\nu }\delta _{\gamma }^{\alpha ^{\prime }}\varepsilon _{\delta
]\varphi \eta \varsigma \tau }, \label{i18c}
\end{eqnarray}
\begin{eqnarray}
&&(\tilde{Z}_{\left( \varphi \right) })\frac{\delta
(\tilde{Z}_{\left( B\right) }^{\mu \nu })_{\alpha \beta \gamma
\delta }}{\delta \varphi }+( \tilde{Z}_{\left( K\right) }^{\omega
\theta \pi })\frac{\delta (\tilde{Z} _{\left( B\right) }^{\mu \nu
})_{\alpha \beta \gamma \delta }}{\delta K^{\omega \theta \pi }}
+(\tilde{Z}_{\left( \phi \right) }^{\omega \theta })\frac{\delta
(\tilde{Z} _{\left( B\right) }^{\mu \nu })_{\alpha \beta \gamma
\delta }}{\delta \phi
^{\omega \theta }}=  \nonumber \\
&&=\tfrac{1}{8}gW_{3}\delta _{[\alpha }^{\alpha ^{\prime }}\delta
_{\beta }^{\beta ^{\prime }}\delta _{\gamma }^{\gamma ^{\prime
}}\delta _{\delta ]}^{\delta ^{\prime }}(\tilde{Z}_{\left(
B\right) }^{\mu \nu })_{\alpha ^{\prime }\beta ^{\prime }\gamma
^{\prime }\delta ^{\prime }}+\tfrac{1}{2}g\frac{dW_{3}}{d\varphi
}\delta _{[\alpha }^{\alpha ^{\prime }}\delta _{\beta }^{\beta
^{\prime }}\phi _{\gamma \delta ]}(\tilde{Z} _{\left( B\right)
}^{\mu \nu })_{\alpha ^{\prime }\beta ^{\prime
}}-\nonumber \\
&&\quad -6gW_{4}\varepsilon _{\alpha \beta \gamma \delta \alpha
^{\prime }}(\tilde{Z}_{\left( B\right) }^{\mu \nu })^{\alpha
^{\prime }} +6g\frac{dW_{4}}{d\varphi }\varepsilon _{\alpha \beta
\gamma \delta \alpha ^{\prime }}K^{\alpha ^{\prime }\beta ^{\prime
}\gamma ^{\prime }}(\tilde{Z} _{\left( B\right) }^{\mu \nu
})_{\beta ^{\prime }\gamma ^{\prime }}, \label{i18d}
\end{eqnarray}
\begin{equation}
(\tilde{Z}_{\left( K\right) }^{\omega \theta \pi })_{\alpha \beta
\gamma } \frac{\delta (\tilde{Z}_{\left( B\right) }^{\mu \nu
})^{\delta }}{\delta K^{\omega \theta \pi }}+(\tilde{Z}_{\left(
\phi \right) }^{\omega \theta })_{\alpha \beta \gamma
}\frac{\delta (\tilde{Z}_{\left( B\right) }^{\mu \nu })^{\delta
}}{\delta \phi ^{\omega \theta }}=
\tfrac{1}{2}g\frac{dW_{2}}{d\varphi }\delta _{[\alpha }^{\alpha
^{\prime }}\delta _{\beta }^{\beta ^{\prime }}\delta _{\gamma
]}^{\delta }(\tilde{Z} _{\left( B\right) }^{\mu \nu })_{\alpha
^{\prime }\beta ^{\prime }}, \label{i18e}
\end{equation}
\begin{eqnarray}
&&(\tilde{Z}_{\left( K\right) }^{\omega \theta \pi })_{\alpha
\beta \gamma
\delta }\frac{\delta (\tilde{Z}_{\left( B\right) }^{\mu \nu })^{\varepsilon }%
}{\delta K^{\omega \theta \pi }}+(\tilde{Z}_{\left( \phi \right)
}^{\omega \theta })_{\alpha \beta \gamma \delta }\frac{\delta
(\tilde{Z}_{\left( B\right) }^{\mu \nu })^{\varepsilon }}{\delta
\phi ^{\omega \theta }}
-(\tilde{Z}_{\left( K\right) }^{\omega \theta \pi })^{\varepsilon }\frac{%
\delta (\tilde{Z}_{\left( B\right) }^{\mu \nu })_{\alpha \beta
\gamma \delta }}{\delta K^{\omega \theta \pi }}-(\tilde{Z}_{\left(
\phi \right) }^{\omega \theta })^{\varepsilon }\frac{\delta
(\tilde{Z}_{\left( B\right) }^{\mu \nu })_{\alpha \beta \gamma
\delta }}{\delta \phi ^{\omega \theta }}=  \nonumber
\\
&&=\tfrac{1}{3}gW_{3}\delta _{[\alpha }^{\alpha ^{\prime }}\delta
_{\beta }^{\beta ^{\prime }}\delta _{\gamma }^{\gamma ^{\prime
}}\delta _{\delta ]}^{\varepsilon }(\tilde{Z}_{\left( B\right)
}^{\mu \nu })_{\alpha ^{\prime }\beta ^{\prime }\gamma ^{\prime }}
-\tfrac{1}{2}g\frac{dW_{3}}{d\varphi }\delta _{[\alpha }^{\alpha
^{\prime }}\delta _{\beta }^{\beta ^{\prime }}\delta _{\gamma
}^{\varepsilon }A_{\delta ]}(\tilde{Z}_{\left( B\right) }^{\mu \nu
})_{\alpha ^{\prime
}\beta ^{\prime }} - \nonumber \\
&& \quad -g\frac{dW_{3}}{d\varphi }\frac{\delta \tilde{S}}{\delta
H^{\alpha ^{\prime }}}\delta _{[\alpha }^{\mu }\delta _{\beta
}^{\nu }\delta _{\gamma }^{\varepsilon }\delta _{\delta ]}^{\alpha
^{\prime }}, \label{i18f}
\end{eqnarray}
\begin{equation}
(\tilde{Z}_{\left( K\right) }^{\omega \theta \pi })_{\varphi \eta
\varsigma }\frac{\delta (\tilde{Z}_{\left( B\right) }^{\mu \nu
})_{\alpha \beta \gamma \delta }}{\delta K^{\omega \theta \pi
}}+(\tilde{Z}_{\left( \phi \right) }^{\omega \theta })_{\varphi
\eta \varsigma }\frac{\delta (\tilde{Z}_{\left( B\right) }^{\mu
\nu })_{\alpha \beta \gamma \delta }}{\delta \phi ^{\omega \theta
}}=\tfrac{1}{2}g\frac{dW_{6}}{d\varphi }\delta _{[\alpha }^{\alpha
^{\prime }}\delta _{\beta }^{\beta ^{\prime }}\varepsilon _{\gamma
\delta ]\varphi \eta \varsigma }(\tilde{Z}_{\left( B\right) }^{\mu
\nu })_{\alpha ^{\prime }\beta ^{\prime }},  \label{i18g}
\end{equation}
\begin{eqnarray}
&&(\tilde{Z}_{\left( K\right) }^{\omega \theta \pi })^{\alpha }\frac{\delta (%
\tilde{Z}_{\left( B\right) }^{\mu \nu })^{\beta }}{\delta
K^{\omega \theta
\pi }}+(\tilde{Z}_{\left( \phi \right) }^{\omega \theta })^{\alpha }\frac{%
\delta (\tilde{Z}_{\left( B\right) }^{\mu \nu })^{\beta }}{\delta
\phi ^{\omega \theta }}
-(\tilde{Z}_{\left( K\right) }^{\omega \theta \pi })^{\beta }\frac{\delta (%
\tilde{Z}_{\left( B\right) }^{\mu \nu })^{\alpha }}{\delta
K^{\omega \theta
\pi }}-(\tilde{Z}_{\left( \phi \right) }^{\omega \theta })^{\beta }\frac{%
\delta (\tilde{Z}_{\left( B\right) }^{\mu \nu })^{\alpha }}{\delta
\phi
^{\omega \theta }}=  \nonumber \\
&&=\tfrac{1}{3}gW_{5}\varepsilon ^{\alpha \beta \alpha ^{\prime
}\beta ^{\prime }\gamma ^{\prime }}(\tilde{Z}_{\left( B\right)
}^{\mu \nu })_{\alpha ^{\prime }\beta ^{\prime }\gamma ^{\prime }}
+\tfrac{1}{2}g\frac{dW_{5}}{d\varphi }\varepsilon ^{\alpha \beta
\alpha
^{\prime }\beta ^{\prime }\gamma ^{\prime }}A_{\alpha ^{\prime }}(\tilde{Z}%
_{\left( B\right) }^{\mu \nu })_{\beta ^{\prime }\gamma ^{\prime
}}+ \nonumber \\
&&\quad +g\frac{dW_{5}}{d\varphi }\varepsilon ^{\mu \nu \alpha
\beta \alpha ^{\prime }}\frac{\delta \tilde{S}}{\delta H^{\alpha
^{\prime }}}. \label{i18h}
\end{eqnarray}

Finally, the remaining nonvanishing commutators that involve the
gauge generators of the field $H^{\mu }$ are expressed like
\begin{eqnarray}
&&(\tilde{Z}_{\left( \varphi \right) })\frac{\delta
(\tilde{Z}_{\left( H\right) }^{\mu })^{\alpha }}{\delta \varphi
}+(\tilde{Z}_{\left( A\right) }^{\omega })\frac{\delta
(\tilde{Z}_{\left( H\right) }^{\mu })^{\alpha }}{ \delta A^{\omega
}}+(\tilde{Z}_{\left( K\right) }^{\omega \theta \pi })\frac{\delta
(\tilde{Z} _{\left( H\right) }^{\mu })^{\alpha }}{\delta K^{\omega
\theta \pi }}+(\tilde{Z}_{\left( \phi \right) }^{\omega \theta
})\frac{\delta (\tilde{Z} _{\left( H\right) }^{\mu })^{\alpha
}}{\delta \phi ^{\omega \theta }}-
\nonumber \\
&&-(\tilde{Z}_{\left( K\right) }^{\omega \theta \pi })^{\alpha
}\frac{\delta (\tilde{Z}_{\left( H\right) }^{\mu })}{\delta
K^{\omega \theta \pi }}-(\tilde{Z}_{\left( \phi \right) }^{\omega
\theta })^{\alpha }\frac{\delta (\tilde{Z}_{\left( H\right) }^{\mu
})}{\delta \phi ^{\omega \theta }}-(\tilde{Z}_{\left( H\right)
}^{\omega })^{\alpha }\frac{\delta (\tilde{Z}_{\left(
H\right) }^{\mu })}{\delta H^{\omega }}=  \nonumber \\
&&=-3gW_{3}\delta _{\alpha ^{\prime }}^{\alpha }(\tilde{Z}_{\left(
H\right) }^{\mu })^{\alpha ^{\prime }}+3g\frac{dW_{3}}{d\varphi
}K^{\alpha \alpha ^{\prime }\beta ^{\prime }}(\tilde{Z}_{\left(
H\right) }^{\mu })_{\alpha ^{\prime }\beta ^{\prime }}
-6g\frac{dW_{3}}{d\varphi }\frac{\delta \tilde{S}}{\delta \phi
_{\mu \alpha }} -\nonumber \\
&&\quad -\tfrac{1}{4}gW_{5}\varepsilon ^{\alpha \alpha ^{\prime
}\beta ^{\prime }\gamma ^{\prime }\delta ^{\prime
}}(\tilde{Z}_{\left( H\right) }^{\mu })_{\alpha ^{\prime }\beta
^{\prime }\gamma ^{\prime }\delta ^{\prime }}-
\tfrac{1}{2}g\frac{dW_{5}}{d\varphi }\varepsilon ^{\alpha \alpha
^{\prime }\beta ^{\prime }\gamma ^{\prime }\delta ^{\prime }}\phi
_{\alpha ^{\prime }\beta ^{\prime }}(\tilde{Z}_{\left( H\right)
}^{\mu })_{\gamma ^{\prime }\delta
^{\prime }} +\nonumber \\
&&\quad +6g\frac{d^{2}W_{3}}{d\varphi ^{2}}\frac{\delta
\tilde{S}}{\delta H^{\alpha ^{\prime }}}K^{\mu \alpha \alpha
^{\prime }}-g\frac{dW_{5}}{d\varphi }\frac{\delta
\tilde{S}}{\delta K^{\alpha ^{\prime }\beta ^{\prime }\gamma
^{\prime }}}\varepsilon ^{\mu \alpha \alpha ^{\prime }\beta
^{\prime }\gamma ^{\prime }}-g\frac{d^{2}W_{5}}{d\varphi ^{2}
}\frac{\delta \tilde{S}}{\delta H^{\alpha ^{\prime }}}\varepsilon
^{\mu \alpha \alpha ^{\prime }\beta ^{\prime }\gamma ^{\prime
}}\phi _{\beta ^{\prime }\gamma ^{\prime }},  \label{i19a}
\end{eqnarray}
\begin{eqnarray}
&&(\tilde{Z}_{\left( \varphi \right) })\frac{\delta
(\tilde{Z}_{\left( H\right) }^{\mu })_{\alpha \beta \gamma \delta
}}{\delta \varphi }+(\tilde{Z} _{\left( A\right) }^{\omega
})\frac{\delta (\tilde{Z}_{\left( H\right) }^{\mu })_{\alpha \beta
\gamma \delta }}{\delta A^{\omega }}+(\tilde{Z}_{\left( K\right)
}^{\omega \theta \pi })\frac{\delta (\tilde{Z} _{\left( H\right)
}^{\mu })_{\alpha \beta \gamma \delta }}{\delta K^{\omega \theta
\pi }}+(\tilde{Z}_{\left( \phi \right) }^{\omega \theta
})\frac{\delta (\tilde{Z}_{\left( H\right) }^{\mu })_{\alpha \beta
\gamma \delta }}{\delta \phi ^{\omega \theta }} - \nonumber \\
&&-(\tilde{Z}_{\left( K\right) }^{\omega \theta \pi })_{\alpha
\beta \gamma \delta }\frac{\delta (\tilde{Z}_{\left( H\right)
}^{\mu })}{\delta K^{\omega \theta \pi }}-(\tilde{Z}_{\left( \phi
\right) }^{\omega \theta })_{\alpha \beta \gamma \delta
}\frac{\delta (\tilde{Z}_{\left( H\right) }^{\mu })}{ \delta \phi
^{\omega \theta }}-(\tilde{Z}_{\left( H\right) }^{\omega
})_{\alpha \beta \gamma \delta }\frac{\delta (\tilde{Z}_{\left(
H\right) }^{\mu })}{\delta H^{\omega }}=
\nonumber \\
&&=\tfrac{1}{8}gW_{3}\delta _{[\alpha }^{\alpha ^{\prime }}\delta
_{\beta }^{\beta ^{\prime }}\delta _{\gamma }^{\gamma ^{\prime
}}\delta _{\delta ]}^{\delta ^{\prime }}(\tilde{Z}_{\left(
H\right) }^{\mu })_{\alpha ^{\prime }\beta ^{\prime }\gamma
^{\prime }\delta ^{\prime }}+\tfrac{1}{2}g\frac{dW_{3}}{d\varphi
}\delta _{[\alpha }^{\alpha ^{\prime }}\delta _{\beta }^{\beta
^{\prime }}\phi _{\gamma \delta ]}(\tilde{Z}_{\left( H\right)
}^{\mu })_{\alpha ^{\prime }\beta ^{\prime }} - \nonumber \\
&&\quad -6gW_{4}\varepsilon _{\alpha \beta \gamma \delta \alpha
^{\prime }}(\tilde{Z}_{\left( H\right) }^{\mu })^{\alpha ^{\prime
}}+6g\frac{dW_{4}}{d\varphi }\varepsilon _{\alpha \beta \gamma
\delta \varepsilon }K^{\varepsilon \alpha ^{\prime }\beta ^{\prime
}}(\tilde{Z}_{\left( H\right) }^{\mu })_{\alpha ^{\prime }\beta
^{\prime }} -g\frac{d^{2}W_{3}}{d\varphi ^{2}}\delta _{[\alpha
}^{\mu }\delta _{\beta }^{\varepsilon }\phi _{\gamma
\delta ]}\frac{\delta \tilde{S}}{\delta H^{\varepsilon }} -\nonumber \\
&&\quad -\tfrac{1}{2}g\frac{dW_{3}}{d\varphi }\delta _{[\alpha
}^{\mu }\delta _{\beta }^{\beta ^{\prime }}\delta _{\gamma
}^{\gamma ^{\prime }}\delta _{\delta ]}^{\delta ^{\prime
}}\frac{\delta \tilde{S}}{\delta K^{\beta ^{\prime }\gamma
^{\prime }\delta ^{\prime }}}-12g\frac{dW_{4}}{d\varphi
}\frac{\delta \tilde{S}}{\delta \phi _{\mu \varepsilon
}}\varepsilon _{\alpha \beta \gamma \delta \varepsilon }-\nonumber
\\
&&\quad -12g\frac{d^{2}W_{4}}{d\varphi ^{2}}\frac{\delta
\tilde{S}}{\delta H^{\alpha ^{\prime }}}K^{\mu \alpha ^{\prime
}\beta ^{\prime }}\varepsilon _{\alpha \beta \gamma \delta \beta
^{\prime }}, \label{i19b}
\end{eqnarray}
\begin{eqnarray}
&&(\tilde{Z}_{\left( A\right) }^{\omega })^{\alpha }\frac{\delta
(\tilde{Z} _{\left( H\right) }^{\mu })^{\beta }}{\delta A^{\omega
}}+(\tilde{Z}_{\left( B\right) }^{\omega \theta })^{\alpha
}\frac{\delta (\tilde{Z}_{\left( H\right) }^{\mu })^{\beta
}}{\delta B^{\omega \theta }}+(\tilde{Z}_{\left( K\right)
}^{\omega \theta \pi })^{\alpha }\frac{\delta (\tilde{Z}_{\left(
H\right) }^{\mu })^{\beta }}{\delta K^{\omega \theta \pi
}}+(\tilde{Z}_{\left( \phi \right) }^{\omega \theta })^{\alpha
}\frac{\delta (\tilde{Z}_{\left( H\right) }^{\mu })^{\beta
}}{\delta \phi ^{\omega \theta }}-
\nonumber \\
&&-(\tilde{Z}_{\left( A\right) }^{\omega })^{\beta }\frac{\delta
(\tilde{Z} _{\left( H\right) }^{\mu })^{\alpha }}{\delta A^{\omega
}}-(\tilde{Z} _{\left( B\right) }^{\omega \theta })^{\beta
}\frac{\delta (\tilde{Z} _{\left( H\right) }^{\mu })^{\alpha
}}{\delta B^{\omega \theta }}-(\tilde{Z}_{\left( K\right)
}^{\omega \theta \pi })^{\beta }\frac{\delta (\tilde{Z}_{\left(
H\right) }^{\mu })^{\alpha }}{\delta K^{\omega \theta \pi
}}-(\tilde{Z}_{\left( \phi \right) }^{\omega \theta })^{\beta
}\frac{\delta (\tilde{Z}_{\left( H\right) }^{\mu })^{\alpha
}}{\delta \phi ^{\omega \theta }}= \nonumber \\
&&=\tfrac{1}{3}gW_{5}\varepsilon ^{\alpha \beta \alpha ^{\prime
}\beta ^{\prime }\gamma ^{\prime }}(\tilde{Z}_{\left( H\right)
}^{\mu })_{\alpha
^{\prime }\beta ^{\prime \gamma ^{\prime }}}+\tfrac{1}{2}g\frac{dW_{5}}{%
d\varphi }\varepsilon ^{\alpha \beta \gamma \alpha ^{\prime }\beta
^{\prime }}A_{\gamma }(\tilde{Z}_{\left( H\right) }^{\mu
})_{\alpha ^{\prime }\beta
^{\prime }} -\nonumber \\
&&\quad -g\frac{dW_{5}}{d\varphi }\frac{\delta \tilde{S}}{\delta
B^{\alpha ^{\prime }\beta ^{\prime }}}\varepsilon ^{\mu \alpha
\beta \alpha ^{\prime
}\beta ^{\prime }}-g\frac{d^{2}W_{5}}{d\varphi ^{2}}\frac{\delta \tilde{S}}{%
\delta H^{\alpha ^{\prime }}}A_{\beta ^{\prime }}\varepsilon ^{\mu
\alpha \beta \alpha ^{\prime }\beta ^{\prime }},  \label{i19c}
\end{eqnarray}
\begin{eqnarray}
&&(\tilde{Z}_{\left( A\right) }^{\omega })_{\alpha \beta \gamma
\delta } \frac{\delta (\tilde{Z}_{\left( H\right) }^{\mu
})^{\varepsilon }}{\delta A^{\omega }}+(\tilde{Z}_{\left( B\right)
}^{\omega \theta })_{\alpha \beta \gamma \delta }\frac{\delta
(\tilde{Z}_{\left( H\right) }^{\mu })^{\varepsilon }}{\delta
B^{\omega \theta }}+(\tilde{Z}_{\left( K\right) }^{\omega \theta
\pi })_{\alpha \beta \gamma \delta }\frac{\delta
(\tilde{Z}_{\left( H\right) }^{\mu })^{\varepsilon }}{\delta
K^{\omega \theta \pi }}+(\tilde{Z}_{\left( \phi \right) }^{\omega
\theta })_{\alpha \beta \gamma \delta }\frac{\delta
(\tilde{Z}_{\left( H\right) }^{\mu })^{\varepsilon }}{\delta \phi
^{\omega \theta }} - \nonumber
\\
&&-(\tilde{Z}_{\left( A\right) }^{\omega })^{\varepsilon
}\frac{\delta (\tilde{Z}_{\left( H\right) }^{\mu })_{\alpha \beta
\gamma \delta }}{\delta A^{\omega }}-(\tilde{Z}_{\left( B\right)
}^{\omega \theta })^{\varepsilon }\frac{\delta (\tilde{Z}_{\left(
H\right) }^{\mu })_{\alpha \beta \gamma \delta }}{\delta B^{\omega
\theta }}-(\tilde{Z}_{\left( K\right) }^{\omega \theta \pi
})^{\varepsilon }\frac{\delta (\tilde{Z}_{\left( H\right) }^{\mu
})_{\alpha \beta \gamma \delta }}{\delta K^{\omega \theta \pi
}}-(\tilde{Z}_{\left( \phi \right) }^{\omega \theta
})^{\varepsilon }\frac{\delta (\tilde{Z}_{\left( H\right) }^{\mu
})_{\alpha \beta \gamma \delta }}{\delta \phi ^{\omega \theta }}=
\nonumber
\\
&&=\tfrac{1}{3}gW_{3}\delta _{[\alpha }^{\alpha ^{\prime }}\delta
_{\beta }^{\beta ^{\prime }}\delta _{\gamma }^{\gamma ^{\prime
}}\delta _{\delta ]}^{\varepsilon }(\tilde{Z}_{\left( H\right)
}^{\mu })_{\alpha ^{\prime
}\beta ^{\prime \gamma ^{\prime }}}-\tfrac{1}{2}g\frac{dW_{3}}{d\varphi }%
\delta _{[\alpha }^{\alpha ^{\prime }}\delta _{\beta }^{\beta
^{\prime }}\delta _{\gamma }^{\varepsilon }A_{\delta
]}(\tilde{Z}_{\left( H\right)
}^{\mu })_{\alpha ^{\prime }\beta ^{\prime }} - \nonumber \\
&&\quad -g\frac{dW_{3}}{d\varphi }\frac{\delta \tilde{S}}{\delta
B^{\alpha ^{\prime }\beta ^{\prime }}}\delta _{[\alpha }^{\alpha
}\delta _{\beta }^{\varepsilon }\delta _{\gamma }^{\alpha ^{\prime
}}\delta _{\delta
]}^{\beta ^{\prime }}-g\frac{d^{2}W_{3}}{d\varphi ^{2}}\frac{\delta \tilde{S}%
}{\delta H^{\alpha ^{\prime }}}A_{\beta ^{\prime }}\delta
_{[\alpha }^{\alpha }\delta _{\beta }^{\varepsilon }\delta
_{\gamma }^{\alpha ^{\prime }}\delta _{\delta ]}^{\beta ^{\prime
}},  \label{i19d}
\end{eqnarray}
\begin{eqnarray}
&&(\tilde{Z}_{\left( \varphi \right) })\frac{\delta
(\tilde{Z}_{\left( H\right) }^{\mu })_{\alpha \beta }}{\delta
\varphi }+(\tilde{Z}_{\left( A\right) }^{\omega })\frac{\delta
(\tilde{Z}_{\left( H\right) }^{\mu })_{\alpha \beta }}{\delta
A^{\omega }}-(\tilde{Z}_{\left( H\right) }^{\omega })_{\alpha
\beta }\frac{\delta (\tilde{Z}_{\left( H\right) }^{\mu
})}{\delta H^{\omega }}=  \nonumber \\
&&=-\tfrac{1}{2}g\frac{dW_{1}}{d\varphi }\delta _{[\alpha
}^{\alpha ^{\prime }}\delta _{\beta ]}^{\beta ^{\prime
}}(\tilde{Z}_{\left( H\right) }^{\mu })_{\alpha ^{\prime }\beta
^{\prime }}+g\frac{d^{2}W_{1}}{d\varphi ^{2}}\frac{\delta
\tilde{S}}{\delta H^{\alpha ^{\prime }}}\delta _{[\alpha }^{\mu
}\delta _{\beta ]}^{\alpha ^{\prime }},  \label{i19e}
\end{eqnarray}
\begin{eqnarray}
&&(\tilde{Z}_{\left( B\right) }^{\omega \theta })_{\alpha \beta \gamma }%
\frac{\delta (\tilde{Z}_{\left( H\right) }^{\mu })^{\varepsilon
}}{\delta B^{\omega \theta }}+(\tilde{Z}_{\left( K\right)
}^{\omega \theta \pi })_{\alpha \beta \gamma }\frac{\delta
(\tilde{Z}_{\left( H\right) }^{\mu })^{\varepsilon }}{\delta
K^{\omega \theta \pi }}+(\tilde{Z}_{\left( \phi
\right) }^{\omega \theta })_{\alpha \beta \gamma }\frac{\delta (\tilde{Z}%
_{\left( H\right) }^{\mu })^{\varepsilon }}{\delta \phi ^{\omega
\theta }}-
\nonumber \\
&&-(\tilde{Z}_{\left( K\right) }^{\omega \theta \pi })^{\varepsilon }\frac{%
\delta (\tilde{Z}_{\left( H\right) }^{\mu })_{\alpha \beta \gamma
}}{\delta K^{\omega \theta \pi }}-(\tilde{Z}_{\left( \phi \right)
}^{\omega \theta })^{\varepsilon }\frac{\delta (\tilde{Z}_{\left(
H\right) }^{\mu })_{\alpha
\beta \gamma }}{\delta \phi ^{\omega \theta }}=  \nonumber \\
&&=\tfrac{1}{2}g\frac{dW_{2}}{d\varphi }\delta _{[\alpha }^{\alpha
^{\prime
}}\delta _{\beta }^{\beta ^{\prime }}\delta _{\gamma ]}^{\varepsilon }(%
\tilde{Z}_{\left( H\right) }^{\mu })_{\alpha ^{\prime }\beta ^{\prime }}+g%
\frac{d^{2}W_{2}}{d\varphi ^{2}}\frac{\delta \tilde{S}}{\delta
H^{\alpha ^{\prime }}}\delta _{[\alpha }^{\mu }\delta _{\beta
}^{\varepsilon }\delta _{\gamma ]}^{\alpha ^{\prime }},
\label{i19f}
\end{eqnarray}
\begin{eqnarray}
&&(\tilde{Z}_{\left( K\right) }^{\omega \theta \pi })_{\alpha
\beta \gamma \delta }\frac{\delta (\tilde{Z}_{\left( H\right)
}^{\mu })_{\varphi \eta \zeta }}{\delta K^{\omega \theta \pi
}}+(\tilde{Z}_{\left( \phi \right) }^{\omega \theta })_{\alpha
\beta \gamma \delta }\frac{\delta (\tilde{Z} _{\left( H\right)
}^{\mu })_{\varphi \eta \zeta }}{\delta \phi ^{\omega \theta
}}-(\tilde{Z}_{\left( B\right) }^{\omega \theta })_{\varphi \eta
\zeta }\frac{\delta (\tilde{Z}_{\left( H\right) }^{\mu })_{\alpha
\beta \gamma \delta }}{\delta B^{\omega \theta
}}- \nonumber \\
&&-(\tilde{Z}_{\left( K\right) }^{\omega \theta \pi })_{\varphi
\eta \zeta }\frac{\delta (\tilde{Z}_{\left( H\right) }^{\mu
})_{\alpha \beta \gamma \delta }}{\delta K^{\omega \theta \pi }}
-(\tilde{Z}_{\left( \phi \right) }^{\omega \theta })_{\varphi \eta
\zeta }\frac{\delta (\tilde{Z}_{\left( H\right) }^{\mu })_{\alpha
\beta \gamma
\delta }}{\delta \phi ^{\omega \theta }}=  \nonumber \\
&&=-g\frac{dW_{6}}{d\varphi }\delta _{[\alpha }^{\alpha ^{\prime
}}\delta _{\beta }^{\beta ^{\prime }}\varepsilon _{\gamma \delta
]\varphi \eta \zeta
}(\tilde{Z}_{\left( H\right) }^{\mu })_{\alpha ^{\prime }\beta ^{\prime }}+g%
\frac{d^{2}W_{6}}{d\varphi ^{2}}\frac{\delta \tilde{S}}{\delta
H^{\alpha ^{\prime }}}\delta _{[\varphi }^{\mu }\delta _{\eta
}^{\alpha ^{\prime }}\varepsilon _{\zeta ]\alpha \beta \gamma
\delta },  \label{i19g}
\end{eqnarray}
\begin{eqnarray}
&&(\tilde{Z}_{\left( A\right) }^{\omega })_{\alpha \beta \gamma \delta }%
\frac{\delta (\tilde{Z}_{\left( H\right) }^{\mu })_{\varphi \eta \zeta \tau }%
}{\delta A^{\omega }}+(\tilde{Z}_{\left( B\right) }^{\omega \theta
})_{\alpha \beta \gamma \delta }\frac{\delta (\tilde{Z}_{\left(
H\right) }^{\mu })_{\varphi \eta \zeta \tau }}{\delta B^{\omega
\theta }}+(\tilde{Z}_{\left( K\right) }^{\omega \theta \pi
})_{\alpha \beta \gamma \delta }\frac{\delta (\tilde{Z}_{\left(
H\right) }^{\mu })_{\varphi \eta \zeta \tau }}{\delta K^{\omega
\theta \pi }}+\nonumber \\
&&+(\tilde{Z}_{\left( \phi \right) }^{\omega \theta })_{\alpha
\beta \gamma \delta }\frac{\delta (\tilde{Z}_{\left( H\right)
}^{\mu })_{\varphi \eta \zeta \tau }}{\delta \phi ^{\omega \theta
}}-(\tilde{Z}_{\left( A\right) }^{\omega })_{\varphi \eta \zeta
\tau }\frac{\delta \left( \tilde{Z}_{\left( H\right) }^{\mu
}\right) _{\alpha \beta \gamma \delta }}{\delta A^{\omega
}}-(\tilde{Z}_{\left( B\right) }^{\omega \theta })_{\varphi \eta
\zeta \tau }\frac{\delta (\tilde{Z}_{\left( H\right) }^{\mu
})_{\alpha \beta \gamma \delta }}{\delta B^{\omega \theta
}}-\nonumber \\
&&-(\tilde{Z}_{\left( K\right) }^{\omega \theta \pi })_{\varphi
\eta \zeta \tau }\frac{\delta (\tilde{Z}_{\left( H\right) }^{\mu
})_{\alpha \beta \gamma \delta }}{\delta K^{\omega \theta \pi
}}-(\tilde{Z}_{\left( \phi \right) }^{\omega \theta })_{\varphi
\eta \zeta \tau }\frac{\delta (\tilde{Z}_{\left( H\right) }^{\mu
})_{\alpha \beta \gamma \delta }}{\delta \phi
^{\omega \theta }}=  \nonumber \\
&&=\tfrac{1}{3}gW_{4}\delta _{[\alpha }^{\alpha ^{\prime }}\delta
_{\beta }^{\beta ^{\prime }}\delta _{\gamma }^{\gamma ^{\prime
}}\varepsilon _{\delta ]\varphi \eta \zeta \tau
}(\tilde{Z}_{\left( H\right) }^{\mu })_{\alpha ^{\prime }\beta
^{\prime \gamma ^{\prime }}}+\tfrac{1}{2}g\frac{dW_{4}}{d\varphi
}\delta _{[\alpha }^{\alpha ^{\prime }}\delta _{\beta }^{\beta
^{\prime }}A_{\gamma }\varepsilon _{\delta ]\varphi \eta \zeta
\tau }(\tilde{Z}_{\left( H\right) }^{\mu })_{\alpha ^{\prime
}\beta ^{\prime }} \nonumber \\
&&\quad -2g\frac{dW_{4}}{d\varphi }\frac{\delta \tilde{S}}{\delta
B^{\alpha ^{\prime }\beta ^{\prime }}}\delta _{[\alpha }^{\mu
}\delta _{\beta }^{\alpha ^{\prime }}\delta _{\gamma }^{\beta
^{\prime }}\varepsilon _{\delta ]\varphi \eta \zeta \tau
}-2g\frac{d^{2}W_{4}}{d\varphi ^{2}}\frac{\delta \tilde{S}}{\delta
H^{\alpha ^{\prime }}}A_{\beta ^{\prime }}\delta _{[\alpha }^{\mu
}\delta _{\beta }^{\alpha ^{\prime }}\delta _{\gamma }^{\beta
^{\prime }}\varepsilon _{\delta ]\varphi \eta \zeta \tau }.
\label{i19h}
\end{eqnarray}

\section{Reducibility of the interacting model \label{defred}}

In what follows we focus on the reducibility functions and
relations corresponding to the resulting interacting BF model. In
view of this we maintain the condensed notations introduced in the
Appendix \ref{defgauge} and observe that the deformed solution
(\ref{soldef}) contains pieces of antighost numbers $k=2,3,4$,
generically written like
\begin{equation*}\sum_{k=2}^{4}\left(
c_{k}\eta _{\alpha _{k-2}}^{*}\Phi _{\beta _{0}}^{*}C_{\alpha
_{k}}^{\alpha _{k-2}\beta _{0}}+\eta _{\alpha
_{k-1}}^{*}Z_{\;\;\alpha _{k}}^{\alpha _{k-1}}\right) \eta
^{\alpha _{k}},
\end{equation*}
where $c_{2}=\frac{1}{2}$, $c_{3}=-1$, and $c_{4}=1$. The
functions $\left( Z_{\;\;\alpha _{k}}^{\alpha _{k-1}}\right)
_{k=2,3,4}$ represent the deformed reducibility functions of order
$\left( k-1\right) $ and the above terms produce the reducibility
relations $Z_{\;\;\alpha _{k-1}}^{\alpha _{k-2}}Z_{\;\;\alpha
_{k}}^{\alpha
_{k-1}}=C_{\alpha _{k}}^{\alpha _{k-2}\beta _{0}}\frac{\delta \tilde{S}}{%
\delta \Phi ^{\beta _{0}}}$ of order $\left( k-1\right) $, with
$k=2,3,4$,
associated with the interacting BF theory. If at least one coefficient $%
C_{\alpha _{k}}^{\alpha _{k-2}\beta _{0}}$ is nonvanishing, we say
that the reducibility relations of order $\left( k-1\right) $ take
place on-shell, while in the opposite situation we say that the
$\left( k-1\right) $-level reducibility holds off-shell. In order
to keep the number of relations at minimum we completely omit
off-shell reducibility relations.

Analyzing these kinds of terms, we obtain that the deformed
first-order reducibility functions ($k=2$) read as
\begin{equation}
(\tilde{Z}_{1})=2gW_{2}\left( \varphi \right), \quad
(\tilde{Z}_{1})_{\mu \nu \rho \lambda \sigma }=-2g\varepsilon
_{\mu \nu \rho \lambda \sigma }W_{6}\left( \varphi \right) , \quad
(\tilde{Z}_{1}^{\mu })=D^{(-)\mu },  \label{rf3}
\end{equation}
\begin{equation}
(\tilde{Z}_{1}^{\mu })_{\alpha \beta \gamma \delta \varepsilon
}=6g\varepsilon _{\alpha \beta \gamma \delta \varepsilon
}W_{4}\left( \varphi \right) A^{\mu }, \quad (\tilde{Z}_{1}^{\mu
})^{\nu \rho \lambda \sigma }=-3g\varepsilon ^{\mu \nu \rho
\lambda \sigma }W_{6}\left( \varphi \right) ,  \label{rf4}
\end{equation}
\begin{equation}
(\tilde{Z}_{1}^{\mu \nu })_{\alpha \beta \gamma
}=-\tfrac{1}{2}D_{[\alpha }\delta _{\beta }^{\mu }\delta _{\gamma
]}^{\nu }, \quad (\tilde{Z}_{1}^{\mu \nu \rho })=2gW_{3}\left(
\varphi \right) -\tfrac{1}{3}gW_{5}\left( \varphi \right)
\varepsilon ^{\mu \nu \rho \lambda \sigma }\phi _{\lambda \sigma
}, \label{rf6}
\end{equation}
\begin{equation}
(\tilde{Z}_{1}^{\mu \nu })=-g\left( \frac{ dW_{2}}{d\varphi
}\left( \varphi \right) B^{\mu \nu }-3\left(
\frac{dW_{3}}{d\varphi }\left( \varphi \right) K^{\mu \nu \rho
}-\tfrac{1}{6}\frac{dW_{5}}{d\varphi }\left( \varphi \right)
\varepsilon ^{\mu \nu \rho \lambda \sigma }\phi _{\lambda \sigma
}\right) A_{\rho }\right),  \label{rf7}
\end{equation}
\begin{equation}
(\tilde{Z}_{1}^{\mu \nu })_{\alpha \beta \gamma \delta
}=\tfrac{1}{2}g\left( \frac{dW_{2}}{d\varphi }\left( \varphi
\right) \delta _{[\alpha }^{\mu }\delta _{\beta }^{\nu }\phi
_{\gamma \delta ]} +6\frac{dW_{3}}{d\varphi }\left( \varphi
\right) K^{\mu \nu \rho }\varepsilon _{\rho \alpha \beta \gamma
\delta }\right) , \label{rf8}
\end{equation}
\begin{eqnarray}
&&(\tilde{Z}_{1}^{\mu \nu })_{\alpha \beta \gamma \delta
\varepsilon }=-\tfrac{1}{2}g\left( \frac{dW_{3}}{d\varphi } \left(
\varphi \right) \delta _{[\alpha }^{\mu }\delta _{\beta }^{\nu
}A_{\gamma }\phi _{\delta \varepsilon ]}+2\varepsilon _{\alpha
\beta \gamma
\delta \varepsilon }\times \right.  \nonumber \\
&&\left. \times \left( 6\frac{dW_{4}}{d\varphi }\left( \varphi
\right) K^{\mu \nu \rho }A_{\rho }- \frac{dW_{6}}{d\varphi }\left(
\varphi \right) B^{\mu \nu }\right) \right) , \label{rf9}
\end{eqnarray}
\begin{equation}
(\tilde{Z}_{1}^{\mu \nu \rho })_{\alpha \beta \gamma
}=\tfrac{1}{3}gW_{1}\left( \varphi \right) \delta _{[\alpha }^{\mu
}\delta _{\beta }^{\nu }\delta _{\gamma ]}^{\rho }, \quad
(\tilde{Z}_{1}^{\mu \nu \rho })_{\alpha \beta \gamma \delta
}=(Z_{1}^{\mu \nu \rho })_{\alpha \beta \gamma \delta },
\end{equation}
\begin{equation}
(\tilde{Z}_{1}^{\mu \nu \rho })_{\alpha \beta \gamma \delta
\varepsilon }=-4gW_{4}\left( \varphi \right) K^{\mu \nu \rho
}\varepsilon _{\alpha \beta \gamma \delta \varepsilon
}-\tfrac{1}{3}gW_{3}\left( \varphi \right) \delta _{[\alpha }^{\mu
}\delta _{\beta }^{\nu }\delta _{\gamma }^{\rho }\phi _{\delta
\varepsilon ]},  \label{rf12}
\end{equation}
\begin{equation}
(\tilde{Z}_{1}^{\mu \nu \rho \lambda })=-\tfrac{1}{4}gW_{5}\left(
\varphi \right) \varepsilon ^{\mu \nu \rho \lambda \sigma
}A_{\sigma } ,\quad (\tilde{Z}_{1}^{\mu \nu \rho \lambda
})_{\alpha \beta \gamma \delta }=\tfrac{1}{8}gW_{2}\left( \varphi
\right) \delta _{[\alpha }^{\mu }\delta _{\beta }^{\nu }\delta
_{\gamma }^{\rho }\delta _{\delta ]}^{\lambda }, \label{rf14}
\end{equation}
\begin{equation}
(\tilde{Z}_{1}^{\mu \nu \rho \lambda })_{\alpha \beta \gamma
\delta \varepsilon }=-\tfrac{1}{24}D_{[\alpha }^{(+)}\delta
_{\beta }^{\mu }\delta _{\gamma }^{\nu }\delta _{\delta }^{\rho
}\delta _{\varepsilon ]}^{\lambda },  \label{rf15}
\end{equation}
while the on-shell, first-order reducibility relations are given
by
\begin{equation}
(\tilde{Z}_{(A)}^{\mu })(\tilde{Z}_{1})+(\tilde{Z}_{(A)}^{\mu })_{\alpha }(%
\tilde{Z}_{1}^{\alpha })+(\tilde{Z}_{(A)}^{\mu })_{\alpha \beta
\gamma \delta }(\tilde{Z}_{1}^{\alpha \beta \gamma \delta
})=2g\frac{dW_{2}}{d\varphi }\frac{\delta \tilde{S}}{\delta H_{\mu
}}, \label{red1a}
\end{equation}
\begin{eqnarray}
&&(\tilde{Z}_{(A)}^{\mu })(\tilde{Z}_{1})_{\alpha \beta \gamma
\delta \varepsilon }+(\tilde{Z}_{(A)}^{\mu })_{\mu ^{\prime
}}(\tilde{Z}_{1}^{\mu
^{\prime }})_{\alpha \beta \gamma \delta \varepsilon }+  \nonumber \\
&&+(\tilde{Z}_{(A)}^{\mu })_{\mu ^{\prime }\nu ^{\prime }\rho
^{\prime }\lambda ^{\prime }}(\tilde{Z}_{1}^{\mu ^{\prime }\nu
^{\prime }\rho ^{\prime }\lambda ^{\prime }})_{\alpha \beta \gamma
\delta \varepsilon }=-2g\varepsilon _{\alpha \beta \gamma \delta
\varepsilon }\frac{dW_{6}}{d\varphi }\frac{\delta \tilde{S}}{\delta H_{\mu }}%
,  \label{red1b}
\end{eqnarray}
\begin{eqnarray}
&&(\tilde{Z}_{(B)}^{\mu \nu })_{\alpha }(\tilde{Z}_{1}^{\alpha })+(\tilde{Z}%
_{(B)}^{\mu \nu })_{\alpha \beta }(\tilde{Z}_{1}^{\alpha \beta })+(\tilde{Z}%
_{(B)}^{\mu \nu })_{\alpha \beta \gamma }(\tilde{Z}_{1}^{\alpha
\beta \gamma
})+(\tilde{Z}_{(B)}^{\mu \nu })_{\alpha \beta \gamma \delta }(\tilde{Z}%
_{1}^{\alpha \beta \gamma \delta })=\nonumber \\
&&=6gW_{3}\frac{\delta \tilde{S}}{\delta
\phi _{\mu \nu }}+g\varepsilon ^{\mu \nu \rho \lambda \sigma }W_{5}\frac{%
\delta \tilde{S}}{\delta K^{\rho \lambda \sigma }}+g\frac{\delta
\tilde{S}}{\delta H^{\rho }}\left( -6\frac{dW_{3}}{d\varphi
}K^{\mu \nu \rho }+\frac{dW_{5}}{d\varphi }\varepsilon ^{\mu \nu
\rho \lambda \sigma }\phi _{\lambda \sigma }\right) ,
\label{red2a}
\end{eqnarray}
\begin{equation}
(\tilde{Z}_{(B)}^{\mu \nu })_{\alpha ^{\prime }\beta ^{\prime }}(\tilde{Z}%
_{1}^{\alpha ^{\prime }\beta ^{\prime }})_{\alpha \beta \gamma }+(\tilde{Z}%
_{(B)}^{\mu \nu })_{\alpha ^{\prime }\beta ^{\prime }\gamma ^{\prime }}(%
\tilde{Z}_{1}^{\alpha ^{\prime }\beta ^{\prime }\gamma ^{\prime
}})_{\alpha \beta \gamma }=-g\frac{dW_{1}}{d\varphi }\frac{\delta
\tilde{S}}{\delta H_{\rho }}\delta _{[ \alpha }^{\mu }\delta
_{\beta }^{\nu }\delta _{\gamma ]}^{\rho }, \label{red2b}
\end{equation}
\begin{eqnarray}
&&(\tilde{Z}_{(B)}^{\mu \nu })_{\alpha ^{\prime
}}(\tilde{Z}_{1}^{\alpha ^{\prime }})_{\alpha \beta \gamma \delta
\varepsilon }+(\tilde{Z}_{(B)}^{\mu \nu })_{\alpha ^{\prime }\beta
^{\prime }}(\tilde{Z}_{1}^{\alpha ^{\prime }\beta ^{\prime
}})_{\alpha \beta \gamma \delta \varepsilon
}+(\tilde{Z}_{(B)}^{\mu \nu })_{\alpha ^{\prime }\beta ^{\prime
}\gamma ^{\prime }}(\tilde{Z}_{1}^{\alpha ^{\prime }\beta ^{\prime
}\gamma ^{\prime }})_{\alpha \beta \gamma \delta \varepsilon }+\nonumber \\
&&+(\tilde{Z}_{(B)}^{\mu \nu })_{\alpha ^{\prime }\beta ^{\prime
}\gamma ^{\prime }\delta ^{\prime }}(\tilde{Z}_{1}^{\alpha
^{\prime }\beta ^{\prime }\gamma ^{\prime }\delta ^{\prime
}})_{\alpha \beta \gamma \delta \varepsilon }=\tfrac{1}{2}g\left(
W_{3}\frac{\delta \tilde{S}}{\delta K^{\rho \lambda \sigma
}}+\frac{dW_{3}}{d\varphi }\frac{\delta \tilde{S}}{\delta H^{\rho
}}\phi _{\lambda \sigma }\right) \delta _{[\alpha }^{\mu }\delta
_{\beta }^{\nu }\delta _{\gamma }^{\rho }\delta _{\delta
}^{\lambda }\delta _{\varepsilon ]}^{\sigma
}-\nonumber \\
&&-12g\left( W_{4}\frac{\delta \tilde{S}}{\delta \phi _{\mu \nu
}}-\frac{dW_{4}}{d\varphi }\frac{\delta \tilde{S}}{\delta H^{\rho
}}K^{\mu \nu \rho }\right) \varepsilon _{\alpha \beta \gamma
\delta \varepsilon }, \label{red3b}
\end{eqnarray}
\begin{eqnarray}
&&(\tilde{Z}_{(\phi )}^{\mu \nu
})(\tilde{Z}_{1})+(\tilde{Z}_{(\phi )}^{\mu \nu })_{\alpha
}(\tilde{Z}_{1}^{\alpha })+(\tilde{Z}_{(\phi )}^{\mu \nu
})_{\alpha \beta \gamma }(\tilde{Z}_{1}^{\alpha \beta \gamma
})+(\tilde{Z}_{(\phi )}^{\mu \nu })_{\alpha \beta \gamma \delta }(\tilde{Z}%
_{1}^{\alpha \beta \gamma \delta })=\nonumber
\\
&&=-6g\left( W_{3}\frac{\delta \tilde{S}}{\delta B_{\mu \nu
}}+\frac{dW_{3}}{d\varphi }\frac{\delta \tilde{S}}{\delta
H_{\alpha }}A^{\beta }\delta _{\alpha }^{[\mu }\delta _{\beta
}^{\nu ]}\right) ,  \label{red4a}
\end{eqnarray}
\begin{eqnarray}
&&(\tilde{Z}_{(\phi )}^{\mu \nu })_{\alpha ^{\prime }\beta
^{\prime }\gamma ^{\prime }}(\tilde{Z}_{1}^{\alpha ^{\prime }\beta
^{\prime }\gamma ^{\prime }})_{\alpha \beta \gamma \delta
}+(\tilde{Z}_{(\phi )}^{\mu \nu })_{\alpha ^{\prime }\beta
^{\prime }\gamma ^{\prime }\delta ^{\prime
}}(\tilde{Z}_{1}^{\alpha ^{\prime }\beta ^{\prime }\gamma ^{\prime
}\delta ^{\prime }})_{\alpha \beta \gamma \delta
}+(\tilde{Z}_{(\phi )}^{\mu \nu })_{\alpha ^{\prime
}}(\tilde{Z}_{1}^{\alpha
^{\prime }})_{\alpha \beta \gamma \delta }=\nonumber \\
&&=-3g\frac{dW_{6}}{d\varphi }\frac{\delta \tilde{S}}{\delta
H_{\lambda }}\delta _{\lambda }^{[\mu }\sigma ^{\nu ]\rho
}\varepsilon _{\rho \alpha \beta \gamma \delta },  \label{red4b}
\end{eqnarray}
\begin{eqnarray}
&&(\tilde{Z}_{(\phi )}^{\mu \nu })_{\alpha ^{\prime
}}(\tilde{Z}_{1}^{\alpha ^{\prime }})_{\alpha \beta \gamma \delta
\varepsilon }+(\tilde{Z}_{(\phi )}^{\mu \nu })_{\alpha ^{\prime
}\beta ^{\prime }\gamma ^{\prime }}(\tilde{Z}_{1}^{\alpha ^{\prime
}\beta ^{\prime }\gamma ^{\prime }})_{\alpha \beta \gamma \delta
\varepsilon }+(\tilde{Z}_{(\phi )}^{\mu \nu })_{\alpha ^{\prime
}\beta ^{\prime }\gamma ^{\prime }\delta ^{\prime
}}(\tilde{Z}_{1}^{\alpha ^{\prime }\beta ^{\prime }\gamma ^{\prime
}\delta ^{\prime }})_{\alpha \beta
\gamma \delta \varepsilon }+\nonumber \\
&&+(\tilde{Z}_{(\phi )}^{\mu \nu })(\tilde{Z}_{1})_{\alpha \beta
\gamma \delta \varepsilon }=12g\left( W_{4}\frac{\delta
\tilde{S}}{\delta B_{\mu \nu }}+\frac{dW_{4}}{d\varphi
}\frac{\delta \tilde{S}}{\delta H_{\rho }}A^{\lambda }\delta
_{\rho }^{[\mu }\delta _{\lambda }^{\nu ]}\right) \varepsilon
_{\alpha \beta \gamma \delta \varepsilon }, \label{red4c}
\end{eqnarray}
\begin{eqnarray}
&&(\tilde{Z}_{(K)}^{\mu \nu \rho
})(\tilde{Z}_{1})+(\tilde{Z}_{(K)}^{\mu \nu \rho })_{\alpha
}(\tilde{Z}_{1}^{\alpha })+(\tilde{Z}_{(K)}^{\mu \nu \rho
})_{\alpha \beta \gamma }(\tilde{Z}_{1}^{\alpha \beta \gamma
})+(\tilde{Z}_{(K)}^{\mu \nu \rho })_{\alpha \beta \gamma \delta
}(\tilde{Z}_{1}^{\alpha \beta \gamma \delta })=\nonumber
\\
&&=-g\left( W_{5}\frac{\delta \tilde{S}}{%
\delta B^{\lambda \sigma }}+\frac{dW_{5}}{d\varphi }\frac{\delta
\tilde{S}}{\delta H^{\lambda }}A_{\sigma }\right) \varepsilon
^{\mu \nu \rho \lambda \sigma },  \label{red5a}
\end{eqnarray}
\begin{eqnarray}
&&(\tilde{Z}_{(K)}^{\mu \nu \rho })_{\alpha ^{\prime }\beta
^{\prime }\gamma ^{\prime }}(\tilde{Z}_{1}^{\alpha ^{\prime }\beta
^{\prime }\gamma ^{\prime }})_{\alpha \beta \gamma \delta
}+(\tilde{Z}_{(K)}^{\mu \nu \rho })_{\alpha ^{\prime }\beta
^{\prime }\gamma ^{\prime }\delta ^{\prime
}}(\tilde{Z}_{1}^{\alpha ^{\prime }\beta ^{\prime }\gamma ^{\prime
}\delta ^{\prime }})_{\alpha \beta \gamma \delta
}+(\tilde{Z}_{(K)}^{\mu \nu \rho })_{\alpha ^{\prime }}
(\tilde{Z}_{1}^{\alpha ^{\prime }})_{\alpha \beta \gamma \delta }=\nonumber \\
&&=\tfrac{1}{2}g\frac{%
dW_{2}}{d\varphi }\frac{\delta \tilde{S}}{\delta H^{\lambda
}}\delta _{[\alpha }^{\mu }\delta _{\beta }^{\nu }\delta _{\gamma
}^{\rho }\delta _{\delta ]}^{\lambda }, \label{red5b}
\end{eqnarray}
\begin{eqnarray}
&&(\tilde{Z}_{(K)}^{\mu \nu \rho })_{\alpha ^{\prime
}}(\tilde{Z}_{1}^{\alpha ^{\prime }})_{\alpha \beta \gamma \delta
\varepsilon }+(\tilde{Z}_{(K)}^{\mu \nu \rho })_{\alpha ^{\prime
}\beta ^{\prime }\gamma ^{\prime }}(\tilde{Z}_{1}^{\alpha ^{\prime
}\beta ^{\prime }\gamma ^{\prime }})_{\alpha \beta \gamma \delta
\varepsilon }+(\tilde{Z}_{(K)}^{\mu \nu \rho })_{\alpha ^{\prime
}\beta ^{\prime }\gamma ^{\prime }\delta ^{\prime
}}(\tilde{Z}_{1}^{\alpha ^{\prime }\beta ^{\prime }\gamma ^{\prime
}\delta ^{\prime }})_{\alpha \beta
\gamma \delta \varepsilon }+\nonumber \\
&&+(\tilde{Z}_{(K)}^{\mu \nu \rho })(\tilde{Z}_{1})_{\alpha \beta
\gamma \delta \varepsilon }= -\tfrac{1}{2}g\left(
W_{3}\frac{\delta \tilde{S}}{\delta B^{\lambda \sigma
}}+\frac{dW_{3}}{d\varphi }\frac{\delta \tilde{S}}{\delta H^{\lambda }}%
A_{\sigma }\right) \delta _{[\alpha }^{\mu }\delta _{\beta }^{\nu
}\delta _{\gamma }^{\rho }\delta _{\delta }^{\lambda }\delta
_{\varepsilon ]}^{\sigma }, \label{red5c}
\end{eqnarray}
\begin{eqnarray}
&&(\tilde{Z}_{(H)}^{\mu })(\tilde{Z}_{1})+(\tilde{Z}_{(H)}^{\mu
})_{\alpha }(\tilde{Z}_{1}^{\alpha })+(\tilde{Z}_{(H)}^{\mu
})_{\alpha \beta }(\tilde{Z}_{1}^{\alpha \beta
})+(\tilde{Z}_{(H)}^{\mu })_{\alpha \beta \gamma
}(\tilde{Z}_{1}^{\alpha \beta \gamma })+(\tilde{Z}_{(H)}^{\mu
})_{\alpha \beta \gamma \delta }
(\tilde{Z}_{1}^{\alpha \beta \gamma \delta })=\nonumber \\
&&=-2g\frac{dW_{2}}{d\varphi }%
\frac{\delta \tilde{S}}{\delta A_{\mu }}-g\frac{\delta
\tilde{S}}{\delta H^{\nu }}\left( 2\frac{d^{2}W_{2}}{d\varphi
^{2}}B^{\mu \nu }-6\frac{d^{2}W_{3}}{d\varphi ^{2}}K^{\mu \nu \rho
}A_{\rho }+\frac{d^{2}W_{5}}{d\varphi ^{2}}\varepsilon ^{\mu \nu
\rho
\lambda \sigma }A_{\rho }\phi _{\lambda \sigma }\right) + \nonumber \\
&&\quad +g\frac{\delta \tilde{S}}{\delta B^{\nu \rho }}\left( 6\frac{dW_{3}}{%
d\varphi }K^{\mu \nu \rho }-\frac{dW_{5}}{d\varphi }\varepsilon
^{\mu \nu \rho \lambda \sigma }\phi _{\lambda \sigma }\right) +\nonumber \\
&&\quad +g\frac{dW_{5}}{d\varphi }\frac{\delta \tilde{S}}{\delta
K^{\rho \lambda
\sigma }}\varepsilon ^{\mu \nu \rho \lambda \sigma }A_{\nu }+6g\frac{dW_{3}}{%
d\varphi }\frac{\delta \tilde{S}}{\delta \phi _{\mu \nu }}A_{\nu
}, \label{red6a}
\end{eqnarray}
\begin{equation}
(\tilde{Z}_{(H)}^{\mu })_{\alpha ^{\prime }\beta ^{\prime }}(\tilde{Z}%
_{1}^{\alpha ^{\prime }\beta ^{\prime }})_{\alpha \beta \gamma }+(\tilde{Z}%
_{(H)}^{\mu })_{\alpha ^{\prime }\beta ^{\prime }\gamma ^{\prime }}(\tilde{Z}%
_{1}^{\alpha ^{\prime }\beta ^{\prime }\gamma ^{\prime }})_{\alpha
\beta \gamma }=g\left( \frac{dW_{1}}{d\varphi }\frac{\delta
\tilde{S}}{\delta B^{\nu \rho
}}+\frac{d^{2}W_{1}}{d\varphi ^{2}}\frac{\delta \tilde{S}}{\delta H^{\nu }}%
A_{\rho }\right) \delta _{[\alpha }^{\mu }\delta _{\beta }^{\nu
}\delta _{\gamma ]}^{\rho },  \label{red6b}
\end{equation}
\begin{eqnarray}
&&(\tilde{Z}_{(H)}^{\mu })_{\alpha ^{\prime }\beta ^{\prime
}}(\tilde{Z}_{1}^{\alpha ^{\prime }\beta ^{\prime }})_{\alpha
\beta \gamma \delta }+(\tilde{Z}_{(H)}^{\mu })_{\alpha ^{\prime
}\beta ^{\prime }\gamma ^{\prime }}(\tilde{Z}_{1}^{\alpha ^{\prime
}\beta ^{\prime }\gamma ^{\prime }})_{\alpha \beta \gamma \delta
}+(\tilde{Z}_{(H)}^{\mu })_{\alpha ^{\prime
}}(\tilde{Z}_{1}^{\alpha ^{\prime }})_{\alpha \beta \gamma \delta
}+
\nonumber \\
&&+(\tilde{Z}_{(H)}^{\mu })_{\alpha
^{\prime }\beta ^{\prime }\gamma ^{\prime }\delta ^{\prime }}(\tilde{Z}%
_{1}^{\alpha ^{\prime }\beta ^{\prime }\gamma ^{\prime }\delta
^{\prime }})_{\alpha \beta \gamma \delta }=g\frac{\delta
\tilde{S}}{\delta H^{\nu }}\left( \tfrac{1}{2}\frac{
d^{2}W_{2}}{d\varphi ^{2}}\phi _{\rho \lambda }\delta _{[\alpha
}^{\mu }\delta _{\beta }^{\nu }\delta _{\gamma }^{\rho }\delta
_{\delta ]}^{\lambda }+6\frac{d^{2}W_{6}}{d\varphi ^{2}}K^{\mu \nu
\rho }\varepsilon _{\rho \alpha \beta \gamma \delta }\right) +
\nonumber
\\
&&+\tfrac{1}{2}g\frac{dW_{2}}{d\varphi }\frac{\delta
\tilde{S}}{\delta K^{\nu \rho \lambda }}\delta _{[\alpha }^{\mu
}\delta _{\beta }^{\nu }\delta _{\gamma }^{\rho }\delta _{\delta
]}^{\lambda }+6\frac{\delta W_{6}}{\delta \varphi }\frac{\delta
\tilde{S}}{\delta \phi _{\mu \nu }}\varepsilon _{\nu \alpha \beta
\gamma \delta },  \label{red6c}
\end{eqnarray}
\begin{eqnarray}
&&(\tilde{Z}_{(H)}^{\mu })(\tilde{Z}_{1})_{\alpha \beta \gamma
\delta \varepsilon }+(\tilde{Z}_{(H)}^{\mu })_{\alpha ^{\prime
}}(\tilde{Z}_{1}^{\alpha ^{\prime }})_{\alpha \beta \gamma \delta
\varepsilon }+(\tilde{Z}_{(H)}^{\mu })_{\alpha ^{\prime }\beta
^{\prime }}(\tilde{Z}_{1}^{\alpha ^{\prime }\beta ^{\prime
}})_{\alpha \beta \gamma \delta \varepsilon }+
\nonumber \\
&&+(\tilde{Z}_{(H)}^{\mu })_{\alpha ^{\prime }\beta ^{\prime
}\gamma ^{\prime }}(\tilde{Z}_{1}^{\alpha ^{\prime }\beta ^{\prime
}\gamma ^{\prime }})_{\alpha \beta \gamma \delta \varepsilon
}+(\tilde{Z}_{(H)}^{\mu })_{\alpha ^{\prime }\beta ^{\prime
}\gamma ^{\prime }\delta ^{\prime }}(\tilde{Z}_{1}^{\alpha
^{\prime }\beta ^{\prime }\gamma ^{\prime }\delta ^{\prime
}})_{\alpha \beta \gamma \delta \varepsilon }=\nonumber \\
&&=g\frac{\delta \tilde{S}}{\delta H^{\nu }}\left( 2\left(
\frac{d^{2}W_{6}}{d\varphi ^{2}}B^{\mu \nu
}-6\frac{d^{2}W_{4}}{d\varphi ^{2}}K^{\mu \nu \rho }A_{\rho
}\right) \varepsilon _{\alpha \beta \gamma \delta \varepsilon
}-\right.  \nonumber \\
&&\quad \left. -\tfrac{1}{2}\frac{d^{2}W_{3}}{d\varphi
^{2}}A_{\rho }\phi _{\lambda \sigma }\delta _{[\alpha }^{\mu
}\delta _{\beta }^{\nu }\delta _{\gamma }^{\rho }\delta _{\delta
}^{\lambda }\delta _{\varepsilon ]}^{\sigma }\right)
+2g\frac{dW_{6}}{d\varphi }\frac{\delta \tilde{S}}{\delta A_{\mu
}}\varepsilon _{\alpha \beta \gamma \delta
\varepsilon } - \nonumber \\
&&\quad -g\frac{\delta \tilde{S}}{\delta B^{\nu \rho }}\left(
12\frac{dW_{4}}{d\varphi }K^{\mu \nu \rho }\varepsilon _{\alpha
\beta \gamma \delta \varepsilon
}+\tfrac{1}{2}\frac{dW_{3}}{d\varphi }\phi _{\lambda \sigma
}\delta _{[\alpha }^{\mu }\delta _{\beta }^{\nu }\delta _{\gamma
}^{\rho }\delta _{\delta }^{\lambda }\delta
_{\varepsilon ]}^{\sigma }\right) - \nonumber \\
&&\quad -12g\frac{dW_{4}}{d\varphi }\frac{\delta \tilde{S}}{\delta
\phi _{\mu \nu }}A_{\nu }\varepsilon _{\alpha \beta \gamma \delta
\varepsilon }+\tfrac{1}{2}g\frac{dW_{3}}{d\varphi }\frac{\delta
\tilde{S}}{\delta K^{\rho \lambda \sigma }}A_{\nu }\delta
_{[\alpha }^{\mu }\delta _{\beta }^{\nu }\delta _{\gamma }^{\rho
}\delta _{\delta }^{\lambda }\delta _{\varepsilon ]}^{\sigma }.
\label{red6d}
\end{eqnarray}

In the next step we determine the deformed second-order
reducibility functions ($k=3$) of the form
\begin{equation}
(\tilde{Z}_{2})_{\alpha \beta \gamma \delta \varepsilon
}=-3g\varepsilon _{\alpha \beta \gamma \delta \varepsilon
}W_{6}\left( \varphi \right) , \quad (\tilde{Z}_{2}^{\mu \nu \rho
})_{\alpha \beta \gamma \delta }=-\tfrac{1}{6}D_{[\alpha }\delta
_{\beta }^{\mu }\delta _{\gamma }^{\nu }\delta _{\delta ]}^{\rho
}, \label{redu2}
\end{equation}
\begin{equation}
(\tilde{Z}_{2}^{\mu \nu \rho })_{\alpha \beta \gamma \delta
\varepsilon }=-g\left( \frac{dW_{6}}{d\varphi }\left( \varphi
\right) K^{\mu \nu \rho }\varepsilon _{\alpha \beta \gamma \delta
\varepsilon }+\tfrac{1}{6}\frac{dW_{2}}{d\varphi }\left( \varphi
\right) \delta _{[\alpha }^{\mu }\delta _{\beta }^{\nu }\delta
_{\gamma }^{\rho }\phi _{\delta \varepsilon ]} \right) ,
\label{redu3}
\end{equation}
\begin{equation}
(\tilde{Z}_{2}^{\mu \nu \rho \lambda })_{\alpha \beta \gamma
\delta }=-\tfrac{1}{12}gW_{1}\left( \varphi \right) \delta
_{[\alpha }^{\mu }\delta _{\beta }^{\nu }\delta _{\gamma }^{\rho
}\delta _{\delta ]}^{\lambda }, \quad (\tilde{Z}_{2}^{\mu \nu \rho
\lambda })_{\alpha \beta \gamma \delta \varepsilon }=(Z_{2}^{\mu
\nu \rho \lambda })_{\alpha \beta \gamma \delta \varepsilon },
\label{redu5}
\end{equation}
\begin{equation}
(\tilde{Z}_{2}^{\mu \nu \rho \lambda \sigma })_{\alpha \beta
\gamma \delta \varepsilon }=-\tfrac{1}{40}gW_{2}\left( \varphi
\right) \delta _{[\alpha }^{\mu }\delta _{\beta }^{\nu }\delta
_{\gamma }^{\rho }\delta _{\delta }^{\lambda }\delta _{\varepsilon
]}^{\sigma }, \label{6}
\end{equation}
such that the corresponding on-shell, second-order reducibility
relations are expressed by
\begin{eqnarray}
&&(\tilde{Z}_{1}^{\mu })(\tilde{Z}_{2})_{\alpha \beta \gamma
\delta \varepsilon }+(\tilde{Z}_{1}^{\mu })_{\alpha ^{\prime
}\beta ^{\prime }\gamma ^{\prime }\delta ^{\prime
}}(\tilde{Z}_{2}^{\alpha ^{\prime }\beta ^{\prime }\gamma ^{\prime
}\delta ^{\prime }})_{\alpha \beta \gamma \delta \varepsilon
}+(\tilde{Z}_{1}^{\mu })_{\alpha ^{\prime }\beta ^{\prime }\gamma
^{\prime }\delta ^{\prime }\varepsilon ^{\prime
}}(\tilde{Z}_{2}^{\alpha ^{\prime }\beta ^{\prime }\gamma ^{\prime
}\delta ^{\prime }\varepsilon ^{\prime
}})_{\alpha \beta \gamma \delta \varepsilon }=\nonumber \\
&&=-3g\frac{dW_{6}}{d\varphi }\frac{\delta \tilde{S}}{\delta
H_{\mu }}\varepsilon _{\alpha \beta \gamma \delta \varepsilon },
\label{redu7a}
\end{eqnarray}
\begin{equation}
(\tilde{Z}_{1}^{\mu \nu \rho })_{\alpha ^{\prime }\beta ^{\prime
}\gamma ^{\prime }}(\tilde{Z}_{2}^{\alpha ^{\prime }\beta ^{\prime
}\gamma ^{\prime }})_{\alpha \beta \gamma \delta
}+(\tilde{Z}_{1}^{\mu \nu \rho })_{\alpha
^{\prime }\beta ^{\prime }\gamma ^{\prime }\delta ^{\prime }}(\tilde{Z}%
_{2}^{\alpha ^{\prime }\beta ^{\prime }\gamma ^{\prime }\delta
^{\prime }})_{\alpha \beta \gamma \delta
}=-\tfrac{1}{3}g\frac{dW_{1}}{d\varphi }\frac{\delta
\tilde{S}}{\delta H^{\lambda }}\delta _{[\alpha }^{\mu }\delta
_{\beta }^{\nu }\delta _{\gamma }^{\rho }\delta _{\delta
]}^{\lambda }, \label{redu7b}
\end{equation}
\begin{eqnarray}
&&(\tilde{Z}_{1}^{\mu \nu \rho \lambda })(\tilde{Z}_{2})_{\alpha
\beta \gamma \delta \varepsilon }+(\tilde{Z}_{1}^{\mu \nu \rho
\lambda })_{\alpha ^{\prime }\beta ^{\prime }\gamma ^{\prime
}\delta ^{\prime }}(\tilde{Z}_{2}^{\alpha ^{\prime }\beta ^{\prime
}\gamma ^{\prime }\delta ^{\prime }})_{\alpha \beta \gamma \delta
\varepsilon }+(\tilde{Z}_{1}^{\mu \nu \rho \lambda })_{\alpha
^{\prime }\beta ^{\prime }\gamma ^{\prime }\delta ^{\prime
}\varepsilon ^{\prime }}(\tilde{Z}_{2}^{\alpha ^{\prime }\beta
^{\prime }\gamma ^{\prime }\delta ^{\prime }\varepsilon ^{\prime
}})_{\alpha \beta \gamma \delta \varepsilon }=
\nonumber \\
&&=\tfrac{1}{8}g\frac{dW_{2}}{d\varphi }\frac{\delta
\tilde{S}}{\delta H^{\sigma }}\delta _{[\alpha }^{\mu }\delta
_{\beta }^{\nu }\delta _{\gamma }^{\rho }\delta _{\delta
}^{\lambda }\delta _{\varepsilon ]}^{\sigma },  \label{redu7c}
\end{eqnarray}
\begin{eqnarray}
&&(\tilde{Z}_{1}^{\mu \nu })_{\alpha ^{\prime }\beta ^{\prime
}\gamma ^{\prime }}(\tilde{Z}_{2}^{\alpha ^{\prime }\beta ^{\prime
}\gamma ^{\prime }})_{\alpha \beta \gamma \delta
}+(\tilde{Z}_{1}^{\mu \nu })_{\alpha
^{\prime }\beta ^{\prime }\gamma ^{\prime }\delta ^{\prime }}(\tilde{Z}%
_{2}^{\alpha ^{\prime }\beta ^{\prime }\gamma ^{\prime }\delta
^{\prime
}})_{\alpha \beta \gamma \delta }=  \nonumber \\
&&=\tfrac{1}{2}g\left( \frac{d^{2}W_{1}}{d\varphi ^{2}}\frac{\delta \tilde{S}%
}{\delta H^{\rho }}A_{\lambda }+\frac{dW_{1}}{d\varphi }\frac{\delta \tilde{S%
}}{\delta B^{\rho \lambda }}\right) \delta _{[\alpha }^{\mu
}\delta _{\beta }^{\nu }\delta _{\gamma }^{\rho }\delta _{\delta
]}^{\lambda },  \label{redu7d}
\end{eqnarray}
\begin{eqnarray}
&&(\tilde{Z}_{1}^{\mu \nu })(\tilde{Z}_{2})_{\alpha \beta \gamma
\delta \varepsilon }+(\tilde{Z}_{1}^{\mu \nu })_{\alpha ^{\prime
}\beta ^{\prime }\gamma ^{\prime }\delta ^{\prime }\varepsilon
^{\prime }}(\tilde{Z}_{2}^{\alpha ^{\prime }\beta ^{\prime }\gamma
^{\prime }\delta ^{\prime }\varepsilon ^{\prime }})_{\alpha \beta
\gamma \delta \varepsilon }+(\tilde{Z}_{1}^{\mu \nu })_{\alpha
^{\prime }\beta ^{\prime }\gamma ^{\prime }}(\tilde{Z}_{2}^{\alpha
^{\prime }\beta ^{\prime }\gamma ^{\prime }})_{\alpha \beta \gamma
\delta \varepsilon }+
\nonumber \\
&&+(\tilde{Z}_{1}^{\mu \nu })_{\alpha ^{\prime }\beta ^{\prime
}\gamma ^{\prime }\delta ^{\prime }}(\tilde{Z}_{2}^{\alpha
^{\prime }\beta ^{\prime }\gamma ^{\prime }\delta ^{\prime
}})_{\alpha \beta \gamma \delta \varepsilon }=\tfrac{1}{4}g\left(
\frac{d^{2}W_{2}}{d\varphi ^{2}}\frac{\delta \tilde{S}}{\delta
H^{\rho }}\phi _{\lambda \sigma }+\frac{dW_{2}}{d\varphi
}\frac{\delta \tilde{S}}{\delta K^{\rho \lambda \sigma }}\right)
\delta _{[\alpha }^{\mu }\delta _{\beta }^{\nu }\delta _{\gamma
}^{\rho }\delta _{\delta }^{\lambda }\delta _{\varepsilon
]}^{\sigma }-
\nonumber \\
&&\quad -3g\left( \frac{dW_{6}}{d\varphi }\frac{\delta
\tilde{S}}{\delta \phi _{\mu \nu }}-\frac{d^{2}W_{6}}{d\varphi
^{2}}\frac{\delta \tilde{S}}{\delta H^{\rho }}K^{\mu \nu \rho
}\right) \varepsilon _{\alpha \beta \gamma \delta \varepsilon }.
\label{redu7e}
\end{eqnarray}

In a similar manner we obtain the deformed third-order
reducibility functions ($k=4$) like
\begin{equation}
(\tilde{Z}_{3}^{\mu \nu \rho \lambda })_{\alpha \beta \gamma
\delta \varepsilon }=-\tfrac{1}{24}D_{[\alpha }\delta _{\beta
}^{\mu }\delta _{\gamma }^{\nu }\delta _{\delta }^{\rho }\delta
_{\varepsilon ]}^{\lambda }, \quad (\tilde{Z}_{3}^{\mu \nu \rho
\lambda \sigma })_{\alpha \beta \gamma \delta \varepsilon
}=\tfrac{1}{60}gW_{1}\left( \varphi \right) \delta _{[\alpha
}^{\mu }\delta _{\beta }^{\nu }\delta _{\gamma }^{\rho }\delta
_{\delta }^{\lambda }\delta _{\varepsilon ]}^{\sigma },
\label{reduc2}
\end{equation}
together with the accompanying on-shell, third-order reducibility
relations
\begin{eqnarray}
&&(\tilde{Z}_{2}^{\mu \nu \rho \lambda })_{\alpha ^{\prime }\beta
^{\prime }\gamma ^{\prime }\delta ^{\prime
}}(\tilde{Z}_{3}^{\alpha ^{\prime }\beta ^{\prime }\gamma ^{\prime
}\delta ^{\prime }})_{\alpha \beta \gamma \delta \varepsilon
}+(\tilde{Z}_{2}^{\mu \nu \rho \lambda })_{\alpha ^{\prime
}\beta ^{\prime }\gamma ^{\prime }\delta ^{\prime }\varepsilon ^{\prime }}(%
\tilde{Z}_{3}^{\alpha ^{\prime }\beta ^{\prime }\gamma ^{\prime
}\delta ^{\prime }\varepsilon ^{\prime }})_{\alpha \beta \gamma
\delta \varepsilon }=
\nonumber \\
&&=-\tfrac{1}{12}g\frac{dW_{1}}{d\varphi }\frac{\delta
\tilde{S}}{\delta H^{\sigma }}\delta _{[\alpha }^{\mu }\delta
_{\beta }^{\nu }\delta _{\gamma }^{\rho }\delta _{\delta
}^{\lambda }\delta _{\varepsilon ]}^{\sigma }, \label{reduc3a}
\end{eqnarray}
\begin{eqnarray}
&&(\tilde{Z}_{2}^{\mu \nu \rho })_{\alpha ^{\prime }\beta ^{\prime
}\gamma ^{\prime }\delta ^{\prime }}(\tilde{Z}_{3}^{\alpha
^{\prime }\beta ^{\prime }\gamma ^{\prime }\delta ^{\prime
}})_{\alpha \beta \gamma \delta \varepsilon }+(\tilde{Z}_{2}^{\mu
\nu \rho })_{\alpha ^{\prime }\beta
^{\prime }\gamma ^{\prime }\delta ^{\prime }\varepsilon ^{\prime }}(\tilde{Z}%
_{3}^{\alpha ^{\prime }\beta ^{\prime }\gamma ^{\prime }\delta
^{\prime }\varepsilon ^{\prime }})_{\alpha \beta \gamma \delta
\varepsilon }=
\nonumber \\
&&=\tfrac{1}{6}g\left( \frac{d^{2}W_{1}}{d\varphi ^{2}}\frac{\delta \tilde{S}%
}{\delta H^{\lambda }}A_{\sigma }+\frac{dW_{1}}{d\varphi
}\frac{\delta \tilde{S}}{\delta B^{\lambda \sigma }}\right) \delta
_{[\alpha }^{\mu }\delta _{\beta }^{\nu }\delta _{\gamma }^{\rho
}\delta _{\delta }^{\lambda }\delta _{\varepsilon ]}^{\sigma }.
\label{reduc3b}
\end{eqnarray}


\begin{thebibliography}{99}
\bibitem{4a}  B. Voronov and I. V. Tyutin, \textit{Formulation of gauge
theories of general form. I}, Theor. Math. Phys. \textbf{50}
(1982) 218.

\bibitem{4b}  B. Voronov and I. V. Tyutin, \textit{Formulation of gauge
theories of general form. II. Gauge invariant renormalizability
and renormalization structure}, Theor. Math. Phys. \textbf{52}
(1982) 628.

\bibitem{4c}  J. Gomis and S. Weinberg, \textit{Are nonrenormalizable gauge
theories renormalizable?}, Nucl. Phys. \textbf{B469} (1996) 473,
hep-th/9510087.

\bibitem{4d}  S. Weinberg, \textit{The Quantum Theory of Fields}, Cambridge
University Press, Cambridge (1996).

\bibitem{5}  O. Piguet and S. P. Sorella, \textit{Algebraic Renormalization:
Perturbative Renormalization, Symmetries and Anomalies}, Lecture
Notes in Physics, Springer Verlag, Berlin, Vol. \textbf{28}
(1995).

\bibitem{6a}  P. S. Howe, V. Lindstr\"{o}m and P. White, \textit{Anomalies
and renormalization in the BRST-BV framework}, Phys. Lett.
\textbf{B246} (1990) 430.

\bibitem{6b}  W. Troost, P. van Nieuwenhuizen and A. van Proeyen, \textit{%
Anomalies and the Batalin-Vilkovisky Lagrangian formalism}, Nucl.
Phys. \textbf{B333} (1990) 727.

\bibitem{6c}  G. Barnich and M. Henneaux, \textit{Renormalization of gauge
invariant operators and anomalies in Yang-Mills theory}, Phys.
Rev. Lett. \textbf{72} (1994) 1588, hep-th/9312206.

\bibitem{6d}  G. Barnich, \textit{Perturbative gauge anomalies in the
Hamiltonian formalism: a cohomological analysis}, Mod. Phys. Lett. \textbf{A9%
} (1994) 665, hep-th/9310167.

\bibitem{6e}  G. Barnich, \textit{Higher order cohomological restrictions on
anomalies and counterterms}, Phys. Lett. \textbf{B419} (1998) 211,
hep-th/9710162.

\bibitem{7}  F. Brandt, M. Henneaux and A. Wilch, \textit{Global symmetries
in the antifield formalism}, Phys. Lett. \textbf{B387} (1996) 320,
hep-th/9606172.

\bibitem{7a}  R. Arnowitt and S. Deser, \textit{Interaction between gauge
vector fields}, Nucl. Phys. \textbf{49} (1963) 133.

\bibitem{7b}  J. Fang and C. Fronsdal, \textit{Deformation of gauge groups.
Gravitation}, J. Math. Phys. \textbf{20} (1979) 2264.

\bibitem{7c}  F. A. Berends, G. J. H. Burgers and H. Van Dam, \textit{On spin
three selfinteractions}, Z. Phys. \textbf{C24} (1984) 247.

\bibitem{7d}  F. A. Berends, G. J. H. Burgers and H. Van Dam, \textit{On the
theoretical problems in constructing interactions involving higher
spin massless particles}, Nucl. Phys. \textbf{B260} (1985) 295.

\bibitem{7e}  A. K. H. Bengtsson, \textit{On gauge invariance for spin-3
fields}, Phys. Rev. \textbf{D32} (1985) 2031.

\bibitem{8a}  G. Barnich and M. Henneaux, \textit{Consistent couplings
between fields with a gauge freedom and deformations of the master equation}%
, Phys. Lett. \textbf{B311} (1993) 123, hep-th/9304057.

\bibitem{17and5}  M. Henneaux, \textit{Consistent interactions between gauge
fields: the cohomological approach}, Contemp. Math. \textbf{219}
(1998) 93, hep-th/9712226.

\bibitem{8b}  J. D. Stasheff, \textit{Deformation theory and the
Batalin-Vilkovisky master equation}, in Deformation theory and
symplectic geometry, Proceedings of Ascona meeting, June 1996,
Eds. D. Sternheimer, J. Rawnsley, and S. Gutt, Math. Physics
Studies \textbf{20}, 271-284, Kluwer Acad. Publ., Dordrecht
(1997), q-alg/9702012.

\bibitem{8c}  J. D. Stasheff, \textit{The (secret?) homological algebra of
the Batalin-Vilkovisky approach}, in Secondary Calculus and
Cohomological Physics, Proceedings of Moscow meeting, August 1997,
Eds. M. Henneaux, J. Krasil'shchik, A. Vinogradov, Contemporary
Mathematics, vol. \textbf{219}, American Mathematical Society
(1998), hep-th/9712157.

\bibitem{8d}  J. A. Garcia and B. Knaepen, \textit{Couplings between
generalized gauge fields}, Phys. Lett. \textbf{B441} (1998) 198,
hep-th/9807016.

\bibitem{12}  D. Birmingham, M. Blau, M. Rakowski and G. Thompson, \textit{%
Topological field theory}, Phys. Rept. \textbf{209} (1991) 129.

\bibitem{psm1}  N. Ikeda, \textit{Two-dimensional gravity and nonlinear
gauge theory}, Annals Phys. \textbf{235} (1994) 435,
hep-th/9312059.

\bibitem{psm2}  T. Strobl, \textit{Dirac quantization of gravity Yang-Mills
systems in (1+1) dimensions}, Phys. Rev. \textbf{D50} (1994) 7346,
hep-th/9403121.

\bibitem{psm3}  P. Schaller and T. Strobl, \textit{Poisson structure induced
(topological) field theories}, Mod. Phys. Lett. \textbf{A9} (1994)
3129, hep-th/9405110.

\bibitem{psm4}  A. Yu. Alekseev, P. Schaller and T. Strobl, \textit{The
topological G/G WZW model in the generalized momentum
representation}, Phys. Rev. \textbf{D52} (1995) 7146,
hep-th/9505012.

\bibitem{psm5}  T. Kl\"{o}sch and T. Strobl, \textit{Classical and quantum
gravity in 1+1 dimensions: I. A unifying approach}, Class. Quantum Grav. \textbf{%
13} (1996) 965, gr-qc/9508020.

\bibitem{psm6}  T. Kl\"{o}sch and T. Strobl, \textit{Classical and quantum
gravity in 1+1 dimensions: II. The universal coverings}, Class.
Quantum Grav. \textbf{13} (1996) 2395, gr-qc/9511081.

\bibitem{psm7}  T. Kl\"{o}sch and T. Strobl, \textit{Classical and quantum
gravity in 1+1 dimensions: III. Solutions of arbitrary topology},
Class. Quantum Grav. \textbf{14} (1997) 1689, hep-th/9607226.

\bibitem{psm8}  A. S. Cattaneo and G. Felder, \textit{A path integral
approach to the Kontsevich quantization formula}, Commun. Math.
Phys. \textbf{212} (2000) 591, math-QA/9902090.

\bibitem{psm9}  A. S. Cattaneo and G. Felder, \textit{Poisson Sigma models
and deformation quantization}, Mod. Phys. Lett. \textbf{A16}
(2001) 179, hep-th/0102208.

\bibitem{bfaspects}  K. I. Izawa, \textit{On nonlinear gauge theory from a
deformation theory perspective}, Prog. Theor. Phys. \textbf{103}
(2000) 225, hep-th/9910133.

\bibitem{bfaspects1}  N. Ikeda, \textit{A deformation of three-dimensional
BF theory}, J. High Energy Phys. \textbf{0011} (2000) 009,
hep-th/0010096.

\bibitem{bfaspects2}  N. Ikeda, \textit{Deformation of BF theories,
topological open membrane and a generalization of the star
deformation}, J. High Energy Phys. \textbf{0107} (2001) 037,
hep-th/0105286.

\bibitem{bfaspects3}  N. Ikeda, \textit{Topological field theories and
geometry of Batalin-Vilkovisky algebras}, J. High Energy Phys.
\textbf{0210} (2002) 076, hep-th/0209042.

\bibitem{bfaspects4}  N. Ikeda, \textit{Chern-Simons gauge theory coupled
with BF theory}, Int. J. Mod. Phys. \textbf{A18} (2003) 2689,
hep-th/0203043.

\bibitem{BFhamnoPT}  C. Bizdadea, E. M. Cioroianu and S. O. Saliu, \textit{%
Hamiltonian cohomological derivation of four-dimensional nonlinear
gauge theories}, Int. J. Mod. Phys. \textbf{A17} (2002) 2191,
hep-th/0206186.

\bibitem{BFhamnoPT1}  C. Bizdadea, C. C. Ciobirca, E. M. Cioroianu, S. O.
Saliu and S. C. Sararu, \textit{Hamiltonian BRST deformation of a
class of n-dimensional BF-type theories}, J. High Energy Phys.
\textbf{0301} (2003) 049.

\bibitem{gen2}  G. Barnich, F. Brandt and M. Henneaux, \textit{Local BRST
cohomology in the antifield formalism. II. Application to Yang-Mills theory}%
, Commun. Math. Phys. \textbf{174} (1995) 93, hep-th/9405194

\bibitem{gen1}  G. Barnich, F. Brandt and M. Henneaux, \textit{Local BRST
cohomology in the antifield formalism. I. General theorems},
Commun. Math. Phys. \textbf{174} (1995) 57, hep-th/9405109.

\bibitem{gen11}  G. Barnich, F. Brandt and M. Henneaux, \textit{Local BRST
cohomology in gauge theories}, Phys. Rept. \textbf{338} (2000)
439, hep-th/0002245.

\bibitem{dubplb}  M. Dubois-Violette, M. Henneaux, M. Talon and C. M.
Viallet, \textit{Some results on local cohomologies in field
theory}, Phys. Lett. \textbf{B267} (1991) 81.
\end{thebibliography}
\end{document}